\documentclass[11pt,aps,prd, superscritptaddress,showpacs]{revtex4-1}
\def\be{\begin{equation}}
\def\ee{\end{equation}}

\def\la{\label}
\def\bea{\begin{eqnarray}}
\def\eea{\end{eqnarray}}

\def\fr{\frac}

\def\gs{\Upsilon_\star}
\def\sun{\odot}

\def\rbm{\rho_{bdm}}

\def\rc{\rho_c}

\def\rcc{\rho_{core}}
\def\rrc{r_{c}}

\usepackage{makeidx}
\usepackage{ctable}
\usepackage{multirow}
\usepackage{graphicx}
\usepackage{grffile}
\usepackage{epstopdf}
\usepackage{subfig}
\usepackage{subfloat}
\usepackage{colortbl}
\usepackage{appendix}
\usepackage{rotating}
\usepackage{array}
\begin{document}
\title {Core-Cusp revisited and Dark Matter Phase Transition Constrained\\ at $\mathcal{O}(0.1)$ eV with LSB Rotation Curve}
\author {Jorge Mastache}
 \email{mastache@fisica.unam.mx}
\author {Axel de la Macorra}
 \email{macorra@fisica.unam.mx}
 \affiliation{Instituto de Fisica, Universidad Nacional Autonoma de Mexico, Apdo. Postal 20-364, 01000, Mexico, D.F.}
\author{Jorge L. Cervantes-Cota}
 \email{jorge.cervantes@inin.gob.mx}
 \affiliation{Depto. de Fisica, Instituto Nacional de Investigaciones Nucleares, Col. Escandon, Apdo. Postal 18-1027, 11801, D.F. Mexico}

\begin{abstract}

Recently a new particle physics model called Bound Dark Matter (BDM) has been proposed \cite{delaMacorra:2009yb} in which dark matter (DM) particles are massless above a threshold energy ($E_c$) and acquire mass below it due to nonperturbative methods. Therefore, the BDM model describes DM particles which are relativistic, hot dark matter (HDM) in the inner regions of galaxies and describes nonrelativistic, cold dark matter (CDM) where halo density is below $\rho_c\equiv E_c^4$. To realize this idea in galaxies we use a particular DM cored profile that contains three parameters: a typical scale length ($r_s$) and density ($\rho_0$) of the halo, and a core radius ($r_c$) stemming from the relativistic nature of the BDM model. We test this model by fitting rotation curves of seventeen Low Surface Brightness (LSB) galaxies from The HI Nearby Galaxy Survey (THINGS). Since the energy $E_c$ parameterizes the phase transition due to the underlying particle physics model, it is independent on the details of galaxy and/or structure formation and therefore the DM profile parameters
$r_s, r_c, E_c$ are constrained, leaving only two free parameters. The high spatial and velocity resolution of this sample allows to derive the model parameters through the numerical implementation of the $\chi^2$-goodness-of-fit test to the mass models. We compare the fittings with those of  Navarro-Frenk-White (NFW), Burkert, and Pseudo-Isothermal (ISO) profiles. Through the results we conclude that the BDM profile fits better, or equally well, than NFW, Burkert, and ISO profiles and agree with previous results implying that cored profiles are preferred over the N-body motivated cuspy profiles.  We also compute 2D likelihoods of the BDM parameters $r_c$ and $E_c$ for the different galaxies and matter contents, and find an average galaxy core radius $r_c = 300$ pc and a transition energy between hot and cold dark matter at $E_c = 0.11^{+0.21}_{-0.07} {\rm \ eV}$ when the DM halo is the only component, therefore the maximum dark matter contribution in galaxies. In a more realistic analysis, as in Kroupa mass model, we obtain a core $r_c = 1.48$ kpc, and energy $E_c = 0.06^{+0.07}_{-0.03} {\rm \ eV}$.

\end{abstract}


\maketitle

\section{Introduction}\label{sec:introduction}

In the last decades intense work to understand the distribution of dark matter (DM) in galaxies has been published \cite{KrAl78,Tr87,SoRu01, deBlok10}. Among the considered, late-type Low Surface Brightness (LSB) galaxies are of special interest since it is believed they are dominated by DM, and high resolution $HI$, $H_{\alpha}$ and optical data can help to distinguish among the different DM profiles proposed in the literature.  There are essentially two types of profiles, the ones stemming from cosmological $N$-body simulations that have a cusp in its inner region, e.g. Navarro-Frenk-White (NFW) profile \cite{Navarro:1995iw,Navarro:1996gj}. On the other hand, the phenomenological motived cored profiles, such as the Burkert or Pseudo-Isothermal (ISO) profiles \cite{Bu95}. Cuspy and cored profiles can both be fitted to most LSB rotation curves, but with a marked preference for a cored inner region with constant density. Furthermore, cuspy profiles that do fit to galaxies in many cases suffer from a parameter inconsistency, since their concentrations ($c$) are too low and velocities $V_{200}$ are too high in comparison to the ones expected from cosmological simulations \cite{KuMcBl08,deBlok10}. Some other fits indicate that cuspy profiles require to tune fine the observer's line of sight with axisymmetric potentials \cite{KuMcMi09}. However, different systematics may play an important role in the observations such as noncircular motions, resolution of data and other issues \cite{Swater99,vandenBosch:1999ka,Swaters:2000nt, Rhee:2003vw,Simon:2003xu,deBlok:2008wp}. And there are attempts to reconcile both approaches through evolution of DM halo profiles including baryonic processes  \cite{SpGiHa05,Oh:2010mc} to transform cuspy to shallower profiles that follow the solid-body velocity curve ($v \sim r$) observed in late-type LSB. However, fthis issue is a matter of recent debate, see e.g. \cite{Pontzen:2011,Ogiya:2011ta} for a recent discussion.

Alternatively to the above models, recently one of us proposed a profile based on a particle physics model called bound dark matter (BDM) \cite{delaMacorra:2009yb}, in which DM particles at high densities, as in galactic inner regions, are relativistic, i.e. Hot DM (HDM), but in the outer regions they behave as standard CDM. To realize this idea in galaxies we use in the present work a particular DM cored profile that contains three parameters: a typical scale length ($r_s$) and density ($\rho_0$) of the halo and a core radius ($r_c$) stemming from the relativistic nature of the BDM model. The galactic core density is given by $\rho_c\equiv E_c^4$ and the profile properties will determine the energy scale of the particle physics model. We will show  that the LSB rotation curves yield a phase transition energy scale $E_c$, between HDM and CDM for our BDM profile, at $E_c = 0.11^{+0.21}_{-0.07} {\rm \ eV}$, when we consider only DM and $E_c = 0.06^{+0.07}_{-0.03}$ when considering gas, DM halo, and the minimum contribution of stars \cite{delaMacorra:2011df}. The $E_c$ parameter is a new fundamental scale  for  DM which can be theoretically determined using  gauge group dynamics, i.e. it does not depend on the properties of galaxies, once the gauge group is known. However, even though we propose $E_c$  as a new fundamental constant for DM, close related to the mass of our DM particle, and  it is important to note that its value is not yet known and we require observational evidence to determine  it.  The same is true for all masses of the standard model of particles, i.e. their value is not predicted by the standard model and it is the
experimental results that fixed them in a consistent manner. We use here the information on galactic rotation curves to determine $E_c$.
However, the extracting of $E_c$  does depend not only on the quality of the observational data, the mass models used but also
on the choice of DM profile.

To perform this task, we use The HI Nearby Galaxy Survey (THINGS), which collects high spectral resolution data revealing extended measurements of gas rotation velocities and circular baryonic matter trajectories \cite{Walter:2008wy}. Given these properties it is adequate to test the above-mentioned DM profiles with THINGS, which has been used to test  different core/cusp mass profiles. For disk-dominated galaxies the core and cusp profiles fit equally well, however for LSB galaxies there is a clear preference for core profiles over the cuspy models \cite{deBlok:2008wp}. Analysis of different high resolution datasets have confirmed this tendency in past recent years \cite{SpGiHa05,KuMcBl08,Salucci:2010pz,Gentile:2004tb,Salucci:2007tm}.  In this work, we present a study of the rotation curves using BDM, NFW, Burkert and ISO DM profiles, taking into account the contribution from different mass models: i) DM alone; ii) DM and gas; and iii) DM, gas and the stellar disk.  By fitting the models to the data we find that the BDM profile fits equally well or better than the cored profiles and only for a few LSB galaxies our model resembles that of NFW.

The kinematics of stars bring a very challenging problem in the analysis, mainly due to the uncertainty of the mass-to-light ratio ($\gs$) and to its dominant behavior close to the galactic center. Some considerations have been made in order to reduce this uncertainty in the parameters \cite{vanAlbada,Salpeter:1955it,Kroupa:2000iv,Bottema:1997qe}, but still the stellar contribution is not well known. In our case, we study four different $\gs$ models and the most inner part of the data enable us to compute the BDM parameters $r_c$ and $E_c$, both related to the galactic core.  We present the 1$\sigma$ and $2\sigma$ likelihood  contour plots of these parameters for the different galaxies and mass models. We notice that being $r_c$ different for each mass model it is consistent within the $2\sigma$ error for each galaxy. Additionally, we find that the galaxy core radius is the order of $r_c \sim 300 {\rm \ pc}$ and the $E_c \sim \mathcal{O}(0.1 {\rm \ eV})$ which gives the transition between hot and cold DM, this is shown in preliminary study \cite{delaMacorra:2011df}. It is also interesting to note the coincidence in the magnitude of the sum of neutrino masses with the magnitude of $E_c$ obtained in this analysis, and how this  could  open an interesting  connection between the generation of DM and neutrinos masses, but this will be presented elsewhere \cite{NBDM}. The magnitude of  the energy transition $E_c$  is of the same order as the upper limit of the total mass of the neutrinos $\sum m_\nu < 0.58 {\rm \ eV} (95\% {\rm \ CL)}$ from the seven year Wilkinson Microwave Anisotropic Probe (WMAP), Baryon Acoustic Oscillation (BAO), and the Hubble constant ($H_0$) \cite{Larson:2010gs,Komatsu:2010fb}. In addition, the recent analysis of the Atacama Cosmology Telescope \cite{Dunkley:2010ge} that combined CMB data  with measurements of BAO and the Hubble constant reported an excess of the effective number of relativistic degrees of freedom, $N_{eff} = 4.6 \pm 0.8$, in consistency with Ref. \cite{Komatsu:2010fb}, opening the possibility to have extra nonstandard-model relativistic degrees of freedom, as our BDM particle.

We organized this work as follows:  In Sec. \ref{SPM} we explain the particle model behind the BDM profile. In Sec. \ref{modelo} we  present our BDM profile, which has NFW as a limit at low energies. In Sec. \ref{sample} we describe the galaxy sample considered for the rotation curves study. The different mass models and components (gas, stars, DM halo) are presented  in detail in Sec. \ref{MaMo}, while the different hypothesis of the mass-to-light ratio are discussed in subsection \ref{MaMoEsDi}. The results and conclusions are presented in Sec. \ref{results} and \ref{conclusion}, respectively. For convenience, some figures and tables are shown in the Appendices. We show in Appendices \ref{apend: diet-salpeter} and \ref{apendix:iso_burkert_result} the fitting values corresponding to the free parameters for some of the cored profiles for the maximum stellar contribution. Finally, in Appendix \ref{apend:LikelihoodPlot} we present the rotation curves of the different profiles and  galaxies and we include the confidence contours levels, pointing out the considerations made for each galaxy.

\section{Particle Model}\label{SPM}

We now present the physics and motivation behind our BDM model \cite{delaMacorra:2009yb}.
Cosmological evolution of gauge groups, similar to QCD,  have been studied to understand the nature of dark energy \cite{Macorra.DE} and also DM \cite{Macorra.DEDM}.
A particle mass can be generated either by the Higgs mechanism or by
a nonperturbative effect. In the Standard Model (SM) the fundamental  particles
(quarks, electrons, neutrinos) get their mass by the interaction with
the Higgs field that acquires a nonvanishing vacuum
expectation value  at the electroweak scale $E_{\rm ew}=O(100 {\rm \ GeV})$, while at higher energies all SM particles have vanishing masses.
On the other hand, the nonperturbative gauge mechanism is based on the strength of
gauge interaction and the mass of the particles is expected to be at the same order of magnitude as the phase transition scale
as for example protons and neutrons.
For asymptotically free gauge groups, such as the strong Quantum Chromodynamics (QCD) force in the SM, the gauge coupling constant becomes strong  at low energies and binds the elementary particles (quarks) forming neutral particles such as protons, neutrons, and mesons.
The condensation or phase transition scale is defined as
the energy where the gauge coupling constant $g$ becomes strong, $g(\Lambda)\gg 1$,
giving a condensation scale $\Lambda_c = \Lambda_i\,e^{-8\pi^2/bg^2_i}$, where $b$ is the one-loop beta function which depends only on the number of fields in the gauge group. For example for a SUSY gauge group $SU(N_c),N_f$, where $N_c(N_f)$ is the number of colors (flavors), we have  $b=3N_c-N_f$  and $g_i$ is the value of the coupling constant at an initial scale $\Lambda_i$.
The fact that $\Lambda_c $ is exponentially suppressed compared to $\Lambda_i$
allows as to understand why $\Lambda_c $ can be much smaller then
the initial $E_i$ which may be identified with the Planck,
Inflation or Unification scale. In our case the gauge group and elementary fields {\it are not} part of the standard model (SM). Our dark gauge group is assumed to interact with the SM only through gravity and is widely predicted by extensions of the SM, such as brane or string theories. We can relate $E_c$ to $\Lambda_c$ since the energy density depends on the average energy per particle and the particle number density $n$, i.e.   $\rho_c\equiv E_c^4 = \Lambda_c\,n$.\\

For asymptotic free gauge groups, such as QCD, the low energy states
consist of bound gauge singlets, such as baryons and mesons in QCD,
formed by  fundamental (nearly massless) particles, quarks in QCD.
The order of magnitude of the mass of these particles is \cite{delaMacorra:2009yb}
\be\label{mb}
m_{BS}=d\,E_c
\ee
with $d=O(1)$ a proportionality constant.
In QCD one has $E_c\simeq 200\,{\rm MeV}$ with the pion mass $m_\pi\simeq140 {\rm \ MeV}$
while the baryons mass (protons and neutrons) $m_b\simeq 940 {\rm \ MeV}$,
i.e. the proportionality constant is in the range $0.7<d<5$, and
with bound mass much larger than the mass of the quarks ($m_u\simeq (1-3) {\rm \ MeV},
m_d\simeq (3.5-6) {\rm \ MeV}$). Clearly the mass of the bound states is
not the sum of its elementary particles but is due to the nonperturbative
effects of the strong force and is well parameterized by $E_c$.
The dynamical formation of bound states is not completely understood since it involves nonperturbative physics. However, it has been shown in RHIC \cite{Adams:2005dq} that at high density, above the transition scale $E_c$, the QCD quarks do indeed behave as free particles, while at low energies there are no free elementary quarks and all quarks form gauge neutral bound states. Since the interaction strength  increases at lower energies, the formation of bound states is expected to be larger  at the smallest possible particle bound state energy $E_{BS}$ (i.e. $E_{BS}=m_{BS}$) with momentum $p^2=E_{BS}^2-m_{BS}^2 \simeq 0$. The energy  distribution of bound states formation is still under investigation \cite{Bazavov:2009zn} and for simplicity we take here $p=0$ which gives a vanishing particle velocity for the bound states.

It is precisely this nonperturbative gauge mechanism that we have in mind for our bound states dark matter BDM. Of course, in our case the gauge group and elementary fields {\it are not} part of the SM. Our ``dark'' gauge group is assumed to interact with the SM only through gravity and is widely predicted by extensions of the SM, such as brane or string theories. Even though we have motivated our BDM in terms of a well motivated particle physics model we stress the fact that  the cosmological implications of BDM do not depend on its origin. The  BDM
is defined by a DM that at high energy densities, $\rbm>\rc$,  the particles behave as relativistic HDM with a particle velocity $v=c$ while for lower energy densities $\rbm<\rc$, the BDM are cold bound state particles, i.e.
CDM with $v \ll c$.

There are two natural places where one may encounter  high energy densities $\rbm$ for dark matter. One is at early cosmological times and the second place is at the inner regions of galaxies. In the first case, we define  $a_c$ as the transition scale factor where
\be
\rc\equiv E_c^4= \rbm(a=a_c)
\ee
and the Universe is dense and for   $a<a_c$  the BDM particles are relativistic and redshift as radiation while for
$a> a_c$ we have  nonrelativistic BDM particles and they redshift as matter, i.e.
\be
\rc < \rbm(a < a_c)\propto  a^{-4}, \hspace{1cm} \rc > \rbm(a>a_c)  \propto  a^{-3}
\ee
In the second case, away from the center of the galaxies the energy density is smaller and it increases towards
its center and in a NFW profile it blows up $\rho \rightarrow \infty$  when the radius $r\rightarrow 0$. Therefore
we will have for   $\rbm < \rc$ and the BDM particles are cold, i.e. CDM,  but in the inner region once $\rbm\geq \rc$ one has a
transition and the BDM particles become relativistic,
\be
\rc < \rbm(r < r_c)\;\;\;\; \textrm{with} \;\; v\approx c , \hspace{1cm} \rc > \rbm(r>r_c)\;\;\;\;   \textrm{with}\;\; v\approx 0
\ee
where $r_c$ is defined by $\rbm(r=r_c)=\rc$.  The dispersion velocity $v$ of the BDM particles is vanishing for $r$ larger than $r_c$,
the particles are CDM, while  at the galactic center the BDM particles are relativistic and for $r<r_c$ it is $v=c$ it gives a core inner region at $r\leq \rrc$ and with $\rcc=\rc$, the energy density of the galaxy at $\rrc$.

\section{BDM profile}\label{modelo}

The BDM model simply consists of particles that at high energy densities are massless relativistic particles with the speed of light, but at low energy densities they acquire a large mass, due to nonperturbative quantum field effects, and become nonrelativistic with a  vanishing (small) dispersion velocity. The phase transition takes place at an energy density defined as $\rc\equiv E_c^4$ and is related to the mass of the DM particle. The  value of $\rc$ can be determined theoretically, given a gauge group model, or phenomenologically  by consistency with cosmological or astrophysical data. In the present work we estimate its value through the study of the BDM from rotation curves. It is worth pointing out that the mass of the SM particles are also free parameters in the SM and it is the physics in colliders, such as in the LHC at CERN, which defines its value. In our case, since the BDM particles do interact only weakly with the SM we cannot use collider data to measure its mass and we are left with astronomical and cosmological observations.

The average energy densities in galactic halos is of the order $\rho_g \sim 10^{5} \rho_{\rm cr}$ ($\rho_{\rm  cr}$, being the critical Universe's background density) and  as long as $\rho_g<\rc$ we expect a standard CDM galaxy profile, which may be given by the NFW profile, $\rho_{{\rm nfw}}$. The NFW profile has a cuspy inner region with $\rho_{{\rm nfw}}$ diverging in the center of the galaxy. Therefore, once one approaches the center of the galaxy the energy density increases in the NFW profile and once it reaches the point $\rho_g=\rc$  we encounter the phase transition and the  BDM particles  become massless.
Inside $r<\rrc$ the BDM particles are relativistic and the DM energy density $\rrc$ remains constant avoiding a galactic cusp. Of course we would expect a smooth  transition region between these two distinct behaviors but we expect the effect of the thickness of this transition region to be small and we will not consider it here.

Since our BDM behaves as CDM for $\rho < \rho_c$  as long as the density of the galaxy is $\rho_g< \rho_c$ we expect to have a NFW type profile in this limiting case. Therefore, the proposed  BDM profile \cite{delaMacorra:2009yb} is given by a cored CDM  profile
\bea \label{eq:rhobdm}
\rho_{bdm}&&=\frac{\rho_0}{\left( \frac{r_c}{r_s}+\frac{r}{r_s} \right)\left(1+\frac{r}{r_s} \right)^2}\\&&
= \frac{2 \rho_c}{\left( 1+\frac{r}{r_c}\right)\left( 1+\frac{r}{r_s}\right)^2},
\label{eq:fix-BDM}\eea
with $\rrc< r_s$ and $r_s,\rho_0$ are typical NFW  halo dependent parameters.
The BDM profile coincides with $\rho_{{\rm nfw}}$ at large radius but has a core inner region, when the galaxy energy density $\rho_{bdm}$ reaches the value $\rc=E_c^4$ at  $r\simeq r_c $  and for $r_c \ll r_s$ we have
\be \label{eq:relation ec_free_bdm_params}
 \rcc \equiv \rho_{bdm}(r=\rrc)\equiv \rc = \fr{\rho_0 r_s}{2\rrc},
\ee
with the core radius given by $\rrc \equiv \fr{\rho_0 r_s}{2\rc}$. The value of $\rrc$  depends on the galaxy profile parameters $\rho_0$ and $r_s$.

If we assume that the transition energy of the BDM particles is a fundamental parameter in the DM nature that can be constrained by  fitting the rotation curve of galaxies, we can substitute one of the three free parameters by the density $\rho_c$ in the BDM profile, Eq. (\ref{eq:rhobdm}), using the relation (\ref{eq:relation ec_free_bdm_params}). We call ``fixed-BDM''  to the profile that is obtained as a result of substituting the free parameter $\rho_0$ by the fundamental quantity $\rho_c$ as in Eq.(\ref{eq:fix-BDM}). The fixed-BDM profile depends now on the fundamental density $\rho_c$, which is the same for all galaxies, and on the two free parameters $r_s$ and $r_c$ that depend on the morphology of each galaxy.

The slope of BDM profile is
\be\label{sl}
\alpha\equiv - \fr{d\, \log \rho}{d\, \log r}=\fr{r/\rrc(1+3r/r_s+2\rrc/r_s)}{(1+r/\rrc)(1+r/r_s)}
\ee
and takes the values $\alpha=(0,1/2,1)$ for $r=(0,\rrc,\rrc\ll r\ll r_s)$  and $\alpha=(2,3)$ for  $r=(r_s,r_s\ll r)$. If we are interested in the inner region of the galaxy, for values of $r\ll r_s$, Eq. (\ref{sl}) can be approximated for $r_c \ll r_s$ with
\be\label{eq:sl2}
0\leq \alpha  = \fr{y}{1+y} < 1
\ee
in terms of $y\equiv r/\rrc$  with  $0\leq y < \infty$. Given a value of $\alpha$ we can determine the value of $y =r/\rrc= \alpha/(1-\alpha)$.

For future reference (see Sec. \ref{results}) we introduce here a phenomenological ansatz for a model independent DM profile, useful to investigate the central region of a  galaxy,  given by $\rho_\alpha=\rho_0\,r^{-\alpha}$ with $\rho_0, \alpha$ constant parameters.

Fittings to observational data have shown that the NFW halo profile is not a good description for rotation curves, and it is generally preferred a core dominated halo model \cite{Salucci:2007tm}, although some scatter in $\alpha$ could be possible \cite{Simon:2005}. In the present work our objective is to test the realization of the BDM model, Eq. (\ref{eq:rhobdm}), through rotation curves of a wide sample of galaxies. We aim to derive the kinematics properties of the galaxies and, in particular, we constrain the slope $\alpha$ and the transition energy $E_c$.

\section{The Sample}\label{sample}

With the current HI data provided by THINGS high resolution and excellent sensitivity of the velocity fields and rotation curves are available making possible to revisit some of the outstanding questions about DM, and to compare the applicability of both core or cusp models. THINGS galaxies have an observing data sample of 34 nearby galaxies containing a large range of luminosities and Hubble types at the desired sub-{\rm kpc} resolution, but we limit our sample to seventeen low luminous (early type and dwarf) galaxies with smooth, symmetric and extended to large radii rotation curves and small or none bulge, see Table \ref{tab:things}. These properties provide a good estimate of the DM halo in galaxies because it is believed that it dominates over all other components at all radii. For technical details and systematic effects treatment of the observations of the THINGS sample refer to Refs. \cite{Walter:2008wy,Trachternach:2008wv,Oh:2008ww} and for a complete analysis of its rotation curves see \cite{deBlok:2008wp}, hereafter deBlok08.

The mass models (in Sec. \ref{MaMo}) are constructed with the rotation curves extracted with THINGS data and the $3.6 \ {\rm \mu m}$ data from SINGS (Spitzer Infrared Nearby Galaxies Survey) \cite{Kennicutt:2003dc}. We follow the analysis of deBlok08 and McGaugh et al. (2007) \cite{McGaugh:2006vv} for the sample considered here.

\captionsetup{margin=10pt,font=small,labelfont=bf,justification=centerlast,indention=.5cm}

\begin{table}[h]
  \centering
    \scriptsize{
    \begin{tabular}{lcccc}
       \multicolumn{5}{c}{THINGS} \\
       \hline
               &            &            & DIET-SALPETER &     KROUPA \\
       \multicolumn{1}{c}{Galaxy}  & Distance & $R_d $ & $\log_{10} M_{star}$ & $\log_{10} M_{star}$ \\
       \multicolumn{1}{c}{(1)} & (2) & (3) & (4) & (5) \\
       \hline
        NGC 925 &        9.2 &      3.30 &      10.01 &       9.86 \\
       NGC 2366 &        3.4 &    1.76 &       8.41 &       8.26 \\
       NGC 2403 &        3.2 &       1.81 &       9.71 &       9.56 \\
       NGC 2403 &        3.2 &       1.81 &       9.67 &       9.52 \\
       NGC 2841 &       14.1 &    4.22 &      11.04 &      10.88 \\
       NGC 2903 &        8.9 &    2.40 &      10.15 &         10 \\
       NGC 2976 &        3.6 &   0.91 &       9.25 &        9.1 \\
       NGC 3031 &        3.6 &     1.93 &      10.84 &      10.69 \\
       NGC 3198 &       13.8 &    3.18 &       10.4 &      10.25 \\
       NGC 3198 &       13.8 &       3.06 &      10.45 &       10.3 \\
       NGC 3521 &       10.7 &    3.09 &      11.09 &      10.94 \\
       NGC 3621 &        6.6 &       2.61 &      10.29 &      10.14 \\
       NGC 4736 &        4.7 &       1.99 &      10.27 &      10.12 \\
       NGC 5055 &       10.1 &     3.68 &      11.09 &      10.94 \\
       NGC 6946 &        5.9 &       2.97 &      10.77 &      10.62 \\
       NGC 7331 &       14.7 &    2.41 &      11.22 &      11.07 \\
       NGC 7793 &        3.9 &     1.25 &       9.44 &       9.29 \\
        IC 2574 &          4 &    2.56 &       9.02 &       8.87 \\
       DDO 154  &        4.3 &       0.72 &       7.42 &       7.27 \\
    \end{tabular}
    }
  \caption{\footnotesize{Sample of late-type and dwarf galaxies taken from THINGS as presented in Walter et al \cite{Walter:2008wy}. Columns: (1) Galaxy's name. (2) Distance to the galaxy in {\rm Mpc}. (3) Characteristic radius of the stellar disk in {\rm kpc} as given in deBlok08. (4) Logarithm of the stellar mass disk when considering the diet-Salpeter IMF in solar masses ($M_\sun$), and (5) Logarithm of the stellar mass disk when considering the Kroupa IMF ($M_\sun$). }}
  \label{tab:things}
\end{table}

\section{Mass Models} \label{MaMo}
Our mass models include the three main components of a spiral galaxy: thin gaseous disk, $V_G$, a thick stellar disk, $V_{\star}$, and a DM halo, $V_H$. In most cases the stellar disk can be well described by a single exponential disk. When necessary, in a small number of galaxies we have considered an additional central component, a bulge, containing a small fraction of the total luminosity of the galaxy, as described by deBlok08 \cite{deBlok:2008wp}. The gravitational potential of the galaxy is the sum of each mass component, thus the observed rotation velocity is

\be
{V_{obs}}^2 = {V_H}^2 + {V_G}^2 + \gs {V_{\star}}^2.
\label{VelT}
\ee

As an input we need the observational rotation curve, $V_{obs}$, the rotation curve of the gas component, $V_G$, and the stellar component $V_\star$. With this data we can extract information for the DM halo through our theoretical model. We describe in more detail the treatment for the  gas component in Sec. \ref{MaMoGa}, as well for the stellar component in Sec. \ref{MaMoEsDi}. In Sec. \ref{MaMoDM} we describe the DM models used.

 \subsection{Neutral Gas Distribution \label{MaMoGa}}

 For the gas we assume an infinitely thin disk in order to compute the corresponding rotation curve. For more technical details we refer to \emph{deBlok08}. We point out that the case of a disk with sufficient central depression in the mass distribution can yield a net force pointing outwards, and this generates an imaginary rotation velocity and therefore a negative $V_G^2$. An imaginary velocity is just a reflect of the effective force of a test particle caused by a nonspherical mass distribution with a depression mass in the center. We have not include the contribution of the molecular gas since its surface density is only a few percent of the that of the stars, therefore its contribution is reflected in a small increase in $\gs$ \cite{Portas:2009}.

 \subsection{Stellar Distribution \label{MaMoEsDi}}

 To model the stellar disk we use the approximation of a radial exponential profile of zero thickness, the Freeman disk \cite{Freeman:1970mx}, since the disk vertical scale height does not change appreciably with radius and the correction to the velocity is around 5\% in most cases \cite{Burlak:1997}. Thus, the central surface density is $\Sigma(R) = \Sigma_0 e^{-R/Rd}$, where $R_d$ is the scale length of the disk and $\Sigma_0$ is the central surface density with units [$M_\sun pc^{-2}$]. These two parameters are obtained first by fitting the observed surface brightness profile, extracted from the SINGS images at the $3.6 \mu m$ band and synthesized by deBlok08, to the linear formula $\mu(R) = \mu_0 + 1.0857 R/R_d$ where $\mu_0$ is the central surface brightness given in observational units [mag arcsec$^{-2}$], $\mu_0$ and the surface brightness are related by a simple change of units. We get the surface density thanks to the mass-to-light ratio $\gs$, an additional free parameter in the mass model, introduced because we generally can only measure the distribution of the light instead of the mass.

 The rotation velocity of an exponential disk is given by the well know Freeman formula \cite{Freeman:1970mx} $V_\star(y') \propto y'^2 \left[I_0(y')K_0(y') - I_1(y')K_1(y')\right]$, where $y' \equiv R/(2 R_d)$  and $I_n$ and $K_n$ are the modified Bessel functions of order \emph{n} of the first and second kind, respectively.

When we deduce the DM properties, the $\gs$ as well as the Initial Mass Function (IMF) remain as the major sources of uncertainty. The unknown value of $\gs$ makes it hard to pick out between cusp and core profiles. In order to constrain the wide range of DM halo parameters we follow the approach used by deBlok08 instead of the van Albada \& Sancisi \cite{vanAlbada}. We considered a ``diet''-Salpeter IMF \cite{Salpeter:1955it} in which stellar mass population syntheses have proved \cite{Bell:2000jt} to maximize the disk mass contribution (maximum disk) for a given photometric constraint, J-K color band. We also consider the Kroupa IMF \cite{Kroupa:2000iv} based on stellar population studies in the Milky Way that produces lower disk masses that minimizes the baryonic contribution by a factor of 1.4 less massive than the diet-Salpeter ones.

In the numerical analysis we use the $\gs$ from the $3.6\mu m$ obtained in deBlok08 using an empirical approach into the relation between the $3.6\mu m$ emission and $\gs$ using the J-K colors given in the 2MASS Large Galaxy Atlas \cite{Jarrett:2003uy}.

Disk galaxies sometimes show radial color gradients, and it is believed that this provides an indication of stellar population between the inner and outer regions of a galaxy and produce $\gs$ gradients between these two regions of the disk \cite{Taylor:2005sf}. We take the $\gs$ as a function of the radius in order to consider the different stars contribution as it depends on the region that we were analyzing.

The $\gs$ has been modeled (e.g. Salpeter \cite{Salpeter:1955it}, Kroupa \cite{Kroupa:2000iv}, Bottema\cite{Bottema:1997qe}), but the precise value for an individual galaxy is not well known and depends on extinction, star formation history, IMF, among others. Some assumptions have to be made respect to $\gs$ in order to reduce the number of free parameters in the model. We present a disk-halo decomposition using different assumptions for $\gs$ for the galaxy sample considered here.

 \begin{description}
     \item[Minimal disk] This model assumes that the observed rotation curve is only due to DM. This gives the upper limit on how concentrated the DM component can be in the galaxy.
     \item[Minimal disk plus gas] The contribution of the atomic gas and the DM halo is taken into account, but stars have no contribution ($\gs=0$).
     \item[Kroupa] Studies of the stellar population in the Milky Way suggest that the Kroupa IMF produces low disk masses that we consider as the minimal limit for the stellar disk.
     \item[diet-Salpeter] Here $\gs$ is set to be a constant value based on the diet-Salpeter IMF, in which the stellar population synthesis model has proven to give a maximum stellar disk for a given photometric constraint.
     \item[Free $\gs$] Here we ignore a priori any knowledge of the IMF and treat $\gs$ as an extra free parameter in the model, and we let the program to choose the best-fitted value for $\gs$.
 \end{description}

The contribution of the atomic gas is considered in the Kroupa, diet-Salpeter, and Free $\gs$ mass models. In our fit we consider radial color gradient and, when present, the bulge for the stellar disk. In the results section we make emphasis on some difference in the core fit when we do not take into account one or both of these two ingredients in the stellar disk.

 \subsection{DM Halos \label{MaMoDM}}

 We use three well-known models for the DM distribution. $\Lambda$CDM simulations come up with a DM density profile independent of the mass of the halo characterized by a cusp central density, the NFW profile. On the other hand observational determinations of the inner mass density distribution seems to indicate that mass density profiles of DM halos can be better described using an approximately constant-density inner core ($\rho \sim r^\alpha$ with $\alpha \ll 1$). This core has a typical size of order of a {\rm kpc} \cite{Moore:1994yx,de Blok:1996ns}, and examples of these profiles are the ISO and Burkert. Here we also consider our proposed BDM to test it and compare it.

 Strictly speaking, we are dealing with circular rotation velocities of test particles in the plane of the galaxy and we assume spherical halos. For this distribution of matter the circular velocity at radius $r$ is given by $V_H^2(r) = G M(r)/r$. We do not consider the adiabatic contraction of the dark matter halo but be are aware that it predicts the increase of dark matter density in the 5\% of the virial radius \cite{Gnedin:2004cx}.

 \subsubsection{NFW Halo \label{MaMoNFW}}

 The NFW profile takes the form
 \be \label{eq:rhoNFW}
    \rho_{{\rm nfw}} = \frac{\rho_0}{\frac{r}{r_s}\left(1+\frac{r}{r_s}\right)^2},
 \ee
 where $r_s$ is the characteristic radius of the halo, and $\rho_0$ is related to the density of the Universe at the time of collapse of the DM halo. This mass distribution gives rise to a halo rotation curve
 \be
    V_H^2(r) = \frac{4 \pi G r_s^3 \rho_0}{r}\left[-\frac{r}{r+r_s} + \ln\left(\frac{r+r_s}{r_s}\right)\right].
 \ee
This density profile has an inner and outer slope of -1 and -3, respectively. The inner slope implies a density cusp. The halo density can be specified in terms of a concentration parameter $c = r_{200}/r_s$ that indicates the amount of collapse that the DM halo has undergone, where the radius $r_{200}$ is defined as the radii at which the density contrast of the galaxy is 200 times greater than the critical density $\rho_{\rm cr}$, defined as $\rho_{\rm  cr} = 3 H^2/(8 \pi G)$, {\it H} being the Hubble parameter.

 We use the $\rho_0$ and $r_s$ as the free parameters for the NFW model instead of the classical $c$ and $V_{200}$, because these two parameters can be interpreted and compared directly with the free parameters of other DM halo profiles.

   \subsubsection{BDM Halo \label{MaMoBDM}}

   Recalling that the BDM model predicts an inner galaxy core radius determined by the energy scale $E_c$, which corresponds to the phase transition energy scale of the subjacent elementary particle model. The proposed profile is shown in Eq. (\ref{eq:rhobdm}) coincides with $\rho_{{\rm nfw}}$ profile at large radii since $\rho \propto r^{-3}$ but has a core inner region at $r = r_c$ where $\rho \propto \rho_0 r_s r_c^{-1}$. Its circular velocity is given by
   \be
    V_H(r)^2 = \frac{4 \pi G \rho_0 r_s^3}{r}\left[ \frac{r_s}{r_c-r_s}\left(\frac{r}{r_s+r}\right) + \frac{r_s(r_s-2 r_c)}{(r_c - r_s)^2}\ln\left(1+\frac{r}{r_s}\right) + \left(\frac{r_c}{r_c-r_s}\right)^2 \, \ln\left(1+\frac{r}{r_c}\right) \right],
   \ee
where the additional free parameter $r_c$ is the galaxy core radius that demarcates the place where the BDM particles have an energy greater than $E_c$ and behaves as HDM for $r < r_c$. If the radius is greater than $r_c$ the BDM particles acquire a large mass through a nonperturbative mechanism and behave as CDM. $r_s$ is just a characteristic scale for the DM halo.

  \subsubsection{Burkert Halo \label{MaMoBur}}

  The cored Burkert halo profile is given by
  \be \label{eq:rhoBurkert}
    \rho_B =\frac{\rho_0}{(1+\frac{r}{rs})\left[1+(\frac{r}{rs})^2\right]},
  \ee
  where $\rho_0$ is the central density, $r_s$ is the core radius. The rotation curves caused by this halo are given by
  \be
    {V_H(r)}^2 = \frac{\pi G \rho_0 r_s^3}{r}\left(\ln\left[\left(1+\frac{r}{r_s}\right)^2\left(1+\frac{r^2}{r_s^2}\right)\right] - 2 \arctan\left[\frac{r}{r_s}\right] \right).
  \ee

   \subsubsection{Pseudo-Isothermal Halo \label{MaMoIso}}

   The spherical pseudo-isothermal (ISO) halo has a density profile
   \be \label{eq:rhoISO}
    \rho_{ISO} =\frac{\rho_0}{1+\left( \frac{r}{r_s} \right)^2},
   \ee
   where $\rho_0$ here is the central density of the halo and $r_s$ is the core radius of the halo. The corresponding DM rotation curve is
   \be
    V_H(r)^2 = 4 \pi G \rho_0 r_s^2\left[1-\frac{r_s}{r}\arctan\left(\frac{r}{r_s}\right) \right].
   \ee

\subsection{Computing the Mass Models}\label{NuMe}

We use the observed rotation curve, stellar, and gas component as an input for the numerical code, in order to obtain the properties of the DM halo. In order to fit the observational velocity curve with the theoretical model we use the $\chi^2$-goodness-of-fit test ($\chi^2$-test), that tell us how ``close" are the theoretical to the observed values. In general the $\chi^2$-test statistics is of the form:
\be
   \chi^2 = \sum_{i=1}^n \left(\frac{V_{{\rm obs}_i}-V_{{\rm model}_i}(r,\rho_0, r_s, r_c)}{\sigma_i}\right)^2,
\ee
where $\sigma$ is the standard deviation, and $n$ is the number of observations.

We fit the free parameters of the DM halo for the Kroupa minimal disk and  diet-Salpeter maximum disk with a $\chi^2$-test. When we minimize $\chi^2$ we use different methods of minimization, Differential Evolution, NelderMead, and SimulatedAnnealing, in order to be sure that we are not in a local minimal, and at the same time we put constraints on the free parameter in order to have values greater than or equal to zero to obtain physical reasonable values. For BDM we constrain $r_c$ to values smaller than $r_s$, because it makes no sense to consider core radius greater than the characteristic radius of the DM halo.

Comparison of the fits derived can tell us which of the DM models is preferred. More important are the differences between the reduced $\chi^2_{\rm red}=\chi^2/(n-p-1)$ values, where $n$ is the number of observations and $p$ is the number of fitted parameters.

The uncertainties in the rotation velocity are reflected in the uncertainties in the model parameters.

As the ISO and Burkert halo have no basis in standard cosmology, there are no a priori expectations for its model parameters. One can derive them and check for possible trends with other fundamental galaxy parameters, as other authors have done \cite{Kormendy:2004se}.

\setlength{\extrarowheight}{3pt}
\begin{table}[h]
  \centering
  \scriptsize{
    \begin{tabular}{rl|r|lllr|llr|llr}
     \multicolumn{ 13}{c}{MINIMUM DISK} \\
     \hline
               &            &                  \multicolumn{ 5}{c|}{BDM} &             \multicolumn{ 3}{c|}{NFW} &     \multicolumn{ 3}{c}{Fixed-BDM $E_c=0.11$} \\
               &     \multicolumn{1}{c|}{Galaxy} & \multicolumn{1}{c|}{$r_c/r_s$} &     \multicolumn{1}{c}{$r_s$} & \multicolumn{1}{c}{$\log_{10} \rho_0$}  &     \multicolumn{1}{c}{$r_c$} & \multicolumn{1}{c|}{$\chi^2_{\rm red}$} &     \multicolumn{1}{c}{$r_s$} & \multicolumn{1}{c}{$\log_{10} \rho_0$}  & \multicolumn{1}{c|}{$\chi^2_{\rm red}$} &     $r_s$ &     $r_c$ & $\chi^2_{\rm red}$ \\
               &        (1) &        (2) &        (3) &        (4) &        (5) &        (6) &        (7) &        (8) &   (9) &   (10)  &  (11) & (12)   \\
               \hline
    \multirow{ 7}{*}{G.A} &     DDO154 &       0.37 & $3.66^{+0.03}_{-0.03}$ & $7.35^{+5.44}_{-5.44}$ & $1.35^{+0.04}_{-0.04}$ &       0.38 & $14.46^{+0.14}_{-0.14}$ & $6.21^{+4.31}_{-4.3}$ &       1.48 & $12.39^{+0.44}_{-0.44}$ & $0.02^{+0.002}_{-0.002}$ & 1.17 \\
     &    NGC2841 & $10^{-5}$ & $4.67^{+0.01}_{-0.01}$ & $8.61^{+6.33}_{-6.11}$ & $0.0001^{+0.01}_{-0.01}$ &       0.58 & $4.67^{+0.01}_{-0.01}$ & $8.61^{+6.33}_{-6.13}$ &       0.58 & $2.59^{+0.02}_{-0.02}$ & $1.87^{+0.005}_{-0.005}$ &        0.80 \\
     &    NGC3031 &       0.12 & $1.55^{+0.003}_{-0.003}$ & $9.38^{+7.}_{-7.01}$ & $0.19^{+0.01}_{-0.01}$ &       4.24 & $1.86^{+0.004}_{-0.004}$ & $9.17^{+6.79}_{-6.79}$ &        4.20 & $0.69^{+0.02}_{-0.02}$ & $1.62^{+0.003}_{-0.003}$ &       4.21 \\
     &    NGC3621 &       0.01 & $7.79^{+0.02}_{-0.02}$ & $7.5^{+5.21}_{-5.22}$ & $0.01^{+0.01}_{-0.01}$ &       2.04 & $7.9^{+0.02}_{-0.02}$ & $7.49^{+5.19}_{-5.21}$ &       2.02 & $6.53^{+0.04}_{-0.04}$ & $0.29^{+0.001}_{-0.001}$ &       2.38 \\
     &    NGC4736 &       0.18 & $0.27^{+0.001}_{-0.001}$ & $10.71^{+8.61}_{-8.61}$ & $0.05^{+0.01}_{-0.01}$ &       1.69 & $0.33^{+ 0.001}_{- 0.001}$ & $10.47^{+8.36}_{-8.36}$ &       1.67 & $0.02^{+0.001}_{-0.001}$ & $1.06^{+0.003}_{-0.003}$ &       1.72 \\
     &    NGC6946 &       0.02 & $5.95^{+0.01}_{-0.01}$ & $8.01^{+5.64}_{-5.64}$ & $0.12^{+0.02}_{-0.01}$ &       1.37 & $6.61^{+0.02}_{-0.02}$ & $7.9^{+5.54}_{-5.54}$ &       1.38 & $4.33^{+0.03}_{-0.03}$ & $0.7^{+0.003}_{-0.003}$ &       1.42 \\
     &    NGC7793 &       0.01 & $7.21^{+0.05}_{-0.05}$ & $7.41^{+5.39}_{-5.39}$ & $0.06^{+0.01}_{-0.01}$ &       3.69 & $8.68^{+0.06}_{-0.06}$ & $7.27^{+5.25}_{-5.25}$ &       3.75 & $5.4^{+0.1}_{-0.1}$ & $0.22^{+0.002}_{-0.002}$ &        3.80 \\
     \hline
    \multirow{ 7}{*}{G.B} &     IC2574 &          1 & $17.27^{+0.17}_{-0.16}$ & $7.^{+5.16}_{-1.24}$ & $18.28^{+0.34}_{-0.38}$ &       0.43 & $\sim10^4$ & $2.85^{+1.03}_{-1.01}$ &       6.17 &  $ >10^6$ & $0.02^{+0.001}_{-0.001}$ &       5.49 \\
     &    NGC2366 &          1 & $2.25^{+0.00001}_{-0.00001}$ & $7.96^{+6.52}_{-6.4}$ & $2.25^{+0.001}_{-0.001}$ &       2.11 & $18.62^{+0.48}_{-0.47}$ & $6.15^{+4.65}_{-4.65}$ &       4.45 & $11.68^{+1.03}_{-1.03}$ & $0.03^{+0.001}_{-0.001}$ & 3.98 \\
     &    NGC2903 &          1 & $1.90^{+0.002}_{-0.003}$ & $9.26^{+7.07}_{-7.00}$ & $1.90^{+0.02}_{-0.02}$ &       1.72 & $3.91^{+0.01}_{-0.01}$ & $8.38^{+6.19}_{-6.12}$ &       2.36 & $2.44^{+0.02}_{-0.02}$ & $1.04^{+0.004}_{-0.004}$ & 1.75 \\
     &    NGC2976 &          1 & $2.53^{+0.001}_{-0.001}$ & $8.5^{+6.6}_{-6.85}$ & $2.53^{+0.001}_{-0.001}$ &       0.69 & $\sim10^4$ & $3.27^{+1.5}_{-1.5}$ &       2.86 &  $ >10^6$ & $0.12^{+0.001}_{-0.001}$ & 1.23 \\
     &    NGC3198 &          1 & $3.76^{+0.01}_{-0.01}$ & $8.41^{+6.18}_{-6.22}$ & $3.76^{+0.06}_{-0.06}$ &       0.59 & $9.02^{+0.03}_{-0.03}$ & $7.39^{+5.16}_{-5.21}$ &        1.80 & $7.85^{+0.05}_{-0.05}$ & $0.25^{+0.001}_{-0.001}$ & 1.23 \\
     &    NGC3521 &          1 & $2.00^{+0.0001}_{-0.0001}$ & $9.31^{+7.47}_{-7.08}$ & $2.00^{+0.001}_{-0.001}$ &       1.37 & $5.25^{+0.03}_{-0.03}$ & $8.22^{+6.19}_{-6.19}$ &       7.17 & $2.43^{+0.04}_{-0.04}$ & $1.22^{+0.008}_{-0.008}$ & 2.72 \\
     &     NGC925 &          1 & $10.36^{+0.08}_{-0.08}$ & $7.53^{+5.6}_{-5.65}$ & $12.18^{+0.23}_{-0.23}$ &       0.31 & $\sim10^4$ & $2.27^{+0.39}_{-0.39}$ &       1.46 &  $ >10^7$ & $0.04^{+0.001}_{-0.001}$ & 1.32 \\
     \hline
    \multirow{ 3}{*}{G.C} &    NGC2403 &   $<10^{-6}$  & $6.94^{+0.01}_{-0.01}$ & $7.51^{+5.11}_{-5.11}$ & $<0.01$ &       0.79 & $6.94^{+0.01}_{-0.01}$ & $7.51^{+5.11}_{-5.11}$ &       0.79 & $5.59^{+0.02}_{-0.02}$ & $0.27^{+0.001}_{-0.001}$ &       1.24 \\
     &    NGC5055 &          $<10^{-6}$ & $4.03^{+0.01}_{-0.01}$ & $8.38^{+6.05}_{-6.05}$ & $<0.01$ &       1.38 & $4.03^{+0.01}_{-0.01}$ & $8.38^{+6.05}_{-6.05}$ &       1.37 & $2.27^{+0.02}_{-0.02}$ & $1.13^{+0.004}_{-0.004}$ &        2.90 \\
     &    NGC7331 & $<10^{-6}$ & $3.56^{+0.01}_{-0.01}$ & $8.67^{+6.5}_{-6.5}$ & $<0.01$ &       0.85 & $3.56^{+0.01}_{-0.01}$ & $8.67^{+6.5}_{-6.5}$ &       0.85 & $1.83^{+0.02}_{-0.02}$ & $1.58^{+0.006}_{-0.006}$ &       1.03 \\
    \end{tabular}
    }
  \caption{\footnotesize{We present the results discussed in Sec. \ref{results} for BDM (Eq. (\ref{eq:rhobdm})) with its three free parameters ($r_c$, $r_s$, and $\rho_0$) and the NFW (Eq. (\ref{eq:rhoNFW})) profiles with the minimal disk model. Also, in the last three columns, we present the result obtained with the BDM profiles fixing the value of $E_c$ to $0.11$ eV as explained in Sec. \ref{sec:bdm.fixed}. Column (2) shows the ratio between the core radius $r_c$ and the scale distance $r_s$, and galaxies are grouped by de value of this quotient. Columns (3-6) are the fitted results when only DM and the BDM profile are considered in the mass model. (7-9) show the results for NFW with the same mass model. In (10-12) we show the results for the fixed BDM model. All distance scales such as $r_c$ and $r_s$ are given in {\rm kpc}. Logarithm base 10 of the densities $\rho_0$ is given in $M_\sun/{\rm kpc}^3$. (6,9,12) show the value of $\chi^2$ is normalized to the numbers of data points minus the number of free parameter. }}
  \label{tab:onlydm_bdmnfw}
\end{table}

\setlength{\extrarowheight}{3pt}
\begin{table}[h]
  \centering
  \scriptsize{
    \begin{tabular}{rl|r|lllr|llr|llr}
    \multicolumn{ 13}{c}{MINIMUM DISK + GAS} \\
    \hline
               &            &                                       \multicolumn{ 5}{c|}{BDM} &             \multicolumn{ 3}{c|}{NFW} &     \multicolumn{3}{c}{Fixed-BDM  $E_c=0.08$} \\
               &  \multicolumn{1}{c|}{Galaxy} & \multicolumn{1}{c|}{$r_c/r_s$} &     \multicolumn{1}{c}{$r_s$} & \multicolumn{1}{c}{$\log_{10} \rho_0$}  &     \multicolumn{1}{c}{$r_c$} & \multicolumn{1}{c|}{$\chi^2_{\rm red}$} &     \multicolumn{1}{c}{$r_s$} & \multicolumn{1}{c}{$\log_{10} \rho_0$}  & \multicolumn{1}{c|}{$\chi^2_{\rm red}$}  &     $r_s$ &     $r_c$ & $\chi^2_{\rm red}$ \\
               & \multicolumn{1}{c|}{(1)} &\multicolumn{1}{c|}{(2)} &\multicolumn{1}{c}{(3)}&\multicolumn{1}{c}{(4)} &\multicolumn{1}{c}{(5)} &\multicolumn{1}{c|}{(6)} & \multicolumn{1}{c}{(7)} &\multicolumn{1}{c}{(8)} &\multicolumn{1}{c|}{(9)} &       (10) &       (11) &       (12) \\
    \hline
    \multirow{ 7}{*}{G.A} &     DDO154 &      0.138 & $4.79^{+0.04}_{-0.04}$ & $6.97^{+5.15}_{-5.15}$ & $0.66^{+0.04}_{-0.04}$ &       0.28 & $13.41^{+0.15}_{-0.15}$ & $6.18^{+4.34}_{-4.35}$ &      1.00 & $9.35^{+0.332}_{-0.332}$ & $0.08^{+0.001}_{-0.001}$ &       0.52 \\
    &    NGC2841 &          0 & $4.47^{+0.01}_{-0.01}$ & $8.65^{+6.25}_{-6.24}$ & $0.00^{+0.01}$ &        0.50 & $4.47^{+0.01}_{-0.01}$ & $8.65^{+6.25}_{-6.24}$ &        0.5 & $3.61^{+0.046}_{-0.046}$ & $3.76^{+0.014}_{-0.014}$ &       6.44 \\
    &    NGC3031 &          1 & $0.88^{+0.002}_{-0.002}$ & $10.06^{+7.69}_{-7.68}$ & $0.91^{+0.01}_{-0.01}$ &        4.5 & $1.73^{+0.003}_{-0.003}$ & $9.23^{+6.85}_{-6.85}$ &       4.48 & $0.53^{+0.016}_{-0.016}$ & $2.96^{+0.009}_{-0.009}$ &       4.89 \\
    &    NGC3621 &       0.01 & $6.22^{+0.02}_{-0.02}$ & $7.66^{+5.4}_{-5.4}$ & $0.06^{+0.01}_{-0.01}$ &       1.61 & $6.58^{+0.02}_{-0.02}$ & $7.60^{+5.34}_{-5.34}$ &       1.64 & $2.88^{+0.026}_{-0.026}$ & $1.41^{+0.009}_{-0.009}$ &       2.87 \\
    &    NGC4736 &          4 & $0.10^{+0.0003}_{-0.0003}$ & $12.02^{+9.91}_{-9.92}$ & $0.4^{+0.009}_{-0.005}$ &        1.7 & $0.04^{+0.0001}_{-0.0001}$ & $12.79^{+10.86}_{-10.52}$ &      14.59 & $<0.01$ & $1.74^{+0.006}_{-0.006}$ &       1.99 \\
    &    NGC6946 &      0.028 & $5.41^{+0.01}_{-0.01}$ & $8.08^{+5.73}_{-5.72}$ & $0.15^{+0.01}_{-0.01}$ &       1.32 & $5.78^{+0.02}_{-0.01}$ & $8.00^{+5.69}_{-5.61}$ &       1.36 & $1.17^{+0.018}_{-0.018}$ & $2.45^{+0.007}_{-0.007}$ &       4.93 \\
    &    NGC7793 &      0.009 & $6.24^{+0.04}_{-0.05}$ & $7.48^{+5.48}_{-5.47}$ & $0.06^{+0.01}_{-0.01}$ &       3.37 & $7.41^{+0.05}_{-0.05}$ & $7.35^{+5.34}_{-5.34}$ &       3.44 & $2.4^{+0.065}_{-0.065}$ & $1.03^{+0.008}_{-0.008}$ &       4.09 \\
    \hline
    \multirow{ 7}{*}{G.B} &     IC2574 &          1 & $13.57^{+0.28}_{-0.13}$ & $6.93^{+5.21}_{-5.18}$ & $13.57^{+0.26}_{-0.29}$ &       0.39 & $\sim10^7$ & $0.08^{+-2.07}_{--1.27}$ &       1.42 & $  >10^6$ & $0.05^{+0.001}_{-0.001}$ &       3.03 \\
    &    NGC2366 &          1 & $1.99^{+0.13}_{-0.14}$ & $7.93^{+6.6}_{-6.47}$ & $1.99^{+0.13}_{-0.12}$ &       1.76 & $14.77^{+0.5}_{-0.49}$ & $6.14^{+4.78}_{-4.77}$ &       3.24 & $5.9^{+0.457}_{-0.457}$ & $0.12^{+0.005}_{-0.005}$ &       2.51 \\
    &    NGC2903 &          1 & $1.80^{+0.01}_{-0.01}$ & $9.30^{+6.82}_{-7.34}$ & $1.80^{+0.01}_{-0.01}$ &       1.63 & $3.66^{+0.01}_{-0.01}$ & $8.43^{+6.22}_{-6.2}$ &        2.40 & $0.74^{+0.025}_{-0.025}$ & $2.7^{+0.009}_{-0.009}$ &       3.89 \\
    &    NGC2976 &        0.5 & $3.01^{+0.03}_{-0.04}$ & $8.24^{+6.5}_{-6.49}$ & $1.62^{+0.04}_{-0.05}$ &       0.88 & $4.97^{+0.07}_{-0.07}$ & $7.43^{+6.08}_{-5.28}$ &        6.60 & $5.34^{+0.765}_{-0.765}$ & $0.75^{+0.016}_{-0.016}$ &       0.86 \\
    &    NGC3198 &          1 & $3.29^{+0.01}_{-0.01}$ & $8.49^{+6.54}_{-6.03}$ & $3.29^{+0.01}_{-0.01}$ &        0.7 & $7.66^{+0.02}_{-0.02}$ & $7.50^{+5.3}_{-5.31}$ &       2.12 & $3.36^{+0.028}_{-0.028}$ & $1.41^{+0.008}_{-0.008}$ &       1.96 \\
    &    NGC3521 &        1.5 & $1.62^{+0.01}_{-0.01}$ & $9.54^{+7.52}_{-7.52}$ & $2.40^{+0.06}_{-0.05}$ &       2.97 & $2.03^{+0.02}_{-0.01}$ & $9.02^{+6.61}_{-7.38}$ &      15.15 & $1.61^{+0.068}_{-0.068}$ & $2.87^{+0.011}_{-0.011}$ &       2.98 \\
    &     NGC925 &          1 & $9.59^{+0.07}_{-0.07}$ & $7.48^{+5.64}_{-5.63}$ & $9.59^{+0.19}_{-0.2}$ &       0.48 & $4.64^{+0.04}_{-0.04}$ & $7.46^{+5.72}_{-5.57}$ &      13.63 & $  >10^7$ & $0.14^{+0.001}_{-0.001}$ &        1.20 \\
    \hline
    \multirow{ 3}{*}{G.C} &    NGC2403 & $<10^{-6}$ & $6.26^{+0.01}_{-0.01}$ & $7.57^{+5.2}_{-5.19}$ &   $<0.01$ &       0.85 & $6.26^{+0.01}_{-0.01}$ & $7.57^{+5.2}_{-5.19}$ &       0.84 & $2.03^{+0.016}_{-0.016}$ & $1.42^{+0.002}_{-0.002}$ &       2.79 \\
    &    NGC5055 & $<10^{-6}$ & $3.76^{+0.01}_{-0.01}$ & $8.43^{+6.11}_{-6.11}$ &  $<0.002$ &       1.26 & $3.76^{+0.01}_{-0.01}$ & $8.43^{+6.11}_{-6.11}$ &       1.25 & $1.05^{+0.022}_{-0.022}$ & $2.73^{+0.001}_{-0.001}$ &       2.84 \\
    &    NGC7331 & $<10^{-6}$ & $3.11^{+0.01}_{-0.01}$ & $8.78^{+6.64}_{-6.62}$ &   $<0.01$ &       0.68 & $3.11^{+0.01}_{-0.01}$ & $8.78^{+6.64}_{-6.62}$ &       0.67 & $5.58^{+0.072}_{-0.072}$ & $2.41^{+0.009}_{-0.009}$ &      13.8 \\
    \end{tabular}
    }
  \caption{\footnotesize{Here we present the fitted values for the minimal disk plus gas mass model when we considered the $\rho_{bdm}$ and the $\rho_{{\rm nfw}}$ profile and the complete set of data as discussed in Sec. \ref{results}, columns (2-6) and (7-9) respectively. Columns (10-12) present the fitted values when fixing the energy $E_c = 0.08$ eV, cf.\ref{sec:inner_analysis}. Units and the set of galaxies are as show in Table.\ref{tab:onlydm_bdmnfw}.}}
  \label{tab:onlydm+gas_bdmnfw}
\end{table}

\setlength{\extrarowheight}{3pt}
\begin{table}[h]
  \centering
  \scriptsize{
    \begin{tabular}{rl|r|lllr|llr|llr}
    \multicolumn{ 13}{c}{KROUPA} \\
    \hline
               &            &      \multicolumn{ 5}{c|}{BDM} &             \multicolumn{ 3}{c|}{NFW} & \multicolumn{ 3}{c}{Fixed-BDM  $E_c=0.06$} \\
               &  \multicolumn{1}{c|}{Galaxy} & \multicolumn{1}{c|}{$r_c/r_s$} &     \multicolumn{1}{c}{$r_s$} & \multicolumn{1}{c}{$\log_{10} \rho_0$}  &     \multicolumn{1}{c}{$r_c$} & \multicolumn{1}{c|}{$\chi^2_{\rm red}$} &     \multicolumn{1}{c}{$r_s$} & \multicolumn{1}{c}{$\log_{10} \rho_0$}  & \multicolumn{1}{c|}{$\chi^2_{\rm red}$} &     $r_s$ &     $r_c$ & $\chi^2_{\rm red}$ \\
               & \multicolumn{1}{c|}{(1)} &\multicolumn{1}{c|}{(2)} &\multicolumn{1}{c}{(3)}&\multicolumn{1}{c}{(4)} &\multicolumn{1}{c}{(5)} &\multicolumn{1}{c|}{(6)} & \multicolumn{1}{c}{(7)} &\multicolumn{1}{c}{(8)} & \multicolumn{1}{c|}{(9)} &  \multicolumn{1}{c}{(10)} & \multicolumn{1}{c}{(11)} & \multicolumn{1}{c}{(12)} \\
    \hline
    \multirow{ 7}{*}{G.A} &     DDO154 &       0.23 & $4.37^{+0.04}_{-0.04}$ & $7.08^{+5.27}_{-5.26}$ & $0.99^{+0.05}_{-0.05}$ &       0.28 & $15.14^{+0.17}_{-0.18}$ & $6.1^{+4.28}_{-4.28}$ &       1.06 & $7.3^{+0.25}_{-0.25}$ & $0.31^{+0.}_{-0.}$ &       0.35 \\
    &    NGC2841 & $<10^{-6}$ & $6.67^{+0.02}_{-0.02}$ & $8.18^{+5.89}_{-5.89}$ &   $<0.01$ &        1.3 & $6.67^{+0.02}_{-0.02}$ & $8.18^{+5.89}_{-5.89}$ &       1.29 & $3.16^{+0.11}_{-0.11}$ & $5.97^{+0.02}_{-0.02}$ &       1.21 \\
    & NGC3031 & $<10^{-6}$ & $8.35^{+0.06}_{-0.06}$ & $7.48^{+5.56}_{-5.55}$ &   $<0.03$ &       5.01 & $8.35^{+0.06}_{-0.06}$ & $7.48^{+5.56}_{-5.55}$ &       4.96 & $3.09^{+0.21}_{-0.21}$ & $3.2^{+0.02}_{-0.02}$ &       5.02 \\
    &    NGC3621 & $<10^{-6}$ & $17.1^{+0.08}_{-0.08}$ & $6.72^{+4.6}_{-4.6}$ &   $<0.01$ &       1.46 & $17.1^{+0.08}_{-0.08}$ & $6.72^{+4.6}_{-4.6}$ &       1.45 & $20.3^{+0.27}_{-0.27}$ & $0.90^{+0.01}_{-0.01}$ &       1.72 \\
    & NGC4736 &       1.05 & $0.13^{+0.02}_{-0.01}$ & $11.19^{+9.49}_{-9.49}$ & $0.14^{+0.02}_{-0.01}$ &       1.34 & $0.23^{+0.02}_{-0.02}$ & $10.44^{+8.74}_{-8.74}$ &       1.34 & $<0.01$ & $1.90^{+0.02}_{-0.02}$ &       1.65 \\
    & NGC6946 &       0.03 & $35.13^{+0.24}_{-0.27}$ & $6.49^{+4.49}_{-4.49}$ & $1.17^{+0.07}_{-0.07}$ &       1.13 & $96.74^{+0.84}_{-0.85}$ & $5.87^{+3.87}_{-3.87}$ &       1.21 & $31.3^{+0.98}_{-0.98}$ & $1.20^{+0.01}_{-0.01}$ &       1.13 \\
    &    NGC7793 & $<10^{-6}$ & $8.62^{+0.08}_{-0.08}$ & $7.15^{+5.26}_{-5.25}$ &   $<0.01$ &       4.13 & $8.62^{+0.08}_{-0.08}$ & $7.15^{+5.26}_{-5.25}$ &       4.07 &  $40.9^{+8.20}_{-8.20}$ & $0.65^{+0.02}_{-0.02}$ &       31.1 \\
    \hline
    \multirow{ 7}{*}{G.B} &     IC2574 &       0.32 & $48.39^{+0.88}_{-0.89}$ & $6.14^{+4.5}_{-4.51}$ & $15.61^{+0.49}_{-0.46}$ &       0.75 &   $>10^6$ & $0.1^{+-1.62}_{--1.46}$ &        2.4 &  $23.0^{+1.20}_{-1.20}$ &   $<0.01$ &         -- \\
    &    NGC2366 &          1 & $2.03^{+0.30}_{-0.30}$ & $7.87^{+5.90}_{-5.90}$ & $2.03^{+0.2}_{-0.2}$ &       1.71 & $15.5^{+1.1}_{-1.1}$ & $6.07^{+4.81}_{-4.81}$ &          3 & $9.02^{+0.80}_{-0.80}$ & $0.28^{+0.01}_{-0.01}$ &       2.22 \\
    & NGC2903 &          1 & $2.34^{+0.18}_{-0.15}$ & $9.02^{+6.97}_{-6.7}$ & $2.35^{+0.03}_{-0.03}$ &       2.13 & $4.92^{+0.02}_{-0.02}$ & $8.12^{+5.94}_{-5.94}$ &       3.43 & $1.67^{+0.07}_{-0.07}$ & $4.36^{+0.01}_{-0.01}$ &       2.67 \\
    &    NGC2976 &        0.8 & $40.2^{+2.10}_{-2.10}$ & $7.61^{+5.80}_{-5.71}$ & $29.9^{+2.10}_{-2.10}$ &       1.28 &   $>10^5$ &      $<2$ &       6.31 &   $>10^5$ & $4.28^{+0.31}_{-0.36}$ &       4.33 \\
    & NGC3198 &        0.9 & $8.28^{+0.03}_{-0.03}$ & $7.58^{+5.53}_{-5.53}$ & $8.27^{+0.18}_{-0.17}$ &       3.51 & $24.2^{+0.13}_{-0.13}$ & $6.42^{+4.37}_{-4.37}$ &       4.99 & $17.2^{+0.21}_{-0.21}$ & $0.93^{+0.01}_{-0.01}$ &        4.1 \\
    &    NGC3521 &       0.06 & $128^{+5.21}_{-5.67}$ & $5.65^{+4.44}_{-4.36}$ & $7.58^{+1.09}_{-0.98}$ &       5.74 &   $>10^6$ & $1.01^{+0.24}_{-0.24}$ &       9.26 &   $>10^7$ & $0.29^{+0.01}_{-0.01}$ &       9.01 \\
    &     NGC925 &          1 & $48.56^{+0.8}_{-0.93}$ & $6.77^{+5.1}_{-5.15}$ & $48.56^{+1.39}_{-1.26}$ &       1.24 &   $>10^5$ & $1.95^{+0.34}_{-0.32}$ &       3.68 & $>10^7$ & $0.21^{+0.01}_{-0.01}$ &       22.2 \\
    \hline
    \multirow{ 3}{*}{G.C} & NGC2403 &      0.004 & $10.45^{+0.03}_{-0.03}$ & $7.14^{+4.87}_{-4.72}$ & $0.05^{+0.01}_{-0.01}$ &        0.8 & $10.38^{+0.03}_{-0.03}$ & $7.14^{+4.79}_{-4.8}$ &       0.82 & $4.36^{+0.04}_{-0.04}$ & $2.29^{+0.01}_{-0.01}$ &       0.98 \\
    & NGC5055 &        0.4 & $45.9^{+0.39}_{-0.48}$ & $6.31^{+4.50}_{-4.50}$ & $18.24^{+0.59}_{-0.76}$ &       4.35 &   $>10^6$ &      $<2$ &       5.09 &   $>10^3$ & $0.25^{+0.01}_{-0.01}$ &          5.00 \\
    & NGC7331 &          0 &   $>10^5$ & $2.11^{+0.22}_{-0.21}$ & $3.76^{+0.2}_{-0.19}$ &       7.92 &   $>10^6$ &      $<2$ &       8.63 & $580^{+127}_{-127}$ & $0.94^{+0.01}_{-0.01}$ &       8.14 \\
    \end{tabular}
    }
  \caption{\footnotesize{This table show the fitted values when considering the Kroupa IMF for the value of the stellar disk, columns (2-6) show the BDM and columns (7-9) NFW parameters. Columns (10-12) are the parameters obtained with the BDM profile by fixing $E_c = 0.05$ eV, cf. \ref{sec:inner_analysis} . Units and the set of galaxies are as show in Table.\ref{tab:onlydm_bdmnfw}. }}
  \label{tab:kroupa_bdmnfw}
\end{table}

\section{Results}\label{results}

We have analyzed the rotation curves of seventeen galaxies using five disk models (Sec. \ref{MaMoEsDi}) and four different DM profiles: our BDM, NFW, Burkert, and pseudo-isothermal (Sec. \ref{MaMoDM}). Since we have a large number of tables and figures, and because we want to emphasize the main results,  in this section we mainly present three disk models (min.disk, min.disk+gas, Kroupa) and two DM profiles (BDM and NFW). However, each DM profile is analyzed with all the disk models except for the free $\gs$, for which only the BDM profile is tested. The rest of the tables and figures are in the Appendices as explained below. In the following we shall use the term ``mass models'' to refer to different disk mass scenarios for the analysis of the rotation curves as described in Sec. \ref{MaMoEsDi}, and  denote ``min.disk'' and ``min.disk+gas'' as minimal disk and minimal disk plus gas abbreviations, respectively.
It is worth pointing out that including color gradients  we obtain a better fit to the rotation curve in our analysis.
We have also consider the effect of a  bulge when it is present in the galaxy and  found
that the rotation curves are still consistent with  a  core  $r_c \neq 0$. However,  since the bulge contributes significantly to the inner part of the galaxy the value of $r_c$ is smaller when a bulge is taken into account and in some galaxies the $1\sigma$ standard deviation of $r_c$ may also consistent with a vanishing core.
The adequate surface brightness distribution for the bulge has been proven  to be of an exponential form \cite{deBlok:2008wp} which is the model that we consider here. Galaxies with a bulge deserve a closer analysis to the parameter estimation, as we discuss below.

We present the results of NFW and BDM profiles in order to compare them since the BDM  profile that has NFW as a limit when $r_c \rightarrow 0$. The fitted values of the profile parameters are shown in the following mass models order: min.disk, min.disk+gas, and Kroupa are in Tables \ref{tab:onlydm_bdmnfw}, \ref{tab:onlydm+gas_bdmnfw}, and \ref{tab:kroupa_bdmnfw}, respectively. We present the  $1\sigma$ and $2\sigma$ likelihood contours  plots of  $\rho_0$ and $r_c$  in Figures \ref{tab:conflevel_A}, \ref{tab:conflevel_B}, \ref{tab:conflevel_C}, and \ref{tab:conflevel_inner B}.

We leave in Appendix \ref{apend: diet-salpeter} the results for the diet-Salpeter (Table \ref{tab:salpeter_bdmnfw}), the BDM free $\gs$ scenario (Table \ref{tab:free_bdm}), and the fitted values for the Burkert and ISO profiles with different mass models (Tables \ref{tab:burkert}, \ref{tab:iso}). The details of the analysis for each galaxy as well as the figures when considering the different mass scenarios are in Appendix \ref {apend:LikelihoodPlot}.

We have grouped the galaxies into three blocks according to the fitted values of $r_c$ of the BDM profile as shown in column (2) of Table \ref{tab:onlydm_bdmnfw}. The first one, Group A (G.A.), is composed of galaxies with a core radius  $r_s > r_c \neq 0$. The second, Group B (G.B.), is formed by galaxies in which $r_c$ took its upper limit value, $r_c = r_s$, and finally, the Group C (G.C.) is where $r_c$ is negligible, $r_c/r_s < 10^{-6}$. For the sake of simplicity we conserve the same group structure for the forthcoming tables. We shall explain the physical interpretation of the results for each group in the following paragraphs.

Let us firstly discuss G.A. galaxies.  This group comprises galaxies with reasonable fitted values for the BDM parameters $r_c, r_s$, and $\rho_0$, and it is composed of the galaxies DDO 154, NGC 2841, NGC 3031, NGC 3621, NGC 4736, NGC 6946, and NGC 7793. In the galaxies NGC 3031, NGC 4736, and NGC 6946 a bulge is present and it gives a better fit, but the  $\chi^2$ is only slightly reduced. On the other hand, the bulge in galaxy NGC 3031 with a diet-Salpeter mass model is overestimated and the fit is worst when is analyzed without the bulge.

For G.A. galaxies we obtain values for $r_c \leq 200$ {\rm pc} and an average $r_s \sim 4$ {\rm kpc} in the minimal disk analysis that are typical values for a galaxy, except for DDO 154 which has a larger core, $r_c = 1.35$ {\rm kpc}. On the other hand,  the average of the scale radius for the NFW profile is slightly bigger, $r_s \sim 6$ {\rm kpc}, shifting the cusp to larger radii.  In Tables   \ref{tab:onlydm_bdmnfw}, \ref{tab:onlydm+gas_bdmnfw}, and  \ref{tab:kroupa_bdmnfw} we show the fittings for BDM and NFW profiles with the three mass models mentioned above. We see that the BDM core radius decreases,  when more mass components are taking into account. The reason for this is that $r_c$ and the amount of disk and bulge masses are degenerated parameters, and this happens more when e.g. the stellar disk has a more dominant behavior close to the center of the galaxy, preventing us from having a better precision for $r_c$.

In Appendix \ref{apendix:ga} we show the fittings for all DM profiles and mass models and the rotation curves for each galaxy. From the likelihoods (e.g. Fig. (\ref{tab:conflevel_A})) we see that the core radius decreases, and even $r_c \rightarrow 0$ becomes valid, when more mass components are taking into account. The reason for this is that $r_c$ and $\gs$ are degenerated parameters, and this happens more when the stellar disk has a dominant behavior close to the center of the galaxy, preventing us from having a better precision.

We notice that for all the G.A. galaxies the $\chi^2$ in BDM is smaller than NFW's. The extra parameter ($r_c$) in BDM, however, makes the reduced $\chi^2_{\rm red}$ equivalent for both profiles. The $\chi^2_{\rm red}$ in the BDM profile tends to NFW's value when $r_c \rightarrow 0$ as explained in Sec. \ref{modelo}. When the value of $r_c \geq 40$ {\rm pc}, BDM is clearly better fitted than NFW, and also than the other two cored profiles, cf. Tables \ref{tab:onlydm_bdmnfw}, \ref{tab:burkert} and \ref{tab:iso} for the min.disk mass model. Burkert and ISO profiles have difficulties when fitting a couple galaxies (NGC 4736, NGC 3621) having $\chi^2_{\rm red} > 5$ while BDM and NFW each have $\chi^2_{\rm red} \leq 2$. For the galaxy DDO 154 all cored profiles fit equally well for every mass model, since they have a $\chi^2_{\rm red} \leq 1$, meaning that the fits lie inside the uncertainty of the observations. For galaxies NGC 3031 and NGC 7793 none of the DM profiles provide a good fit, being BDM the best fit for both galaxies with a $\chi^2_{\rm red} > 3$ for each of the mass models. Such a large $\chi^2_{\rm red}$ is due to the uncommon velocity drop in the rotation curve at large radii in these galaxies.

We now turn to the analysis of confidence levels of $r_c$ and $\rho_0$ in the BDM profile stemming from  the different mass models for this set of galaxies, shown in Fig. (\ref{tab:conflevel_A}). From this table we deduce that the core average is $r_c \simeq 40$ {\rm pc} and a core energy $E_c \simeq 0.22 {\rm \ eV}$ for the min.disk mass model, where we have taken a logarithm-normal distribution.It is worth noticing that energy $E_c $  has a smaller dispersion than $r_c$ or $r_s$ and we supports the hypothesis that a signal that $E_c$ is a constant parameter. This indicates the fundamental importance of the energy parameter in the BDM  model. The likelihood contour plots also reveal how the central contribution of the gas makes $r_c$ to increase (c.f. min.disk vs. min.disk + gas), but on the contrary, stellar disk fades away the evidence of the core making $r_c \rightarrow 0$, e.g. the galaxies NGC 3031, NGC 3621, and NGC 6946 for the Kroupa mass model are consistent with a null $r_c$ at $1\sigma$. We also observe that even when $r_c$ is different for each mass model its value lies inside the confidence levels obtained from the other mass models for the same galaxy. For example, for the galaxy NGC 3621 the central value of $r_c < 60$ {\rm pc} for min.disk and min.disk+gas, but the confidence levels are still equivalent with a zero core. On the other hand, when stars are considered in Kroupa the central value is $r_c \rightarrow 0$, but confidence contours show that can be well accepted values up to 30 {\rm pc}. This, on the other hand, means that the amount of data in this galaxy is not good enough to clearly discern between a core or cusp profile. We obtain $r_c = r_s$ for the galaxy NGC 4736 (Kroupa mass model), in which the likelihoods restrict quite strongly the $r_c$ value.

Let us now discuss Group B (G.B.) which is composed of the galaxies IC 2574, NGC 2366, NGC 2903, NGC 2976, NGC 3198, NGC 3521, and NGC 925. These galaxies have preferred a fitted values $r_c \simeq r_s$, that makes our BDM profile of a particular type, $\rho \propto (r_c + r)^{-3}$. The $r_c$ value for IC 2574, NGC 2976, and NGC 925 galaxies is of the order of the farthest observation from the galactic center which entails that the density of the DM halo is constant at all observed radii. In G.B. the only galaxies that have a bulge are NGC 2903  and  NGC 3198. The galaxy NGC 3198 presents a sudden reddening in the inner most region that might indicate the presence of a central component, therefore we performed the fits with and without a bulge for this galaxy. The bulge component in both galaxies overestimated the  rotation velocity, indicating that is likely to be dynamically less important than suggested by the mass models. The fits for the G.B. galaxies are in Tables \ref{tab:onlydm_bdmnfw}, \ref{tab:onlydm+gas_bdmnfw}, \ref{tab:kroupa_bdmnfw}, and \ref{tab:salpeter_bdmnfw},  and the rotation curves are in Appendix \ref{apendix:gb}.

It is worth noticing that for this set of galaxies the BDM profile fits much better than NFW's, having the former smaller $\chi_{\rm red}^2$. In fact,
Burkert and ISO profiles also fit better than NFW's, indicating that a core is needed for G.B. galaxies. The BDM profile fits better or equally well than the other two cored profiles in all the analyzed scenarios. The BDM and Burkert profiles are slightly better than ISO's profile when we consider min.disk and min.disk+gas analysis. Otherwise, for Kroupa and diet-Salpeter mass models the BDM and ISO  profiles have equivalent $\chi^2_{\rm red}$ and fit a bit better than Burkert's.

The $1\sigma$ and $2\sigma$  likelihood contour plots of $r_c$ and $\rho_0$ for the G.B. are shown in Fig. (\ref{tab:conflevel_B}). We notice that when more mass components are included in the analysis the contours plots become broader. The BDM profile consistent fits values of $r_c \simeq r_s$ for all mass models have an average $r_c$ value of 6 and 5 {\rm kpc} for min.disk and min.disk+gas, respectively. There is an exception when considering the Kroupa and diet-Salpeter mass models in IC 2574, NGC 2903, and NGC 3521 galaxies for which $r_c \ne r_s$, as seen in Table \ref{tab:kroupa_bdmnfw} and \ref{tab:salpeter_bdmnfw}.

In general, the rotation curves fits prefer cored profiles over NFW's, but for this group of galaxies we found that $r_c \simeq r_s$, that defines the particular cored profile mentioned before. To understand in more detail the parameter estimation in these galaxies we performed an analysis of the inner galactic region, see Sec. \ref{sec:inner_analysis}. The galaxies IC 2574, NGC 2976, and NGC 925 are characterized for having a rotation curve with constant positive slope at all radii, see Figs. \ref{fig:inner_grph1}, \ref{fig:inner_grph2}, and \ref{fig:inner_grph4}. It is important to emphasize that for these galaxies the different results obtained from the BDM and inner analysis are all in good agreement. First, the fit of BDM profile results in $r_c \simeq r_s$ that is of the order of the maximum observed radius. Then, for the inner profile $\rho_{in}$ we obtain $r_s/r_c \geq 10^2$ in most of the cases. Finally, we obtained $\alpha \leq 0.9$ for these galaxies when fitting $\rho_{\alpha}=\rho_0 r^{-\alpha}$ taking into account all the observations, see Table \ref{tab:inner_result}. The different results obtained from the different analysis imply that the DM halo density for these galaxies is constant up to the maximum observed radius. This clearly indicates that DM halos need to have cored profiles for the correct description of these galaxies.

The $1\sigma$ and $2\sigma$ confidence levels of $r_c$ and $\rho_c$ are shown in Fig. (\ref{tab:conflevel_inner B}). With the inner analysis we notice that the length of the core radius is reduced, for example, the average $r_c = 2.2$ {\rm kpc}, $r_c = 3.5$ {\rm kpc} for min.disk and min.disk+gas, respectively, with energy $E_c \sim 0.06 {\rm \ eV}$ in both cases and assuming a logarithm-normal distribution in our sample. We notice since stars are difficult to treat,  for the Kroupa and diet-Salpeter mass models, the confidence levels increase up to an order of magnitude for the $r_c$ value.

The last and smallest group is Group C (G.C.) composed of galaxies NGC 2403, NGC 5055, and NGC 7331.  In these  galaxies is present a bulge in their stellar structure, however, their contribution is not enough to adequately explain the rotation curves. This is probably an indication that the bulge has been dynamically underestimated in the calculations, as suggested by the Kroupa or diet-Salpeter mass models, except for the NGC 2403 galaxy that has a good fit considering a diet-Salpeter mass model. These galaxies have fitted values such that $r_c/r_s < 10^{-6}$, c.f. Table \ref{tab:onlydm_bdmnfw}. The reduced number of data close to galactic center prevents us from finding whether these galaxies actually have or not a core. ISO and Burkert are poorly fitting profiles, meanwhile BDM has a fitted value for $r_c \rightarrow 0$, so, BDM is equivalent to NFW profile and both have equivalent fits. Their values are displayed in Table \ref{tab:onlydm_bdmnfw} for the minimal disk mass model and rotation curves, and confidence contours of $r_c$ and $\rho_0$ are shown in Appendix \ref{apend:LikelihoodPlot} in Fig. (\ref{tab:conflevel_C}). We conclude that NFW/BDM profiles fit better than Burkert's and ISO's profile.

All the mass models are consistent with a null core, except for the diet-Salpeter analysis where the core take values of $r_c > r_s$ which has no physical interpretation in our model and indicates an overestimation of the stellar disk. Galaxies NGC 2403 and NGC 5055 present only few observational points close to the center, therefore we decided to perform an inner analysis that is described in section \ref{sec:inner_analysis}. We compute the confidence level contours of $r_c$ and $\rho_c$, see Fig. (\ref{tab:conflevel_C}). We obtain $0 \leq r_c \leq 50 $ {\rm pc} as a plausible interval that can fit the observations within the $2\sigma$ error and $r_c < 10^{-6} r_s$. Although a cusp profile is preferred for these galaxies, a more definitive conclusion would demand more data close to the center in these galaxies.

To summarize, we have seen that in G.A. and G.B. cored profiles fit better than the cuspy NFW profile. The BDM profile prefers values for $r_c \neq 0$ and fits better or equally well than the other two cored profiles. In the cases where $r_c$ is very small ($r_c < 10^{-6} r_s$) a larger number of observational  data is needed close to the center in order to clearly discern between cored or cuspy profiles. BDM, Burkert, and ISO profiles have good fits for G.B.galaxies, where NFW over-predicts the velocity in the inner parts of the galaxy or does fit with unphysical values for $r_s > \mathcal{O}(10^4)$ or $\rho_0 < \mathcal{O}(10^2) {\rm M_\sun/kpc^{3}}$ . In G.C. galaxies the Burkert and ISO profiles do not have a good fit, and the BDM profile demands a very small   core radius ($r_c < 10^{-6} r_s$), but the number of inner data is poor and it does not give enough information on $r_c$. Assuming these data, BDM results are indistinguishable from NFW. Thus,  BDM-NFW have much better agreement than ISO and Burkert profiles. Confidence level contours show consistency in the fitted values of $r_c$ for the different mass models. Given the richer structure the BDM profile has, through its transition from CDM to HDM, it allows for a better explanation of the rotation curves of the different galaxies.

The free mass model treats $\gs$ as an extra free parameter in the model, and the best value for $\gs$ is chosen by fitting the theory to the observations. The mass of the stellar disk is directly related to $\gs$ with the Freeman equation, refer to Sec. \ref{MaMoEsDi}. The results when analyzing the BDM profile in the free mass model are presented in Table \ref{tab:free_bdm}. We notice from Tables \ref{tab:things} and \ref{tab:free_bdm} that values of the stellar mass in the free mass model are in good agreement with the photometric disk masses derived from the $3.6 \mu m$ images in combination with stellar population synthesis arguments predicting diet-Salpeter as the maximum stellar disk consistent with deBlok08 \cite{deBlok:2008wp}, nevertheless the diet-Salpeter mass scenario overestimated the rotation velocity of most of the galaxies and therefore being less realistic than the Kroupa mass scenario.

\subsection{Inner Analysis}\label{sec:inner_analysis}

\setlength{\extrarowheight}{3pt}
\begin{table}[h]
  \centering
  \scriptsize{
    \begin{tabular}{lr|rrrrr|rrrrr|rrrrr}
    \multicolumn{ 17}{c}{Inner Analysis - $\rho_{in}$} \\
    \hline
     &   &   \multicolumn{ 5}{c|}{Min. Disk.} &      \multicolumn{ 5}{c|}{Min.Disk+gas} &                      \multicolumn{ 5}{c}{Kroupa} \\
               &     Galaxy & $R_{m}$ &  \multicolumn{1}{c}{$r_c$} &  $\log_{10} \rho_c$ & $\chi_{inn} ^2$ & $\chi_{t}^2$ & $R_{m}$ & \multicolumn{1}{c}{$r_c$} & $\log_{10} \rho_c$ & $\chi_{inn} ^2$ & $\chi_{t}^2$ & $R_{m}$ &  \multicolumn{1}{c}{$r_c$} & $\log_{10} \rho_c$ & $\chi_{inn} ^2$ & $\chi_{t}^2$ \\
    \hline
          G.A. &    NGC 3621 &       3.32 & $<0.01$ & $9.89^{+10.0}_{-8.19}$ &       1.24 &       2.43 &       3.32 & $0.03^{+0.01}_{-0.01}$ & $9.46^{+8.87}_{-7.59}$ &       1.43 &       1.97 &       3.32 &   $<0.02$ & $9.41^{+7.6}_{-7.22}$ &       1.54 &    4.13 \\
    \hline
    \multirow{ 7}{*}{G.B} &     IC 2574 &      11.6 & $3.27^{+0.11}_{-0.12}$ & $6.72^{+5.29}_{-5.32}$ &       0.63 &       0.54 &      11.6 & $2.22^{+0.12}_{-0.12}$ & $6.72^{+5.47}_{-5.46}$ &       0.58 &       0.39 &      11.6 & $7.37^{+0.29}_{-0.28}$ & $6.35^{+5.01}_{-5.01}$ &       0.79 &       0.75 \\
    &    NGC 2366 &       2.28 & $7.39^{+0.82}_{-0.70}$ & $7.22^{+6.38}_{-6.32}$ &       0.04 &       1.41 &       2.08 & $9.42^{+1.12}_{-0.98}$ & $7.21^{+6.41}_{-6.37}$ &       0.07 &        1.20 &       0.99 & $24.2^{+5.55}_{-4.10}$ & $7.12^{+6.60}_{-6.52}$ &       0.12 &       1.86 \\
    &    NGC 2903 &       2.11 & $5.07^{+0.88}_{-0.73}$ & $8.06^{+7.41}_{-7.32}$ &       1.61 &       18.0 &       1.81 & $22.0^{+7.33}_{-4.78}$ & $7.84^{+7.47}_{-7.34}$ &       0.73 &      21.4 &       2.72 & $81.3^{+5.10}_{-7.20}$ & $6.44^{+4.80}_{-4.80}$ &      14.7 &       39.3 \\
    &    NGC 2976 &       2.57 & $0.25^{+0.02}_{-0.01}$ & $8.44^{+7.31}_{-6.67}$ &       1.14 &       0.99 &       2.57 & $1.07^{+0.01}_{-0.18}$ & $8.01^{+6.51}_{-7.24}$ &       2.88 &       0.83 &       2.57 & $197^{+6.68}_{-6.07}$ & $7.40^{+6.07}_{-6.01}$ &       1.18 &      423 \\
    &    NGC 3198 &       3.61 & $0.31^{+0.09}_{-0.09}$ & $8.34^{+7.78}_{-7.60}$ &       0.03 &       1.76 &       3.21 & $0.33^{+0.10}_{-0.10}$ & $8.33^{+7.80}_{-7.08}$ &       0.04 &       3.16 &       7.23 & $191^{+18.0}_{-15.2}$ & $6.63^{+5.73}_{-5.69}$ &      10.5 &      15.8 \\
    &    NGC 3521 &       2.18 & $1.95^{+0.12}_{-0.35}$ & $8.47^{+7.36}_{-7.80}$ &      15.2 &      45.7 &       1.87 & $6.19^{+0.67}_{-0.58}$ & $8.40^{+7.63}_{-7.45}$ &      13.4 &      69.3 &        5.60 &   $>10^3$ &      $<2$ &       48.9 &       51.6 \\
    &     NGC 925 &       6.02 & $22.3^{+0.73}_{-0.70}$ & $6.91^{+5.54}_{-5.54}$ &       0.21 &       0.62 &       6.02 & $6.10^{+0.27}_{-0.26}$ & $7.01^{+5.75}_{-5.70}$ &       0.25 &       0.98 &       6.02 &   $>10^3$ & $6.31^{+5.01}_{-5.01}$ &       1.98 &      44.0 \\
    \hline
    \multirow{ 2}{*}{G.C} &    NGC 2403 &       2.04 & $<0.01$ & $10.4^{+10.7}_{-8.77}$ &       1.86 &       1.94 &       2.04 & $<0.004$ & $10.5^{+10.9}_{-8.83}$ &       1.87 &       1.89 &       2.04 &   $>10^5$ & $5.27^{+4.3}_{-4.3}$ &      228 &     211 \\
    &    NGC 5055 &       0.73 & $<0.003$ & $11.3^{+12.4}_{-10.3}$ &         -- &       4.65 &       0.73 & $<0.003$ & $11.3^{+12.4}_{-10.3}$ &         -- &       3.84 &       0.73 &   $>10^5$ & $6.7^{+5.2}_{-5.2}$ &         -- &      10.1 \\
    \hline
    \end{tabular}
  }
  \caption{\footnotesize{We show the values obtained from the inner analysis, for all the mass model (except diet-Salpeter, cf. Appendix \ref{apend: diet-salpeter}), with the profile $\rho_{in}$. $R_{m}$ is the maximum radius to consider when making the inner-core analysis. The core radius $r_c$ is given in [{\rm kpc}] and the logarithm of the central density $\rho_c$ in [$M_\sun/{\rm kpc}^3$], both parameters obtained from the inner analysis in which we fit the profile $\rho_{in} = 2 \rho_c (1+r/r_c)^{-1}$ to the data below $R_{m}$ where we obtained the goodness-of-fit $\chi_{\rm red}^2$. Using the parameters ($r_c$, $r_s$, and $\rho_0$) obtained from the analysis of the complete set of data and the $\rho_{bdm}$, we compute $\chi^2_t$ which is the contribution of the data corresponding to the inner region, $r < R_m$.}}
  \label{tab:inner_dm_dmg_k}
\end{table}

\setlength{\extrarowheight}{3pt}
\begin{table}[h]
  \centering
  \scriptsize{
    \begin{tabular}{r|rcrr|rcrr|rcrr}
    \multicolumn{ 13}{c}{Inner Analysis - $\rho_\alpha$} \\
    \hline
               &                    \multicolumn{ 4}{c|}{Min.Disk} &               \multicolumn{ 4}{c|}{Min.Disk.+gas} &                      \multicolumn{ 4}{c}{Kroupa} \\
        Galaxy &  \multicolumn{1}{c}{$\alpha$} & \multicolumn{1}{c}{$\log_{10} \rho_0$}  & $\chi^2_{red}$ & \multicolumn{1}{c|}{$r_c$} & \multicolumn{1}{c}{$\alpha$} & \multicolumn{1}{c}{$\log_{10} \rho_0$}  & $\chi^2_{red}$ & \multicolumn{1}{c|}{$r_c$} & \multicolumn{1}{c}{$\alpha$} & \multicolumn{1}{c}{$\log_{10} \rho_0$}  & $\chi^2_{red}$ & \multicolumn{1}{c}{$r_c$} \\
    \hline
        IC 2574 &    $0.47$ &    $6.93$ &       1.09 &      11.84 &    $0.54$ &    $6.87$ &       1.03 &      10.22 &    $0.28$ &     $6.6$ &       0.98 &      20.32 \\
       NGC 2366 &    $0.19$ &    $7.45$ &       0.04 &       5.55 &    $0.21$ &    $7.47$ &       0.05 &       4.50 &    $0.11$ &    $7.47$ &       0.16 &       4.22 \\
       NGC 2903 &    $0.13$ &    $8.26$ &       1.03 &       8.49 &    $0.15$ &    $8.46$ &       4.13 &       6.23 &   $<0.01$ &    $6.76$ &       12.6 &      74.79 \\
       NGC 2976 &    $0.74$ &    $8.02$ &       1.17 &       1.13 &    $0.72$ &    $8.06$ &       2.02 &       1.18 &   $<0.01$ &    $5.17$ &        381 &     130.9 \\
       NGC 3198 &    $0.76$ &    $7.99$ &       0.12 &       1.90 &    $0.74$ &    $8.00$ &       0.13 &       1.70 &    $0.03$ &    $6.92$ &       9.77 &     126.5 \\
       NGC 3521 &    $0.01$ &    $8.66$ &       6.77 &      123.8 &    $0.03$ &    $8.46$ &       13.7 &      35.5  &    $0.90$ &      $<1$ &       45.8 &        2.60 \\
        NGC 925 &    $0.22$ &    $7.25$ &       0.15 &       13.9 &    $0.20$ &    $7.22$ &       0.16 &      15.3  &    $1.60$ &    $4.63$ &        36.0 &       1.32 \\
    \end{tabular}
  }
  \caption{\footnotesize{Fits with the profile $\rho = \rho_0 r^{-\alpha}$ c.f. Eq. (\ref{eq:rhoalpha}) for galaxies belonging to group G.B. for all the mass models (except diet-Salpeter, cf. Appendix \ref{apend: diet-salpeter}). The distance until which we consider the observations is given by $R_{m}$ given in Table \ref{tab:inner_dm_dmg_k}. The magnitude of the slope of the rotation curve is given by the dimensionless parameter $\alpha$. The $r_c$ is the core radius, given in Kpc, obtain based on the value of $\alpha$ and Eq. (\ref{eq:alpha_integral}). }}
  \label{tab:inner_result}
\end{table}

In this section we perform  an analysis of the inner region of the galaxies. We  only do it  to galaxies in which we  found a $r_c \simeq r_s$, i.e. in Group G.B. These galaxies have a large number of points in the outer region and the DM  profile is therefore  fixed by these points. However, the estimation of the DM  profile parameters  gives a bad fit to the inner region of these galaxies and gives then no information about the core. We will see that if we use only the galactic inner region we get a much better fit.

To  extract  information of  the galactic central region  and  determine  the core radius, central density with a inner slope of the profiles, and we examine the central region data in two different scenarios:\\
i) The first one by taking the limit $r \ll r_s$ in the BDM profile which turns out to be
\be\la{in1}
\rho_{in} = \frac{2 \rho_c}{ (1+\frac{r}{r_c})};
\ee
ii) The second is with the ansatz profile
\begin{equation} \label{eq:rhoalpha}
\rho_{\alpha} = \rho_0 r^{-\alpha},
\end{equation}
where $\alpha$ is a constant slope. Both approaches are related through Eq. (\ref{eq:sl2}). If the slope takes values $\alpha \leq 1/2$  it implies that one is in the region where $r < r_c$, hence we have a  core behavior.

In case (i), Eq.(\ref{in1}), when the BDM profile is approximated by $\rho_{in}$, we take the data points for $r\leq Minimum(r_c,r_s)$ and determine the parameters $r_c$, $\rho_c$, and the corresponding chi-square, $\chi_{inn}^2$. The values obtained using this method are shown in Table \ref{tab:inner_dm_dmg_k}. The $\chi^2_t$ it is computed using also only the inner points but with the  $r_c$ and $\rho_c$ values of the complete data. We notice that $\chi_{inn}^2$ is much smaller than $\chi_{t}^2$ in most of the cases.

We present the different G.B. galaxies that where analyzed when the DM halo is the only mass component with the inner approximation for the BDM profile $\rho_{in}$ up to a radius $r < Minimum{r_c,r_s}$ in Fig. (\ref{fig:inner_grph1} - \ref{fig:inner_grph4}) with a purple, long, dashed line. From this first approach we obtain the core distance $r_c$ and the central density $\rho_c$. The gray points are the observations with its error bars. Below we show the difference between the observational and the theoretical approach, the purple region represent the error bars and the line the fitted curve. Fitted values are presented in Table \ref{tab:inner_dm_dmg_k} and the confidence levels between $\rho_c$ and $r_c$ in Fig. \ref{tab:conflevel_inner B}.

In case (ii), Eq.(\ref{eq:rhoalpha}), we analyze the rotation curves with $\rho_{\alpha}$ and we fit the free parameters $\alpha$ and $\rho_0$ using the same dataset taken for the inner analysis described in case (i). The values obtained for $\alpha$ are in good agreement, cf. Eq. (\ref{eq:sl2}), with those established for $r_c$ in case i), see Table \ref{tab:inner_result}. For most of the galaxies we obtain values in the interval $y \leq 1.08$, that implies a dominant core behavior. For galaxies NGC 2976 and NGC 3198 we obtain $y \sim 5.67$ which means that their inner data correspond to distances beyond the core region, such that $r_c < r < r_s$,  which can be still accepted for the core analysis since we include the core and the transition region between $r_c$ and $r_s$. Given these results, we can now trust to the $r_c$ values obtained from the $\rho_{in}$ analysis in case i), since we confirm that the set of data points used for inner analysis correctly describes the core region of the galaxy.

It is interesting to notice that we can obtain the value of $r_c$ from the value of $\alpha$. We developed a different approach in order to calculate the core radii from the value of the slope. The constant $\alpha$ obtained from the fit with $\rho_\alpha$ is the average value of the $\alpha(r)$, Eq. (\ref{eq:sl2}), in the interval from $r_{min}$ to $r_{max}$, where $r_{min}$ is the first observation and $r_{max}$ is the maximum radius where we are doing the inner analysis. The average value of $\alpha$ is given by
\begin{equation} \label{eq:alpha_integral}
\alpha = \frac{1}{r_{max} - r_{min}} \int_{r_{min}}^{r_{max}} \frac{r}{rc + r}  \, \mathrm{d} r = 1 + \frac{r_c}{r_{max} - r_{min}}\ln\left[ \frac{r_c + r_{min}}{r_c + r_{max}} \right].
\end{equation}
We can numerically solve the last equation in order to obtain the value of $r_c$. The values obtained from these method are shown in Table \ref{tab:inner_result}, from which we can see that the core radius agree in the order of magnitude with those obtain from the $\rho_{in}$ analysis for most of the galaxies.
\setlength{\extrarowheight}{3pt}
\begin{table}[h]
  \centering
  \scriptsize{
    \begin{tabular}{ll|lll|lll|lll}
                                                                    \multicolumn{ 11}{c}{BDM} \\
    \hline
               &            &   \multicolumn{ 3}{c|}{MINIMUM DISK} & \multicolumn{ 3}{c|}{MINIMUM DISK+GAS} &         \multicolumn{ 3}{c}{KROUPA} \\
               &\multicolumn{ 1}{c|}{Galaxy} &\multicolumn{ 1}{c}{$r_c$} & \multicolumn{ 1}{c}{$\log_{10} \rho_c$} & \multicolumn{ 1}{c|}{$E_c$} & \multicolumn{ 1}{c}{$r_c$} & \multicolumn{ 1}{c}{$\log_{10} \rho_c$} & \multicolumn{ 1}{c|}{$E_c$} & \multicolumn{ 1}{c}{$r_c$} & \multicolumn{ 1}{c}{$\log_{10} \rho_c$} & \multicolumn{ 1}{c}{$E_c$} \\
    \hline
    \multirow{ 7}{*}{G.A} &     DDO154 & $1.35^{+0.04}_{-0.04}$ & $7.20^{+6.18}_{-6.17}$ & $0.05^{+0.001}_{-0.001}$ & $0.66^{+0.04}_{-0.04}$ & $7.42^{+6.54}_{-6.53}$ & $0.05^{+0.002}_{-0.002}$ & $0.99^{+0.05}_{-0.05}$ & $7.24^{+6.07}_{-6.07}$ & $0.05^{+0.0008}_{-0.0008}$ \\
    &    NGC2841 & $0.0001^{+0.01}_{-0.01}$ & $13.0^{+15.0}_{-11.08}$ & $1.29^{+0.004}_{-0.004}$ & $<0.01$ &         -- &         -- &   $<0.01$ &         -- &         -- \\
    &    NGC3031 & $0.19^{+0.01}_{-0.01}$ & $9.90^{+8.74}_{-8.02}$ & $0.22^{+0.004}_{-0.001}$ & $0.91^{+0.01}_{-0.01}$ & $9.14^{+7.74}_{-7.73}$ & $0.14^{+0.001}_{-0.001}$ &   $<0.03$ &         -- &         -- \\
    &    NGC3621 & $0.01^{+0.01}_{-0.01}$ & $9.94^{+9.87}_{-9.83}$ & $0.22^{+0.05}_{-0.04}$ & $0.06^{+0.01}_{-0.01}$ & $9.36^{+8.76}_{-8.74}$ & $0.16^{+0.01}_{-0.01}$ &   $<0.01$ &         -- &         -- \\
    &    NGC4736 & $0.05^{+0.01}_{-0.01}$ & $11.0^{+10.2}_{-8.9}$ & $0.41^{+0.02}_{-0.001}$ & $0.40^{+0.009}_{-0.005}$ & $9.71^{+8.61}_{-7.61}$ & $0.2^{+0.004}_{-0.001}$ & $0.14^{+0.02}_{-0.}$ & $10.3^{+9.61}_{-8.55}$ & $0.27^{+0.02}_{-0.002}$ \\
    &    NGC6946 & $0.12^{+0.02}_{-0.01}$ & $9.37^{+8.53}_{-8.46}$ & $0.16^{+0.01}_{-0.01}$ & $0.15^{+0.01}_{-0.01}$ & $9.30^{+8.33}_{-8.32}$ & $0.16^{+0.004}_{-0.004}$ & $1.17^{+0.07}_{-0.07}$ & $7.64^{+6.46}_{-6.46}$ & $0.06^{+0.001}_{-0.001}$ \\
    &    NGC7793 & $0.06^{+0.01}_{-0.01}$ & $9.18^{+8.51}_{-8.49}$ & $0.15^{+0.007}_{-0.007}$ & $0.06^{+0.01}_{-0.01}$ & $9.21^{+8.56}_{-8.55}$ & $0.15^{+0.01}_{-0.01}$ &   $<0.01$ &         -- &         -- \\
    \hline
    \multirow{ 7}{*}{G.B} &     IC2574 & $3.27^{+0.11}_{-0.12}$ & $6.72^{+5.29}_{-5.32}$ & $0.04^{+0.0003}_{-0.0003}$ & $2.22^{+0.12}_{-0.12}$ & $6.72^{+5.47}_{-5.46}$ & $0.04^{+0.001}_{-0.001}$ & $7.37^{+0.29}_{-0.28}$ & $6.35^{+5.00}_{-5.00}$ & $0.03^{+0.0003}_{-0.0003}$ \\
    &    NGC2366 & $7.40^{+0.82}_{-0.70}$ & $7.22^{+6.38}_{-6.32}$ & $0.05^{+0.002}_{-0.002}$ & $9.42^{+1.12}_{-0.98}$ & $7.21^{+6.41}_{-6.37}$ & $0.05^{+0.002}_{-0.002}$ & $24.2^{+5.55}_{-4.10}$ & $7.12^{+6.60}_{-6.52}$ & $0.04^{+0.003}_{-0.003}$ \\
    &    NGC2903 & $5.07^{+0.88}_{-0.73}$ & $8.06^{+7.41}_{-7.32}$ & $0.08^{+0.004}_{-0.004}$ & $22.0^{+7.33}_{-4.78}$ & $7.84^{+7.47}_{-7.34}$ & $0.07^{+0.01}_{-0.01}$ & $81.3^{+9.10}_{-9.10}$ & $6.44^{+4.80}_{-4.80}$ & $0.03^{+0.001}_{-0.001}$ \\
    &    NGC2976 & $0.25^{+0.02}_{-0.02}$ & $8.44^{+7.31}_{-6.67}$ & $0.09^{+0.002}_{-0.001}$ & $1.07^{+0.01}_{-0.18}$ & $8.01^{+6.51}_{-7.24}$ & $0.07^{+0.001}_{-0.003}$ & $197^{+6.68}_{-6.07}$ & $7.40^{+6.07}_{-6.01}$ & $0.05^{+0.001}_{-0.001}$ \\
    &    NGC3198 & $0.31^{+0.09}_{-0.09}$ & $8.34^{+7.78}_{-7.78}$ & $0.09^{+0.006}_{-0.001}$ & $0.33^{+0.10}_{-0.10}$ & $8.33^{+7.80}_{-7.08}$ & $0.09^{+0.01}_{-0.001}$ & $191.3^{+18.0}_{-15.2}$ & $6.63^{+5.73}_{-5.69}$ & $0.03^{+0.001}_{-0.001}$ \\
    &    NGC3521 & $1.95^{+0.12}_{-0.35}$ & $8.47^{+7.36}_{-7.80}$ & $0.10^{+0.002}_{-0.005}$ & $6.19^{+0.67}_{-0.58}$ & $8.4^{+7.63}_{-7.45}$ & $0.09^{+0.004}_{-0.003}$ & $98.0^{+8.10}_{-8.10}$ &      $<2$ &         -- \\
    &     NGC925 & $22.25^{+0.73}_{-0.70}$ & $6.91^{+5.54}_{-5.54}$ & $0.04^{+0.0004}_{-0.0004}$ & $6.10^{+0.27}_{-0.26}$ & $7.01^{+5.75}_{-5.70}$ & $0.04^{+0.001}_{-0.001}$ &   $>10^3$ & $6.31^{+5.1}_{-5.1}$ & $0.03^{+0.001}_{-0.001}$ \\
    \end{tabular}
  }
  \caption{\footnotesize{We show the main parameters, $r_c$, $\rho_c$ and $E_c$, for the BDM model as defined in Sec. \ref{modelo} for the min.disk, min.disk+gas, and Kroupa mass model. For the G.A. galaxy we present the results obtained for analysis considering the BDM profile, Eq. (\ref{eq:rhobdm}). The values for G.B. galaxies are those obtained from the inner analysis using $\rho_{in}$ as described in the Sec.\ref{results}. The core $r_c$ is given in {\rm kpc}. The logarithm base 10 of the density $\rho_c$ is given in $\rho_c$ in $M_\sun/{\rm kpc}^3$. Finally, the energy $E_c$ in which the DM transition take place is given in {\rm \ eV}. }}
  \label{tab:bdm_odm_dmg_k}
\end{table}

\begin{figure}
  \centering
 \subfloat[\footnotesize{$R_c vs E_c$ : Minimal Disk}]{ \includegraphics[width=0.50\textwidth]{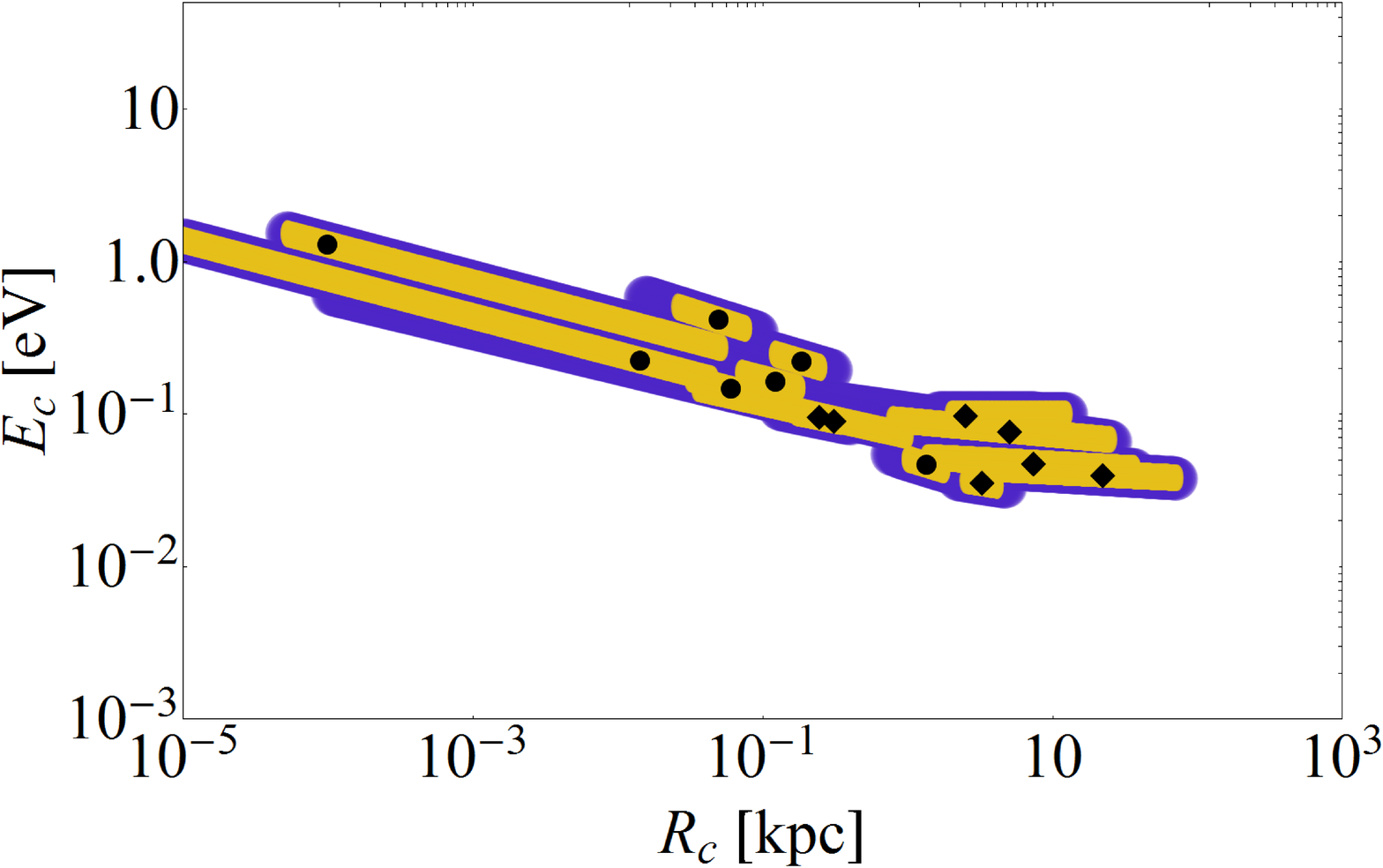}}
    \subfloat[\footnotesize{$r_c$ vs $E_c$ : Min.disk + gas}]{ \includegraphics[width=0.50\textwidth]{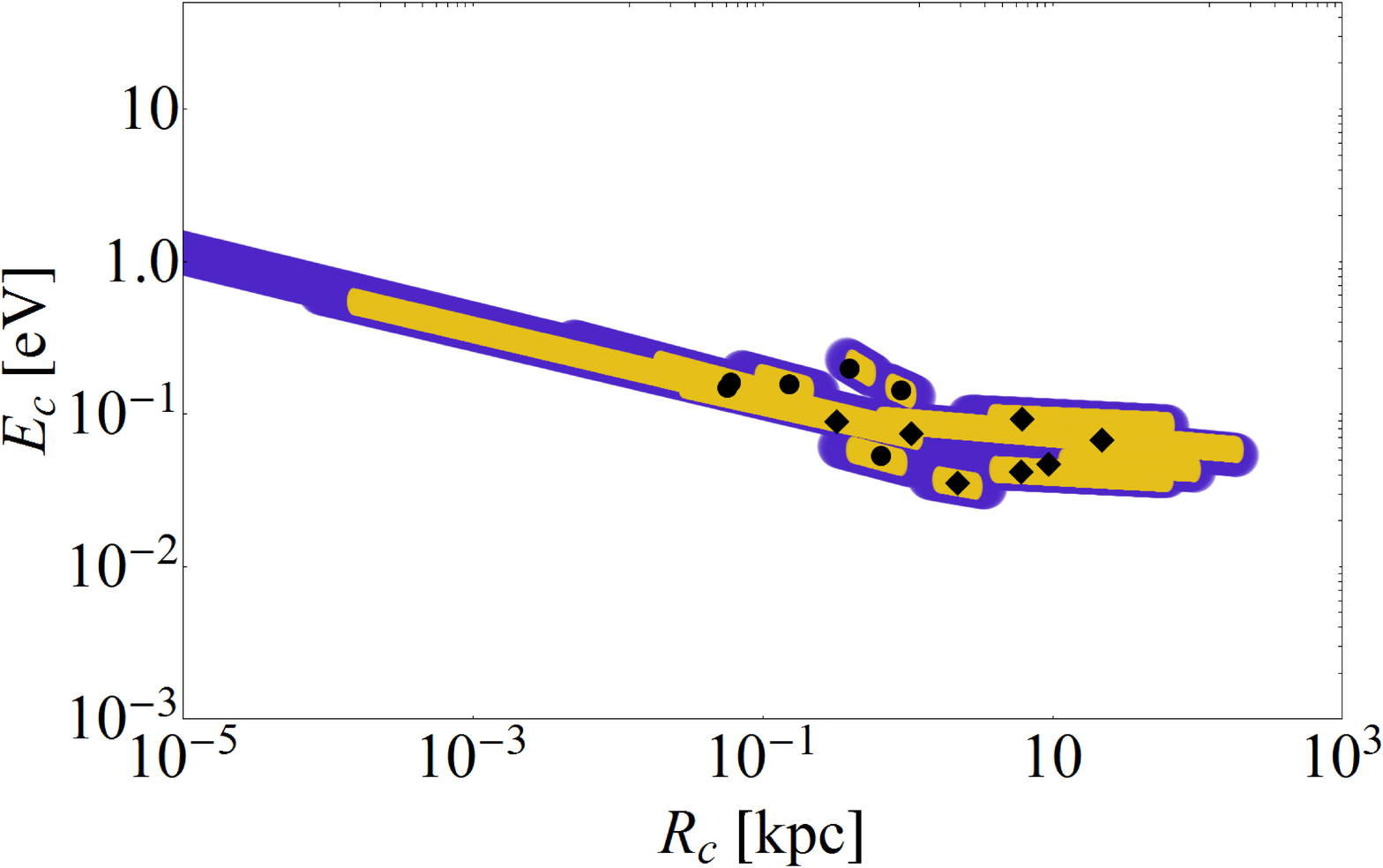}}   \\
    \subfloat[\footnotesize{$r_c$ vs $E_c$ : Kroupa}]{ \includegraphics[width=0.50\textwidth]{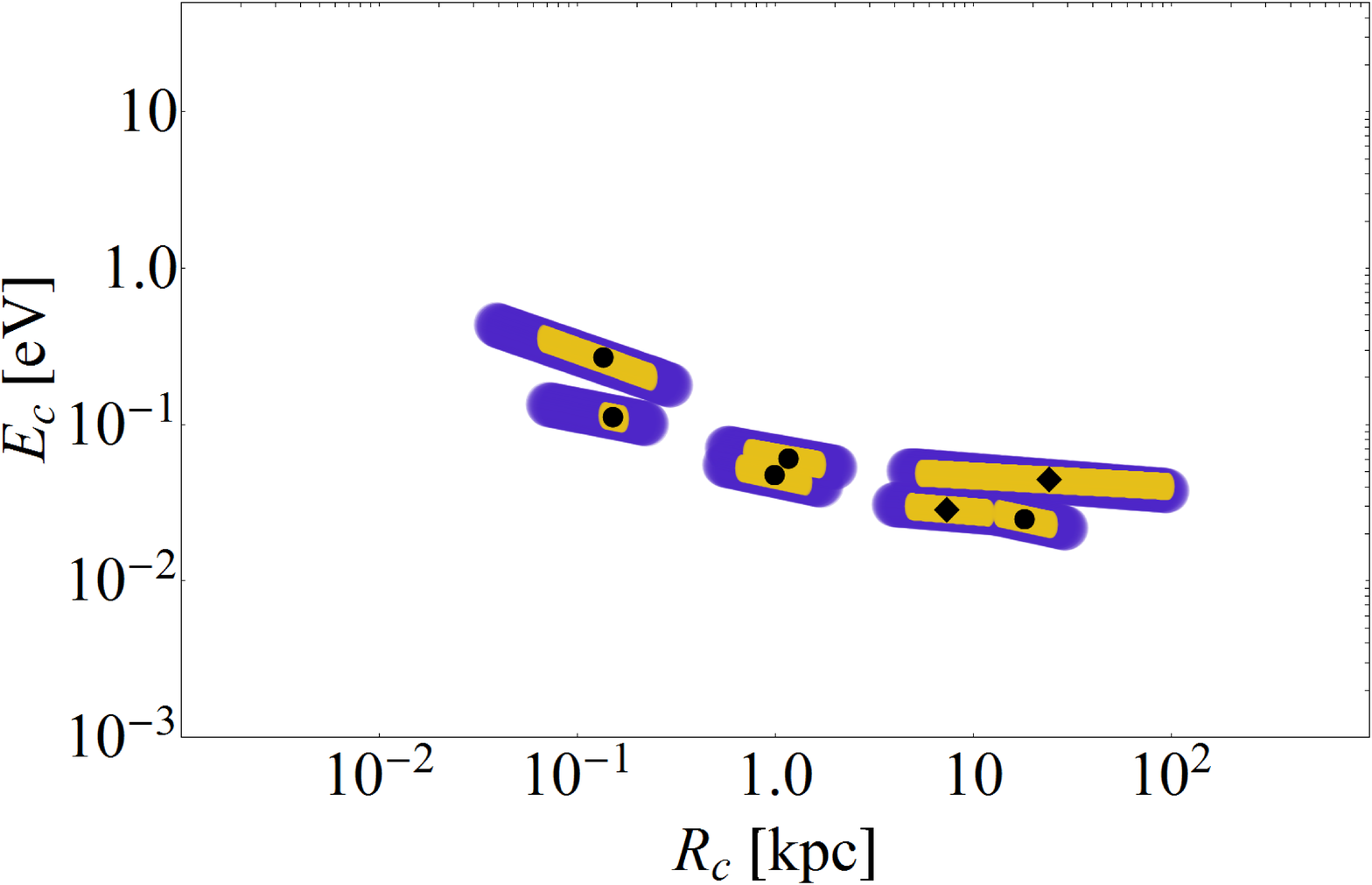}}
    \subfloat[\footnotesize{$r_c$ vs $E_c$ : diet-Salpeter}]{ \includegraphics[width=0.50\textwidth]{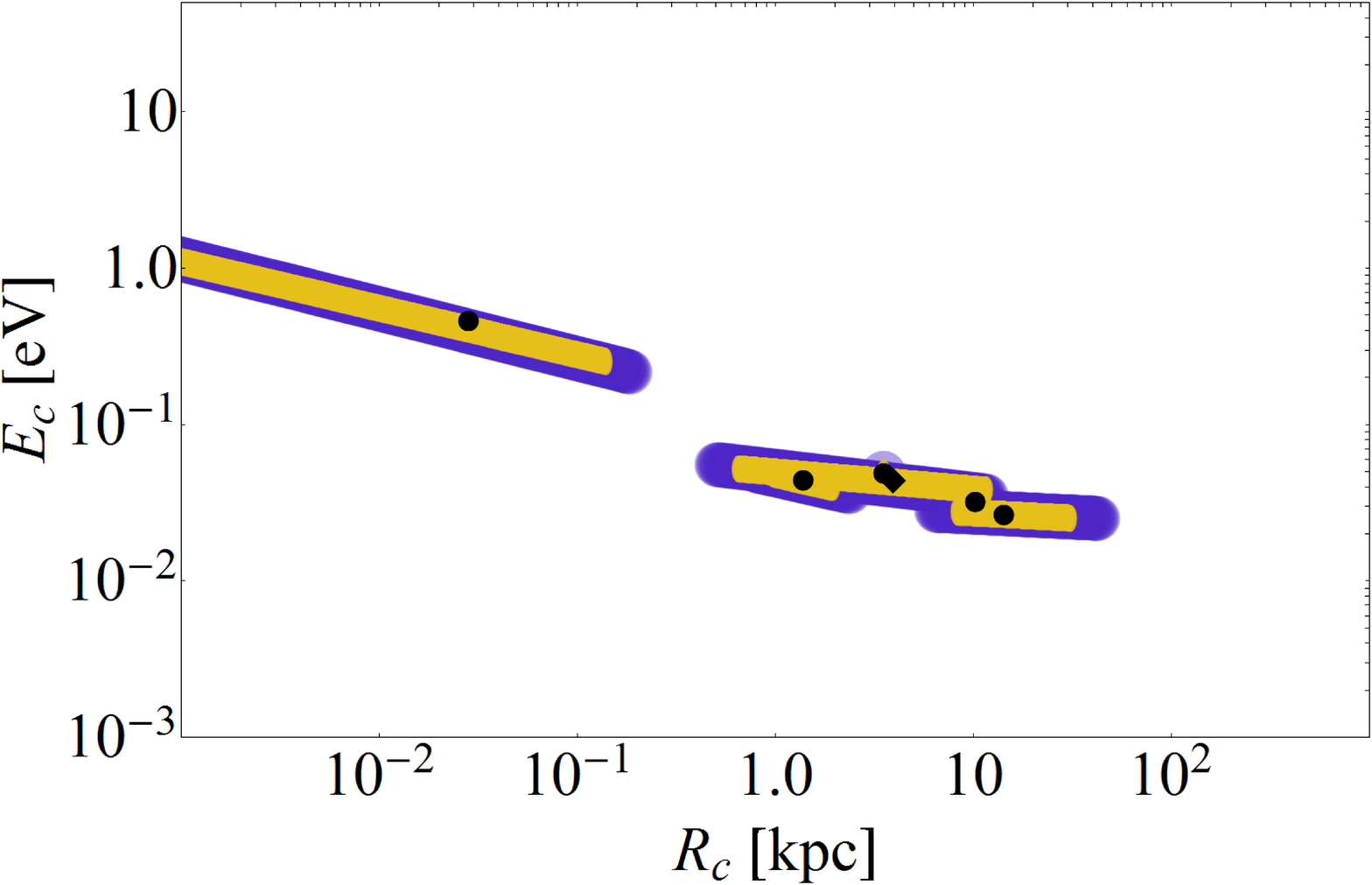}}
    \caption{\footnotesize{
    We plot the values of $r_c$ vs $E_c$ for each galaxy and different mass models. Circles represent galaxies with $r_c\ll r_s$, diamond markers are galaxies corresponding $r_c = r_s$ group, but considering $r_c$ and $E_c$ obtained from the inner analysis in Sec.\ref{sec:inner_analysis}. In all cases the areas in yellow and blue represent the $1\sigma$ and $2\sigma$ confidence levels. }}
    \label{fig:RcvsEcONLYDM}
\end{figure}

\setlength{\extrarowheight}{3pt}
\begin{table}[h]
  \centering
  \scriptsize{
    \begin{tabular}{l|r|rrr|r|c|r|rrr|r}
    \multicolumn{ 12}{c}{BDM Statistics} \\
    \hline
               &  \multicolumn{ 6}{c|}{Energy $E_c$} &  \multicolumn{ 5}{c}{Core $r_c$} \\
   Mass Models & $\tilde{E}_{c_n}$  & $E_{c_{-}}$ & $E_c$ & $E_{c_{+}}$ & $\sigma(E_c)$ & $\log_{10}\rho_c$ &  $\tilde{r}_{c_n}$ & $r_{c_{-}}$ & $\hat{r}_c$ & $r_{c_{+}}$ & $\sigma(r_c)$ \\
    \multicolumn{1}{c|}{(1)} &\multicolumn{1}{c|}{(2)} &\multicolumn{1}{c}{(3)}&\multicolumn{1}{c}{(4)} &\multicolumn{1}{c|}{(5)} &\multicolumn{1}{c|}{(6)} & \multicolumn{1}{c|}{(7)} &\multicolumn{1}{c|}{(8)} & \multicolumn{1}{c}{(9)} &  \multicolumn{1}{c}{(10)} & \multicolumn{1}{c|}{(11)} & \multicolumn{1}{c}{(12)}\\
               \hline
      Min.Disk &       0.21 &       0.04 &       0.11 &       0.33 &  0.46  & 8.85  & 3.02 &       0.01 &       0.30 &       7.02 & 1.37 \\
  Min.Disk+gas &       0.10 &       0.05 &       0.08 &       0.15 &  0.25  & 8.35  & 3.81 &       0.15 &       0.98 &       6.60 & 0.83 \\
        Kroupa &       0.08 &       0.03 &       0.06 &       0.13 &  0.33  & 7.65  & 6.59 &       0.16 &       1.48 &      13.73 & 0.97 \\
 diet-Salpeter &       0.13 &       0.02 &       0.07 &       0.21 &  0.47  & 7.92  & 5.85 &       0.17 &       1.70 &      17.37 & 1.01 \\
    \end{tabular}
    }
  \caption{\footnotesize{We show the statistics of the transition energy $E_c$ and the core $r_c$ for the BDM profile for the different mass models. In column (2) we show the arithmetic mean energy transition from HDM to CDM, $\tilde{E}_{c_n}$. In columns  (3-5) we present the median energy ($E_c$), and the its related uncertainty ($E_{c \pm} \equiv E_c 10^{\pm \sigma}$), where the $\sigma$ is the standard deviation of the $\log E_c$ distribution.  In columns (8-12) we present the same statistical parameters but related with the core distance $r_c$. The energy $E_c$ and the core distance are given in unit of {\rm \ eV} and {\rm kpc}, respectively. In column (7) we have $\rho_c\equiv E_c^4$ in units of $M_\sun/{\rm kpc}^3$. }}
  \label{tab:statistics_BDM}
\end{table}

\subsection{BDM Energy Phase Transition $E_c$}\label{sec.bdm.ec}

In the inner region of the galaxies and for $r\ll r_s$ the relevant parameters are $\rho_0$ and $r_c$  and  we present their $1\sigma$ and $2\sigma$ likelihood contours for the different mass models in Figures \ref{tab:conflevel_A}, \ref{tab:conflevel_B}, \ref{tab:conflevel_C}, and \ref{tab:conflevel_inner B}. Using the fit of the rotation curves for the seventeen LSB galaxies for our BDM halo profile, Eq. (\ref{eq:rhobdm}) presented in Sec.\ref{results} we  determine the properties of the DM halo through its three free parameters ($r_c$, $r_s$, and $\rho_0$) and  constrain the energy value $E_c$, in which the BDM particles acquire a non-perturbative mass and has the transition from CDM to HDM, as explained in Sec. \ref{modelo}. We calculate the maximum, minimum, and central value of $E_c$ and $r_c$ for the four mass  models and present the results in  Table \ref{tab:statistics_BDM}.
The values of the core radius $r_c$, central energy density $\rho_c$ and transition energy $E_c=\rho_c^{1/4}$, as defined in Sec.\ref{modelo}, are shown in Tables \ref{tab:bdm_odm_dmg_k} and \ref{tab:bdm_k+s}  and  we plot $E_{c}$ vs $r_c$  in Figs.\ref{fig:RcvsEcONLYDM}   for the different mass models and galaxies of G.A. and G.B., taking into account the inner analysis of Sec.{\ref{sec:inner_analysis}) and  its respectively $1\sigma$ and $2\sigma$ likelihood contours errors. The circles and diamonds represent the  galaxies from G.A. and G.B., respectively. We use a log-normal distribution for  $E_c,r_c$ since these
quantities are constraint from $(0,\infty)$, and the $1\sigma,2\sigma$ represent a $A\times 10^{\pm\sigma}$ standard deviation from the central value with $A=E_c$ or $A=r_c$.

The average core radius $r_c$ and the energy $E_c$ of the transition between HDM and CDM depends on the choice of mass models used. We
can see that in general the core radius is larger for the min.disk+gas  than for the min.disk but decreases   when bulge and stars are considered as in Kroupa and diet-Salpeter. From Figs (\ref{fig:DDO154} - \ref{fig:NGC7793}) we see that the contribution of the gas is very small at small or negligible  radius, however Kroupa, and even more diet-Salpeter, have a considerable contribution of mass at small radius $\mathcal{O}$(0.1 - 1 kpc) and therefore the required contribution of DM in this range of radius would be smaller than in min.disk or min.disk+gas mass models. Therefore,  it is not surprising that the extraction of the properties of any DM profile for small radius in Kroupa or diet-Salpeter gives a less constraint to $r_c$, i.e. a larger standard deviation.

For galaxies and mass models  where the  central  value of $r_c$ is vanishing, the energy density $\rc$ or $E_c$ can not be calculated because it blows up as seen in Eq.(\ref{eq:relation ec_free_bdm_params}). In these cases we have a blank (i.e. $"-"$) in Tables  \ref{tab:bdm_odm_dmg_k} and \ref{tab:bdm_k+s}}. However, even though the central value of $r_c < r_s\, 10^{-6}$ it is consistent with a non-vanishing value  at a $1\sigma$ and $2\sigma$ deviation. More data and better understanding about the distribution of stars in these galaxies are needed to settled the question of a core DM profile.

The average values for $r_c$ and $E_c$ and the standard deviations $\sigma$ are presented for the different mass models
in Table \ref{tab:statistics_BDM}.  For the min.disk mass scenario we have
\be
E_c=0.11 \times 10^{\pm 0.46}{\rm \ eV} = 0.11^{+0.22}_{-0.07} {\rm \ eV}; \:\, r_c = 300\times 10^{\pm 1.37}\;\; {\rm pc},
\ee
i.e. $13\, {\rm pc}<r_c< 5308\, {\rm pc}$, while in the more realistic approach, the energy and the core for the Kroupa mass model take the values
\be
E_c=0.06 \times 10^{\pm 0.33}{\rm \ eV} = 0.06^{+0.07}_{-0.03} {\rm \ eV}; \:\, r_c = 1.48\times 10^{\pm 0.97}\;\; {\rm kpc},
\ee
The figures of $E_c$ vs $r_c$ for the different mass models are shown in Fig \ref{fig:RcvsEcONLYDM}. We note that when more components are considered, the $1\sigma$ and $2 \sigma$ confidence levels of $r_c$  become larger and the transition energy $E_c$ decreases but are all consistent within a 1 and 2$\sigma$ levels as seen in Table\ref{tab:statistics_BDM}.

Clearly the scale of $E_c = \mathcal{O}(0.1 {\rm \ eV})$ is very close the sum of neutrino masses.  Therefore, an interesting connection could
be further explored  between the phase transition scale $E_c$ of our BDM model, which also sets the mass of the BDM particles, to
the neutrino mass generation mechanism \cite{NBDM}. However, we would like to emphasize that our BDM are not neutrinos, since neutrinos
are HDM,  and a combination of CDM plus neutrinos would have a cuspy NFW profile with CDM dominating in the inner region of the galaxies,
assuming that CDM do indeed have a NFW profile as suggested by N-body  simulations  \cite{Navarro:1995iw,Navarro:1996gj}.
A main difference between neutrinos and BDM is that neutrinos were in thermal equilibrium at $E\gg$ MeV  with the SM particles, but BDM may not have been in thermal equilibrium with SM at all.

\subsection{Fixed-BDM profile}\label{sec:bdm.fixed}

Now we are able to extract further information of our BDM model.The BDM model predicts that the transition scale $E_c$ is an universal constant which depends on  the nature of the interaction of the DM particles \cite{Macorra.DEDM} and as explained in Sec.\ref{modelo} and is  independent of the galaxies details  and therefore it should be the same for all galaxies. However,  the determination on the value of $E_c$ depends on the quality of the data,  the choice of matter content and also on the choice of the galactic profile as the one proposed in Eq.(\ref{eq:rhobdm}).

Our BDM profile  has three  parameters $r_s, r_c$ and $\rc=E^4_c$. In Sec.\ref{sec.bdm.ec} we presented the analysis and conclusions of comparing the BDM profile to the rotation curves as described in Sec. \ref{results}. We now proceed as follow: we obtain an average value of $E_c$ for each of the different mass models  and then  we take this average value and proceed to reanalyze the rotation curves for the BDM profile but with only two free parameters, i.e. we take $\rc$ fixed in Eq.(\ref{eq:fix-BDM}) and we have $r_c$ and $r_s$ as free parameters. Following, we present these results and conclusions.

The results are shown in Tables \ref{tab:onlydm_bdmnfw} - \ref{tab:kroupa_bdmnfw}, and in Table \ref{tab:kroupa_bdmnfw} for each mass model, and Table \ref{tab:statistics_fix-BDM} show the statics values of $r_s$ and $r_c$ for this analysis. We generally find that in most of the galaxies the fixed-BDM profile is better than NFW regardless of the mass model used. The fixed-BDM profile has a smaller $\chi^2_{\rm red} \sim 1$ than NFW, e.g. in DDO 154, NGC 2366, NGC 3198, or because the fit is clearly competitive with that of NFW, $\Delta\chi^2_{\rm red} < 1$, e.g. NGC 4736, NGC 2841, NGC 6946. It is also relevant that the core obtained in the min.disk scenario with the fixed-BDM profile (460 pc) is of the order to the one obtained with the BDM profile (300 pc)

\begin{table}[h]
  \centering
    \begin{tabular}{l|r|rrr|r|r|r|rrr|r}
    \multicolumn{ 12}{c}{Fixed-BDM Statistics} \\
    \hline
    Mass Model & $E_c$ & $\tilde{r_c}$ & ${r_c}_-$ &   $r_c$ & ${r_c}_+$ & $\sigma(r_c)$ & $\tilde{r_s}$ & ${r_s}_-$ &     $r_s$ & ${r_s}_+$ & $\sigma(r_s)$ \\
    \multicolumn{1}{c|}{(1)} &\multicolumn{1}{c|}{(2)} &\multicolumn{1}{c}{(3)}&\multicolumn{1}{c}{(4)} &\multicolumn{1}{c|}{(5)} &\multicolumn{1}{c|}{(6)} & \multicolumn{1}{c|}{(7)} &\multicolumn{1}{c|}{(8)} & \multicolumn{1}{c}{(9)} &  \multicolumn{1}{c}{(10)} & \multicolumn{1}{c|}{(11)} & \multicolumn{1}{c}{(12)}\\
    \hline
    Min.Disk     &   0.11 &       0.81 &       0.11 &       0.46 &       1.85 &       0.61 &       4.71 &       0.51 &       2.64 &      13.44 &       0.71 \\
    Min.Disk+gas &   0.08 &       1.86 &       0.40 &       1.28 &       4.11 &       0.51 &       3.25 &       1.03 &       2.41 &       5.62 &       0.37 \\
    Kroupa       &   0.06 &       2.01 &       0.44 &       1.27 &       3.65 &       0.46 &      13.84 &       2.90 &       8.56 &      25.26 &       0.47 \\
    diet-Salpeter&   0.07 &       1.59 &       0.25 &       0.85 &       3.50 &       0.62 &      11.99 &       2.80 &       7.84 &      22.01 &       0.45 \\
    \end{tabular}
  \caption{\footnotesize{We show the statistics for the core $r_c$ and $r_s$ resulting from the fit of the fixed-BDM profile, Eq. (\ref{eq:fix-BDM}) for the different mass models and the computed $E_c$ shown in Table \ref{tab:statistics_BDM}. We had considered the log-normal distribution. In columns (3) we show the arithmetic mean $\tilde{r_c}$. In columns (4-6) we present the median energy ($r_c$), and the its related uncertainty ($r_{c \pm} \equiv r_c 10^{\pm \sigma}$), where the $\sigma$ is the standard deviation of the $\log r_c$ distribution.  In columns (8-12) we present the same statistical parameters but related with the distance $r_s$. The distance are given in unit of {\rm kpc}.}}
  \label{tab:statistics_fix-BDM}
\end{table}

\section{Conclusions}\label{conclusion}

We have presented the analysis of the rotation curves using four different DM profiles (BDM, NFW, Burkert, and ISO) and five different disk mass components (only DM (min.disk), DM plus gas (min.disk+gas), DM plus gas plus stellar disk (Kroupa, diet-Salpeter, Free $\gs$) for a subsample of the THINGS galaxies as described in Sec. \ref{MaMo}.

The new tested profile, BDM, is introduced in Sec. \ref{modelo} and  corresponds to CDM particles below certain critical energy density $\rc = E_c^4$ while it behaves as HDM above $\rc$ with a core radius $r_c$. Therefore, the resulting profile behaves as NFW's for large radius, but it has a core in the central region.

We have found that fits to the THINGS galaxies favor core profiles over the cuspy NFW in accordance to general results found in other works \cite{deBlok:2008wp,deBlok10} for the same galaxies. We studied the cases when the five different disk mass components are taking into account and conclude that BDM is in good agreement with observations for all the disk mass scenarios and fits better or equally well than Burkert and ISO profiles.  For three galaxies (Group C) the fits to the BDM profile demanded $r_c/r_s <10^{-6}$ and in these cases the $\chi^2$ are almost identical to NFW's.  For these galaxies a cuspy profile fits best because there is not enough observational data close to the galactic center to test it.

The observational resolution of the THINGS sample is of high quality, but one still needs data closer to the center of some galaxies, on scales smaller than 200 {\rm pc}, in order to discern between cored or cuspy profiles.  This is because stars pose a very challenging problem when testing the core-cusp problem, largely due to the uncertainty of the mass-to-light ratio and their dominant behavior close to the center of the galaxy. We have performed an analysis of the inner regions of  Group B and C galaxies, for which $r_c \leq r_s$, extracting information from the rotation curves consistent with a core profile.  Confidence level contours show that the core radius depends on the number of disk mass components taking into account in the analysis.  We found that $r_c$ is highly constrained if the stellar disk has a dominant behavior close to the center of the galaxy.

We computed the 1 and 2$\sigma$ confidence levels for the BDM parameters $r_c$ and $E_c$ for the different galaxies and mass models. We found that the energy of transition between HDM and CDM takes places for most galaxies with an energy $E_c = \mathcal{O}(0.1{\rm \, {\rm eV}})$ and most of the galaxies have a core radius $r_c \sim \mathcal{O}(300 \,$ {\rm pc}), see Table \ref{tab:statistics_BDM}. We found that $r_c$ is highly constrained if the stellar disk has a dominant behavior close to the center of the galaxy. However, we notice that even when $r_c$ is different for each disk  mass model there is an interval of values where $r_c$ is consistent for all mass models within the $1\sigma$ for most galaxies and some within the $2\sigma$ error, as shown in Figures \ref{tab:conflevel_A}, \ref{tab:conflevel_B}, \ref{tab:conflevel_C}, and \ref{tab:conflevel_inner B}.  This provides us with enough tools to later test the Universe background evolution and large scale structure formation in order to continue testing the particle-physics-motivated BDM model. This energy transition is similar to the mean energy of a relativistic neutrino (see e.g. \cite{Larson:2010gs}) which is such that $\sum m_\nu < 0.58 {\rm \ eV} (95\% {\rm \ CL)}$.

The dispersion on $E_c$ is much smaller than that of $r_c$ and this is consistent with our BDM model since $E_c$ is a new fundamental scale for DM model while $r_c$ depends on $r_s,\rho_0$ which depend on the formation and initial conditions for each galaxy. We test the BDM profile with a fixed energy $E_c$ finding that the proposed profile still being competitive compare with NFW profile. BDM profile has a rich -particle physics motivated- structure that, through its transition from CDM to HDM (when $\rho \simeq \rho_c$),  allows for a more physical explanation of the rotation curves of the different galaxies treated here.

\begin{acknowledgments}
We thanks to Prof. Erwin de Blok for providing the observational data of THINGS and Prof. Christiane Frigerio for the help granted in the research. A.M. and J.M.  acknowledge financial support from Conacyt Project 80519 and UNAM PAPIIT Project IN100612, and J.L.C.C. acknowledges support from Conacyt project 84133-F.
\end{acknowledgments}

\newpage

\begin{appendices}
We organize this Appendix as follow, in section \ref{apend: diet-salpeter} are shown the fitted values corresponding to the diet-Salpeter, for the NFW and BDM profile, and the free $\gs$, for BDM profile, mass models. In section \ref{apendix:iso_burkert_result} we show the fitted values for the minimal disk, min.disk+gas, Kroupa and diet-Salpeter when considering Burkert, Table \ref{tab:burkert}, and pseudo-isothermal, Table \ref{tab:iso}, profiles. Finally, in the section \ref{apend:LikelihoodPlot} we show the likelihood contour plots and the rotation curves for the different galaxies and mass models.

\section{diet-Salpeter and Free mass model}\label{apend: diet-salpeter}
In this section we present the results of the diet-Salpeter mass model for the BDM, Eq. (\ref{eq:rhobdm}), and NFW, Eq. (\ref{eq:rhoNFW}), profile. We also present the results for the free $\gs$ mass model for the BDM profile, as explained in the Section \ref{MaMo}.

\begin{table}[h]
  \centering
  \scriptsize{
    \begin{tabular}{rl|r|lllr|llr|llr}
         \multicolumn{ 13}{c}{DIET-SALPETER} \\
         \hline
               &            &              \multicolumn{ 5}{c|}{BDM} &             \multicolumn{ 3}{c|}{NFW} &     \multicolumn{ 3}{c}{$E_c=0.07$} \\
               &     \multicolumn{1}{c|}{Galaxy} & \multicolumn{1}{c|}{$r_c/r_s$} &     \multicolumn{1}{c}{$r_s$} & \multicolumn{1}{c}{$\log_{10} \rho_0$}  &     \multicolumn{1}{c}{$r_c$} & \multicolumn{1}{c|}{$\chi^2_{\rm red}$} &     \multicolumn{1}{c}{$r_s$} & \multicolumn{1}{c}{$\log_{10} \rho_0$}  & \multicolumn{1}{c|}{$\chi^2_{\rm red}$} &     $r_s$ &     $r_c$ & $\chi^2_{\rm red}$ \\
    \hline
    \multirow{ 7}{*}{G.A} &     DDO154 &       0.46 & $3.53^{+0.03}_{-0.03}$ & $7.32^{+5.51}_{-5.5}$ & $1.63^{+0.06}_{-0.06}$ &       0.28 & $15.95^{+0.19}_{-0.19}$ & $6.07^{+4.25}_{-4.25}$ &       1.09 & $9.74^{+0.35}_{-0.35}$ & $0.14^{+0.}_{-0.}$ &       0.51 \\
    &    NGC2841 &       0.04 & $5.36^{+0.20}_{-0.31}$ & $8.47^{+6.30}_{-6.30}$ & $0.22^{+0.02}_{-0.01}$ & 0.49 & $5.84^{+0.43}_{-0.43}$ & $8.37^{+6.51}_{-6.51}$ &  0.49 & $6.09^{+0.09}_{-0.09}$ & $3.32^{+0.02}_{-0.02}$ &       2.21 \\
    & NGC3031 &      $>1$ &   $>10^4$ & $8.73^{+7.56}_{-7.55}$ &   $>10^6$ &      21.29 &   $>10^6$ &      $<2$ &      22.76 & $>10^7$ & $0.06^{+0.01}_{-0.01}$ &       22.6 \\
    &    NGC3621 &          0 & $43.8^{+0.27}_{-0.27}$ & $6.03^{+3.98}_{-3.98}$ &   $<0.01$ &       1.45 & $43.81^{+0.27}_{-0.27}$ & $6.03^{+3.98}_{-3.98}$ &       1.44 & $>10^6$ & $0.14^{+0.01}_{-0.01}$ &       31.7 \\
    & NGC4736 &       0.59 & $0.05^{+0.01}_{-0.01}$ & $12.0^{+10.6}_{-10.6}$ & $0.03^{+0.01}_{-0.01}$ &  1.31 & $0.07^{+0.01}_{-0.01}$ & $11.5^{+10.1}_{-10.2}$ &        1.3 &  $<0.01$ & $1.14^{+0.02}_{-0.02}$ &       1.65 \\
    & NGC6946 &    $>1$ &   $>10^5$ & $8.96^{+7.22}_{-7.18}$ &   $>10^6$ &        2.5 &   $>10^5$ & $2.89^{+1.15}_{-1.14}$ &       4.42 & $>10^5$ & $0.22^{+0.01}_{-0.01}$ &        4.2 \\
    &    NGC7793 &          1 & $3.53^{+0.02}_{-0.02}$ & $8.17^{+6.60}_{-6.60}$ & $3.53^{+0.02}_{-0.02}$ &  4.05 & $30.1^{+2.2}_{-2.2}$ & $6.4^{+4.8}_{-4.8}$ &       5.01 &   $>10^6$ & $0.18^{+0.01}_{-0.01}$ &      75.92 \\
    \hline
    \multirow{ 7}{*}{G.B} &     IC2574 &       0.14 & $143.55^{+3.18}_{-3.2}$ & $5.66^{+4.03}_{-4.04}$ & $20.06^{+0.63}_{-0.61}$ &       1.21 &   $>10^5$ &      $<2$ &       5.49 &  $9.40^{+1.03}_{-1.03}$ &   $<0.01$ &        148 \\
    &    NGC2366 &      0.15 & $4.59^{+0.13}_{-0.13}$ & $7.01^{+5.82}_{-5.56}$ & $0.68^{+0.10}_{-0.09}$ &       1.87 & $15.99^{+0.63}_{-0.62}$ & $6.03^{+4.74}_{-4.71}$ &       2.94 & $>10^3$ &   $<0.02$ &       31.3 \\
    & NGC2903 &          1 & $2.30^{+0.24}_{-0.24}$ & $9.05^{+6.58}_{-7.14}$ & $2.30^{+0.24}_{-0.24}$ & 2.24 & $4.81^{+0.01}_{-0.01}$ & $8.15^{+6.04}_{-5.96}$ & 3.73 & $1.96^{+0.05}_{-0.05}$ & $3.15^{+0.01}_{-0.01}$ &       2.91 \\
    &    NGC2976 &   $>1$ & $300.^{+20.31}_{-12.11}$ & $11.84^{+10.65}_{-10.76}$ &   $>10^6$ &       5.28 &   $>10^5$ & $2.41^{+1.22}_{-1.22}$ &      10.49 &   $>10^6$ &   $<0.01$ &        640 \\
    & NGC3198 &          1 & $11.6^{+0.06}_{-0.05}$ & $7.28^{+5.29}_{-5.27}$ & $11.6^{+0.24}_{-0.25}$ & 6.28 & $41.1^{+0.27}_{-0.27}$ & $6.01^{+4.01}_{-4.01}$ & 7.86 & $34.8^{+0.61}_{-0.61}$ & $0.28^{+0.02}_{-0.02}$ &       7.45 \\
    &    NGC3521 &      $>1$ & $339^{+21.2}_{-18.8}$ & $10.75^{+9.55}_{-9.54}$ & $   >10^7$ & 11.62 &   $>10^6$ &      $<2$ & 12.5 &   $>10^6$ & $0.12^{+0.01}_{-0.01}$ &      12.35 \\
    &     NGC925 &  $>1$ & $>10^3$ & $9.59^{+8.07}_{-8.06}$ &  $ >10^6$ &       2.51 &   $>10^6$ & $0.06^{+0.01}_{-0.01}$ &       36.4 \\
    \hline
    \multirow{ 3}{*}{G.C}  & NGC2403 &       1.38 & $3.66^{+0.01}_{-0.01}$ & $8.32^{+5.98}_{-5.99}$ & $5.05^{+0.05}_{-0.05}$ & 0.87 & $12.53^{+0.03}_{-0.03}$ & $6.99^{+4.65}_{-4.65}$ & 1.08 & $7.32^{+0.05}_{-0.05}$ & $1.05^{+0.03}_{-0.03}$ &       0.84 \\
    & NGC5055 &   0.33 & $417.6^{+7.61}_{-7.54}$ & $5.49^{+3.8}_{-3.79}$ & $135.9^{+3.3}_{-3.54}$ &    13.45 &   $>10^6$ & $<1.51$ &      15.74 & $>10^6$ & $0.09^{+0.01}_{-0.01}$ &       15.7 \\
    & NGC7331 &          0 &   $>10^6$ & $1.51^{+0.41}_{-0.18}$ & $58.1^{+1.53}_{-0.86}$ & 26.2 &   $>10^6$ &      $<2$ & 29.6 &   $>10^8$ & $0.4^{+0.01}_{-0.01}$ &       23.5 \\
    \end{tabular}
    }
  \caption{\footnotesize{It shows the fitted values when considering the diet-Salpeter mass model for the value of the stellar disk. BDM and NFW parameters are show in columns (2-7) and (8-10) respectively. Units and the set of galaxies are as show in Table.\ref{tab:onlydm_bdmnfw}. }}
  \label{tab:salpeter_bdmnfw}
\end{table}

\begin{table}[h]
  \centering
  \scriptsize{
    \begin{tabular}{rr|rrrrr}
                                                          \multicolumn{ 7}{c}{DIET-SALPETER} \\
    \hline
               &            &                                   \multicolumn{ 5}{c}{BDM Inner} \\
               &     Galaxy & $R_{max}$ &     $r_c$ & $\log \rho_0$  & $\chi^2_{\rm red}$ & $\chi_{t}^2$ \\
    \hline
          G.A. &    NGC3621 &       3.32 &   $<0.01$ & $9.47^{+7.38}_{-7.3}$ &       3.09 &     3.66 \\
    \hline
    \multirow{ 7}{*}{G.B} &     IC2574 &      11.64 & $14.35^{+0.49}_{-0.48}$ & $6.21^{+4.83}_{-4.82}$ &       1.22 &       1.21 \\
    &    NGC 2366 &       0.79 & $3.94^{+1.23}_{-0.83}$ & $7.09^{+6.69}_{-6.59}$ &       0.09 &       2.04 \\
    &    NGC 2903 &       2.72 &   $>10^4$ & $7.29^{+7.04}_{-6.89}$ &      17.9 &       39.8 \\
    &    NGC 2976 &       2.57 &   $>10^3$ & $7.20^{+5.6}_{-5.6}$ &       5.56 &     684 \\
    &    NGC 3198 &       6.02 &   $>10^5$ & $5.43^{+3.70}_{-3.70}$ &      25.1 &       52.4 \\
    &    NGC 3521 &       3.74 &   $>10^5$ & $6.44^{+4.7}_{-4.6}$ &     110 &     164 \\
    &     NGC 925 &       6.02 &   $>10^5$ & $5.38^{+5.16}_{-5.16}$ &       4.14 &      77.1 \\
    \hline
    \multirow{ 2}{*}{G.C} &    NGC2403 &       2.04 & $44.5^{+23.88}_{-11.78}$ & $6.87^{+6.68}_{-6.51}$ &       0.71 &       3.11 \\
    &    NGC5055 &       0.73 & $0.45^{+0.24}_{-0.}$ & $9.04^{+8.8}_{-8.44}$ &         -- &       1.87 \\
    \hline
    \end{tabular}
    }
  \caption{\footnotesize{ Here we show the values obtained from the inner analysis for the diet-Salpeter mass model, with the profile $\rho_{in}$. All the parameters are explained in Table \ref{tab:inner_dm_dmg_k}. The dashed (--) means that the $\chi^2_{\rm red}$ could not be computed since the number of data taken to do the analysis were too small. }}
  \label{tab:salpeter_inner}
\end{table}

\setlength{\extrarowheight}{3pt}
\begin{table}[h]
  \centering
  \scriptsize{
    \begin{tabular}{rrrrr}
                          \multicolumn{ 5}{c}{BDM - diet-Salpeter} \\
    \hline
               &     Galaxy &     $r_c$ &  $\rho_c$ &     $E_c$ \\
    \multicolumn{ 1}{c}{G.A} &     DDO154 & $1.63^{+0.20}_{-0.20}$ & $6.80^{+4.80}_{-6.80}$ & $0.04^{+0.001}_{-0.001}$ \\
    &    NGC2841 & $0.22^{+0.01}_{-0.01}$ & $8.83^{+6.80}_{-6.80}$ & $0.12^{+0.01}_{-0.01}$ \\
    &    NGC3031 & $1.44^{+0.03}_{-0.03}$ & $9.08^{+7.80}_{-7.80}$ & $0.14^{+0.01}_{-0.01}$ \\
    &    NGC3621 &  $<0.001$ &      $<2$ &         -- \\
    &    NGC4736 & $0.03^{+0.01}_{-0.01}$ & $11.53^{+11.36}_{-10.15}$ & $0.56^{+0.09}_{-0.01}$ \\
    &    NGC6946 &   $>10^6$ &      $<2$ &         -- \\
    &    NGC7793 & $3.53^{+0.50}_{-0.50}$ & $7.32^{+6.80}_{-6.80}$ & $0.05^{+0.01}_{-0.01}$ \\
    \hline
    \multicolumn{ 1}{c}{G.B} &     IC2574 & $14.4^{+0.49}_{-0.48}$ & $6.21^{+4.83}_{-4.82}$ & $0.03^{+0.0003}_{-0.0003}$ \\
    &    NGC2366 & $3.94^{+1.23}_{-0.83}$ & $7.09^{+6.69}_{-6.59}$ & $0.04^{+0.004}_{-0.003}$ \\
    &    NGC2903 &   $>10^4$ & $7.29^{+7.04}_{-6.89}$ & $0.05^{+0.007}_{-0.005}$ \\
    &    NGC2976 &   $>10^3$ & $7.2^{+5.50}_{-5.50}$ & $0.05^{+0.001}_{-0.001}$ \\
    &    NGC3198 &   $>10^5$ & $5.43^{+3.1}_{-3.1}$ & $0.02^{+0.001}_{-0.001}$ \\
    &    NGC3521 &   $>10^5$ & $6.44^{+4.50}_{-4.50}$ & $0.03^{+0.001}_{-0.001}$ \\
    &     NGC925 &   $>10^5$ & $5.38^{+5.16}_{-5.16}$ & $0.02^{+0.002}_{-0.002}$ \\
    \hline
    \end{tabular}
  }
  \caption{\footnotesize{We show the main parameters, $r_c$, $\rho_c$ and $E_c$, for the BDM model as defined in Sec. \ref{modelo} for the diet-Salpeter mass model. The values for G.A. galaxies are those obtained with the BDM profile, Eq. (\ref{eq:rhobdm}). The values for G.B. galaxies are those obtained from the inner analysis using $\rho_{in}$ as explained in Sec.\ref{results}.  The cases with central value $r_c < 10^{-6}$ the energy value $E_c$ cannot be computed, in this case we show a single dash. Units are as in Table. \ref{tab:bdm_odm_dmg_k} }}
  \label{tab:bdm_k+s}
\end{table}

\setlength{\extrarowheight}{3pt}
\begin{table}[h]
  \centering
  \scriptsize{
    \begin{tabular}{ll|llrrrrr}
                                                                         \multicolumn{ 9}{c}{FREE :: BDM} \\
                                                                         \hline
               &    \multicolumn{1}{c|}{Galaxy} & \multicolumn{1}{c}{$r_c/r_s$} & \multicolumn{1}{c}{$r_s$} & \multicolumn{1}{c}{$\log_{10} \rho_0$}  & \multicolumn{1}{c}{$r_c$} & \multicolumn{1}{c}{$\log_{10} M_{s}$} & \multicolumn{1}{c}{$\gs$} & \multicolumn{1}{c}{$\chi^2_{\rm red}$} \\
               &    \multicolumn{1}{c|}{(1)} &  \multicolumn{1}{c}{(2)} &  \multicolumn{1}{c}{(3)} &  \multicolumn{1}{c}{(4)} &  \multicolumn{1}{c}{(5)} & \multicolumn{1}{c}{(6)} &  \multicolumn{1}{c}{(7)} &  \multicolumn{1}{c}{(8)} \\
        \hline
    \multirow{ 7}{*}{G.A} &     DDO154 &       0.14 & $4.79^{+0.04}_{-0.04}$ & $6.97^{+5.15}_{-5.15}$ & $0.66^{+0.04}_{-0.04}$ &   $<1$ & $<0.01$  &   0.28 \\
      &    NGC2841 &          0 & $4.47^{+0.01}_{-0.01}$ & $8.65^{+6.25}_{-6.24}$ & $0.00^{+0.01}$ & $4.7^{+8.95}_{-4.00}$ &  $<0.01$  &   0.5 \\
      &    NGC3031 &       0.21 & $1.33^{+0.01}_{-0.01}$ & $9.54^{+7.17}_{-7.16}$ & $0.28^{+0.01}_{-0.01}$ & $7.52^{+8.53}_{-7.50}$ &  $0.01$ &     4.5 \\
      &    NGC3621 &          0 & $60.^{+0.45}_{-0.38}$ & $5.83^{+3.79}_{-3.85}$ & $0.00^{+0.01}$ & $10.31^{+8.35}_{-8.35}$ &  $0.24$ &     0.67 \\
      &    NGC4736 &       6    & $0.08^{+0.01}_{-0.01}$ & $12.33^{+10.24}_{-10.24}$ & $0.466^{+0.01}_{-0.01}$ & $8.91^{+8.33}_{-8.5}$ &  $0.01$ &     1.67 \\
      &    NGC6946 &          0 & $7.36^{+0.02}_{-0.02}$ & $7.73^{+5.45}_{-5.45}$ & $0.0^{+0.01}$ & $10.13^{+8.55}_{-8.54}$ &   $0.10$ &    1.24 \\
      &    NGC7793 &      0.009 & $6.24^{+0.04}_{-0.05}$ & $7.48^{+5.48}_{-5.47}$ & $0.06^{+0.01}_{-0.01}$ &   $<1$ &   $<0.01$ &    3.37 \\
      \hline
    \multirow{ 7}{*}{G.B} &     IC2574 &          1 & $13.41^{+0.09}_{-0.13}$ & $6.94^{+5.15}_{-5.26}$ & $13.40^{+0.28}_{-0.19}$ &   $<1$ & $<0.01$ &  0.39 \\
      &    NGC2366 &       3.8  & $1.53^{+0.05}_{-0.05}$ & $8.19^{+7.}_{-7.11}$ & $5.92^{+0.64}_{-0.56}$ & $8.95^{+8.08}_{-7.74}$ &  $0.41$ &     1.63 \\
      &    NGC2903 &          1 & $2.32^{+0.02}_{-0.03}$ & $9.00^{+6.77}_{-6.92}$ & $2.32^{+0.03}_{-0.03}$ & $10.15^{+8.75}_{-8.74}$ &  $0.22$ &      1.29 \\
      &    NGC2976 &        0.5 & $4.26^{+0.06}_{-0.06}$ & $8.01^{+6.34}_{-6.33}$ & $2.09^{+0.07}_{-0.07}$ & $8.61^{+7.6}_{-7.58}$ &  $0.03$ &   0.85 \\
      &    NGC3198 &       0.98 & $5.17^{+0.02}_{-0.02}$ & $8.00^{+5.9}_{-5.9}$ & $5.08^{+0.11}_{-0.11}$ & $10.17^{+8.62}_{-8.61}$ &  $0.21$  &     0.47 \\
      &    NGC3521 &          1 & $1.91^{+0.02}_{-0.01}$ & $9.34^{+7.21}_{-7.43}$ & $1.91^{+0.005}_{-0.005}$ & $9.59^{+9.02}_{-8.98}$ &  $0.01$ &     1.1 \\
      &     NGC925 &          1 & $9.32^{+0.08}_{-0.08}$ & $7.52^{+5.67}_{-5.68}$ & $10.14^{+0.22}_{-0.22}$ &   $<1$ &   $<0.01$ &    0.49 \\
      \hline
    \multirow{ 3}{*}{G.C} &    NGC2403 &      0 & $13.2^{+0.04}_{-0.04}$ & $6.92^{+4.61}_{-4.61}$ & $0.0^{+0.001}$ & $9.75^{+8.17}_{-7.95}$ & $0.19$ & 0.82 \\
      &    NGC5055 &          0 & $3.76^{+0.01}_{-0.01}$ & $8.43^{+6.11}_{-6.11}$ & $0.0^{+0.01}$ &   $<1$ &   $<0.01$ &    1.26 \\
      &    NGC7331 &     292    & $0.12^{+0.001}_{-0.001}$ & $13.41^{+11.3}_{-11.3}$ & $35.643^{+0.34}_{-0.33}$ & $10.24^{+9.1}_{-8.91}$ &   $0.02$ &    0.43 \\
    \end{tabular}
    }
  \caption{\footnotesize{We show the fitted values when we let the mass-to-light ratio of the stellar disk as a free parameter. Columns (2-5) are the BDM parameters with units given in {\rm kpc} for distances and $M_\sun/{\rm kpc}^3$ for the $\rho_0$. Column (6) is the logarithm of the fitted mass of the stellar disk and in (7) its associated mass-to-light. We can see that some galaxies are consistent with the minimal disk+gas, and must of them get values closer to stellar mass suggested by Kroupa instead of the Salpeter IMF.  The set of galaxies are as shown in Table.\ref{tab:onlydm_bdmnfw}. }}
  \label{tab:free_bdm}
\end{table}

\clearpage
\newpage
\section{ISO and Burkert results}\label{apendix:iso_burkert_result}
In this section we present the tables where we show the results of the fit using the pseudo-isothermal (Eq.(\ref{eq:rhoISO})) and Burkert profile (Eq.(\ref{eq:rhoBurkert})) for all the for mass models as explained in Section \ref{MaMo}. The values for BDM and NFW for min.disk, min.disk+gas and Kroupa mass model were presented in the Sec. \ref{results} in Tables \ref{tab:onlydm_bdmnfw}, \ref{tab:onlydm+gas_bdmnfw}, \ref{tab:kroupa_bdmnfw}.

\setlength{\extrarowheight}{3pt}
\begin{table}[h]
  \centering
  \scriptsize{
    \begin{tabular}{rr|lrr|rrr|lrr|lrr}
    \multicolumn{ 14}{c}{BURKERT} \\
    \hline
           & & \multicolumn{ 3}{c|}{MIN. DISK} & \multicolumn{ 3}{c|}{MIN. DISK + GAS} & \multicolumn{ 3}{c|}{KROUPA} &  \multicolumn{ 3}{c}{DIET-SALPETER} \\
               &     Galaxy & \multicolumn{1}{c}{$r_s$} & $\log_{10} \rho_0$  & $\chi^2_{\rm red}$ & \multicolumn{1}{c}{$r_s$} & $\log_{10} \rho_0$  & $\chi^2_{\rm red}$ &     \multicolumn{1}{c}{$r_s$} & $\log_{10} \rho_0$  & $\chi^2_{\rm red}$ & \multicolumn{1}{c}{$r_s$} & $\log_{10} \rho_0$  & $\chi^2_{\rm red}$ \\
               & \multicolumn{1}{c|}{(1)} &    \multicolumn{1}{c}{(2)} &\multicolumn{1}{c}{(3)}&\multicolumn{1}{c|}{(4)}&\multicolumn{1}{c}{(5)}&\multicolumn{1}{c}{(6)}&\multicolumn{1}{c|}{(7)}&\multicolumn{1}{c}{(8)} &\multicolumn{1}{c}{(9)}&\multicolumn{1}{c|}{(10)}&\multicolumn{1}{c}{(11)}&\multicolumn{1}{c}{(12)}& \multicolumn{1}{c}{(13)} \\
    \hline
    \multirow{ 7}{*}{G.A} &     DDO 154 & $2.44^{+0.02}_{-0.02}$ & $7.52^{+5.62}_{-5.61}$ &       0.57 & $2.31^{+0.02}_{-0.02}$ & $7.48^{+5.64}_{-5.65}$ &       0.53 & $2.47^{+0.02}_{-0.02}$ & $7.43^{+5.61}_{-5.6}$ &       0.43 & $2.53^{+0.03}_{-0.03}$ & $7.41^{+5.59}_{-5.59}$ &       0.39 \\
    &    NGC 2841 & $3.23^{+0.01}_{-0.01}$ & $8.95^{+6.66}_{-6.46}$ &       1.53 & $2.01^{+0.002}_{-0.006}$ & $9.39^{+7.42}_{-6.55}$ &       5.79 & $4.08^{+0.01}_{-0.01}$ & $8.62^{+6.34}_{-6.33}$ &       3.02 & $3.85^{+0.03}_{-0.03}$ & $8.75^{+7.02}_{-7.01}$ &    0.93 \\
    &    NGC 3031 & $1.35^{+0.003}_{-0.003}$ & $9.45^{+7.07}_{-7.08}$ &       4.34 & $1.02^{+0.004}_{-0.001}$ & $9.71^{+7.28}_{-7.39}$ &       5.45 & $3.5^{+0.02}_{-0.03}$ & $8.22^{+6.29}_{-6.3}$ &       5.12 &   $1.90^{+0.01}_{-0.01}$ & $9.06^{+7.82}_{-7.81}$ &    4.37 \\
    &    NGC 3621 & $3.33^{+0.01}_{-0.01}$ & $8.26^{+5.96}_{-5.98}$ &       7.94 & $0.89^{+0.003}_{-0.001}$ & $9.43^{+7.32}_{-7.06}$ &       55.7 & $5.77^{+0.02}_{-0.02}$ & $7.63^{+5.51}_{-5.51}$ &       5.59 & $9.41^{+0.05}_{-0.05}$ & $7.21^{+5.15}_{-5.17}$ &       3.64 \\
    &    NGC 4736 & $0.29^{+0.001}_{-0.001}$ & $10.6^{+8.47}_{-8.47}$ &       1.73 & $0.04^{+0.0001}_{-0.0001}$ & $12.8^{+10.84}_{-10.52}$ &       15.8 & $0.22^{+0.05}_{-0.05}$ & $10.5^{+8.76}_{-8.76}$ &        1.30 & $0.55^{+0.05}_{-0.05}$ & $9.80^{+7.40}_{-7.40}$ &       3.33 \\
    &    NGC 6946 & $2.86^{+0.01}_{-0.01}$ & $8.63^{+6.27}_{-6.27}$ &       2.11 & $1.77^{+0.002}_{-0.006}$ & $9.04^{+6.51}_{-6.87}$ &       7.13 & $8.12^{+0.06}_{-0.06}$ & $7.52^{+5.52}_{-5.52}$ &       1.14 & $309^{+322}_{-105}$ & $6.65^{+4.89}_{-4.9}$ &       2.49 \\
    &    NGC 7793 & $1.90^{+0.01}_{-0.01}$ & $8.47^{+6.43}_{-6.44}$ &       6.11 & $1.77^{+0.01}_{-0.01}$ & $8.49^{+6.49}_{-6.49}$ &       5.68 & $1.88^{+0.01}_{-0.02}$ & $8.35^{+6.44}_{-6.45}$ &       7.88 & $3.09^{+0.03}_{-0.03}$ & $7.98^{+6.14}_{-6.13}$ &        4.30 \\
    \hline
    \multirow{ 7}{*}{G.B} &     IC 2574 & $10.5^{+0.19}_{-0.19}$ & $6.90^{+5.06}_{-5.1}$ &       0.29 & $8.71^{+0.17}_{-0.17}$ & $6.86^{+5.13}_{-5.18}$ &       0.31 & $14.7^{+0.57}_{-0.54}$ & $6.62^{+4.96}_{-4.95}$ &       0.69 & $20.8^{+1.27}_{-1.14}$ & $6.51^{+4.89}_{-4.89}$ &       1.16 \\
    &    NGC 2366 & $1.92^{+0.04}_{-0.03}$ & $7.81^{+6.43}_{-6.3}$ &       1.63 & $0.76^{+0.01}_{-0.03}$ & $8.33^{+7.16}_{-6.82}$ &       3.94 & $1.75^{+0.04}_{-0.05}$ & $7.72^{+6.37}_{-6.37}$ &       1.34 & $1.78^{+0.05}_{-0.05}$ & $7.69^{+6.36}_{-6.35}$ &       1.34 \\
    &    NGC 2903 & $2.55^{+0.01}_{-0.01}$ & $8.77^{+6.59}_{-6.52}$ &       1.43 & $2.43^{+0.01}_{-0.01}$ & $8.81^{+6.60}_{-6.58}$ &       1.31 & $3.06^{+0.01}_{-0.01}$ & $8.55^{+6.37}_{-6.45}$ & 2.01 & $2.85^{+0.01}_{-0.01}$ & $8.63^{+6.44}_{-6.43}$ & 1.23 \\
    &    NGC 2976 & $1.77^{+0.03}_{-0.03}$ & $8.40^{+6.62}_{-6.63}$ &       0.95 & $0.32^{+0.003}_{-0.003}$ & $9.41^{+7.3}_{-8.07}$ &       36.3 &   $>10^4$ & $7.69^{+6.19}_{-6.19}$ & 1.18 & $23.7^{+3.06}_{-3.06}$ & $7.50^{+5.39}_{-5.39}$ & 5.29 \\
    &    NGC 3198 & $4.50^{+0.01}_{-0.01}$ & $8.02^{+5.79}_{-5.83}$ &       0.67 & $4.00^{+0.01}_{-0.01}$ & $8.09^{+5.90}_{-5.91}$ &       0.42 & $8.76^{+0.04}_{-0.04}$ & $7.27^{+5.22}_{-5.23}$ & 2.92 & $9.74^{+0.05}_{-0.05}$ & $7.17^{+5.16}_{-5.13}$ & 0.93 \\
    &    NGC 3521 & $2.18^{+0.01}_{-0.01}$ & $9.00^{+6.87}_{-7.05}$ &       1.25 & $2.14^{+0.01}_{-0.01}$ & $9.01^{+6.98}_{-6.98}$ &       1.32 & $24.5^{+1.32}_{-1.25}$ & $6.66^{+5.39}_{-5.39}$ & 8.64 & $162^{+135}_{-50.1}$ & $6.22^{+5.02}_{-5.01}$ & 11.5 \\
    &     NGC 925 & $7.98^{+0.09}_{-0.09}$ & $7.35^{+5.55}_{-5.47}$ &       0.19 & $2.14^{+0.01}_{-0.03}$ & $8.15^{+5.86}_{-6.77}$ &       13.7 & $21.2^{+1.02}_{-0.95}$ & $6.77^{+5.14}_{-5.13}$ &       1.19 & $61.1^{+18.5}_{-8.61}$ & $6.54^{+4.78}_{-5.23}$ &       2.54 \\
    \hline
    \multirow{ 3}{*}{G.C} &    NGC 2403 & $2.78^{+0.01}_{-0.01}$ & $8.32^{+5.83}_{-5.91}$ &       4.33 & $0.84^{+0.001}_{-0.001}$ & $9.42^{+7.45}_{-6.66}$ &       27.8 & $3.86^{+0.01}_{-0.01}$ & $7.98^{+5.63}_{-5.63}$ &  1.6 & $3.9^{+0.01}_{-0.01}$ & $7.96^{+5.59}_{-5.62}$ & 2.25 \\
    &    NGC 5055 & $2.26^{+0.01}_{-0.01}$ & $8.91^{+6.59}_{-6.58}$ &       6.34 & $0.07^{+0.0001}_{-0.0001}$ & $12.5^{+10.4}_{-10.0}$ &       94.4 & $23.6^{+0.27}_{-0.26}$ & $6.6^{+4.79}_{-4.8}$ & 4.11 & $2.52^{+0.03}_{-0.03}$ & $8.79^{+6.89}_{-6.89}$ & 3.47 \\
    &    NGC 7331 & $2.36^{+0.01}_{-0.01}$ & $9.05^{+6.88}_{-6.88}$ &       1.47 & $2.16^{+0.01}_{-0.01}$ & $9.12^{+6.96}_{-6.96}$ &       1.08 &   $17.4^{+0.19}_{-0.19}$ & $7.23^{+5.33}_{-5.34}$ & 7.06 & $27.58^{+0.49}_{-0.48}$ & $7.00^{+5.15}_{-5.14}$ & 15.57 \\
    \end{tabular}
    }
  \caption{\footnotesize{We present the fitted values of the free Burkert parameters, Eq. (\ref{eq:rhoBurkert}), for all the mass model. Each column. The $r_c$ is given in {\rm kpc}. Logarithm base 10 of the central densities $\rho_0$ is given in $M_\sun/{\rm kpc}^3$. The value of $\chi^2$ is normalized to the numbers of data points minus the number of free parameter. The set of galaxies are as show in Table.\ref{tab:onlydm_bdmnfw}.}}
  \label{tab:burkert}
\end{table}

\setlength{\extrarowheight}{3pt}
\begin{table}[h]
  \centering
  \scriptsize{
    \begin{tabular}{rr|lrr|lrr|lrr|lrr}
    \multicolumn{ 14}{c}{PSEUDO-ISOTHERMAL}\\
    \hline
    &  &  \multicolumn{ 3}{c}{MIN.DISK} &   \multicolumn{ 3}{c}{MIN.DISK+GAS} &         \multicolumn{ 3}{c}{KROUPA} &  \multicolumn{ 3}{c}{DIET-SALPETER} \\
               &     Galaxy & \multicolumn{1}{c}{$r_s$} & $\log\rho_0$  & $\chi^2_{\rm red}$ & \multicolumn{1}{c}{$r_s$} & $\log\rho_0$  & $\chi^2_{\rm red}$ & \multicolumn{1}{c}{$r_s$} & $\log \rho_0$  & $\chi^2_{\rm red}$ & \multicolumn{1}{c}{$r_s$} & $\log \rho_0$  & $\chi^2_{\rm red}$ \\
               &\multicolumn{1}{c|}{(1)}&\multicolumn{1}{c}{(2)}&\multicolumn{1}{c}{(3)}&\multicolumn{1}{c|}{(4)}&\multicolumn{1}{c}{(5)}&\multicolumn{1}{c}{(6)}&\multicolumn{1}{c|}{(7)}&\multicolumn{1}{c}{(8)} &\multicolumn{1}{c}{(9)}&\multicolumn{1}{c|}{(10)}&\multicolumn{1}{c}{(11)}&\multicolumn{1}{c}{(12)}& \multicolumn{1}{c}{(13)} \\
    \hline
    \multirow{ 7}{*}{G.A} &     DDO 154 & $1.30^{+0.01}_{-0.01}$ & $7.55^{+5.64}_{-5.64}$ &        0.50 & $1.21^{+0.01}_{-0.01}$ & $7.52^{+5.68}_{-5.69}$ &        0.40 & $1.31^{+0.01}_{-0.01}$ & $7.45^{+5.64}_{-5.63}$ &       0.36 & $1.36^{+0.01}_{-0.01}$ & $7.43^{+5.61}_{-5.61}$ &       0.35 \\
    &    NGC 2841 & $0.14^{+0.01}_{-0.01}$ & $11.0^{+8.61}_{-8.53}$ &       5.36 & $0.82^{+0.01}_{-0.01}$ & $9.47^{+7.32}_{-6.81}$ &       10.9 & $<0.01$ & $13.6^{+11.3}_{-11.3}$ &       0.72 & $<0.01$ & $14.95^{+12.4}_{-12.3}$ &       3.14 \\
    &    NGC 3031 & $\sim10^{-7}$ & $22.80^{+20.4}_{-20.5}$ &       6.98 & $1.07^{+0.02}_{-0.03}$ & $9.08^{+7.14}_{-6.26}$ &       25.6 & $1.42^{+0.01}_{-0.01}$ & $8.42^{+6.50}_{-6.49}$ &       5.14 &   $>10^8$ & $6.54^{+4.79}_{-5.76}$ &       20.4 \\
    &    NGC3621 & $1.01^{+0.01}_{-0.01}$ & $8.65^{+6.37}_{-6.35}$ &       0.86 & $0.58^{+0.01}_{-0.01}$ & $9.05^{+7.15}_{-6.43}$ &       1.89 & $1.99^{+0.01}_{-0.01}$ & $7.94^{+5.83}_{-5.82}$ &        2.8 & $4.56^{+0.03}_{-0.03}$ & $7.29^{+5.23}_{-5.24}$ &       3.01 \\
    &    NGC 4736 & $2.63^{+0.01}_{-0.03}$ & $8.10^{+6.10}_{-5.99}$ &        168 & $0.38^{+0.01}_{-0.01}$ & $9.51^{+7.64}_{-7.22}$ &       47.2 &   $<0.01$ & $21.0^{+19.3}_{-19.3}$ &    5.96 & $<0.01$ & $12.0^{+10.7}_{-10.6}$ &       4.29 \\
    &    NGC 6946 & $1.06^{+0.02}_{-0.02}$ & $8.88^{+6.52}_{-6.52}$ &       1.66 & $1.68^{+0.06}_{-0.03}$ & $8.53^{+6.27}_{-6.06}$ &       4.09 & $4.79^{+0.04}_{-0.04}$ & $7.49^{+5.49}_{-5.49}$ &        1.10 & $41.6^{+7.14}_{-4.87}$ & $6.66^{+4.91}_{-4.91}$ &       2.48 \\
    &    NGC 7793 & $0.95^{+0.01}_{-0.01}$ & $8.52^{+6.48}_{-6.49}$ &       5.74 & $0.86^{+0.01}_{-0.01}$ & $8.56^{+6.56}_{-6.56}$ &       5.22 & $0.87^{+0.01}_{-0.01}$ & $8.44^{+6.54}_{-6.55}$ &       7.02 & $1.86^{+0.02}_{-0.02}$ & $7.94^{+6.09}_{-6.09}$ &       4.78 \\
    \hline
    \multirow{ 7}{*}{G.B} &     IC 2574 & $6.62^{+0.12}_{-0.11}$ & $6.84^{+5.01}_{-5.04}$ &       0.26 & $5.47^{+0.11}_{-0.11}$ & $6.81^{+5.07}_{-5.12}$ &       0.27 & $8.95^{+0.30}_{-0.31}$ & $6.57^{+4.91}_{-4.9}$ &       0.64 & $11.7^{+0.64}_{-0.57}$ & $6.48^{+4.86}_{-4.86}$ &       1.12 \\
    &    NGC 2366 & $1.16^{+0.02}_{-0.02}$ & $7.75^{+6.24}_{-6.23}$ &       2.07 & $0.06^{+0.01}_{-0.02}$ & $9.35^{+8.04}_{-8.35}$ &       11.8 & $1.36^{+0.02}_{-0.02}$ & $7.47^{+5.83}_{-6.45}$ & 1.79 & $1.06^{+0.03}_{-0.03}$ & $7.62^{+6.34}_{-6.29}$ &       1.64 \\
    &    NGC 2903 & $0.29^{+0.01}_{-0.02}$ & $9.96^{+7.73}_{-7.77}$ &       6.47 & $0.51^{+0.02}_{-0.02}$ & $9.46^{+7.35}_{-7.13}$ &       7.58 & $0.59^{+0.02}_{-0.02}$ & $9.31^{+7.13}_{-7.12}$ &  6.04 & $0.56^{+0.02}_{-0.02}$ & $9.35^{+7.16}_{-7.24}$ & 6.46 \\
    &    NGC 2976 & $1.12^{+0.02}_{-0.02}$ & $8.35^{+6.56}_{-6.57}$ &       1.03 & $0.04^{+0.01}_{-0.01}$ & $10.4^{+9.08}_{-8.36}$ &       36.7 &   $>10^6$ & $7.69^{+5.61}_{-5.56}$ & 1.15 & $3.00^{+0.01}_{-0.01}$ & $7.56^{+6.20}_{-6.15}$ & 6.01 \\
    &    NGC 3198 & $1.37^{+0.02}_{-0.02}$ & $8.40^{+6.17}_{-6.22}$ &       2.79 & $0.36^{+0.01}_{-0.01}$ & $9.47^{+6.97}_{-7.58}$ &       8.04 & $3.80^{+0.02}_{-0.02}$ & $7.43^{+5.39}_{-5.38}$ &       3.42 & $6.70^{+0.04}_{-0.04}$ & $6.96^{+4.68}_{-4.98}$ &        6.40 \\
    &    NGC 3521 & $0.90^{+0.01}_{-0.01}$ & $9.16^{+7.12}_{-7.12}$ &        5.70 & $0.63^{+0.04}_{-0.02}$ & $9.43^{+7.03}_{-7.76}$ &       7.19 & $15.38^{+0.83}_{-0.79}$ & $6.6^{+5.34}_{-5.34}$ & 8.62 & $46.3^{+11.7}_{-7.39}$ & $6.24^{+5.03}_{-5.03}$ & 11.49 \\
    &     NGC 925 & $5.01^{+0.06}_{-0.06}$ & $7.29^{+5.49}_{-5.41}$ &        0.20 & $1.03^{+0.01}_{-0.01}$ & $8.20^{+6.25}_{-6.47}$ &       10.3 & $12.4^{+0.54}_{-0.50}$ & $6.73^{+5.10}_{-5.09}$ &       1.13 &   $>10^6$ & $6.47^{+0.60}_{-0.60}$ &       2.47 \\
    \hline
    \multirow{ 3}{*}{G.C} &    NGC 2403 & $0.84^{+0.01}_{-0.01}$ & $8.72^{+6.23}_{-6.31}$ &       1.04 & $0.57^{+0.01}_{-0.01}$ & $9.00^{+6.86}_{-6.38}$ &       1.14 & $1.53^{+0.01}_{-0.01}$ & $8.19^{+5.85}_{-5.84}$ & 0.85 & $1.91^{+0.01}_{-0.01}$ & $8.01^{+5.67}_{-5.68}$ & 0.8 \\
    &    NGC 5055 & $0.14^{+0.01}_{-0.01}$ & $10.6^{+8.24}_{-8.24}$ &       2.79 & $<0.002$ & $14.2^{+11.9}_{-11.9}$ &       3.43 & $15.04^{+0.18}_{-0.18}$ & $6.53^{+4.72}_{-4.72}$ & 4.14 & $45.3^{+1.69}_{-1.58}$ & $5.99^{+4.29}_{-4.29}$ & 13.3 \\
    &    NGC 7331 & $0.35^{+0.01}_{-0.01}$ & $10.0^{+7.91}_{-7.73}$ &        1.80 & $0.20^{+0.01}_{-0.01}$ & $10.5^{+8.35}_{-8.23}$ &       2.18 &   $>10^7$ & $6.73^{+4.66}_{-5.02}$ & 10.2 &   $>10^8$ & $6.64^{+4.98}_{-4.70}$ & 25.9 \\
    \end{tabular}
    }
  \caption{\footnotesize{We present the fitted values of the free pseudo-isothermal parameters, Eq. (\ref{eq:rhoISO}), for all the mass model. Each column. The $r_c$ is given in {\rm kpc}. Logarithm base 10 of the central densities $\rho_0$ is given in $M_\sun/{\rm kpc}^3$. The value of $\chi^2$ is normalized to the numbers of data points minus the number of free parameter. The set of galaxies are as show in Table.\ref{tab:onlydm_bdmnfw}.}}
  \label{tab:iso}
\end{table}

\begin{figure}
  \includegraphics[width=0.65\textwidth]{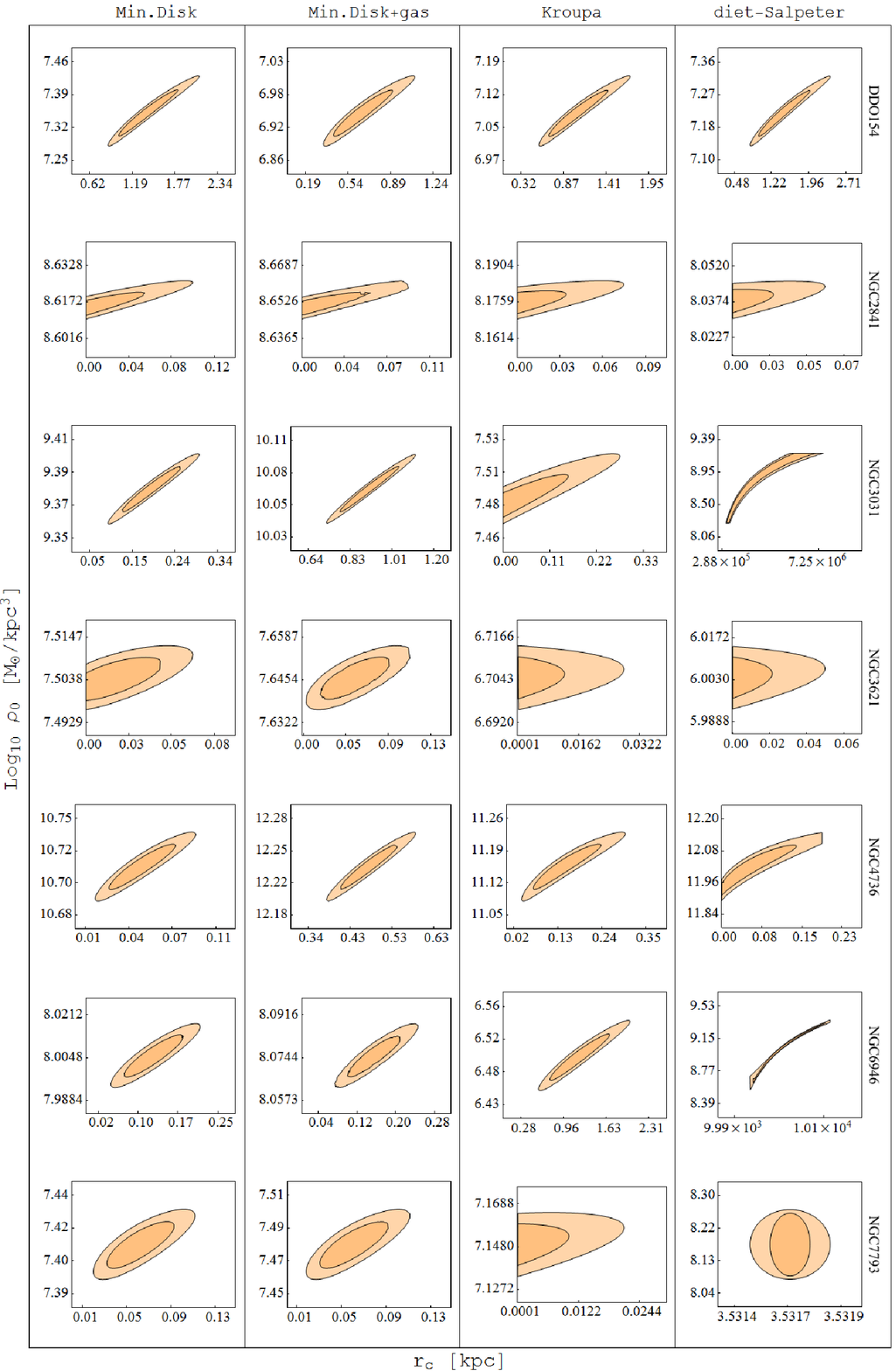}\\
  \caption{\footnotesize{We present the 2D likelihood contours for the BDM parameters $\rho_0$ and $r_c$ for galaxies with $r_c\ll r_s$, group A. The two regions in each figure correspond to $1\sigma$ (68\%) and $2\sigma$ (95\%) confidence level. Columns from left to right correspond to minimal disk, minimal disk + gas, Kroupa and diet-Salpeter mass models. This clearly shows how the stars has an important contribution close to the center of the galaxy and erase trace of the core, we can see that even if the value of $r_c$ that minimize $\chi^2$ is zero, is consistent with values different from zero up to $1\sigma$.}}
  \label{tab:conflevel_A}
\end{figure}

\section{Rotation curves and likelihoods contours}\label{apend:LikelihoodPlot}
In this Sec. we present the figures where we plot all the different profiles, presented in the mass model Sec. \ref{NuMe}, the confidence level contours obtained from the numerical fitting as well as the consideration made for each galaxy. The groups are described in the Sec. \ref{results}. In each subsection we present the rotation curve for each galaxy and mass model, at the end of each subsection we show the respectively confidence level contours. The G.A. galaxies are presented in Sec. \ref{apendix:ga}, in subsection Sec. \ref{apendix:gb} and \ref{apendix:gb_inner} are presented the galaxies included in G.B for the BDM and the BDM-inner analysis. Finally, in subsection \ref{apendix:gc} we present G.C galaxies.

\begin{figure}
  \includegraphics[width=0.75\textwidth]{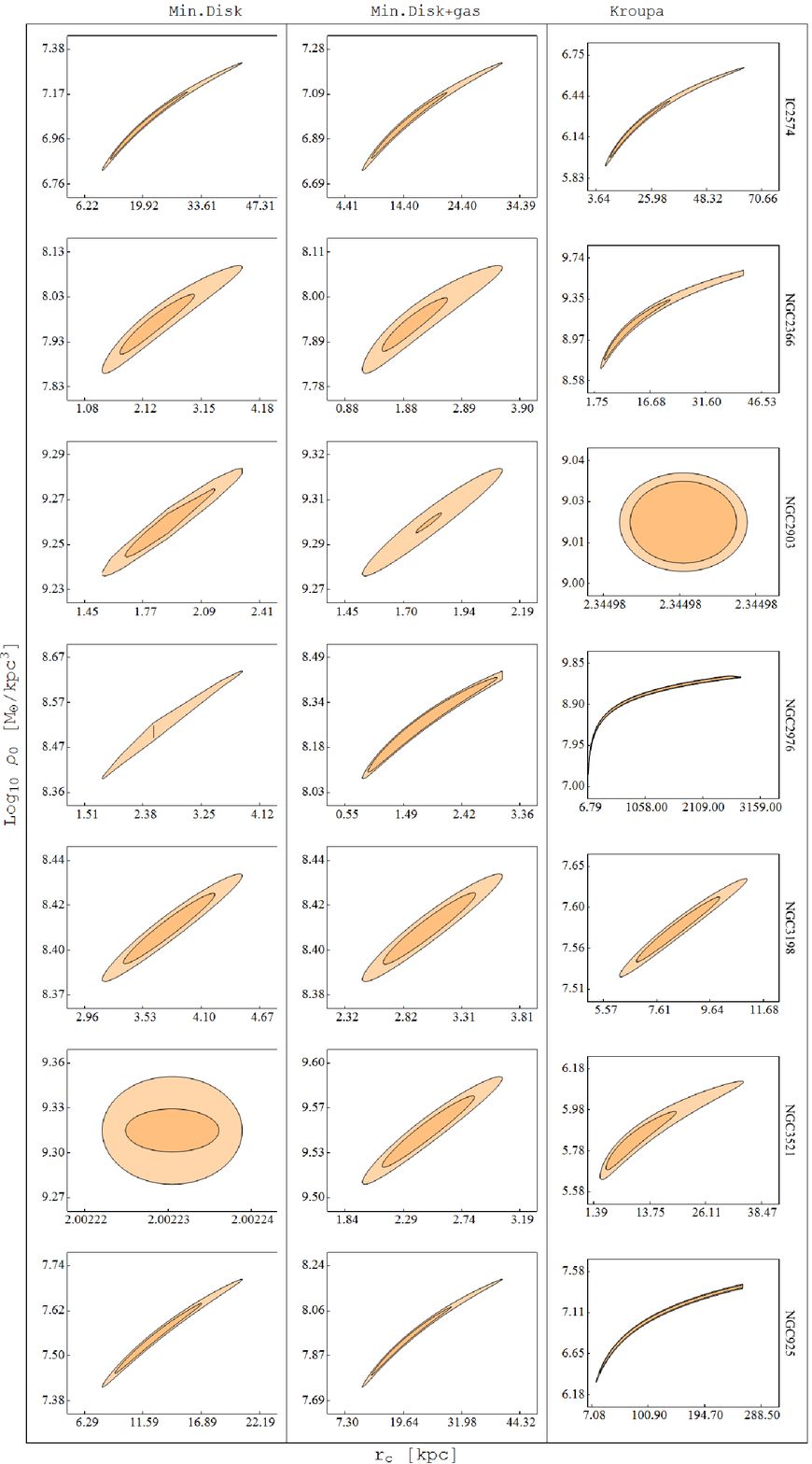}\\
  \caption{\footnotesize{This table shows the 2D likelihood contours for the BDM parameters $\rho_0$ and $r_c$ for galaxies with $r_c = r_s$, group B. The different colors in each figure represent the $1\sigma$ and $2\sigma$ confidence levels. Columns from left to right corresponds to minimal disk, minimal disk + gas, and Kroupa. When more mass components are included in the analysis the contours of the confidence levels become larger, resulting in the difficulty of finding the minimal fitted value for $r_c$ because stars vanish the traces of the core. We obtain uncommon confidence level contours for diet-Salpeter that suggest an overestimation for the stellar component. For this mass model we obtain values for $r_c < r_s$ and $r_c$ unconstrained inside the $2\sigma$ error for the other analysis and we decide not to show it here.}}
  \label{tab:conflevel_B}
\end{figure}

\begin{figure}
  \includegraphics[width=\textwidth]{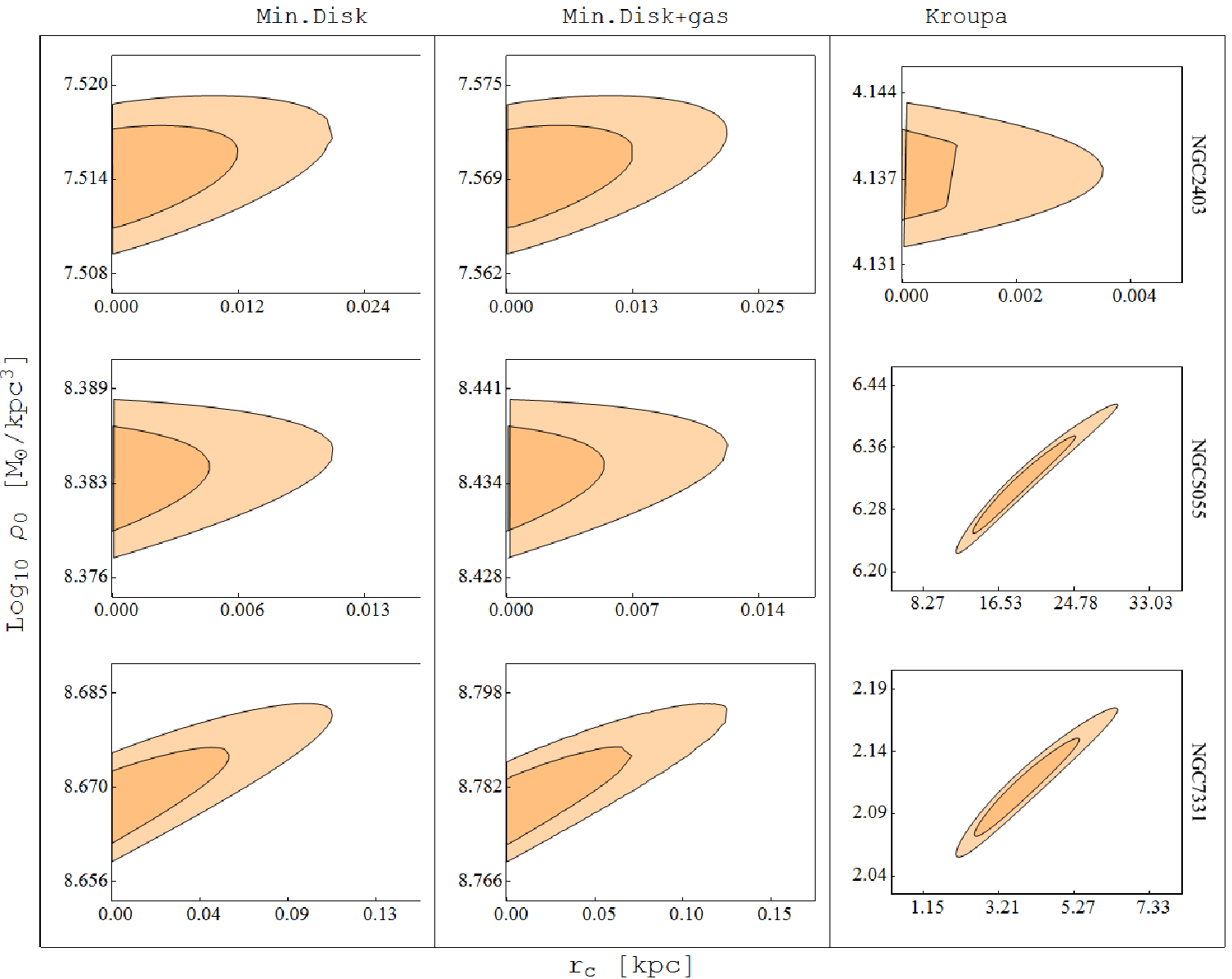}\\
  \caption{\footnotesize{This table shows the 2D likelihood contours between the BDM parameters $\rho_0$ and $r_c$, corresponding to $1\sigma$ and $2\sigma$ for galaxies with $r_c = 0$, group C. Columns from left to right correspond to minimal disk, min.disk+gas, Kroupa, and diet-Salpeter mass models.}}
  \label{tab:conflevel_C}
\end{figure}

\newpage
\subsection{Inner analysis}\label{apendix:gb_inner}
\begin{figure}[!h]
      \subfloat[\footnotesize{Fit and Difference}]{
       	\includegraphics[width=0.4\textwidth]{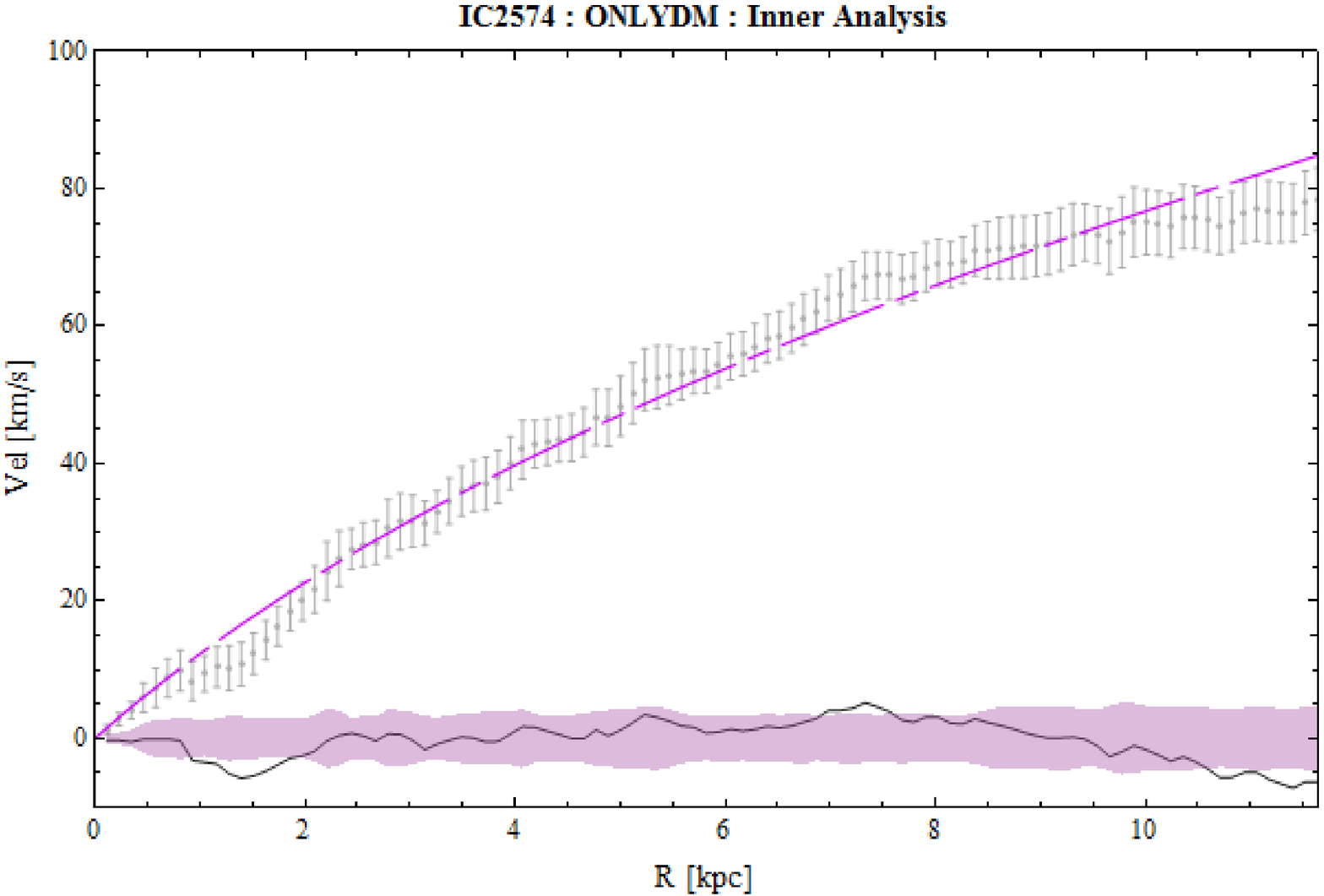} }
       \subfloat[\footnotesize{Fit and Difference}]{
       	\includegraphics[width=0.4\textwidth]{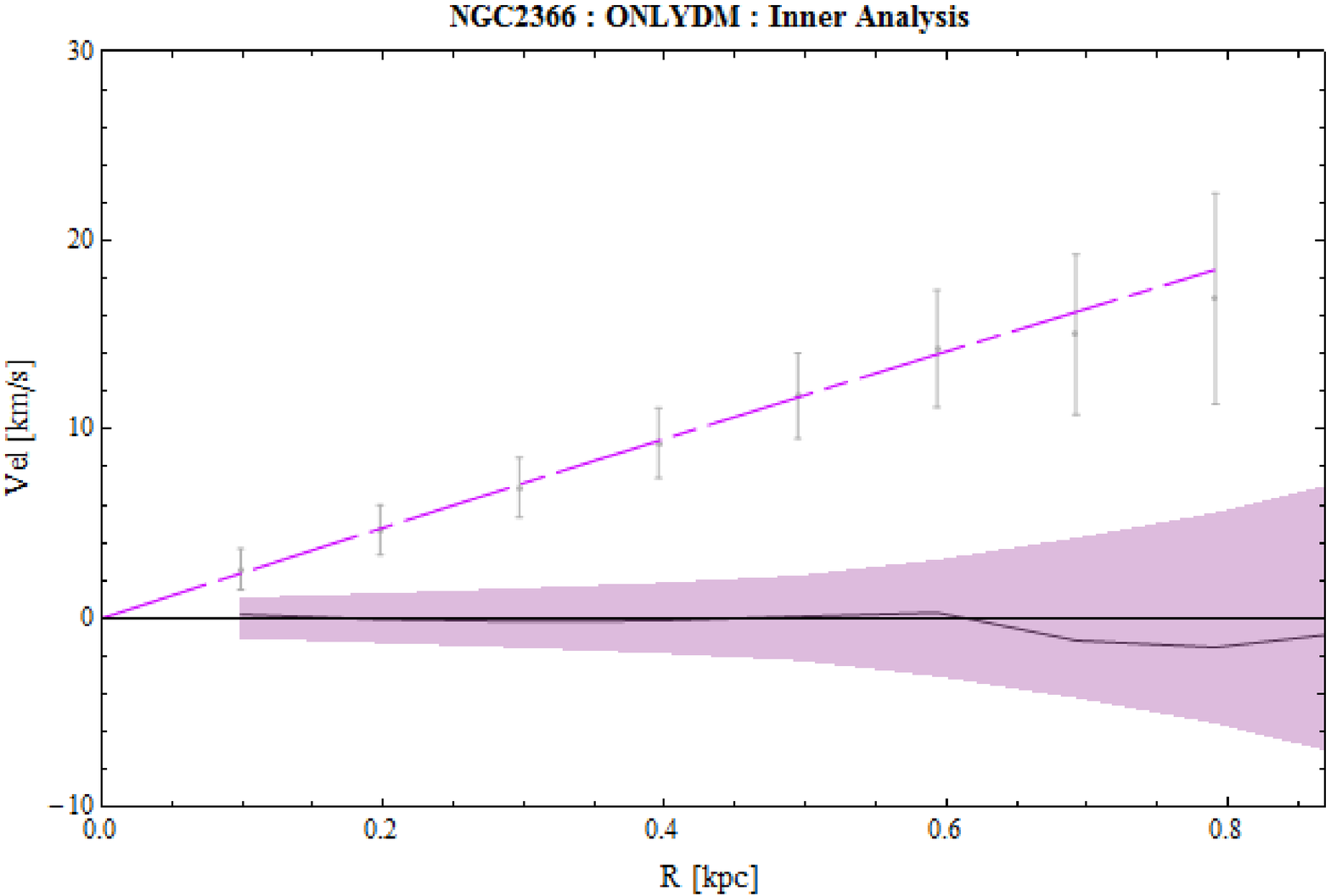} }
      \caption{IC 2574 and NGC 2366 inner analysis. The gray points are the observations with its error bars. Below we show the difference between the observational and the theoretical approach, the purple region represent the error bars and the line the fitted curve.}\label{fig:inner_grph1}
\end{figure}

\begin{figure}[!h]
      \subfloat[\footnotesize{Fit and Difference}]{
       	\includegraphics[width=0.4\textwidth]{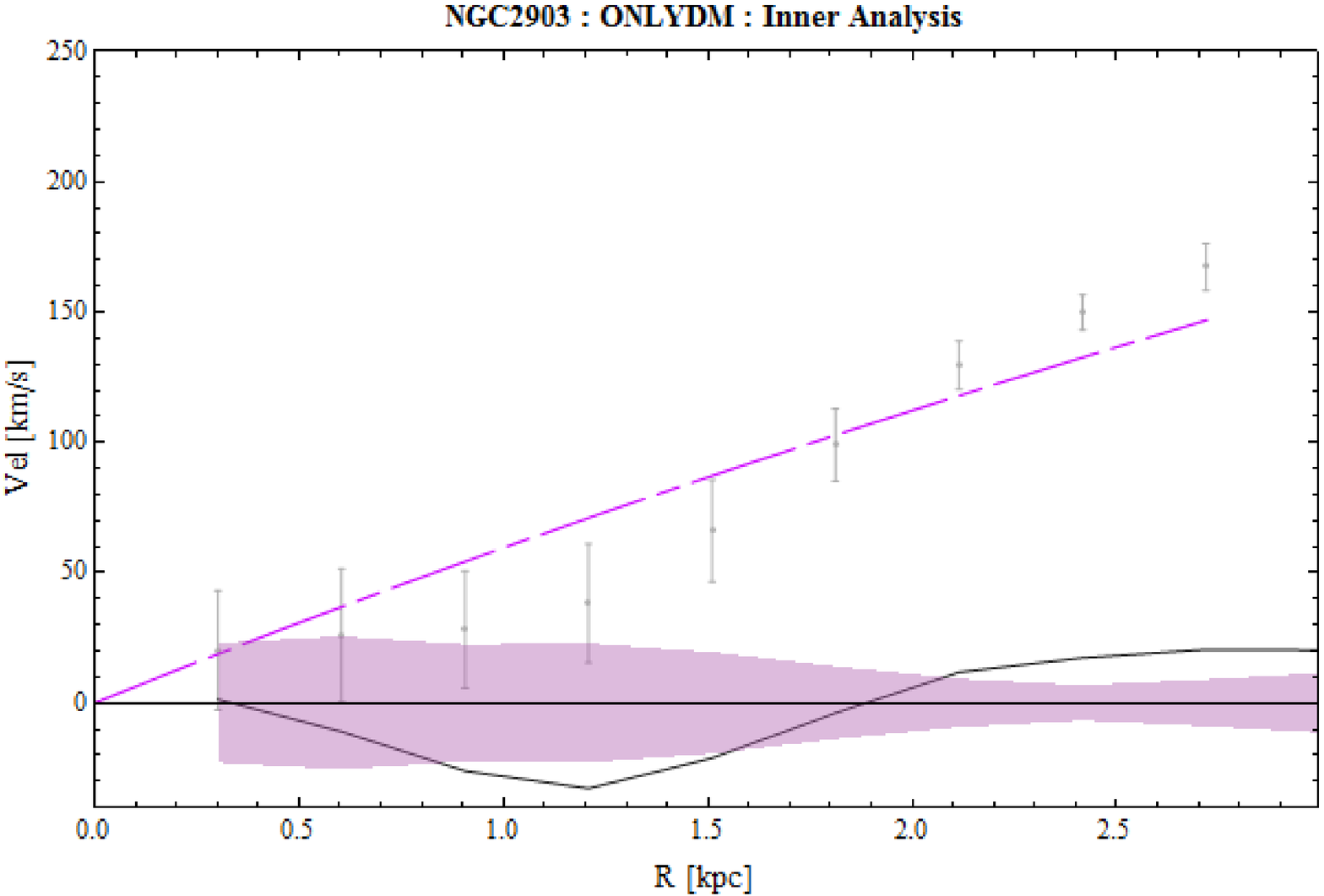}}
      \subfloat[\footnotesize{Fit and Difference}]{
       	\includegraphics[width=0.4\textwidth]{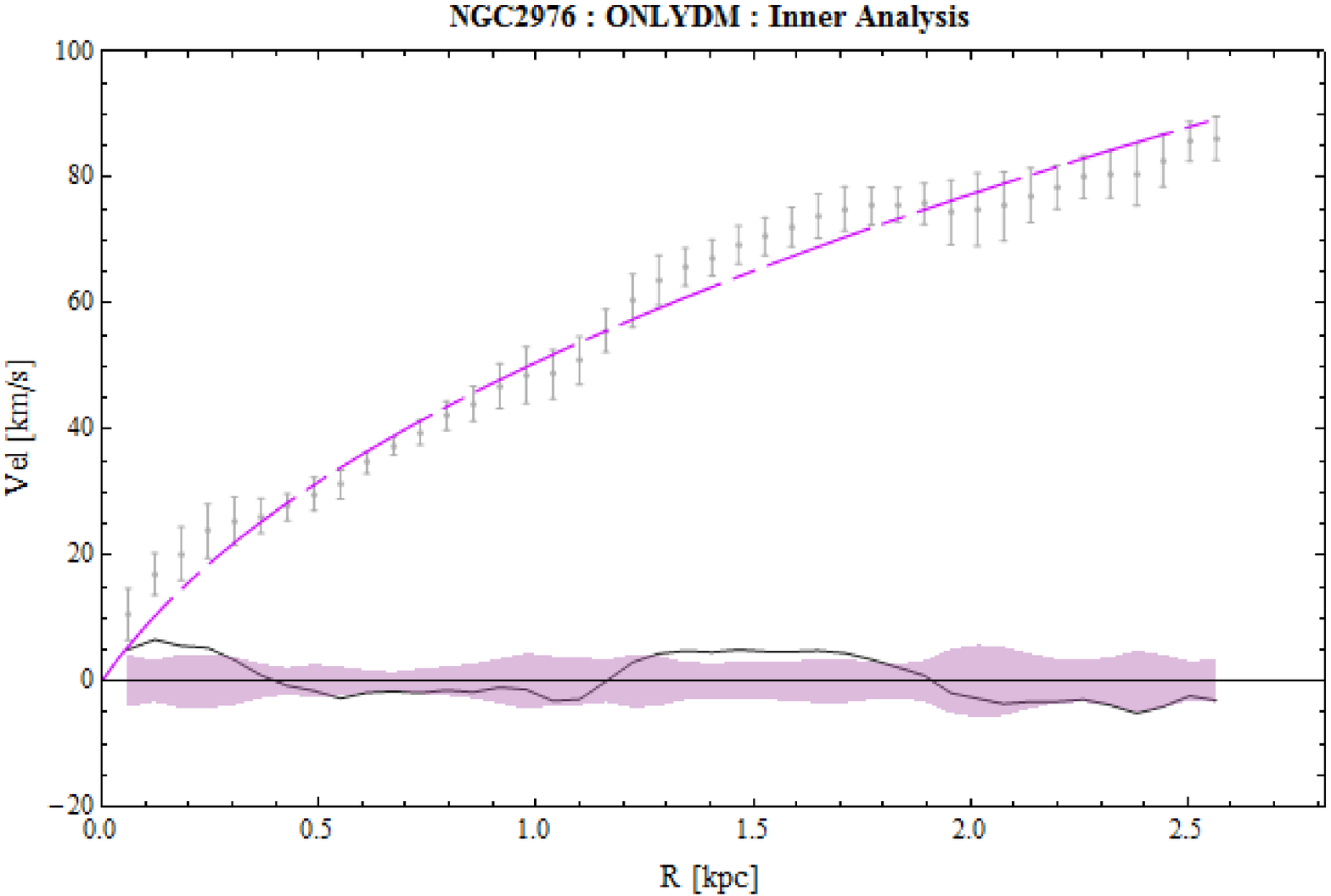} }
      \caption{NGC2976 inner analysis}\label{fig:inner_grph2}
    \end{figure}

\begin{figure}[!h]
      \subfloat[\footnotesize{Fit and Difference}]{
       	\includegraphics[width=0.4\textwidth]{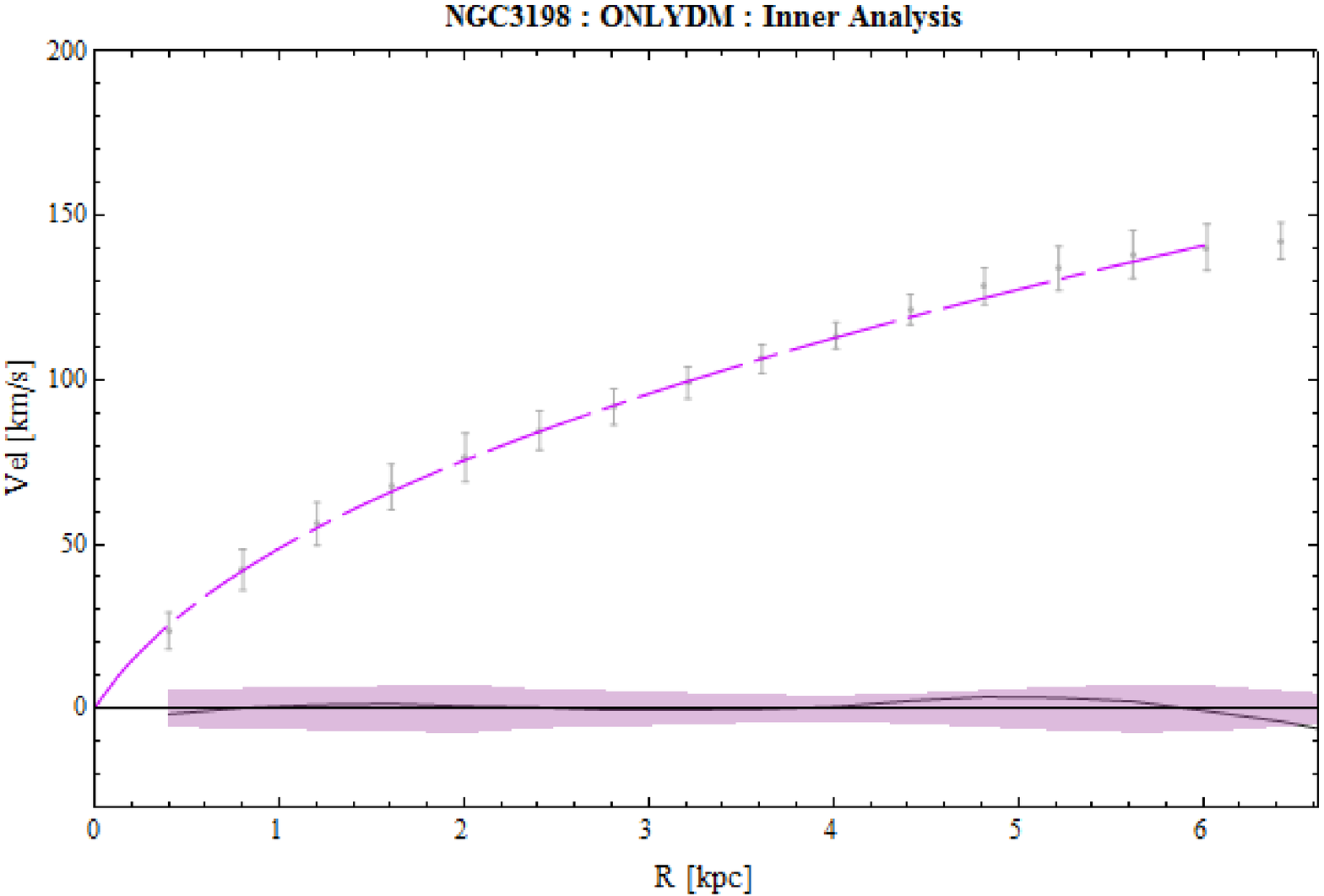} }
     \subfloat[\footnotesize{Fit and Difference}]{
       	\includegraphics[width=0.4\textwidth]{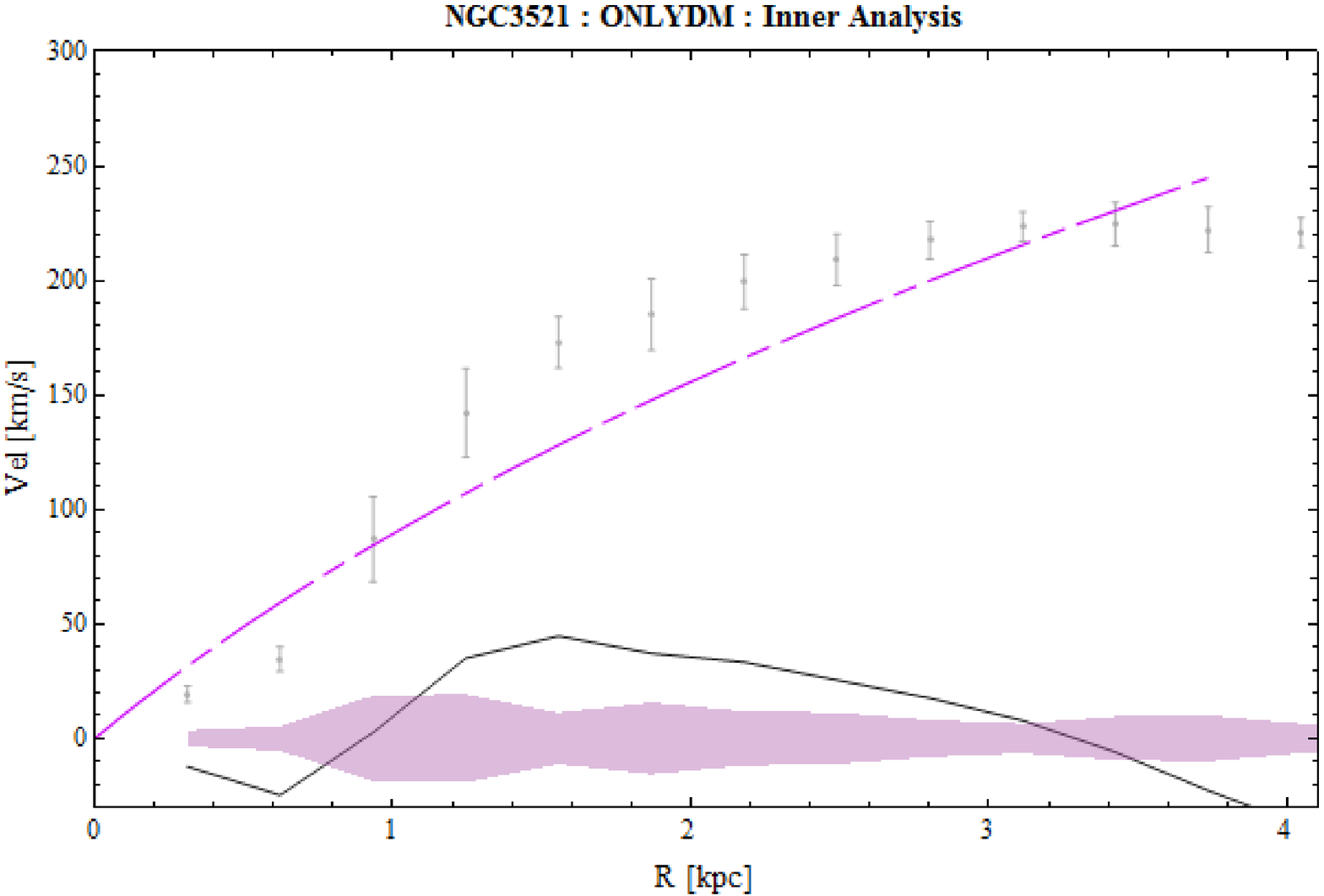} }
     \caption{NGC3198 inner analysis}\label{fig:inner_grph3}
\end{figure}

\begin{figure}[!h]
      \subfloat[\footnotesize{Fit and Difference}]{
       	\includegraphics[width=0.4\textwidth]{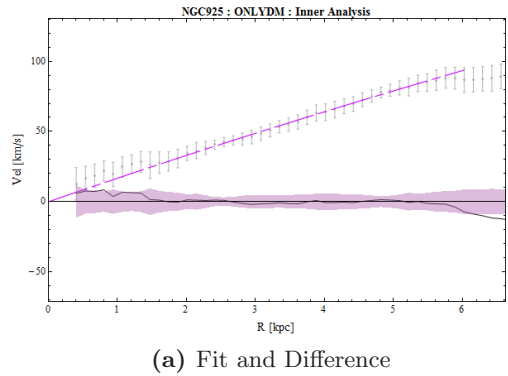} }
      \caption{NGC925 inner analysis}\label{fig:inner_grph4}
\end{figure}

We present the different G.B. galaxies that where analyzed when the DM halo is the only mass component with the inner approximation for the BDM profile $\rho_{in} = 2 \rho_c (1+r/r_c)^{-1}$ up to a radius $r < r_s$ with a constant positive slope shown in Fig. (\ref{fig:inner_grph1} - \ref{fig:inner_grph4}) with a purple, long, dashed line. From this first approach we obtain the the core distance $r_c$ and the central density $\rho_c$, then we use these parameters to obtain $\rho_0$ and $r_s$ from the fit of BDM profile to all the set of observational data (thick black line). Fitted values are presented in Table. \ref{tab:inner_dm_dmg_k} and the confidence levels between $\rho_c$ and $r_c$ in Fig. \ref{tab:conflevel_B}.

\newcolumntype{V}{>{\centering\arraybackslash} m{0.22\textwidth} }
\newcolumntype{W}{>{\centering\arraybackslash} m{0.21\textwidth} }
\begin{figure}
  \includegraphics[width=0.75\textwidth]{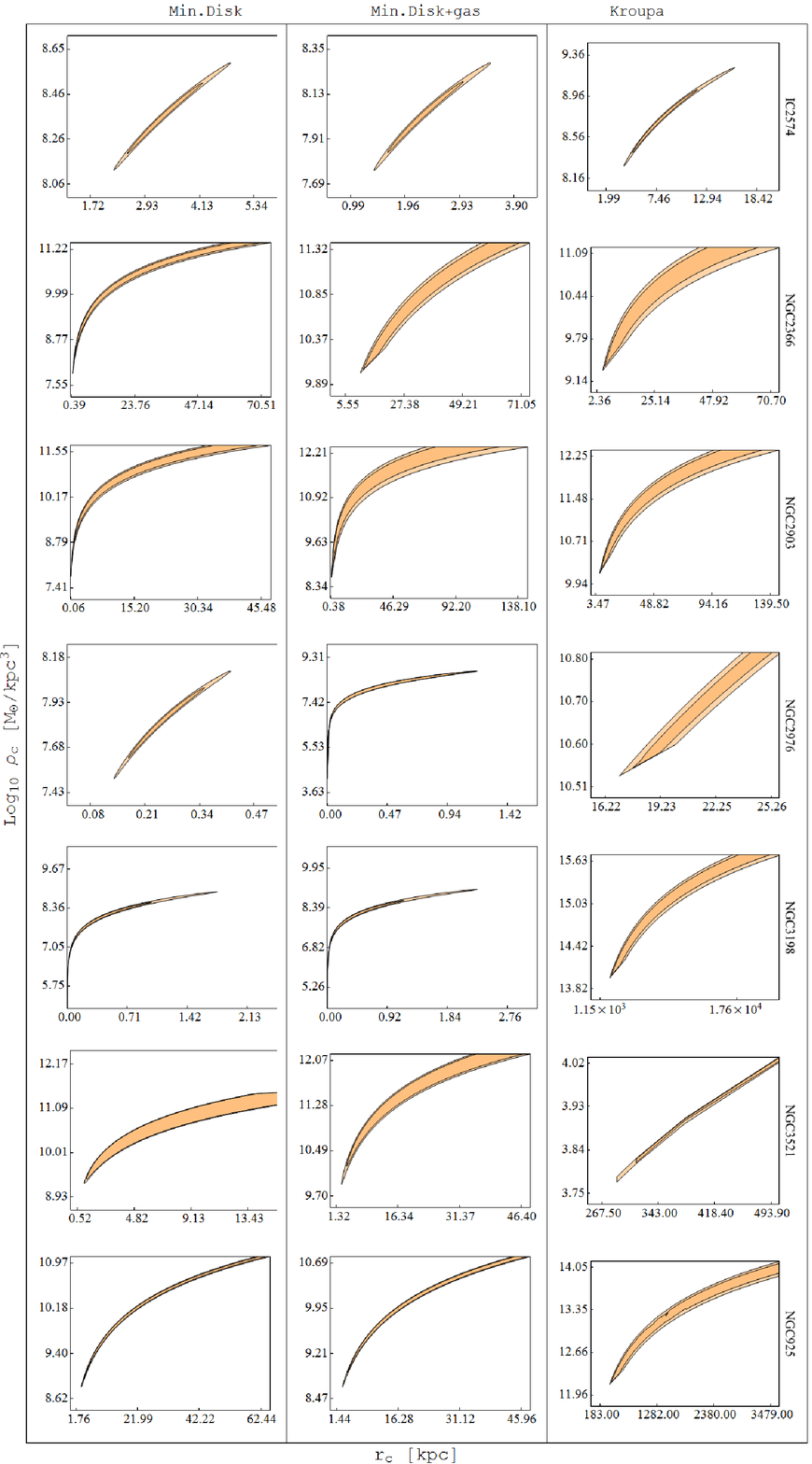}\\
  \caption{\footnotesize{The 2D likelihood contours for the BDM parameters $\rho_0$ and $r_c$ for galaxies with $r_c = r_s$, group B. The different colors in each figure represent the $1\sigma$ and $2\sigma$ confidence levels. Columns from left to right corresponds to minimal disk, minimal disk + gas, and Kroupa mass models obtained with the inner analysis.}}
  \label{tab:conflevel_inner B}
\end{figure}

\begin{subappendices}
\subsection{GROUP A}\label{apendix:ga}

In this section we present the considerations made in each one of the galaxies for the numerical analysis, this galaxy has in particular fitted values of $r_c < r_s$ when analyzed with the minimal disk model, we have named this set of galaxies as G.A. The conclusion are presented in Sec. \ref{conclusion}. At the end of the section can be seen the confidence level for each galaxy in every mass model.

\begin{figure}[h!]
    \subfloat[\footnotesize{Minimal disk}]{
    \begin{tabular}[b]{c}
    \includegraphics[width=0.35\textwidth]{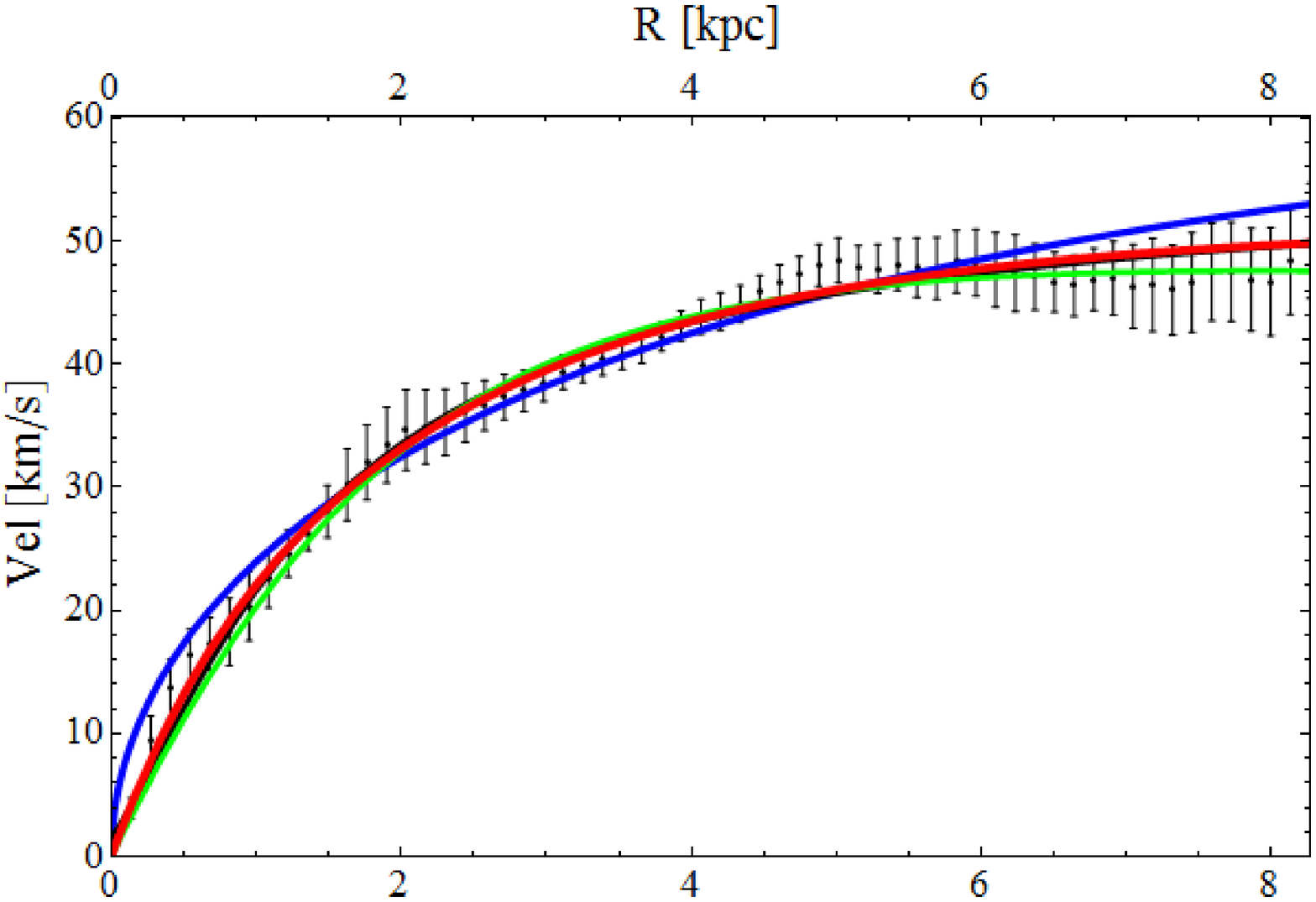} \\
    \includegraphics[width=0.35\textwidth]{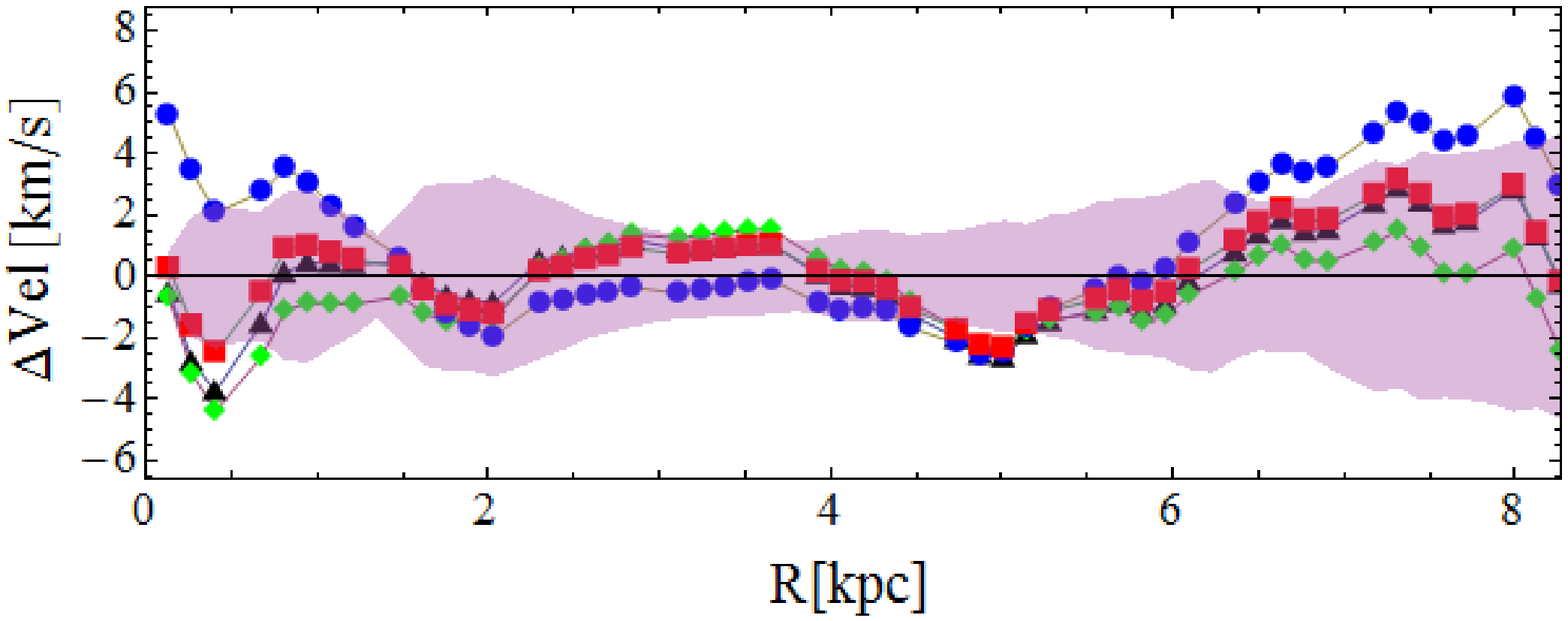}
    \end{tabular}  }
    \subfloat[\footnotesize{Min. disk + Gas}]{
    \begin{tabular}[b]{c}
    \includegraphics[width=0.35\textwidth]{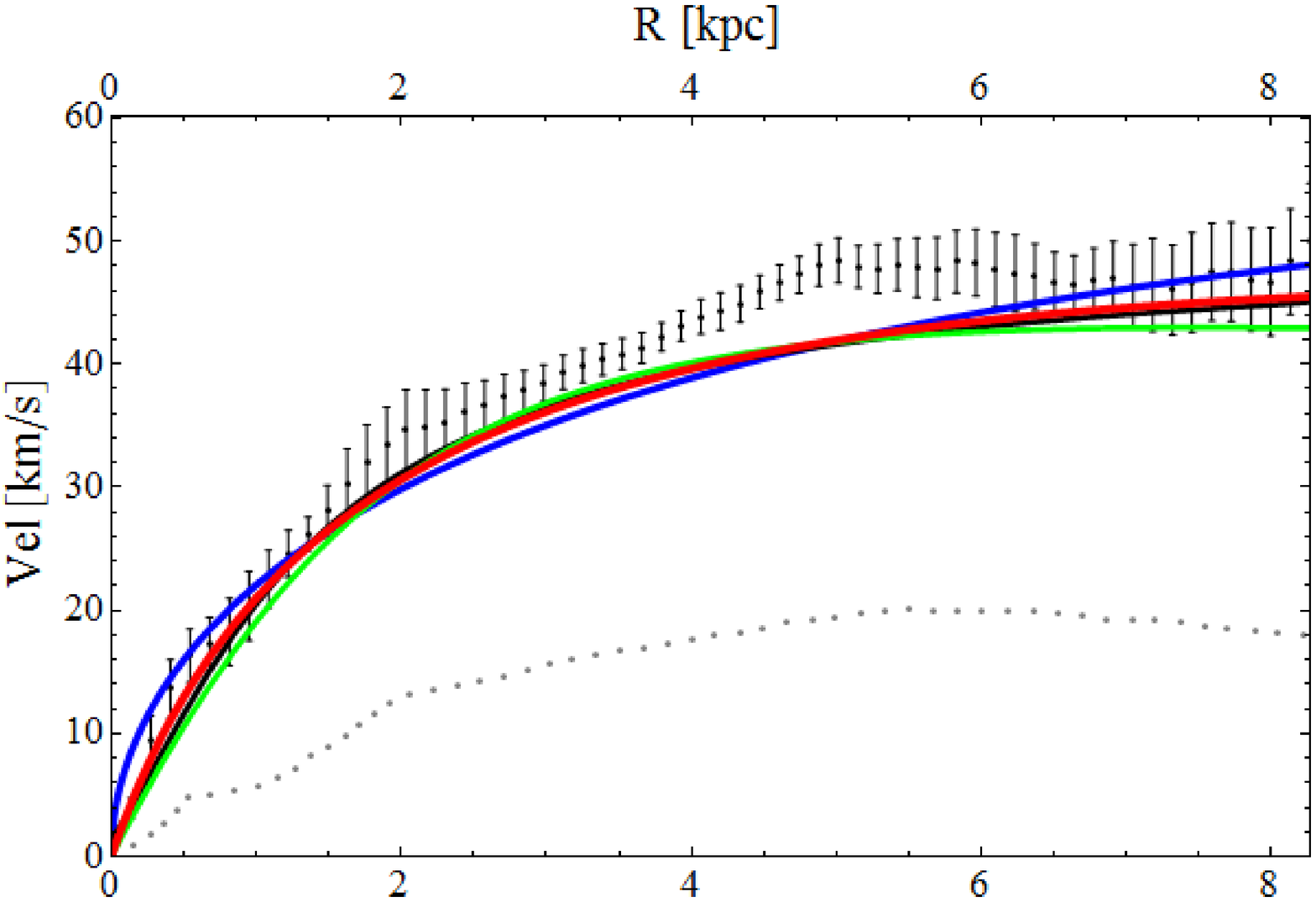} \\
    \includegraphics[width=0.35\textwidth]{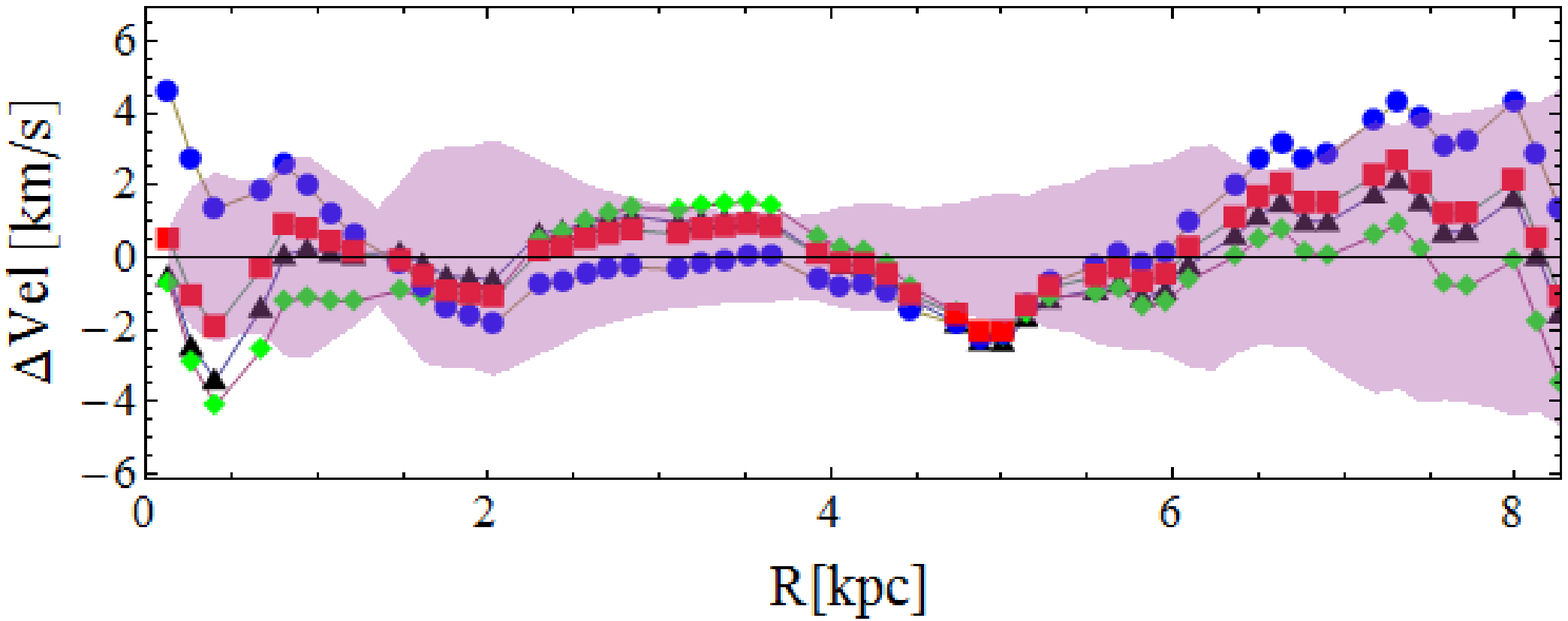}
    \end{tabular}  }  \\
    \subfloat[\footnotesize{Kroupa}]{
    \begin{tabular}[b]{c}
    \includegraphics[width=0.35\textwidth]{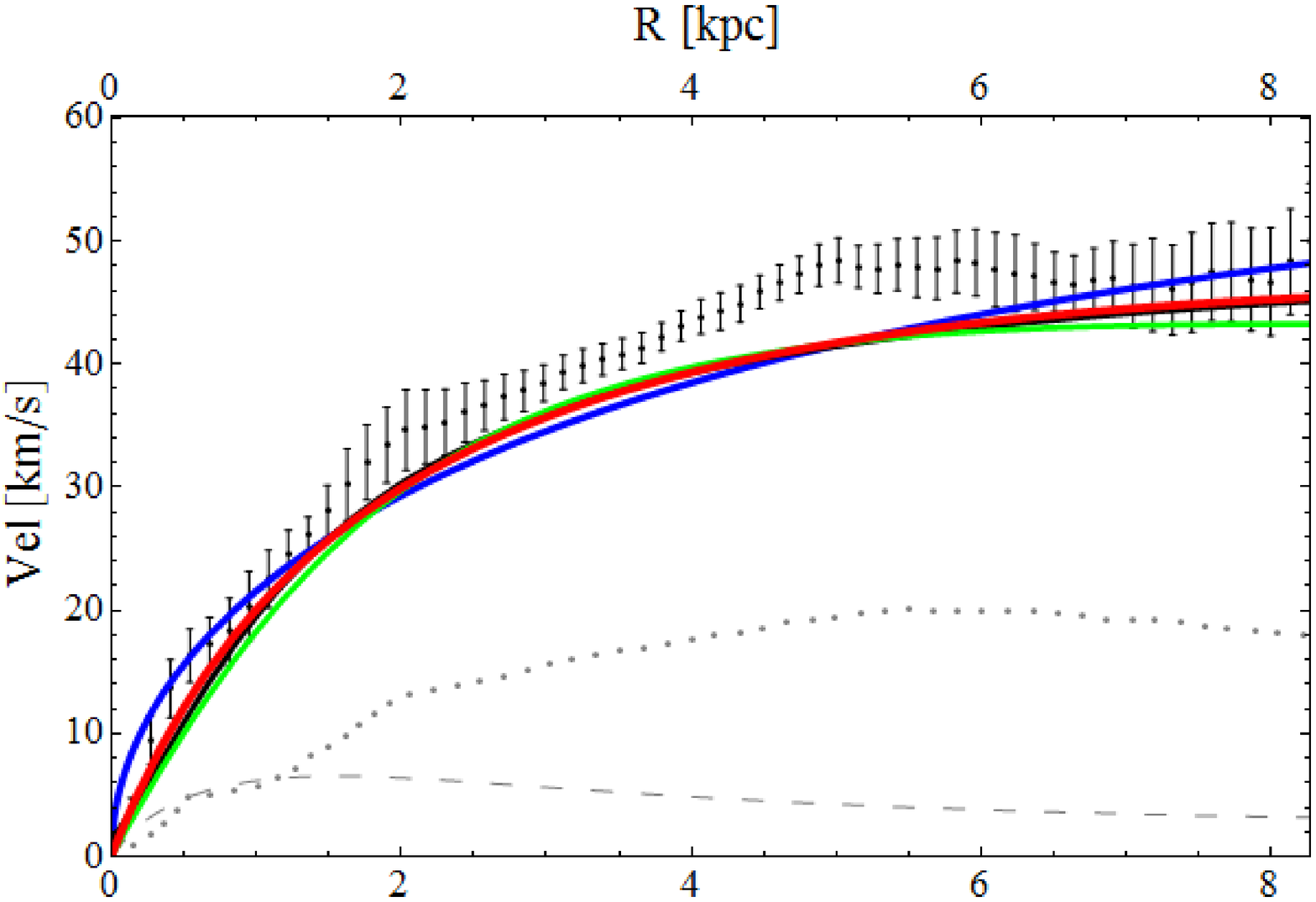} \\
    \includegraphics[width=0.35\textwidth]{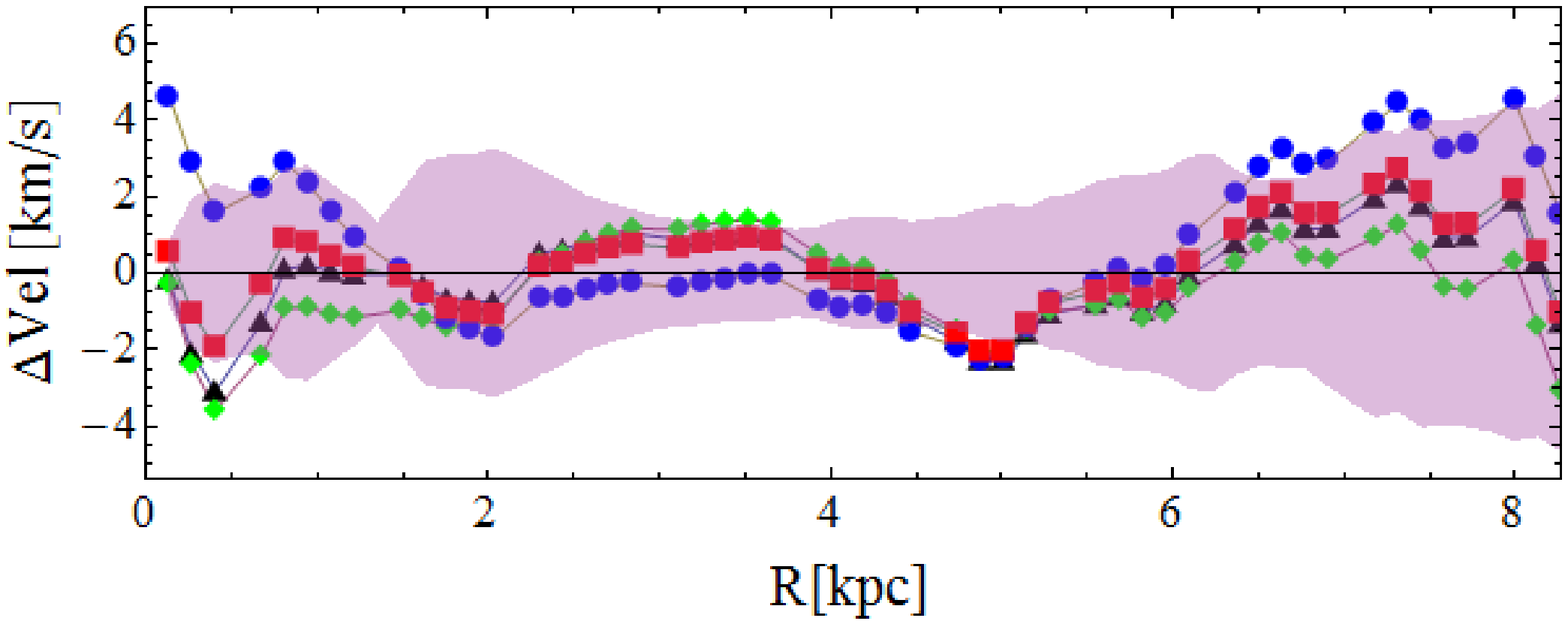}
    \end{tabular}  }
    \subfloat[\footnotesize{diet-Salpeter}]{
    \begin{tabular}[b]{c}
    \includegraphics[width=0.35\textwidth]{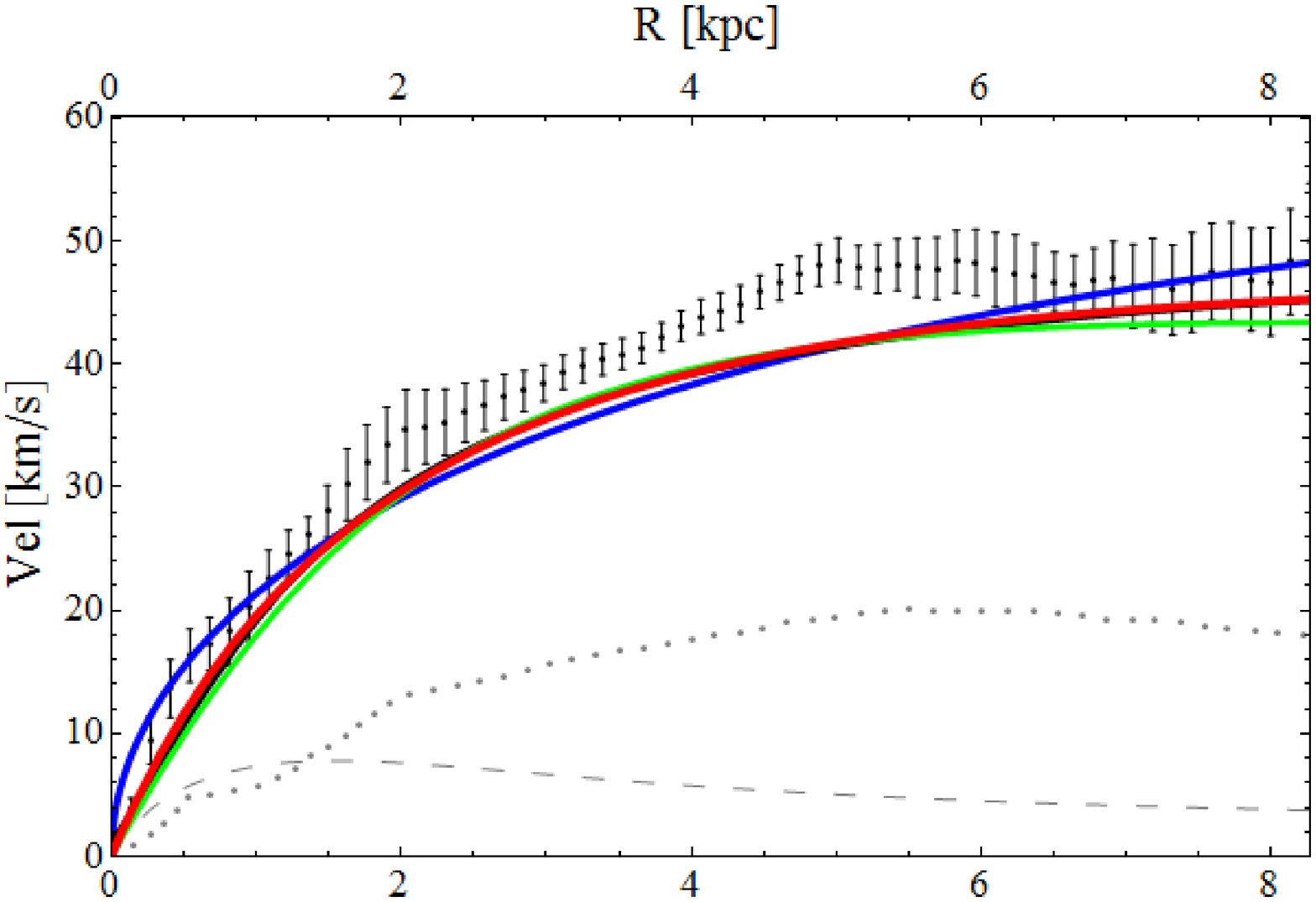} \\
    \includegraphics[width=0.35\textwidth]{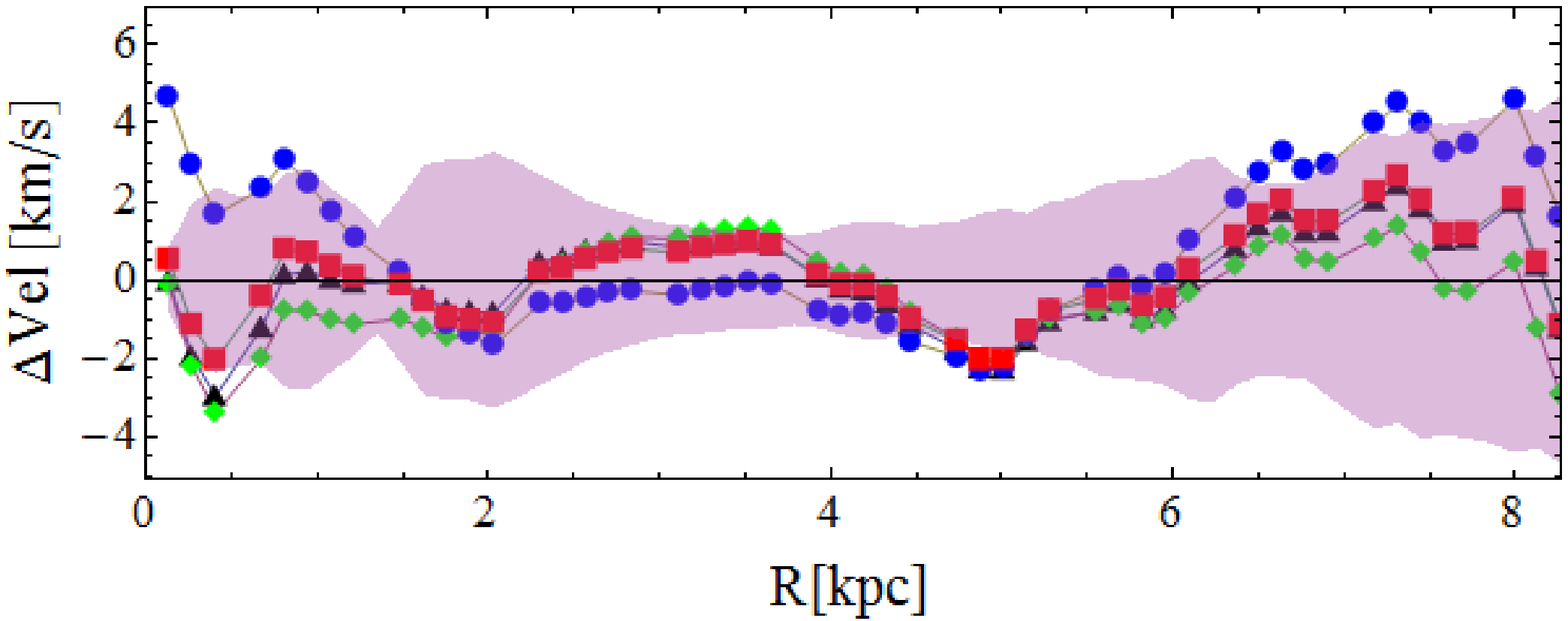}
    \end{tabular}  }
  \caption{\footnotesize{Graphical visualization of the rotation curves for the galaxies DDO 154 belonging to the group G.A. where $r_c < r_s$ for all mass models. Is a dwarf galaxy, one of the first galaxies used to illustrate the conflict with the theoretical prediction of the $\lambda$CDM [17]. No significant color gradient in the colors (B-V) and (B-R) is detected, constant color as function of radius in our models is well assumed with the value for $\gs = 0.32$. For a single exponential disk we use value of $\mu_0 = 20.8$ {\rm mag $arcsec^2$} and $R_d = 0.72$ {\rm kpc}. BDM is the best option since it predict a value of $\gs$. very close to the minimal disk and gives the best fit. The values for $r_s$ and $r_c$ are very reasonable except when we considered the Kroupa IMF. From left to right and top to bottom, images from the fit considering minimal disk, minimal disk+gas, Kroupa, diet-Salpeter. In the top images of each galaxy are represent the four DM profiles: BDM, red squares; NFW, blue circles; Burkert, green diamond; Isothermal, black triangles. The dotted line is the gas component, the short dashed line is the stellar contribution and the tick points are the observational data with its respectively error bars. At the bottom images we plot the difference between the observed and predicted curves where the purple region represents the error of the observations and each different line represent the fitted curve of the different profiles with the observations taking into account already all the mass contributions.}}
  \label{fig:DDO154}
\end{figure}

\begin{figure}[h!]
    \subfloat[\footnotesize{Minimal disk}]{
    \begin{tabular}[b]{c}
    \includegraphics[width=0.35\textwidth]{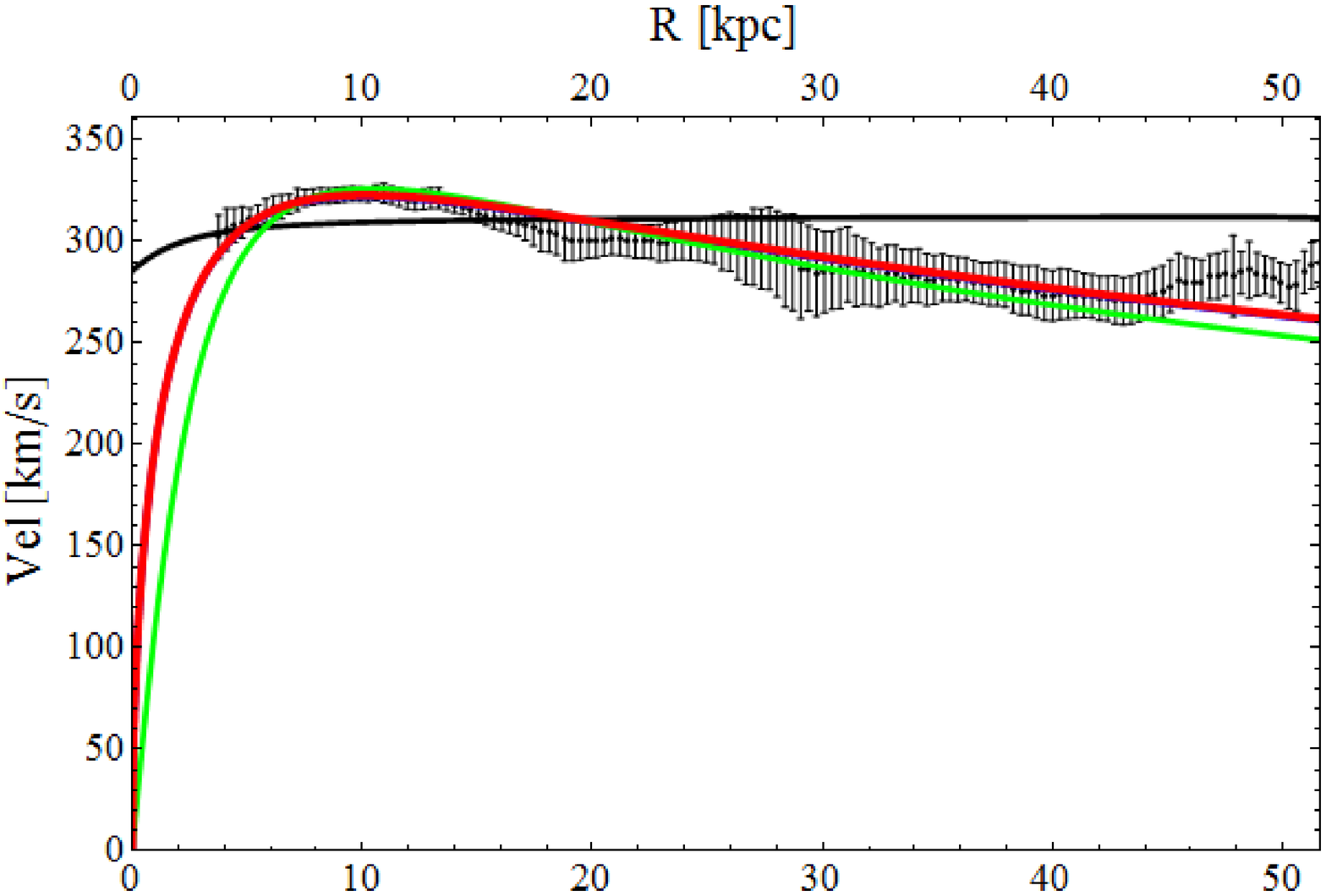} \\
    \includegraphics[width=0.35\textwidth]{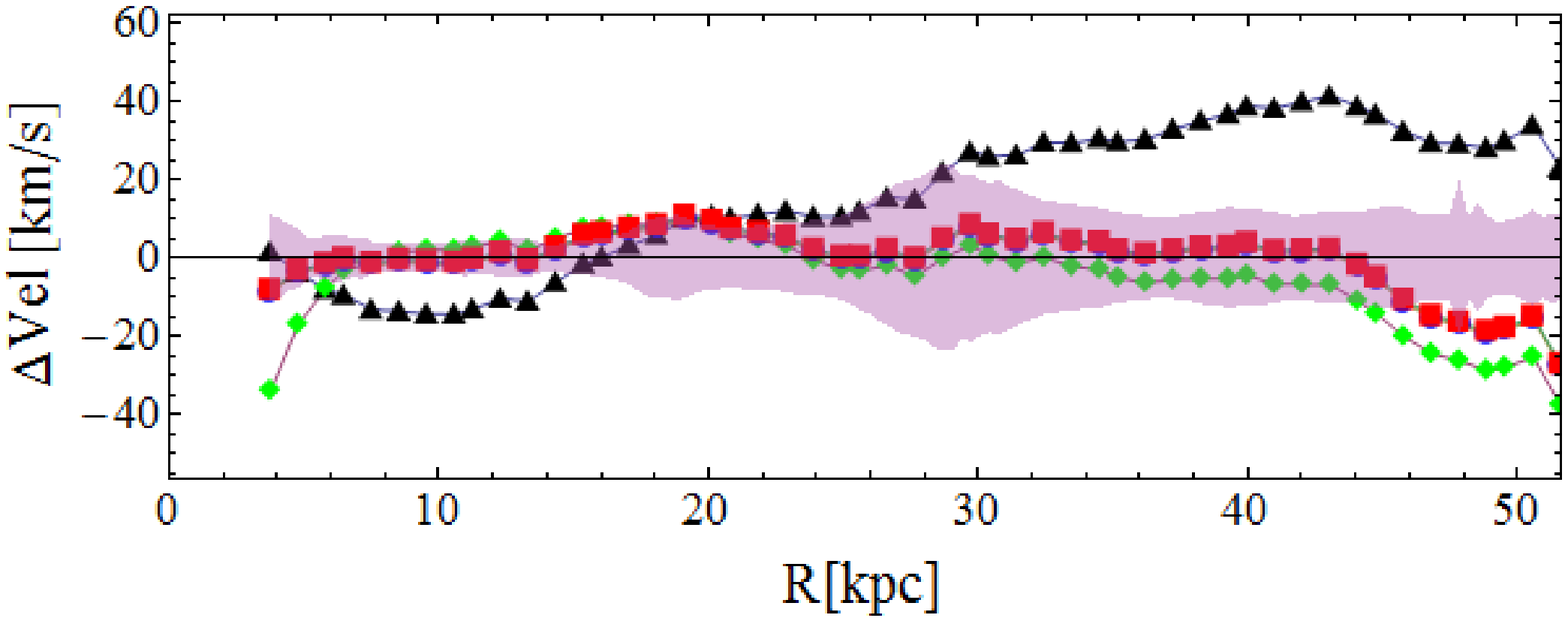}
    \end{tabular}  }
    \subfloat[\footnotesize{Min. disk + Gas}]{
    \begin{tabular}[b]{c}
    \includegraphics[width=0.35\textwidth]{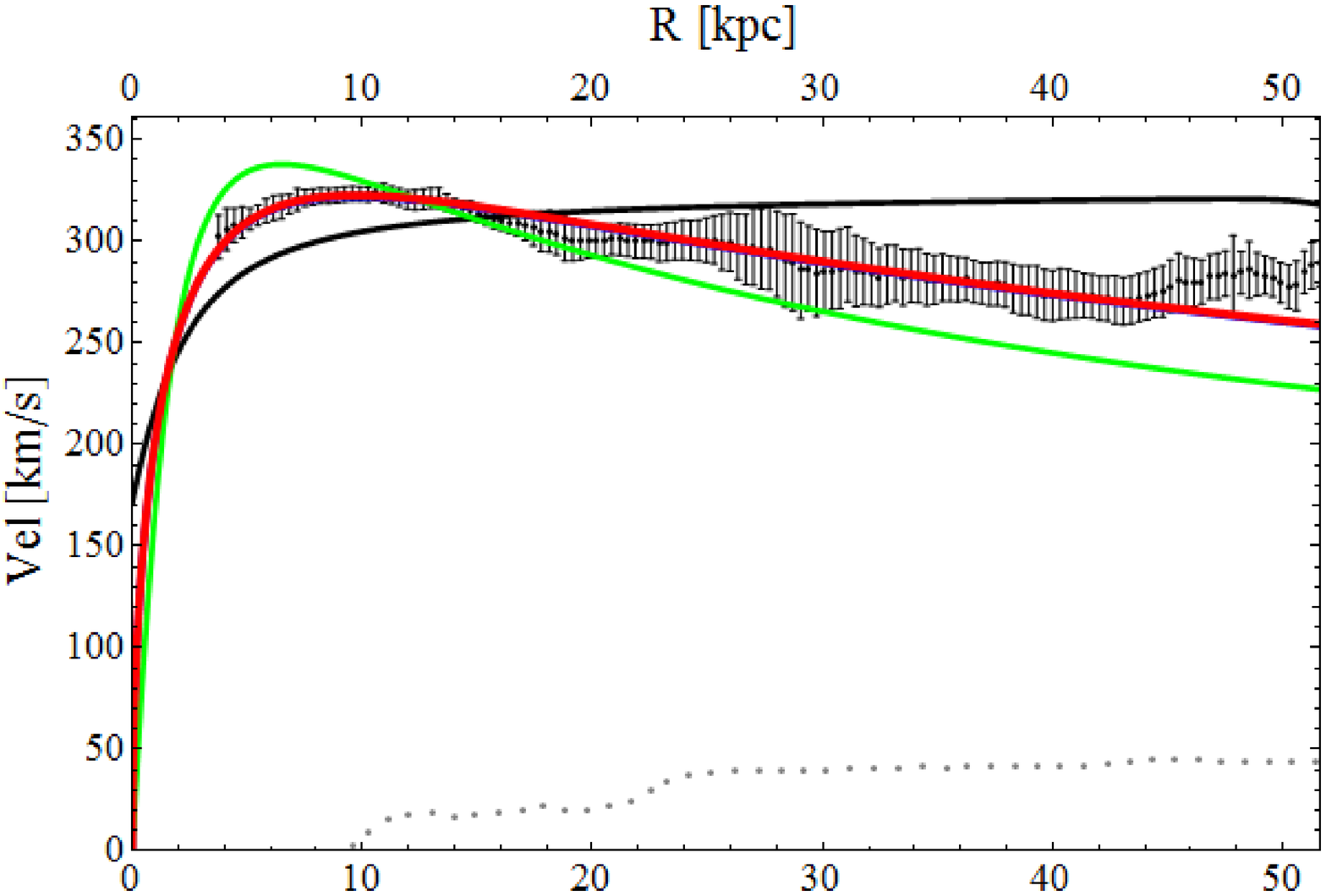} \\
    \includegraphics[width=0.35\textwidth]{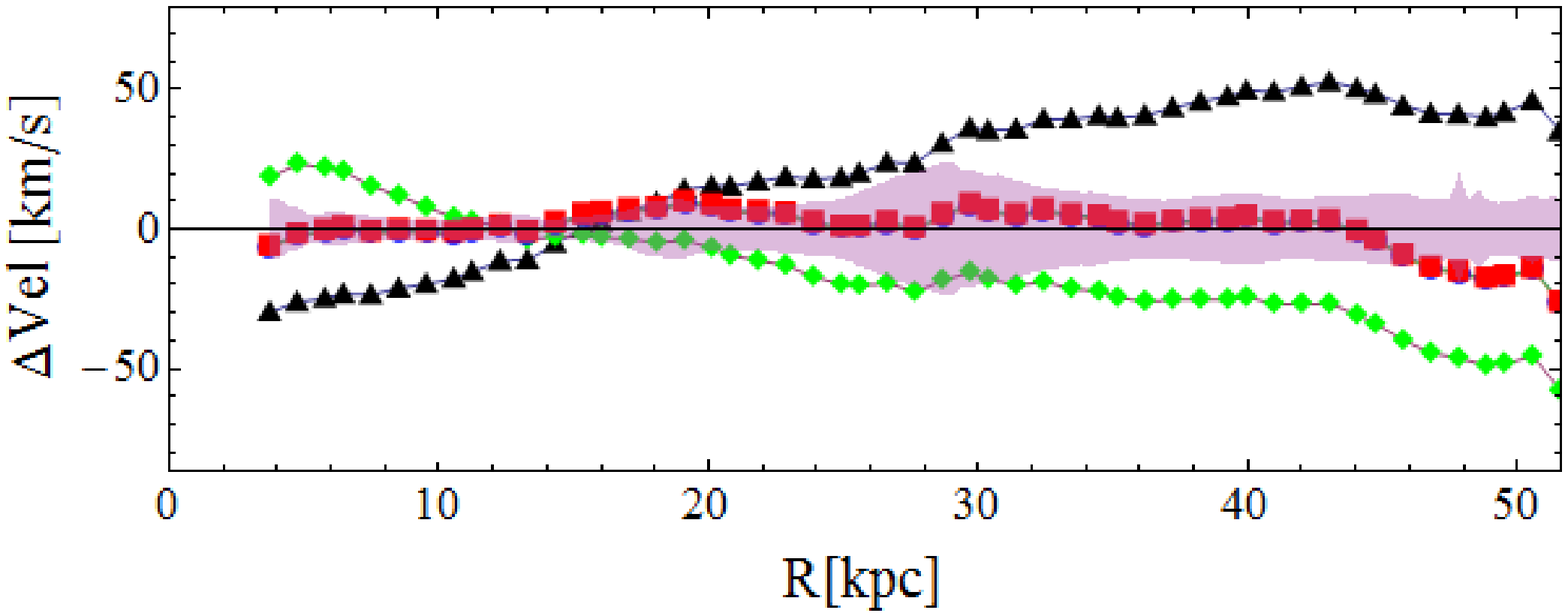}
    \end{tabular}  }  \\
    \subfloat[\footnotesize{Kroupa}]{
    \begin{tabular}[b]{c}
    \includegraphics[width=0.35\textwidth]{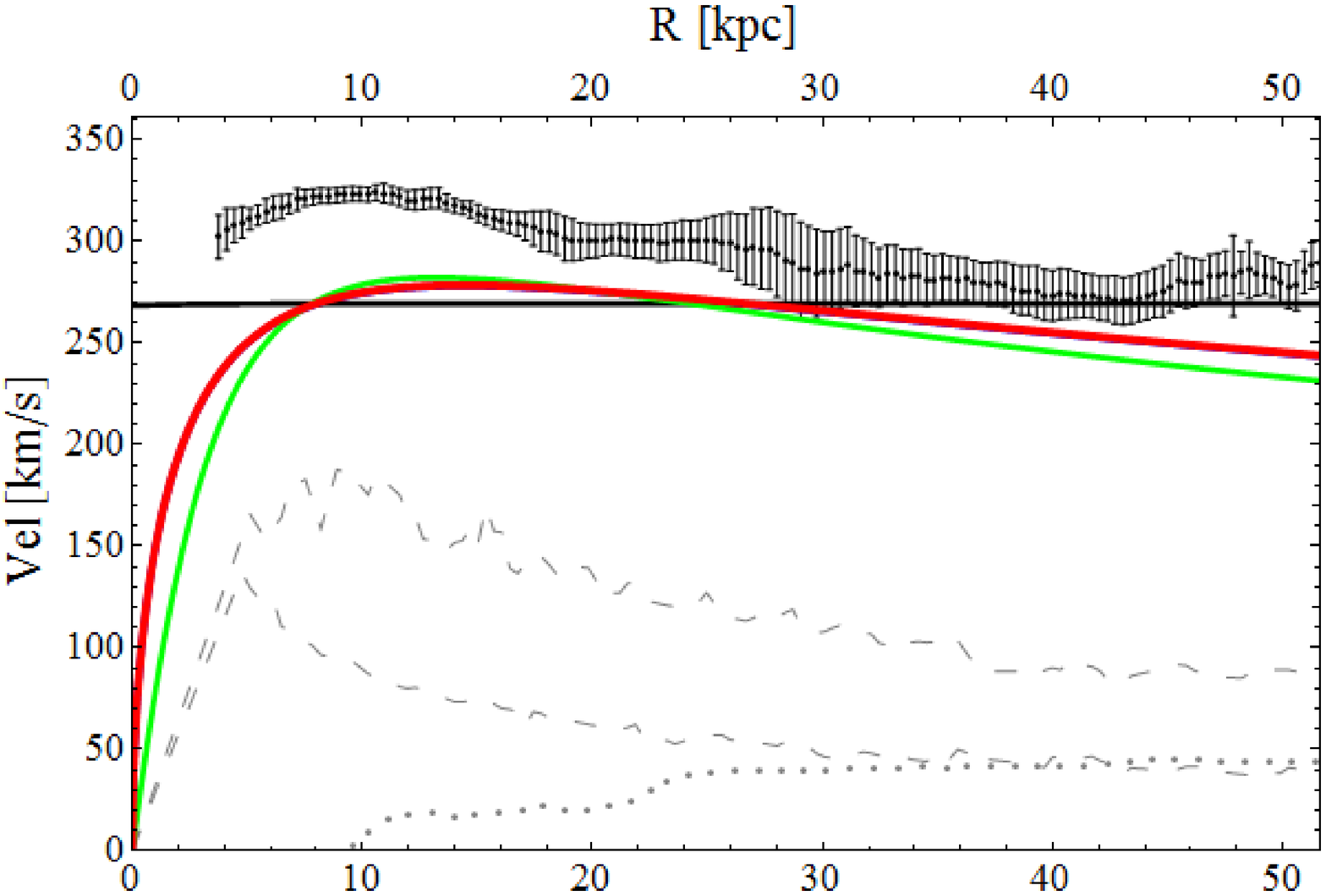} \\
    \includegraphics[width=0.35\textwidth]{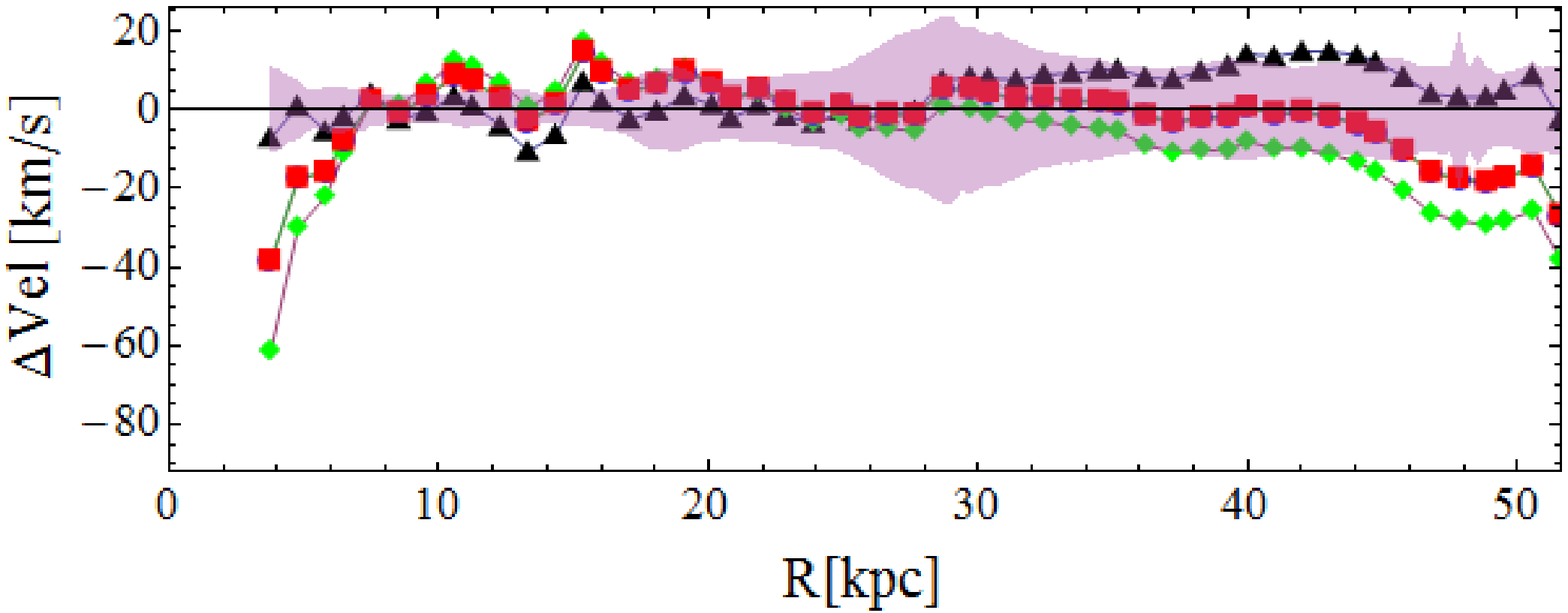}
    \end{tabular}  }
    \subfloat[\footnotesize{diet-Salpeter}]{
    \begin{tabular}[b]{c}
    \includegraphics[width=0.35\textwidth]{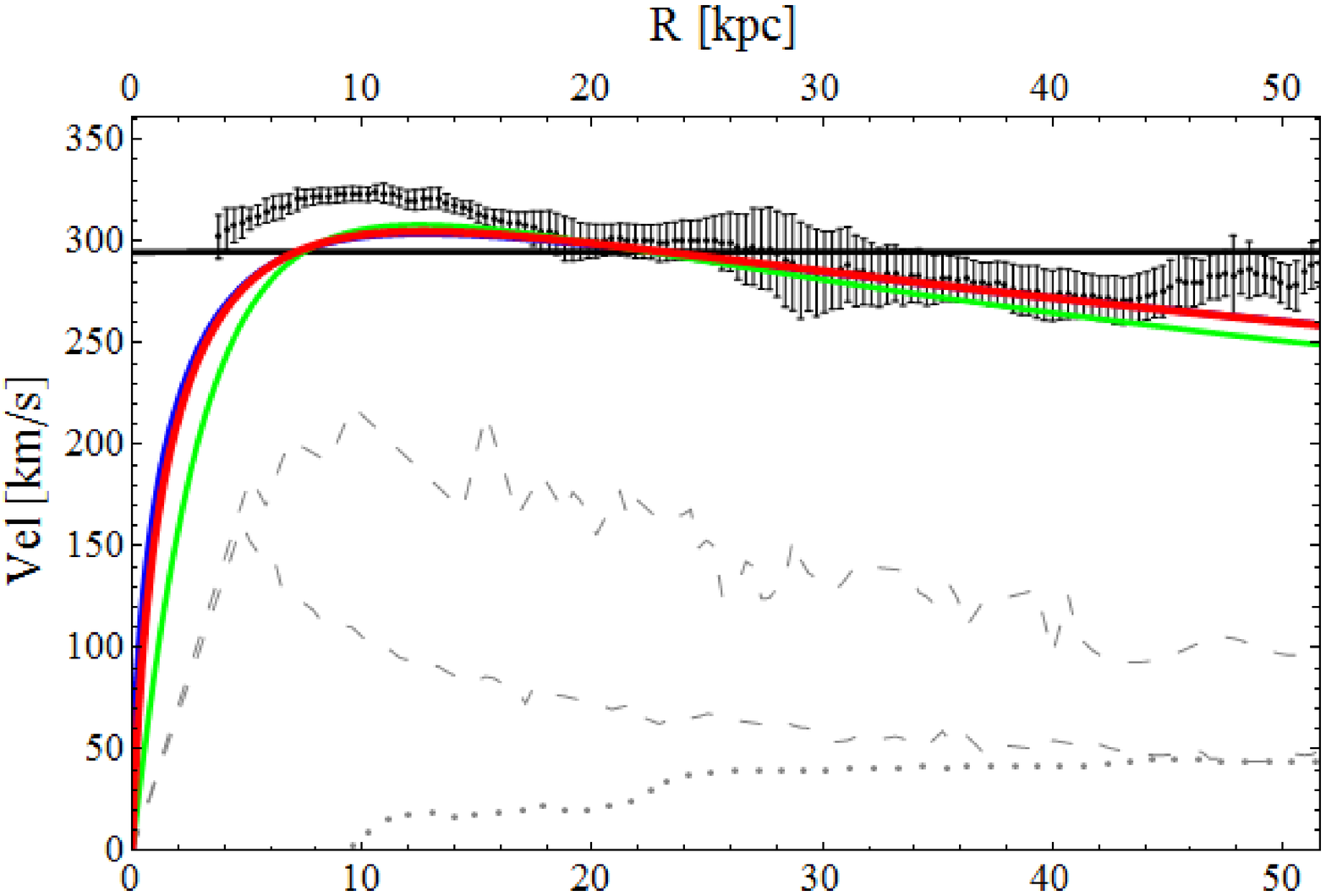} \\
    \includegraphics[width=0.35\textwidth]{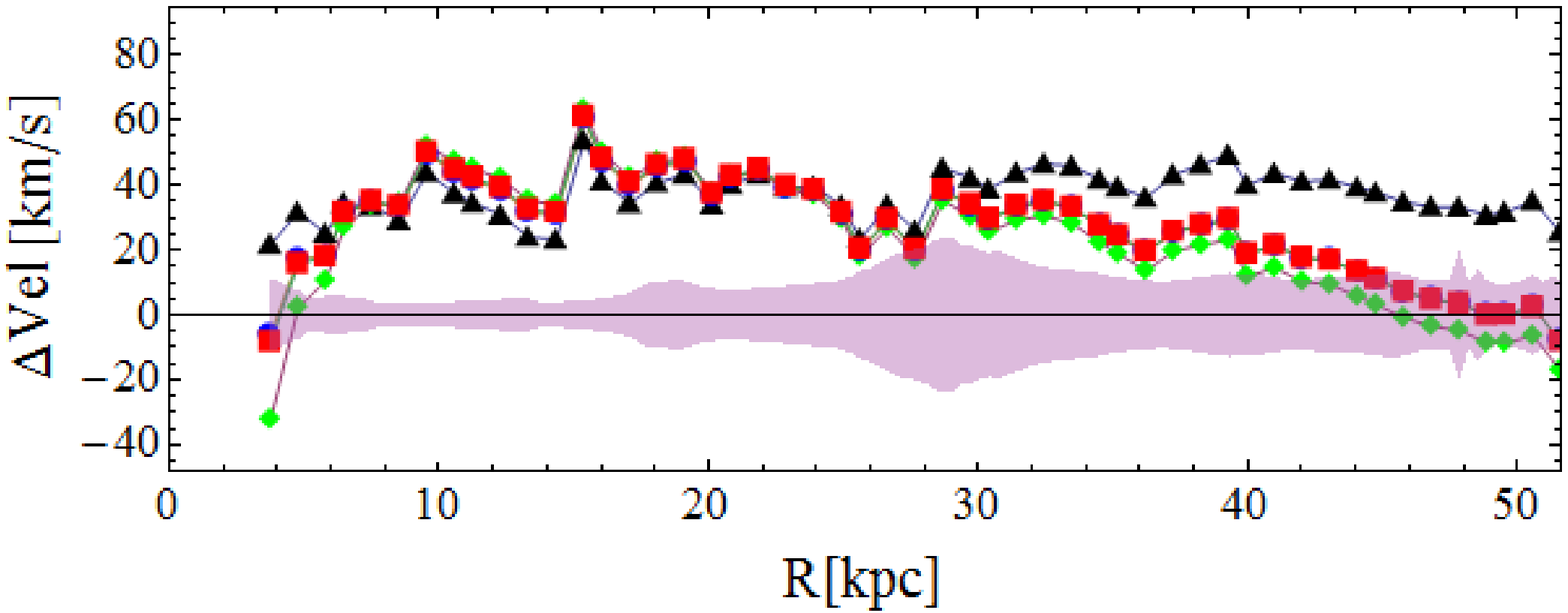}
    \end{tabular}  }
  \caption{\footnotesize{We present the rotation curves for the galaxy NGC 2841. Colors and symbols are as in Fig.\ref{fig:DDO154}. This galaxy shows a inner component of characteristic radius of $R_d = 0.72 {\rm kpc}$ but we consider it as one-component with total mass of the star disk of $M_\star=10^11.04 M_\sun$ and $R_d=0.72 {\rm kpc}$. BDM and NFW presents the best fitting when considering the DM as the only one which contributes to the rotation curve. BDM does not differ much from NFW since $r_c$ is very close to zero. Fits when we considered other mass elements tends to zero the value of the core, but are consistent with $1\sigma$ confidence intervals. From left to right, images from the fit of the galaxy NGC 2841 considering minimal disk, minimal disk+gas, Kroupa, diet-Salpeter. }}
  \label{fig:NGC2841}
\end{figure}

\begin{figure}[h!]
    \subfloat[\footnotesize{Minimal disk}]{
    \begin{tabular}[b]{c}
    \includegraphics[width=0.35\textwidth]{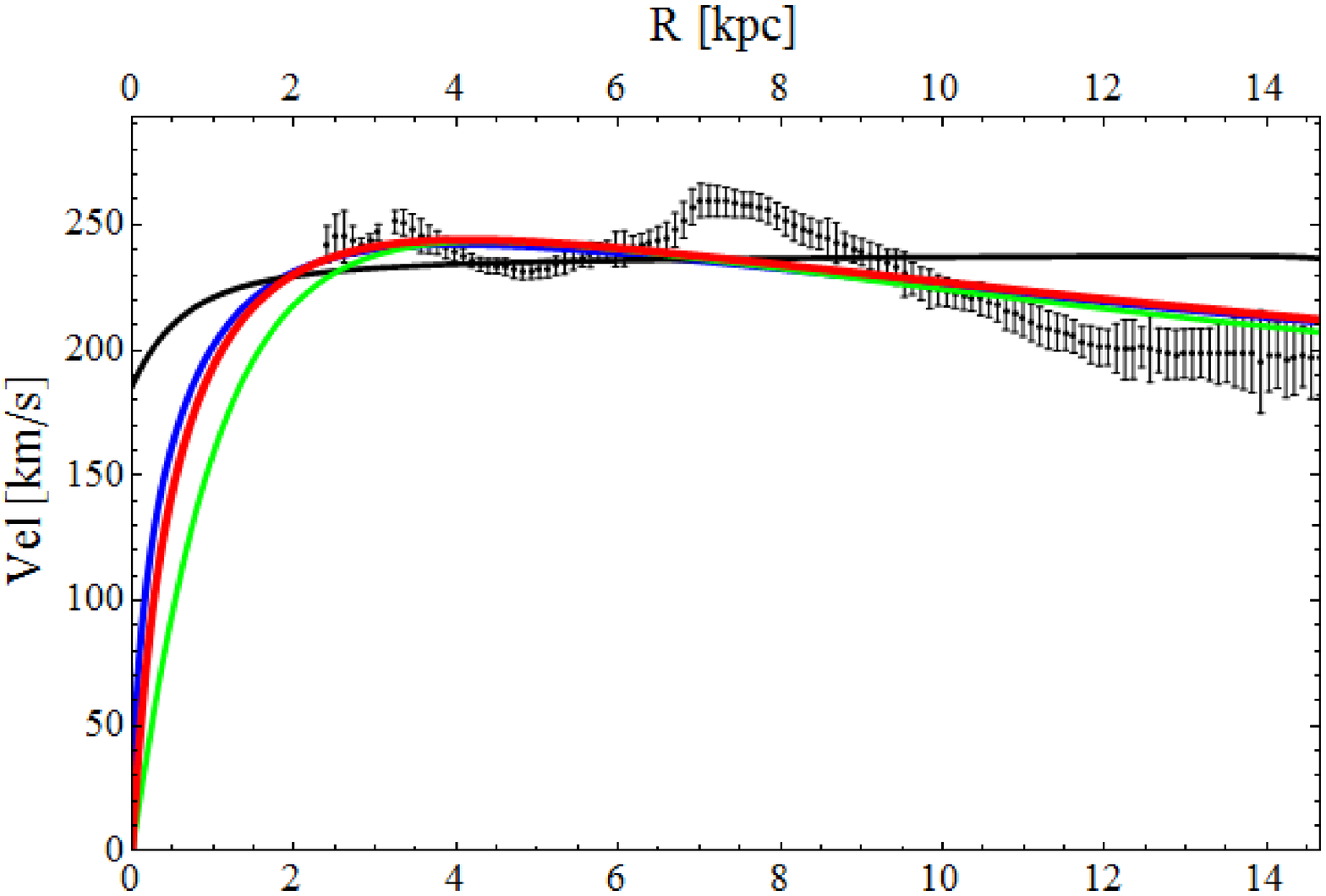} \\
    \includegraphics[width=0.35\textwidth]{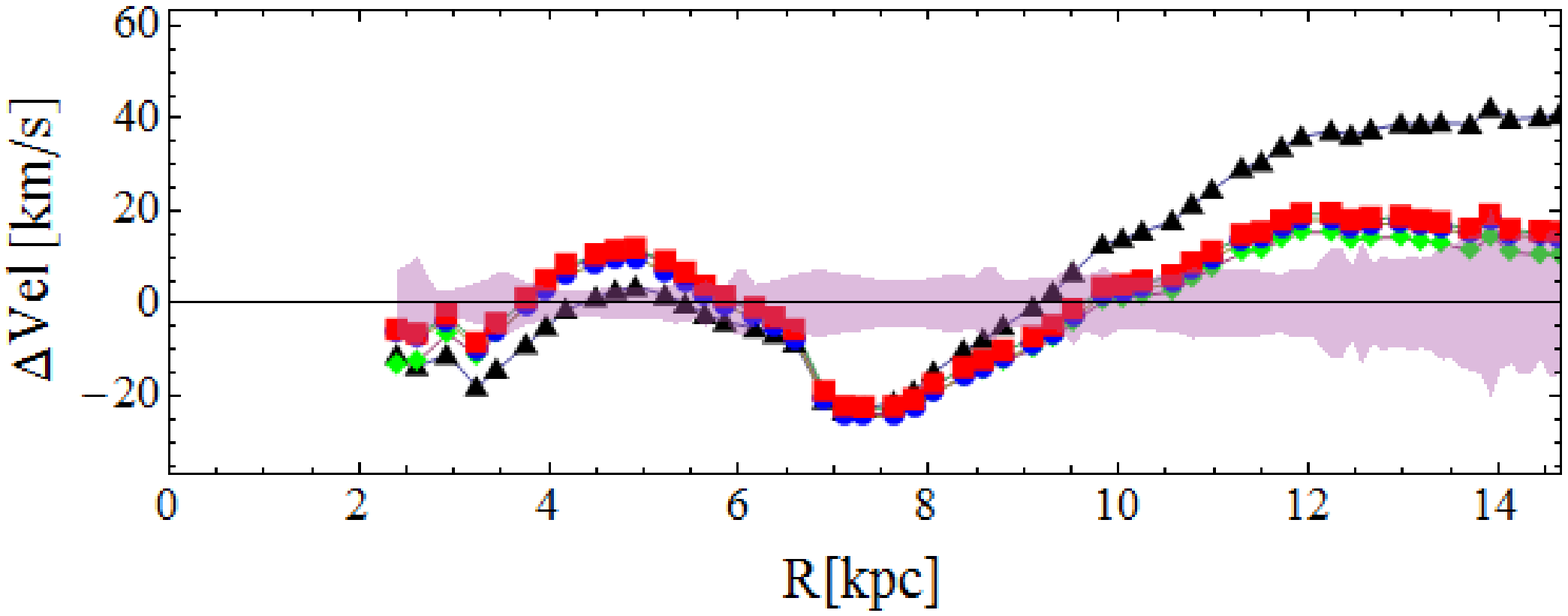}
    \end{tabular}  }
    \subfloat[\footnotesize{Min. disk + Gas}]{
    \begin{tabular}[b]{c}
    \includegraphics[width=0.35\textwidth]{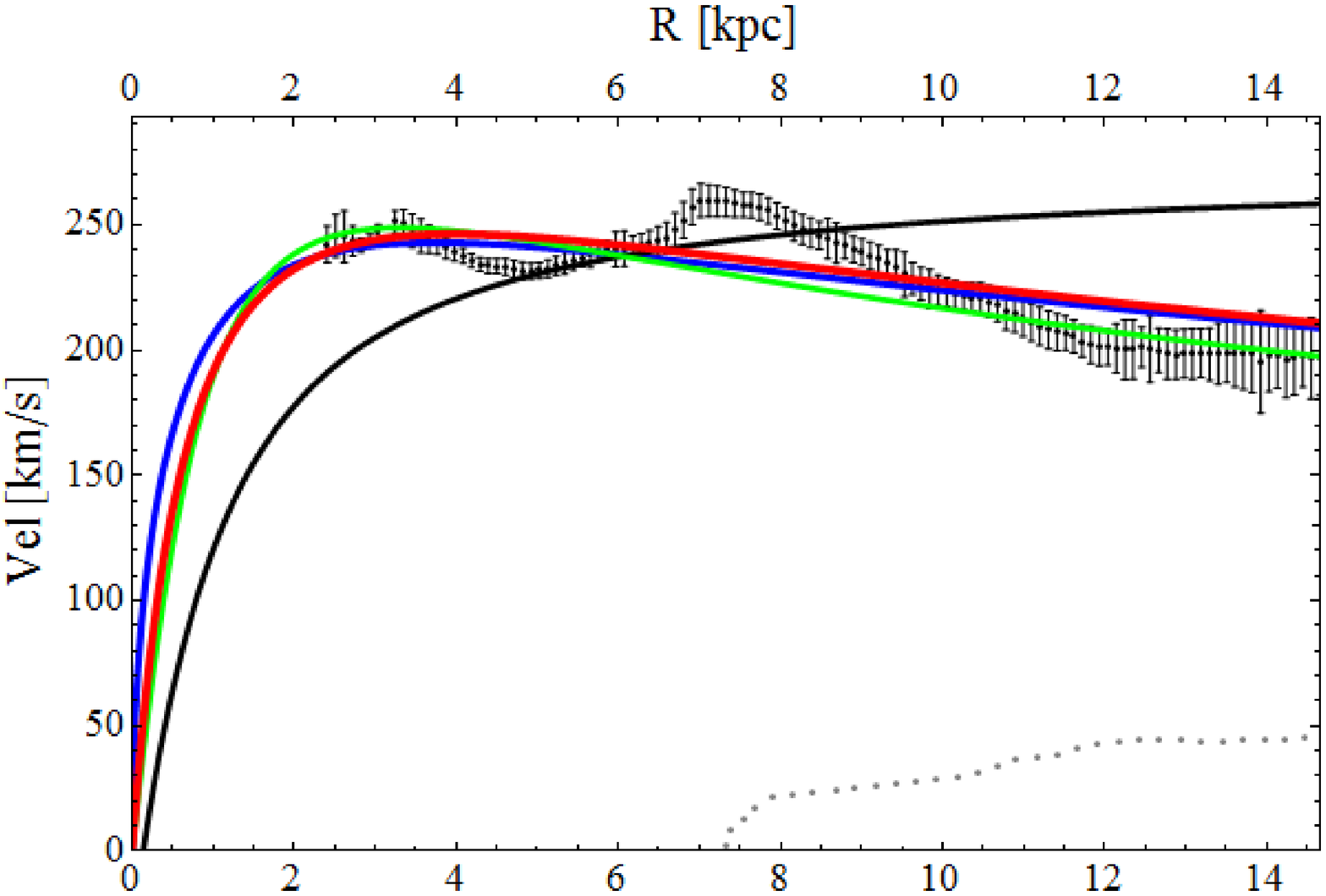} \\
    \includegraphics[width=0.35\textwidth]{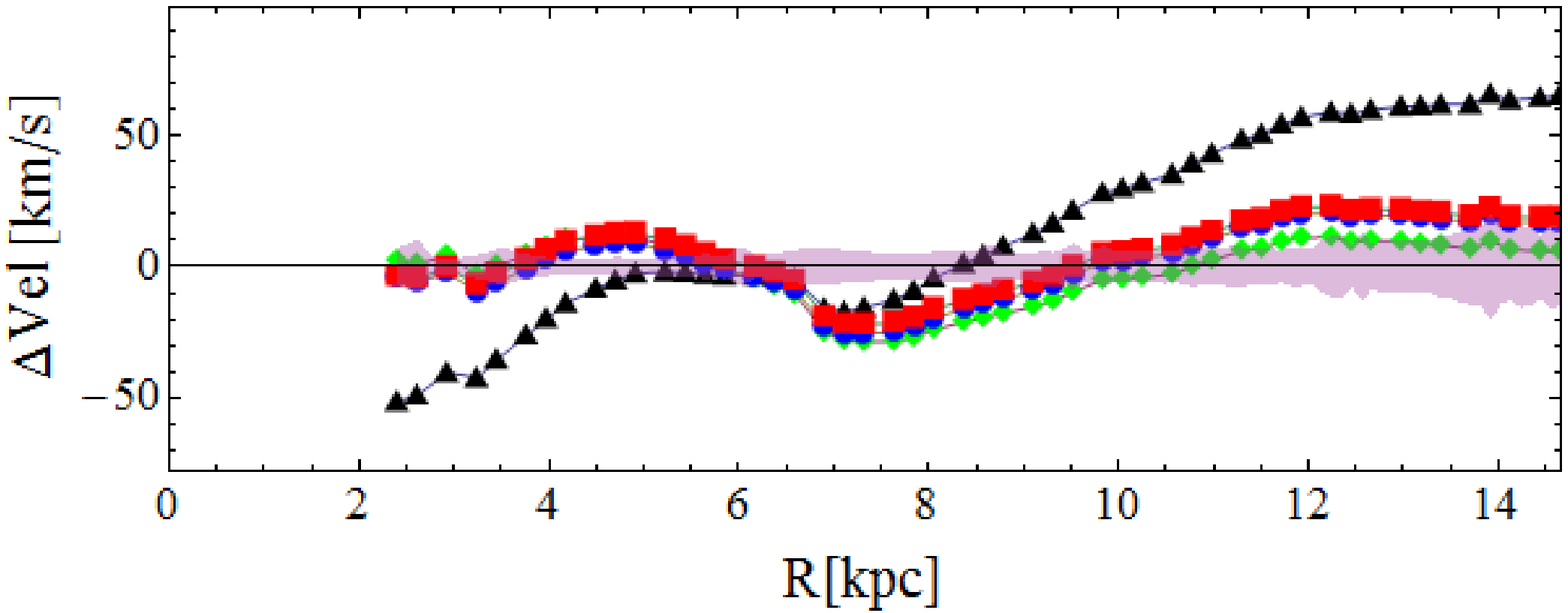}
    \end{tabular}  }  \\
    \subfloat[\footnotesize{Kroupa}]{
    \begin{tabular}[b]{c}
    \includegraphics[width=0.35\textwidth]{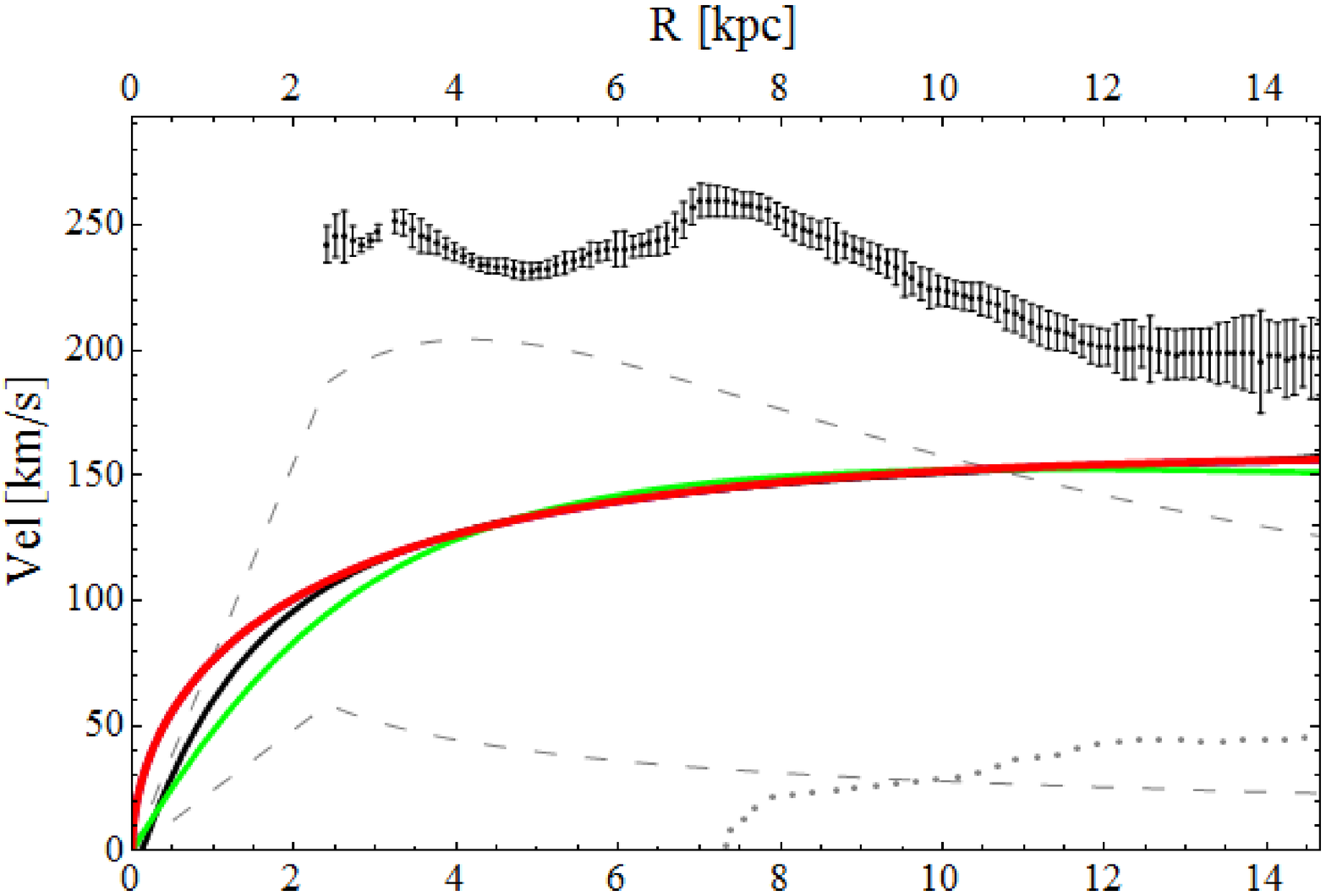} \\
    \includegraphics[width=0.35\textwidth]{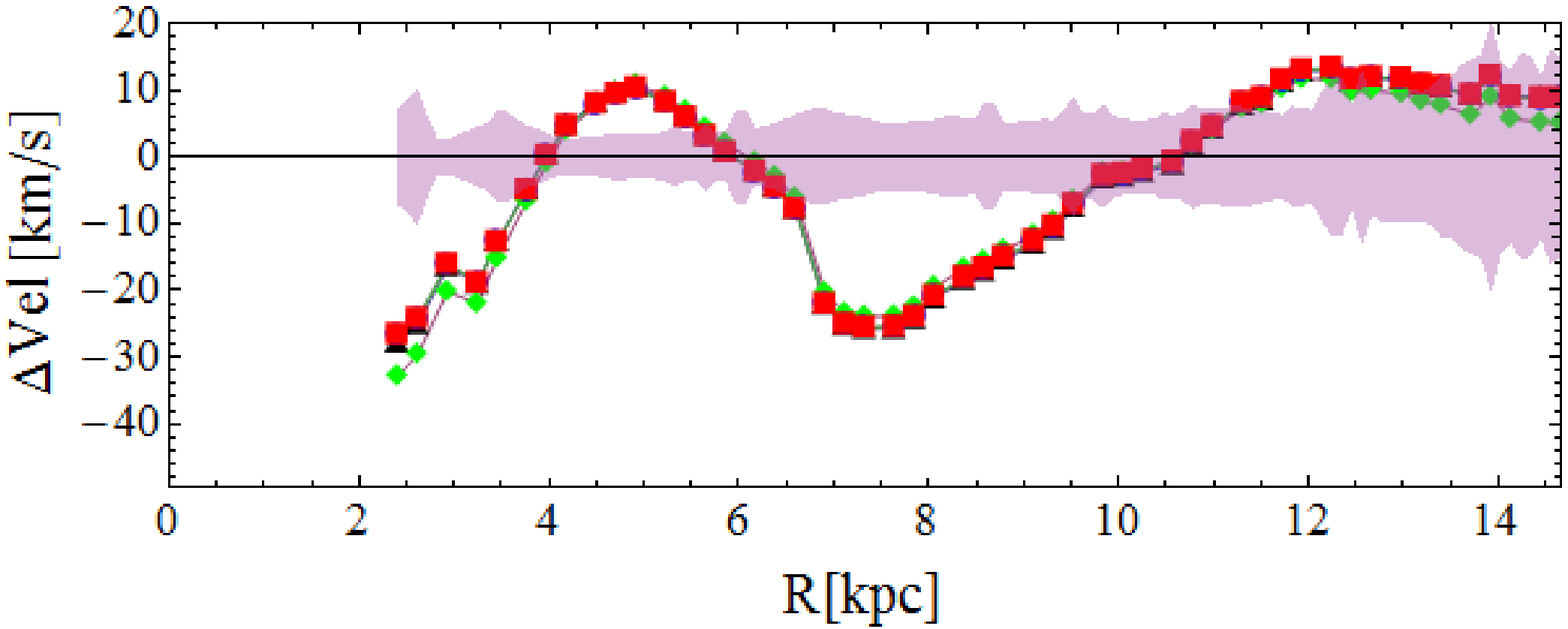}
    \end{tabular}  }
    \subfloat[\footnotesize{diet-Salpeter}]{
    \begin{tabular}[b]{c}
    \includegraphics[width=0.35\textwidth]{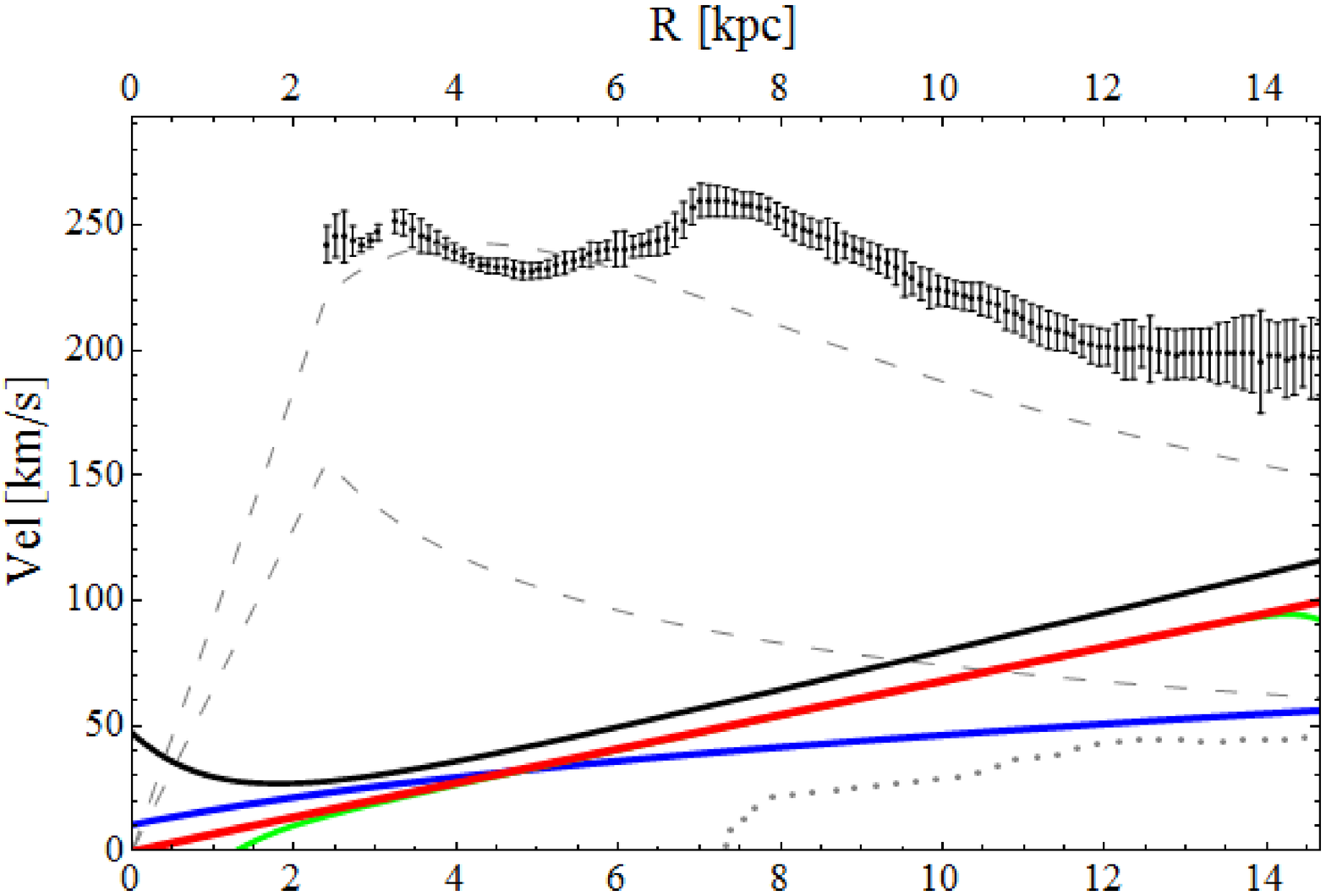} \\
    \includegraphics[width=0.35\textwidth]{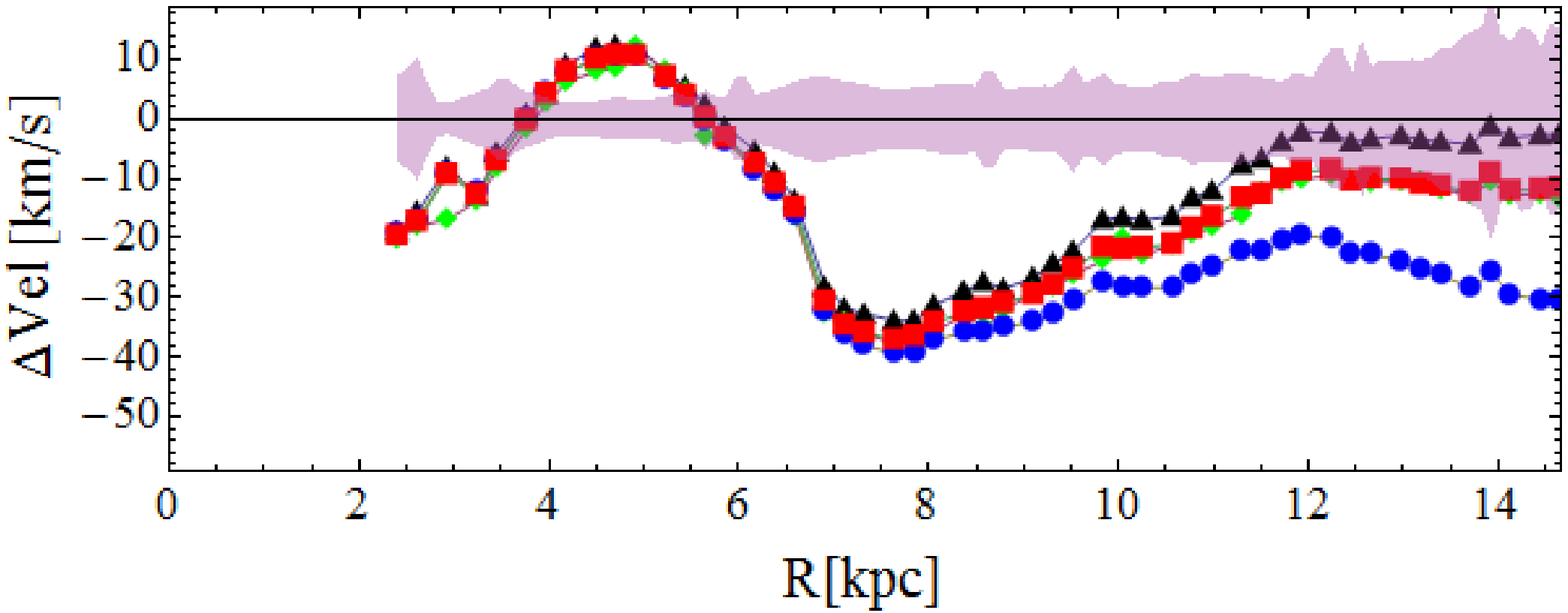}
    \end{tabular}  }
  \caption{\footnotesize{We display the rotation curves for the galaxy NGC 3031. Colors and symbols are as in Fig.\ref{fig:DDO154}. This galaxy is the proto-typical grand design spiral. The profiles show evidence for a central small component. To describe the outer disk we use the parameters $\mu_0=15.9$ and $R_d = 1.94$. The (J - K) color profile shows no significant color gradient. We assume a constant $\gs = 0.8$ for the disk. We can not extract information of these galaxy since we do not have observational points close to the center. From left to right, images from the fit of the galaxy NGC 3031 considering minimal disk, minimal disk+gas, Kroupa, diet-Salpeter. }}
  \label{fig:NGC3031}
\end{figure}

\begin{figure}[h!]
    \subfloat[\footnotesize{Minimal disk}]{
    \begin{tabular}[b]{c}
    \includegraphics[width=0.35\textwidth]{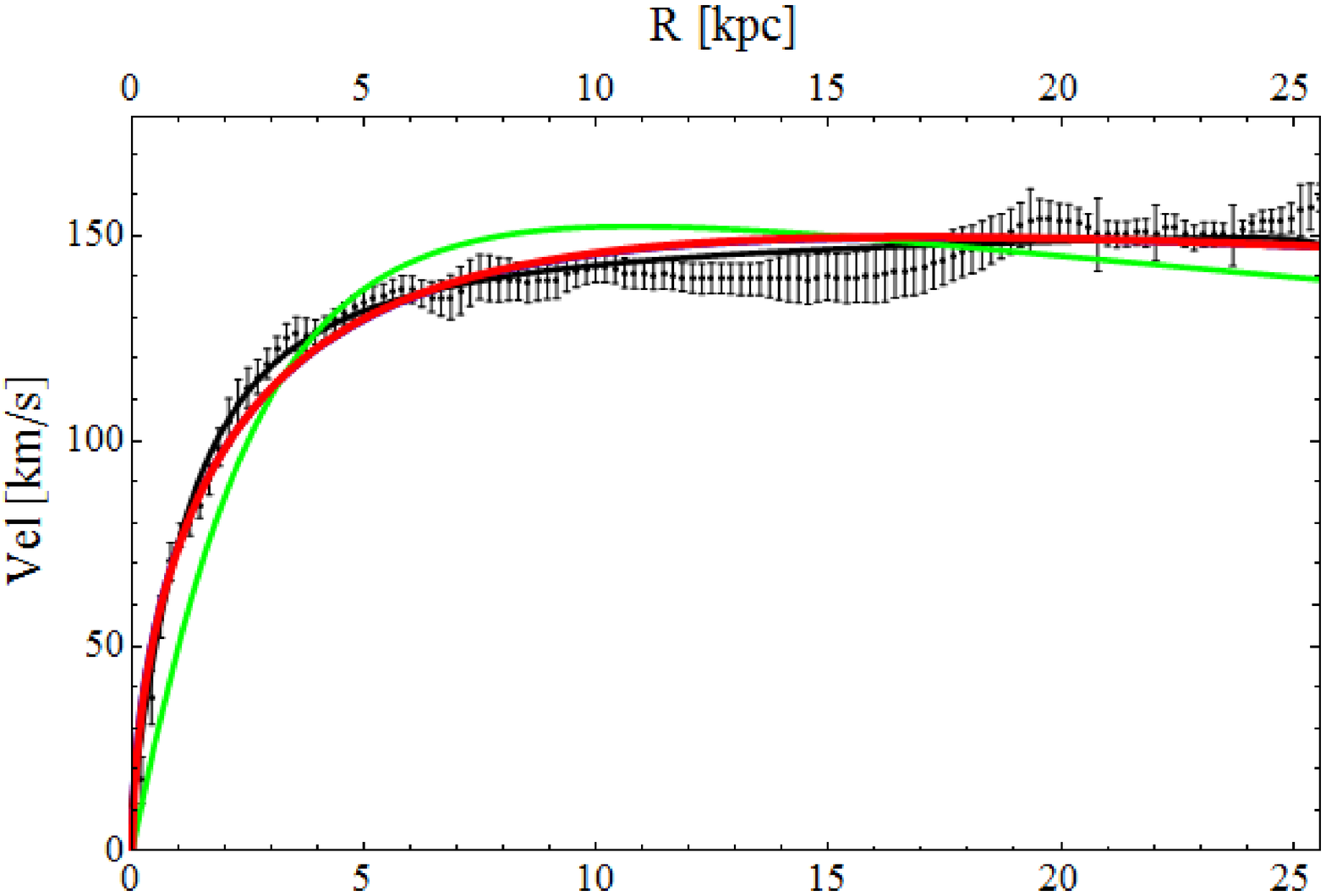} \\
    \includegraphics[width=0.35\textwidth]{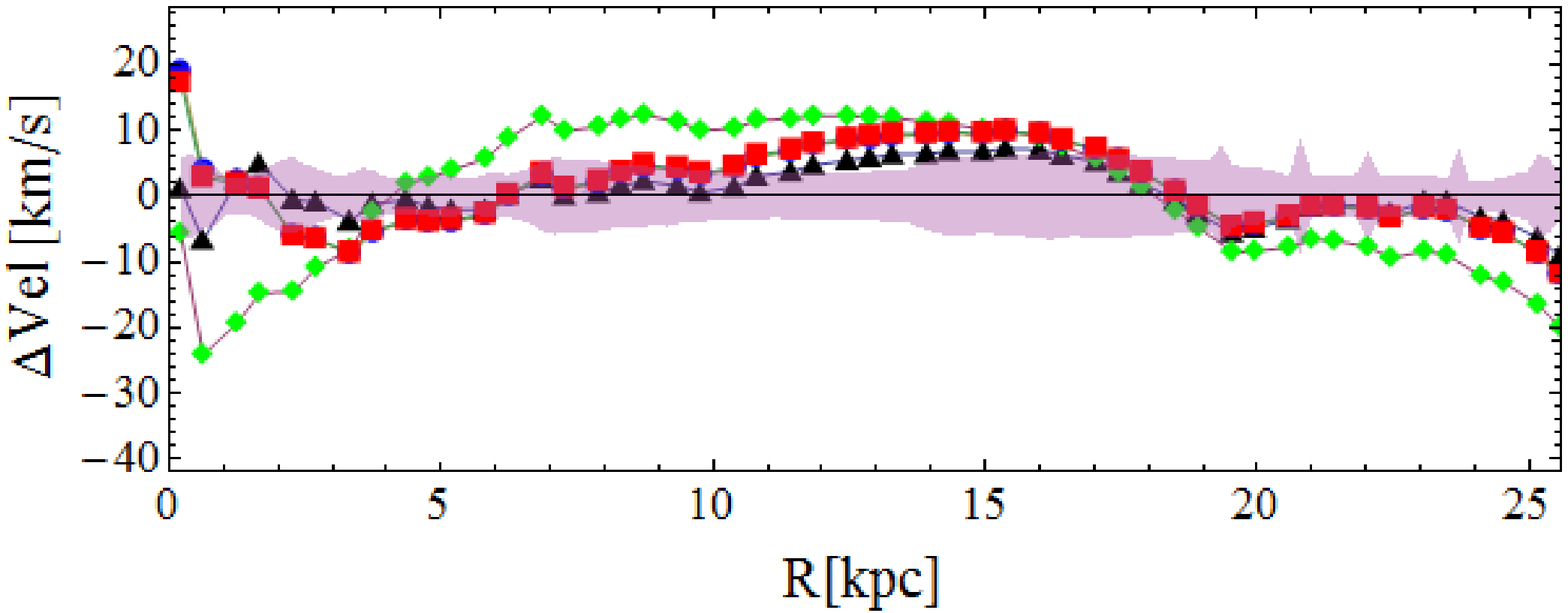}
    \end{tabular}  }
    \subfloat[\footnotesize{Min. disk + Gas}]{
    \begin{tabular}[b]{c}
    \includegraphics[width=0.35\textwidth]{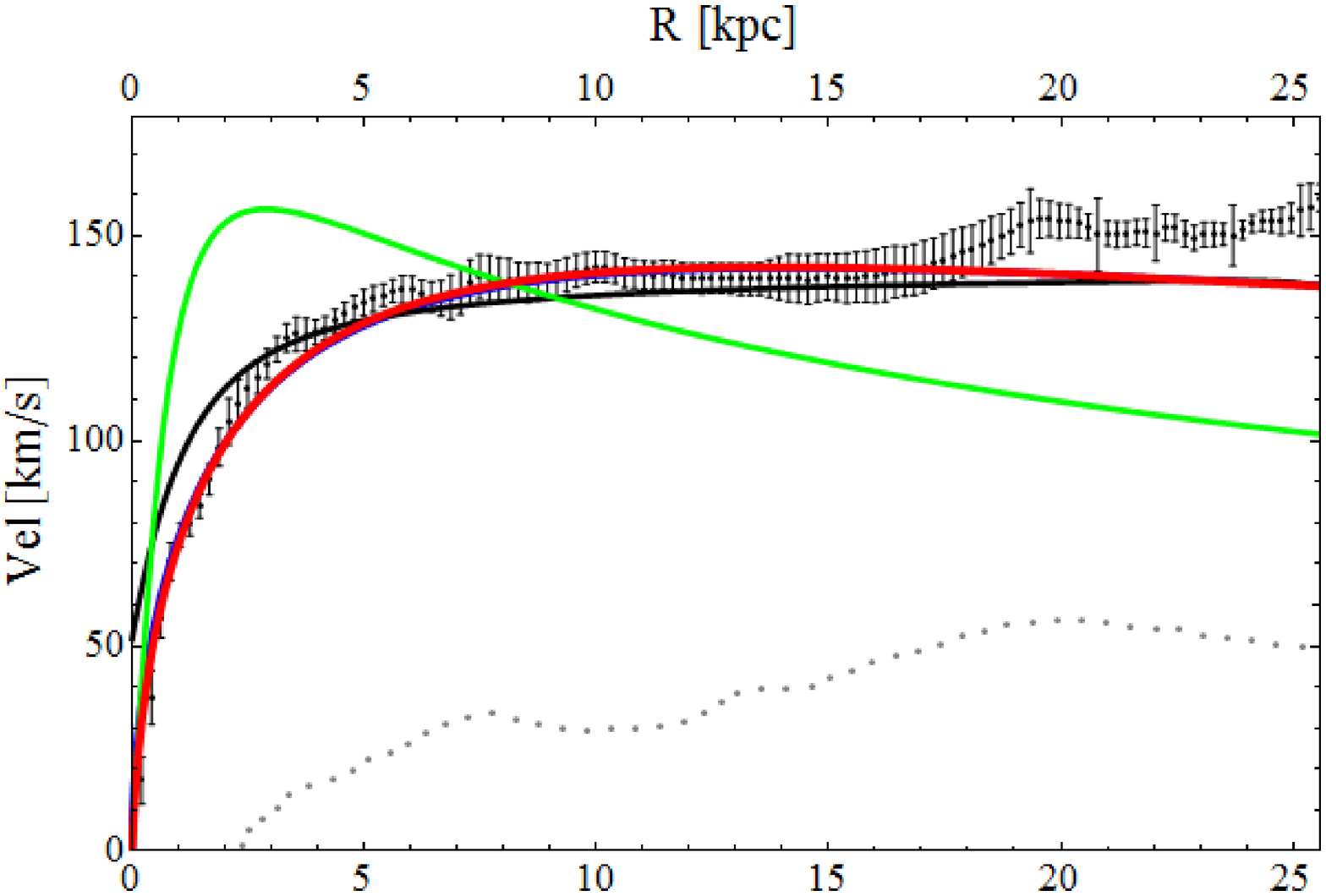} \\
    \includegraphics[width=0.35\textwidth]{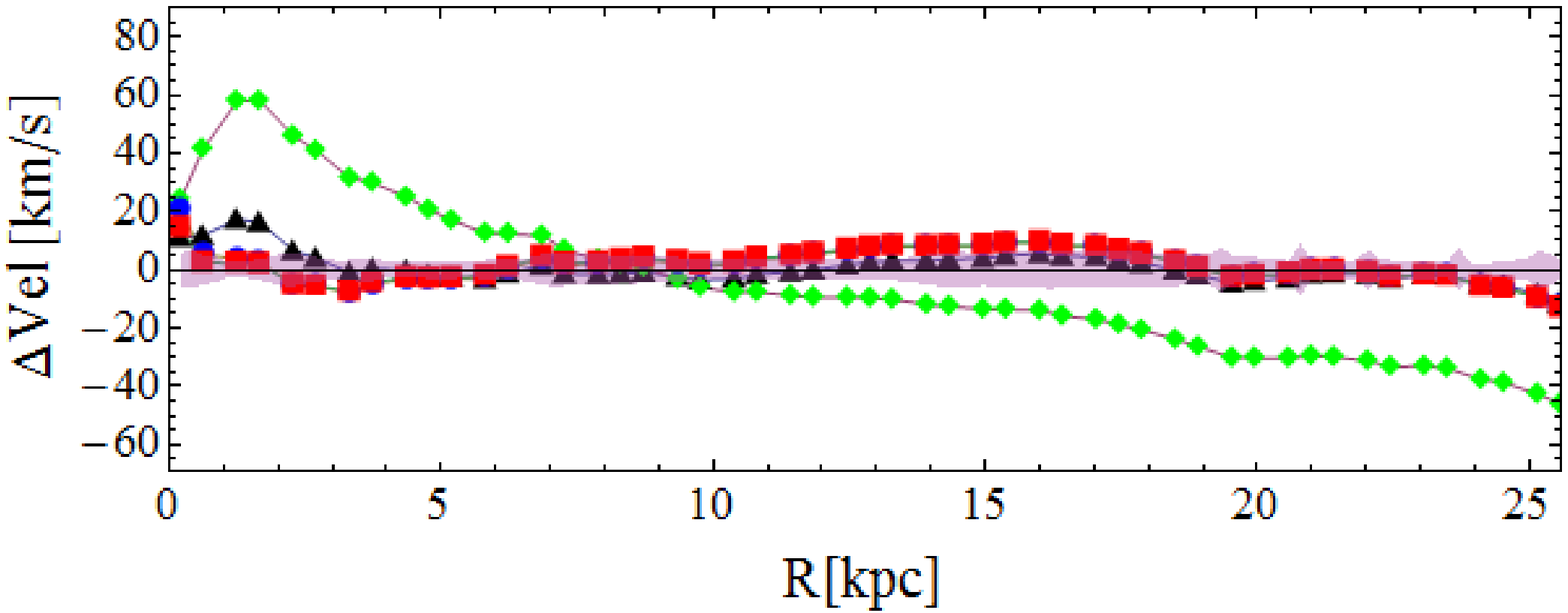}
    \end{tabular}  }  \\
    \subfloat[\footnotesize{Kroupa}]{
    \begin{tabular}[b]{c}
    \includegraphics[width=0.35\textwidth]{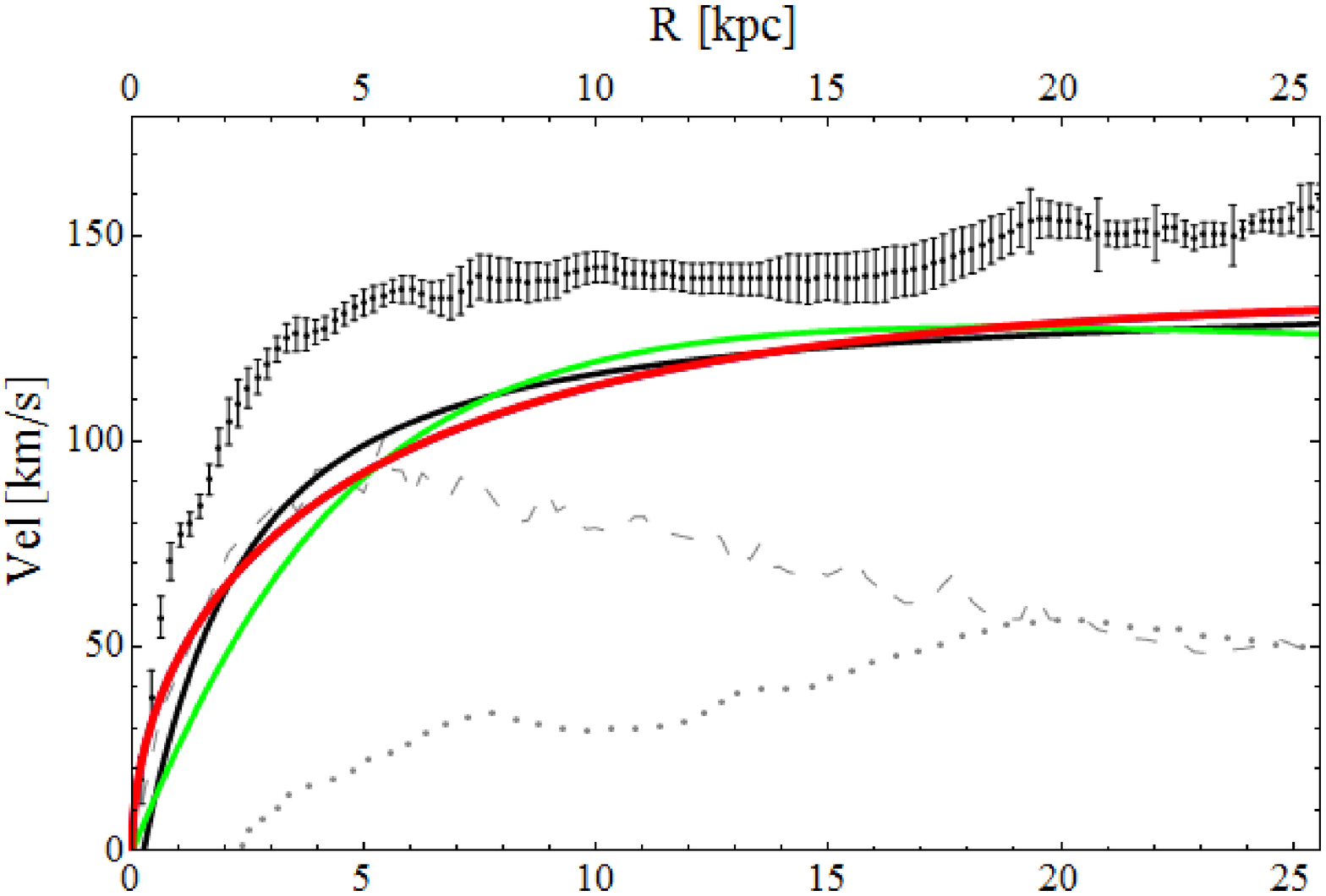} \\
    \includegraphics[width=0.35\textwidth]{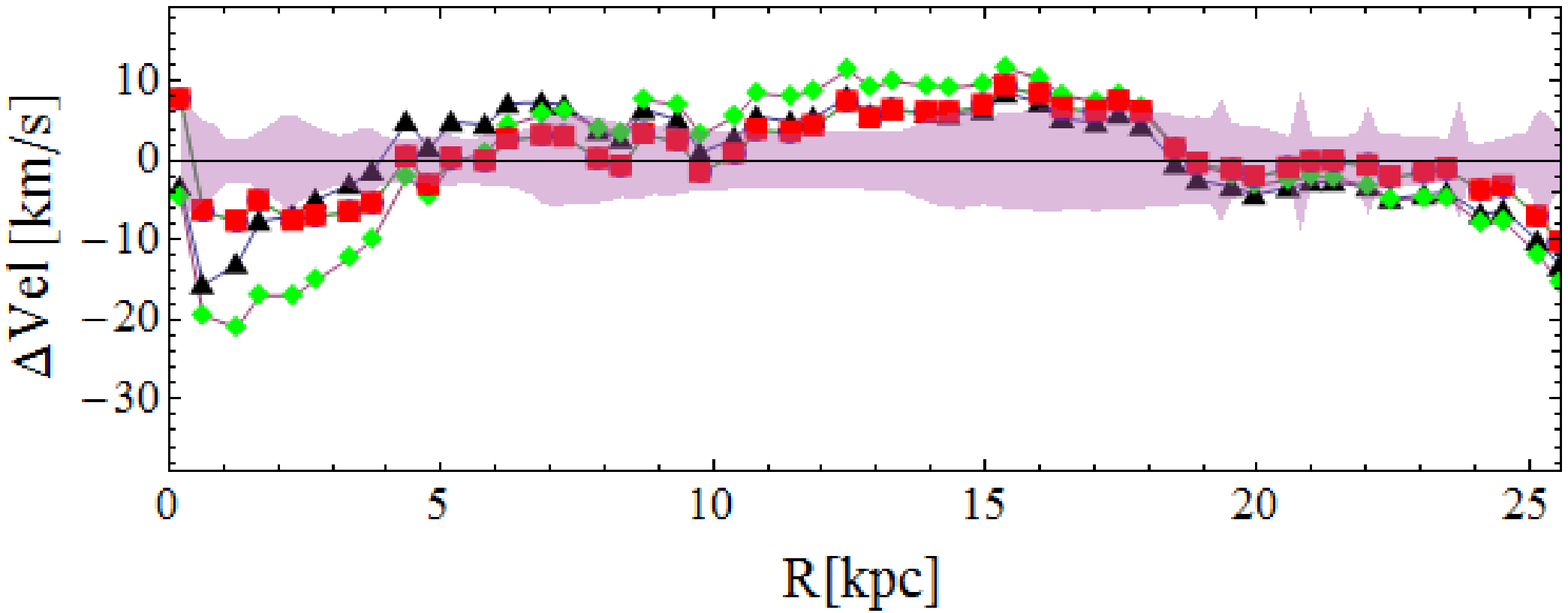}
    \end{tabular}  }
    \subfloat[\footnotesize{Diet-Salpeter}]{
    \begin{tabular}[b]{c}
    \includegraphics[width=0.35\textwidth]{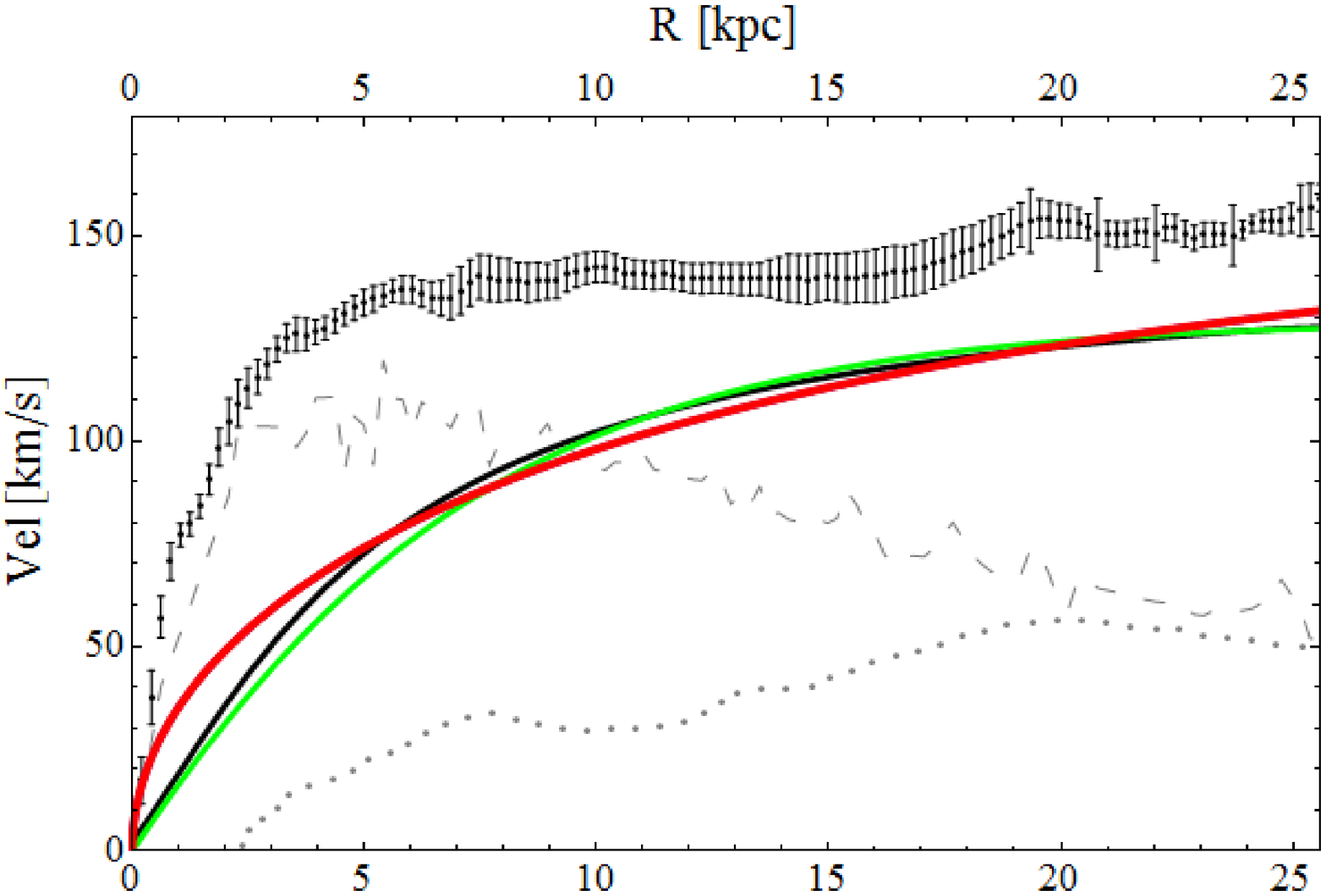} \\
    \includegraphics[width=0.35\textwidth]{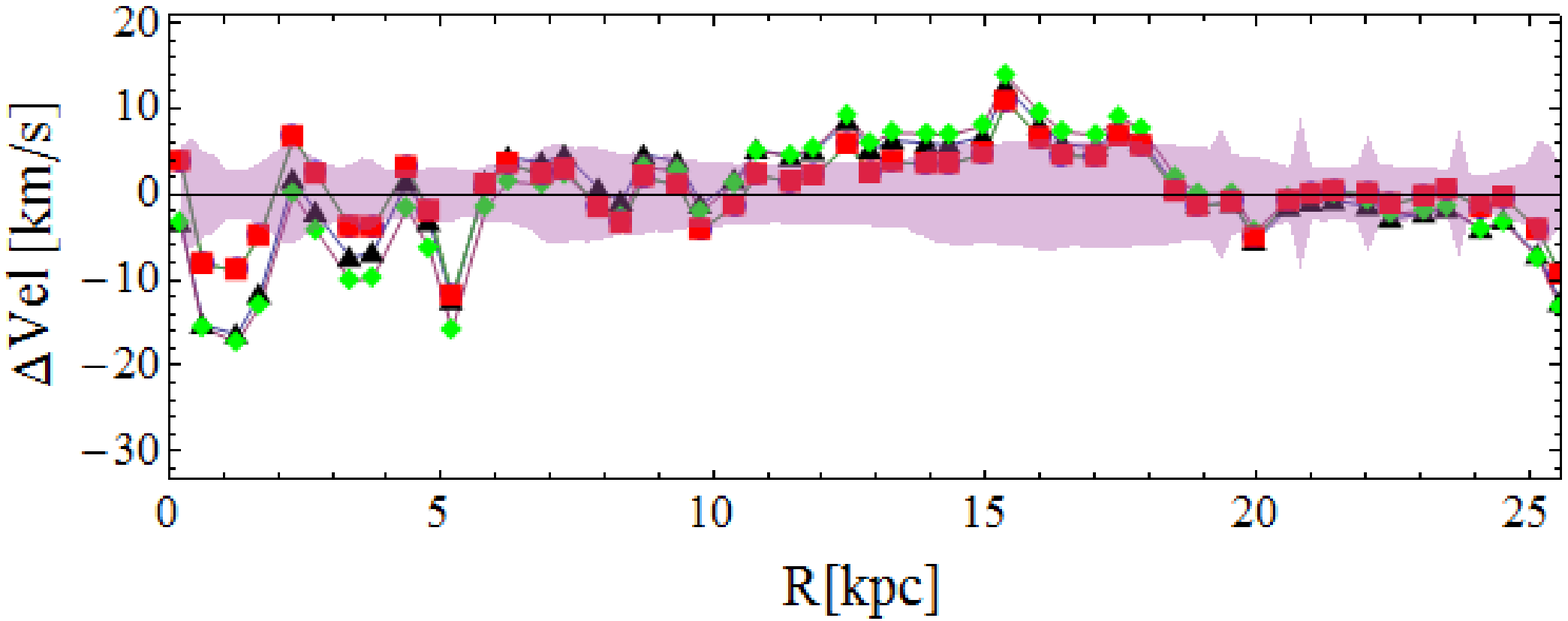}
    \end{tabular}  }
  \caption{\footnotesize{The rotation curves for the galaxy NGC 3621. Colors and symbols are as in Fig.\ref{fig:DDO154}. There is no a convincible evidence to believe that this galaxy has a bulge, neither the color gradient, morphology or deviation for a single exponential disk. Therefore we assume a single-component disk to larger radii using an exponential of mass $M_\star = 1.9 10^{10} M_\sun$  and $R_d = 2.61$ {\rm kpc}. It is worth noticing how close are the predicted values of the BDM and NFW to the expected values for the  maximum stellar disk and at the same time they provided the best fit to this galaxy. The value of the core is underestimated the more mass components are considered. Since the galaxy present sufficient data we proceed with the inner analysis obtaining consistent results with the minimal disk analysis.From left to right, images from the fit of the galaxy NGC 3621 considering minimal disk, minimal disk+gas, Kroupa, diet-Salpeter. }}
  \label{fig:NGC3621}
\end{figure}

\begin{figure}[h!]
    \subfloat[\footnotesize{Minimal disk}]{
    \begin{tabular}[b]{c}
    \includegraphics[width=0.35\textwidth]{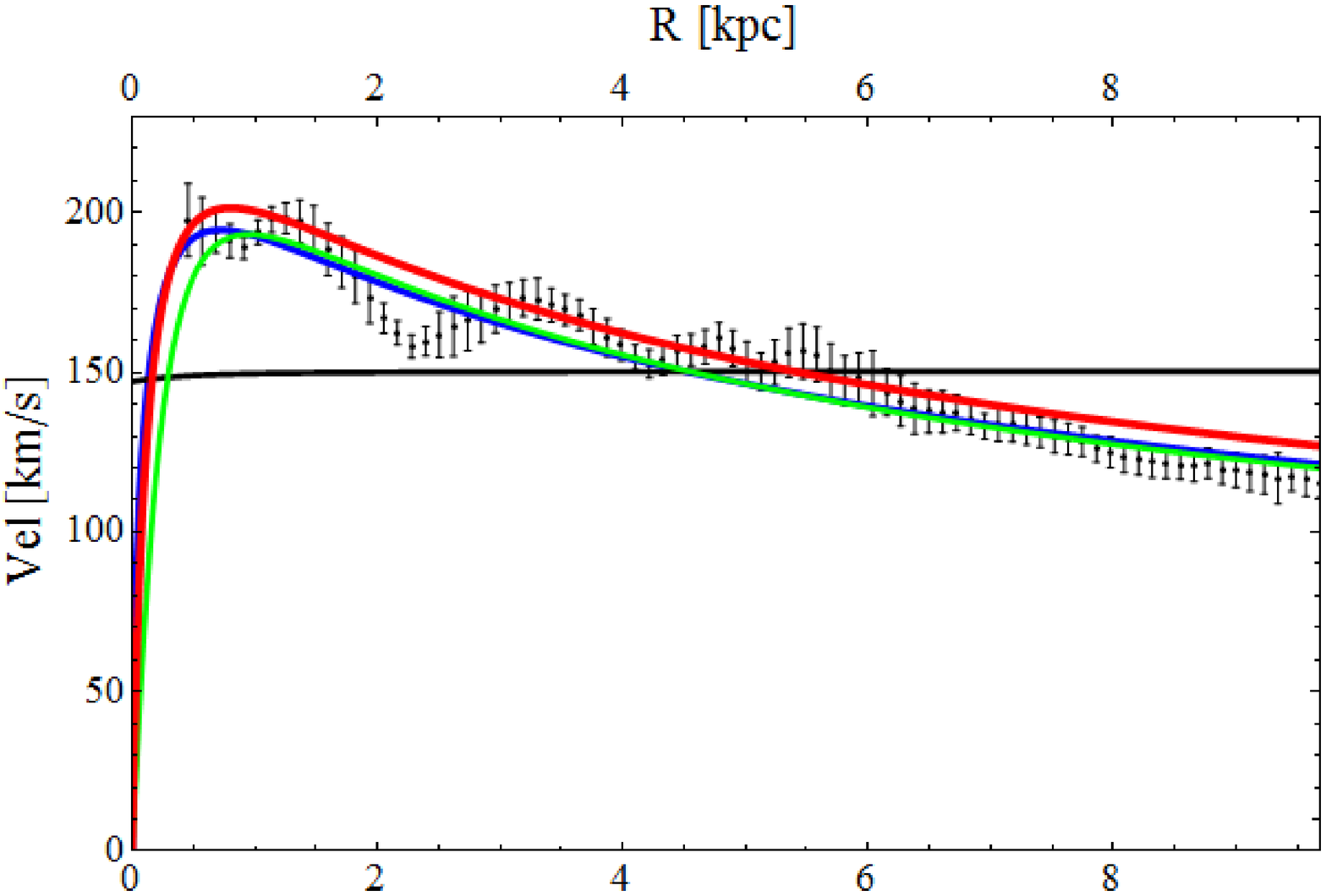} \\
    \includegraphics[width=0.35\textwidth]{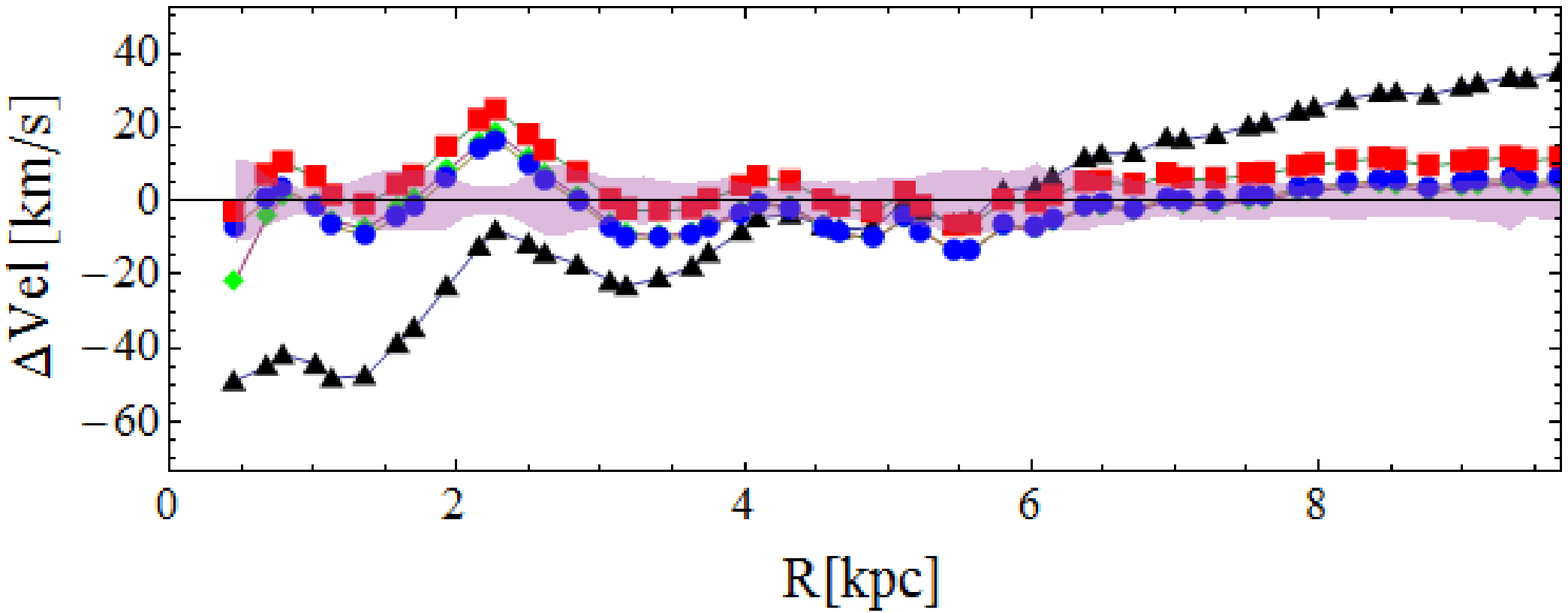}
    \end{tabular}  }
    \subfloat[\footnotesize{Min. disk + Gas}]{
    \begin{tabular}[b]{c}
    \includegraphics[width=0.35\textwidth]{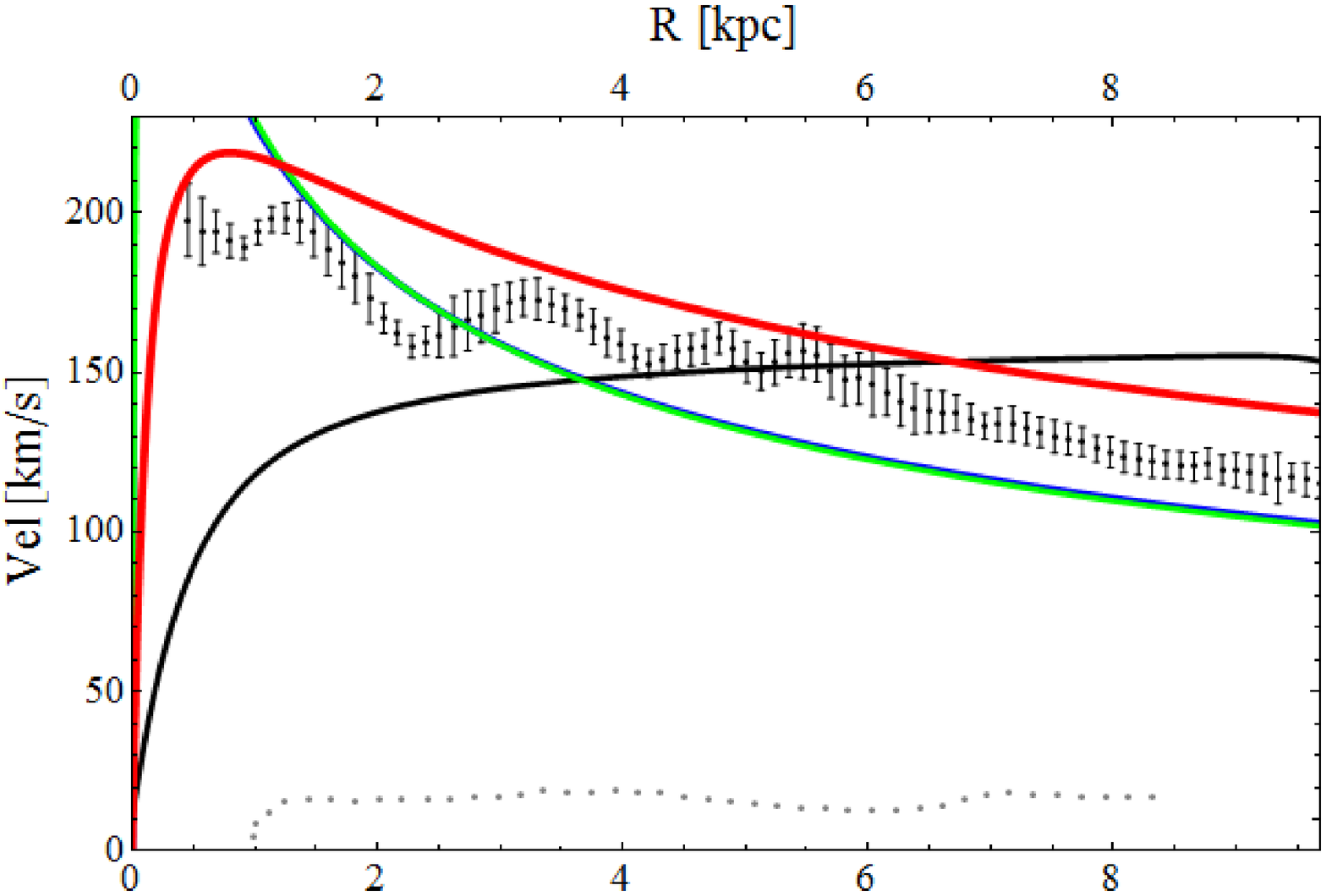} \\
    \includegraphics[width=0.35\textwidth]{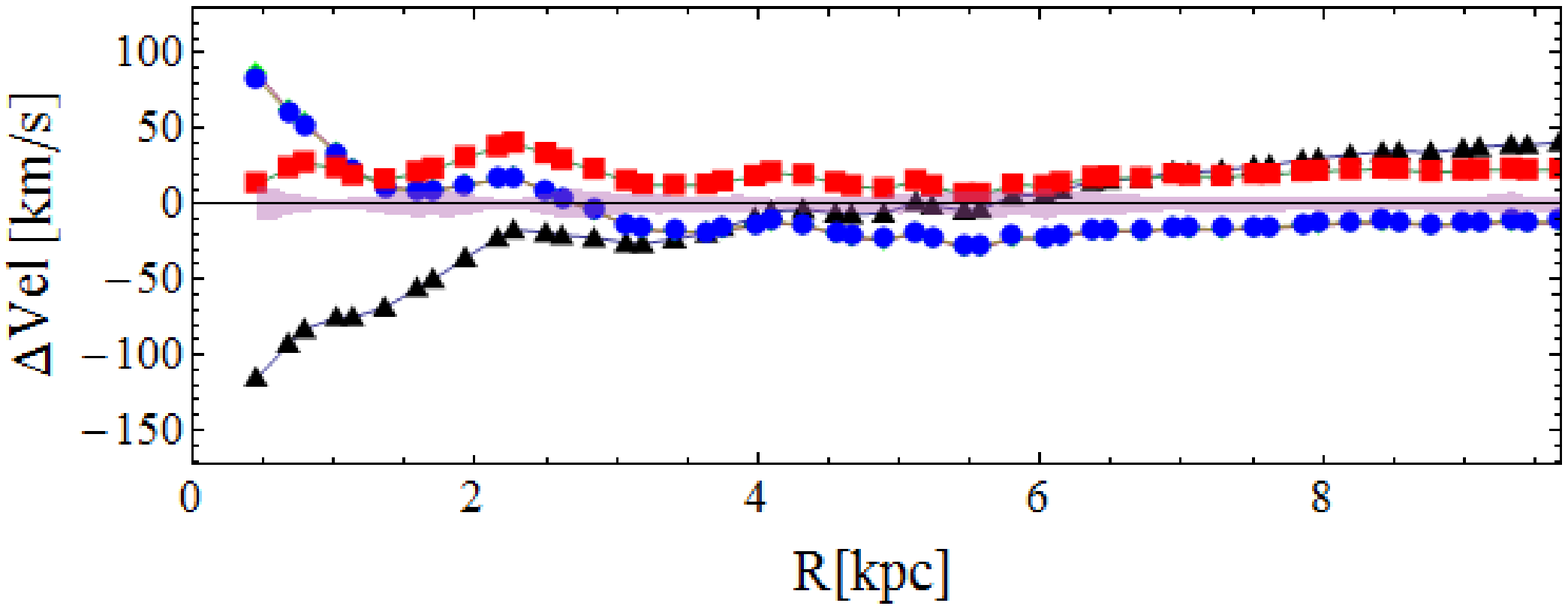}
    \end{tabular}  }  \\
    \subfloat[\footnotesize{Kroupa}]{
    \begin{tabular}[b]{c}
    \includegraphics[width=0.35\textwidth]{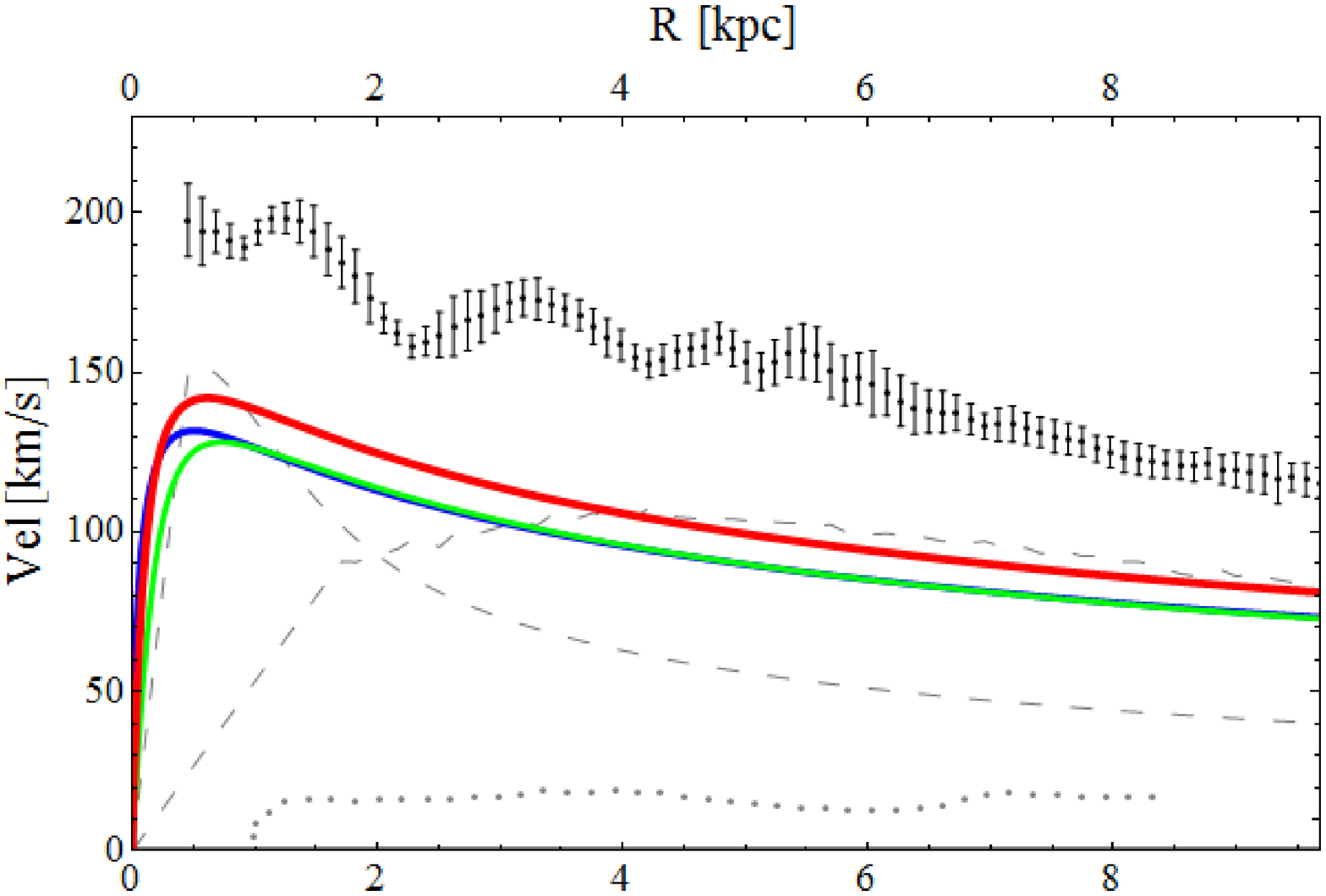} \\
    \includegraphics[width=0.35\textwidth]{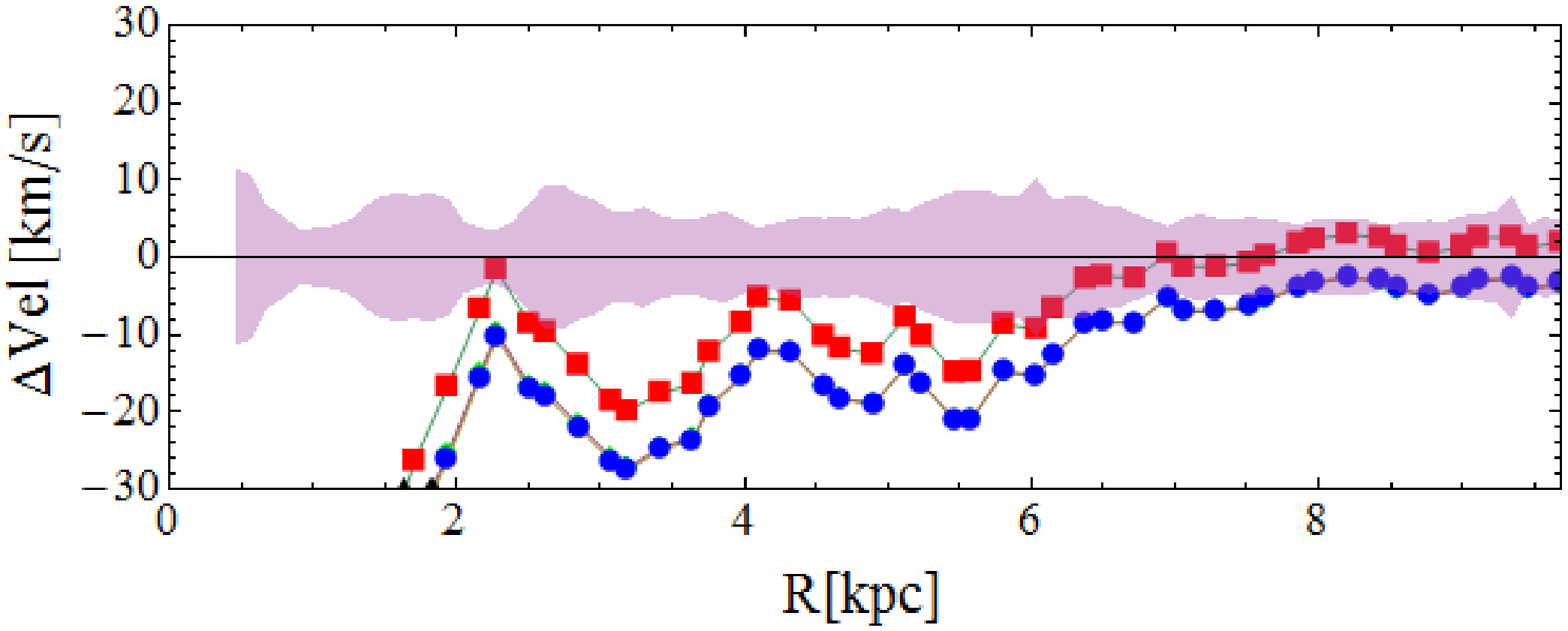}
    \end{tabular}  }
    \subfloat[\footnotesize{diet-Salpeter}]{
    \begin{tabular}[b]{c}
    \includegraphics[width=0.35\textwidth]{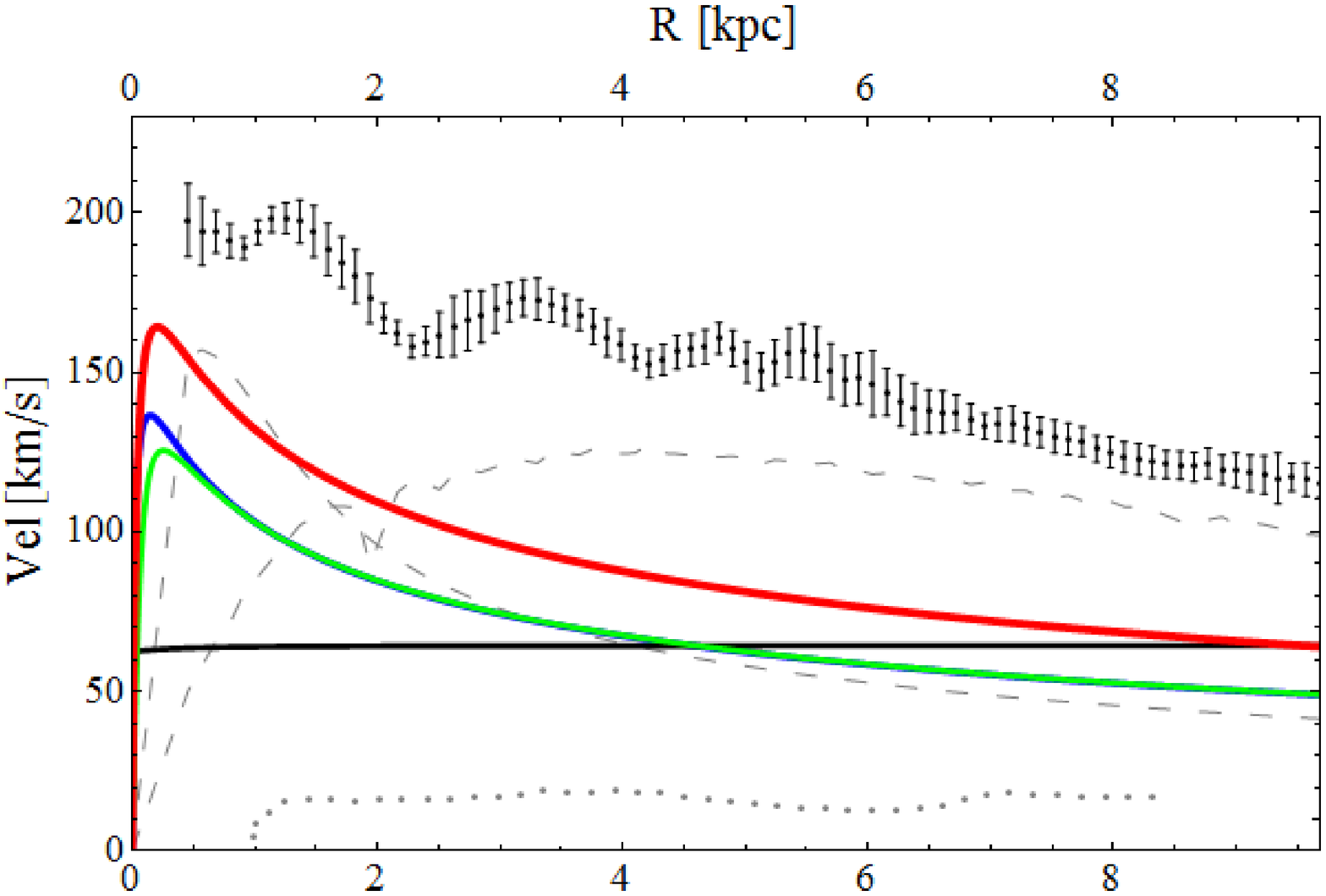} \\
    \includegraphics[width=0.35\textwidth]{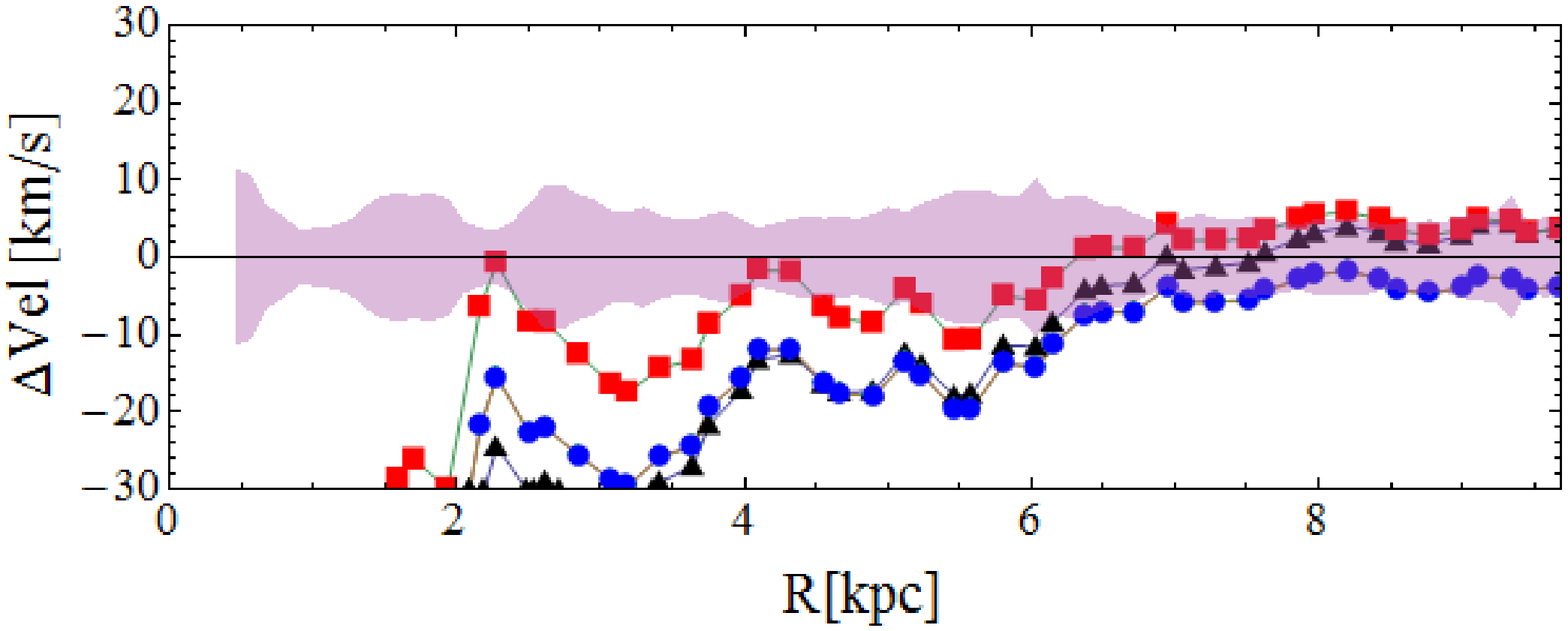}
    \end{tabular}  }
  \caption{\footnotesize{The rotation curves for the galaxy NGC 4736. Colors and symbols are as in Fig.\ref{fig:DDO154}. The stellar component of the galaxy has a complex structure, we analyzed it as one and two component stellar galaxy, where its central component has a the parameters $\mu_0= 11.8 mag arcsec^{-2}$ y $h=0.26 {\rm kpc}$. For the outer disk we assume a mass of  $M_\star=1.8*10^{10}$ $M_\sun$ and $R_d = 1.99$ {\rm kpc}. The galaxy exhibits a small color gradient and we assume a constant $\gs = 0.6$ and $\gs = 0.9$ for the outer and inner disk respectively. The best fit is given by the minimal analysis and is consistent with Kroupa IMF without bulge. The analysis with bulge does not improve significantly the value of $\chi^2$. From left to right, images from the fit of the galaxy NGC 4736 considering minimal disk, minimal disk+gas, Kroupa, diet-Salpeter. }}
  \label{fig:NGC4736}
\end{figure}

\begin{figure}[h!]
    \subfloat[\footnotesize{Minimal disk}]{
    \begin{tabular}[b]{c}
    \includegraphics[width=0.35\textwidth]{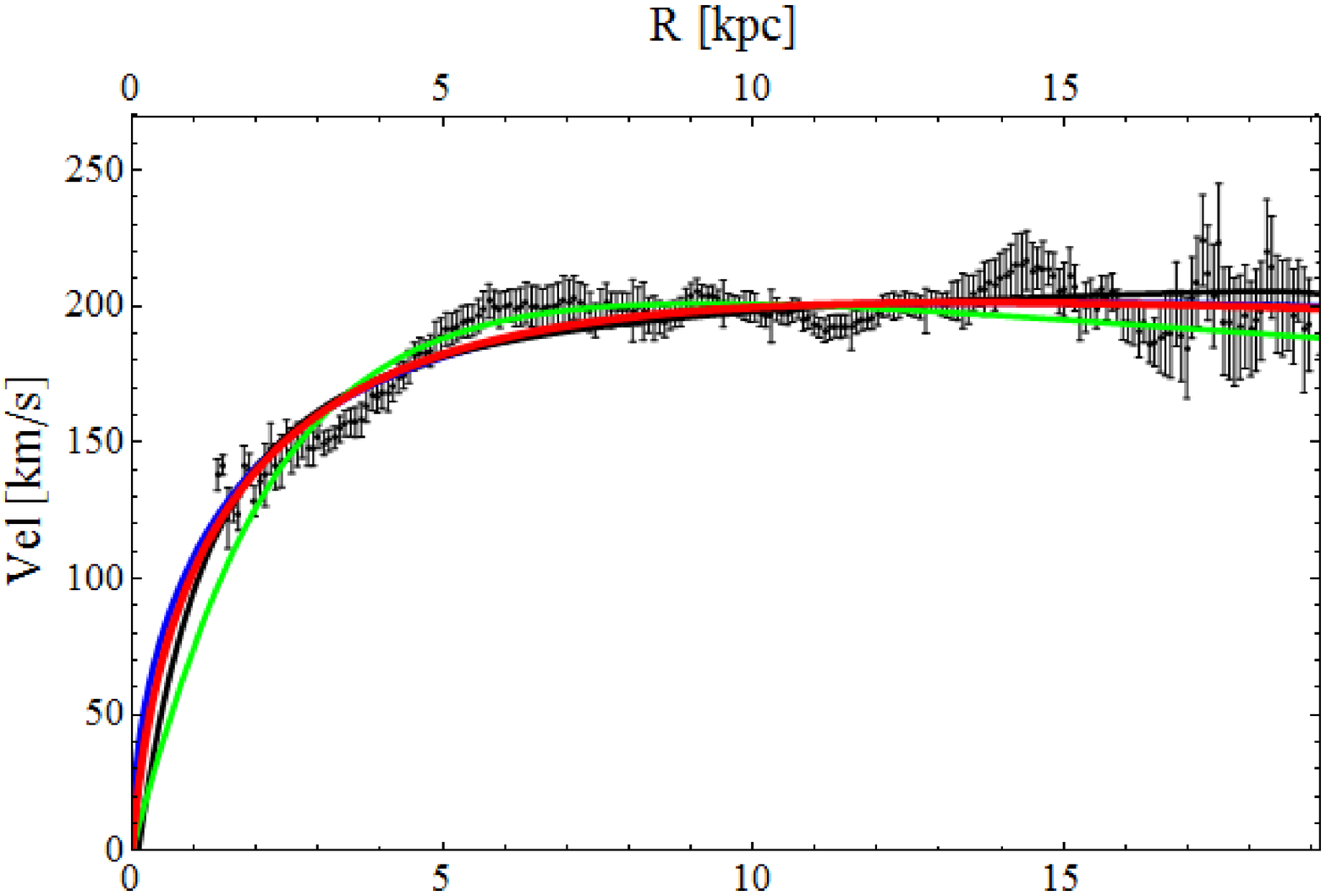} \\
    \includegraphics[width=0.35\textwidth]{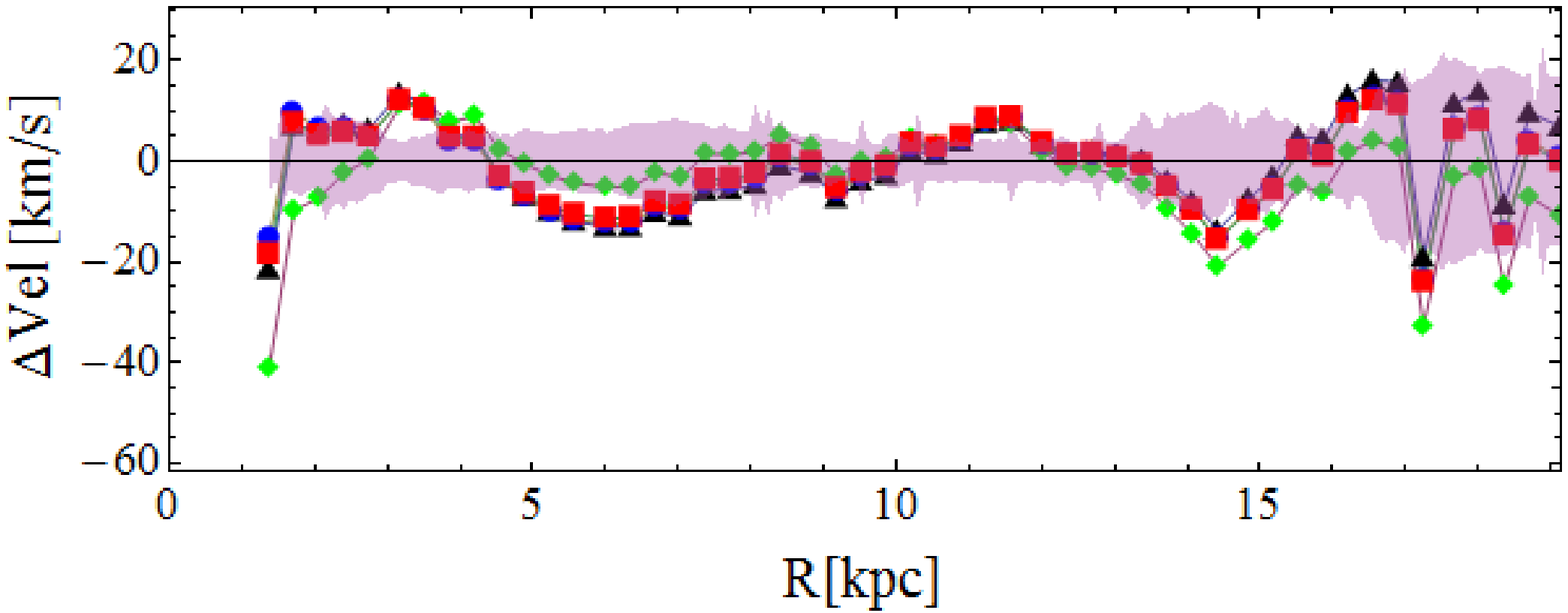}
    \end{tabular}  }
    \subfloat[\footnotesize{Min. disk + Gas}]{
    \begin{tabular}[b]{c}
    \includegraphics[width=0.35\textwidth]{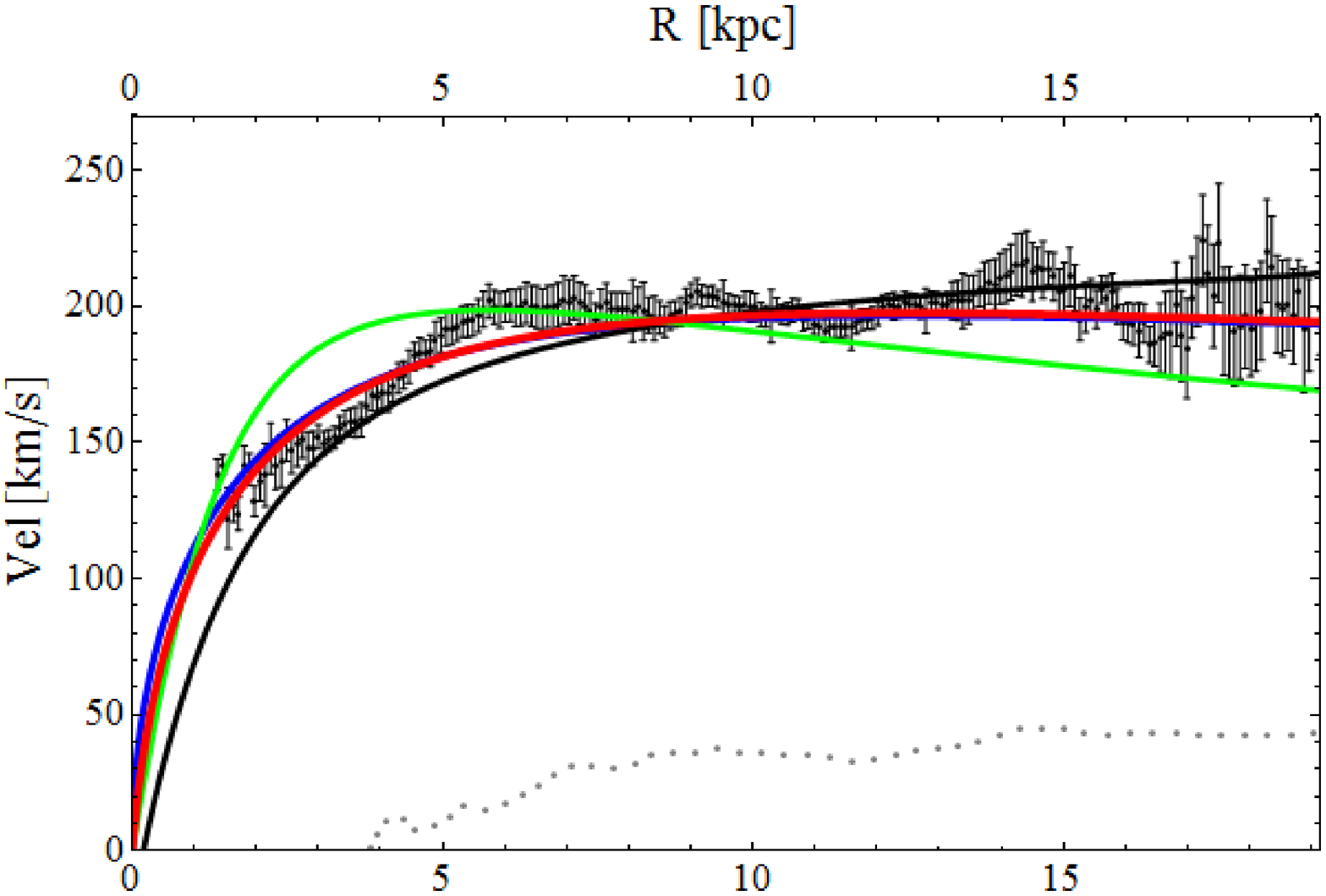} \\
    \includegraphics[width=0.35\textwidth]{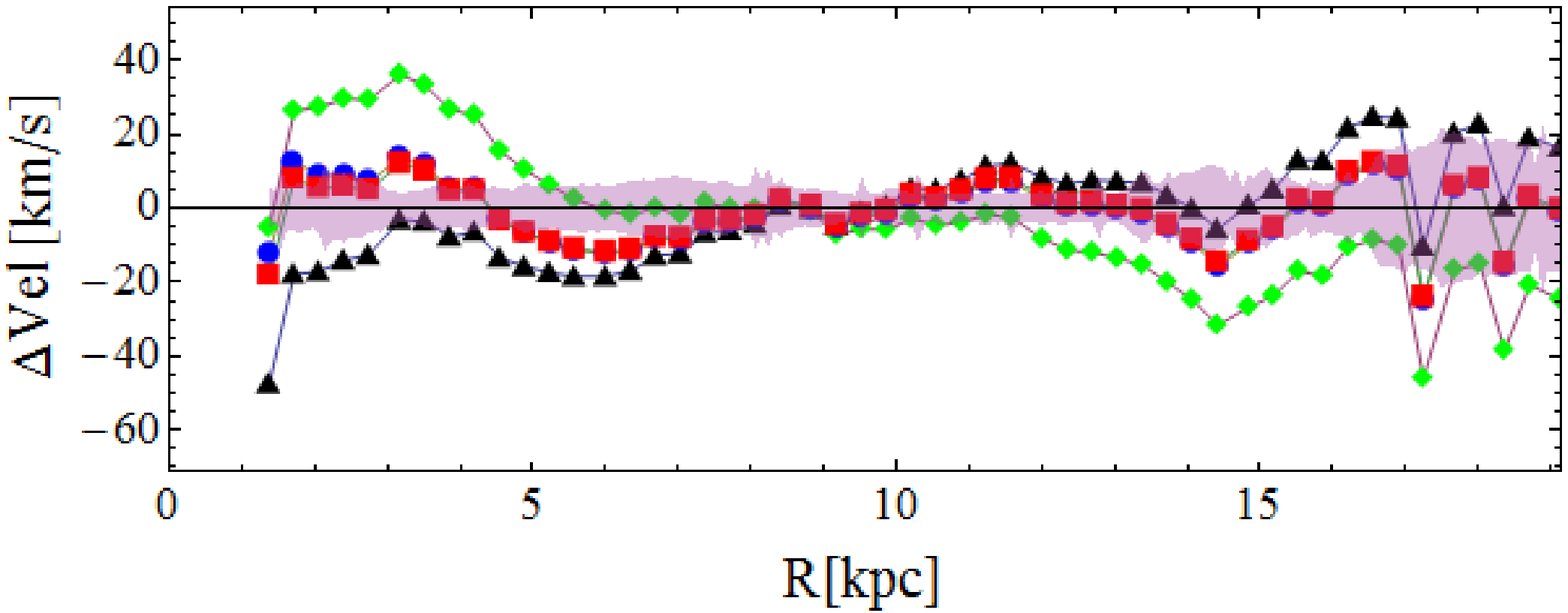}
    \end{tabular}  }  \\
    \subfloat[\footnotesize{Kroupa}]{
    \begin{tabular}[b]{c}
    \includegraphics[width=0.35\textwidth]{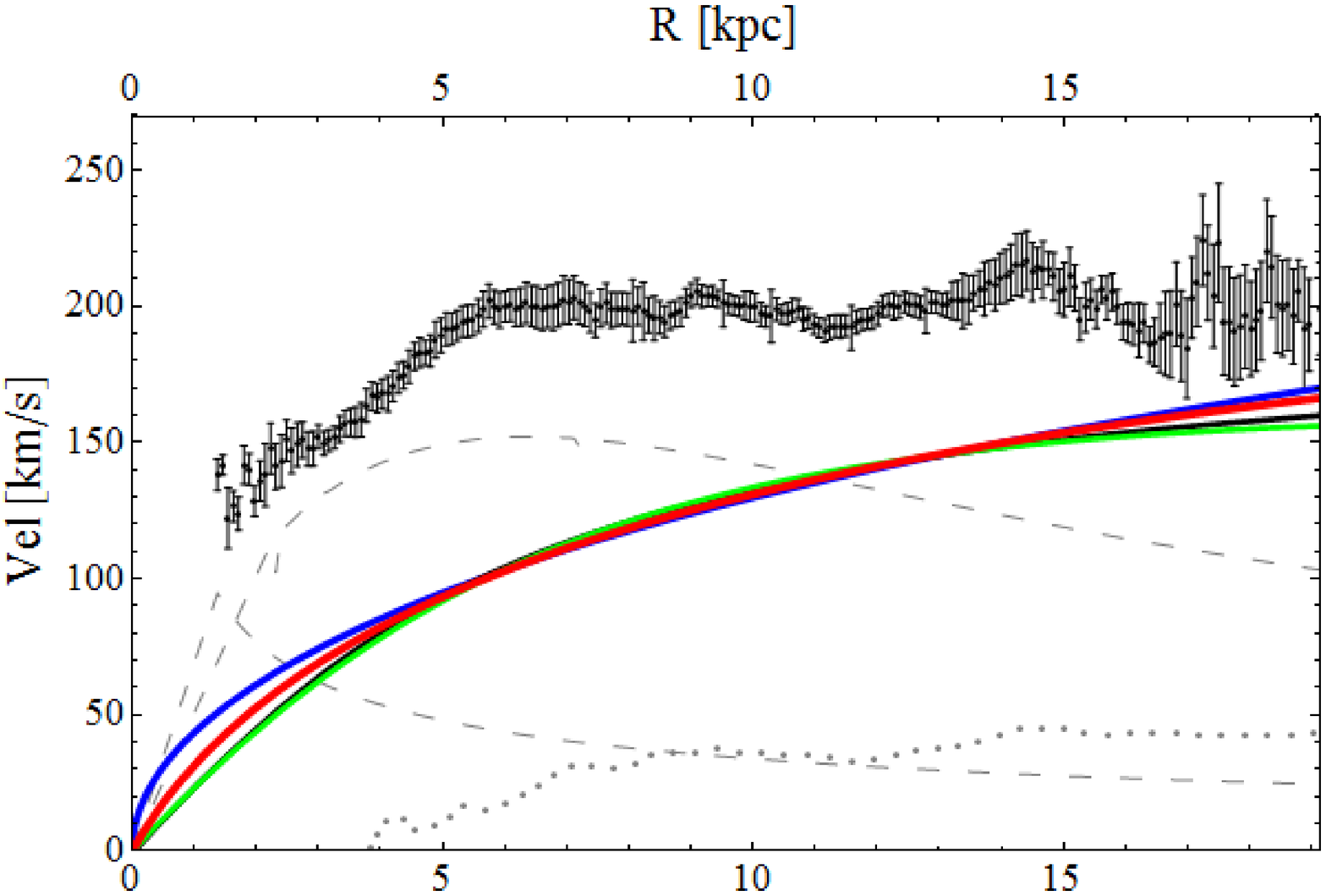} \\
    \includegraphics[width=0.35\textwidth]{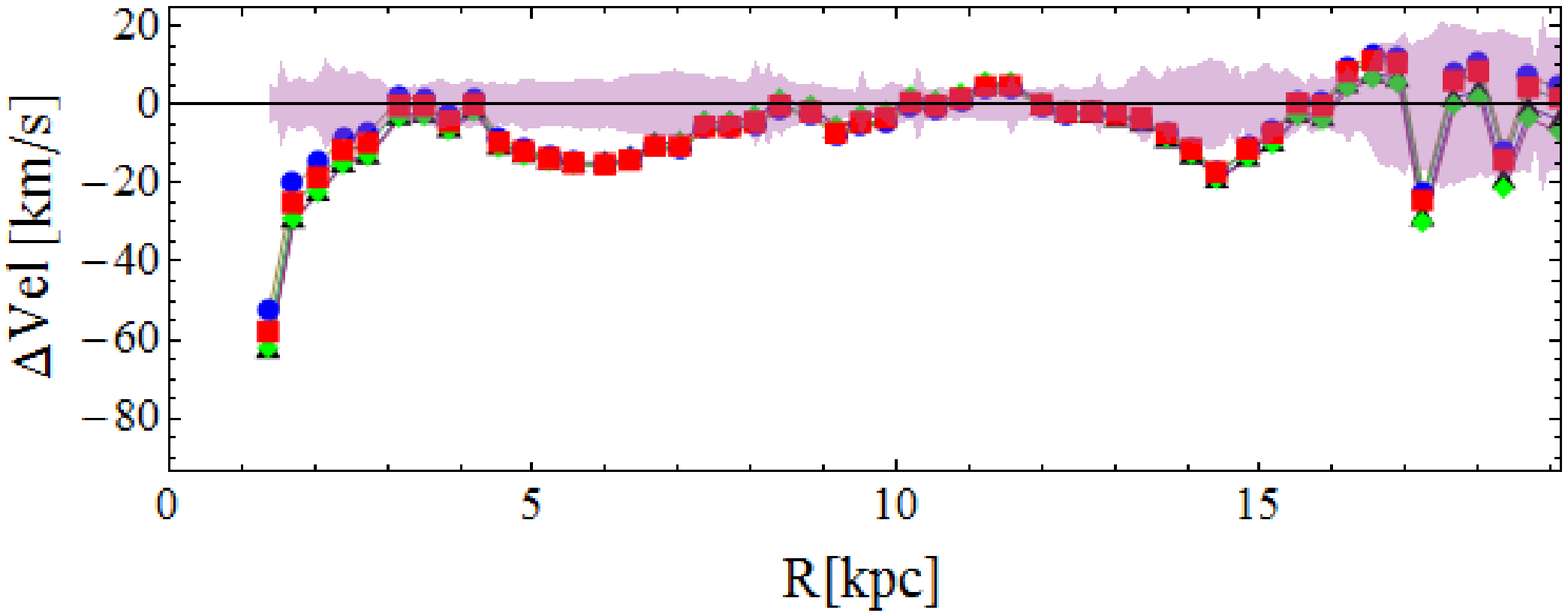}
    \end{tabular}  }
    \subfloat[\footnotesize{diet-Salpeter}]{
    \begin{tabular}[b]{c}
    \includegraphics[width=0.35\textwidth]{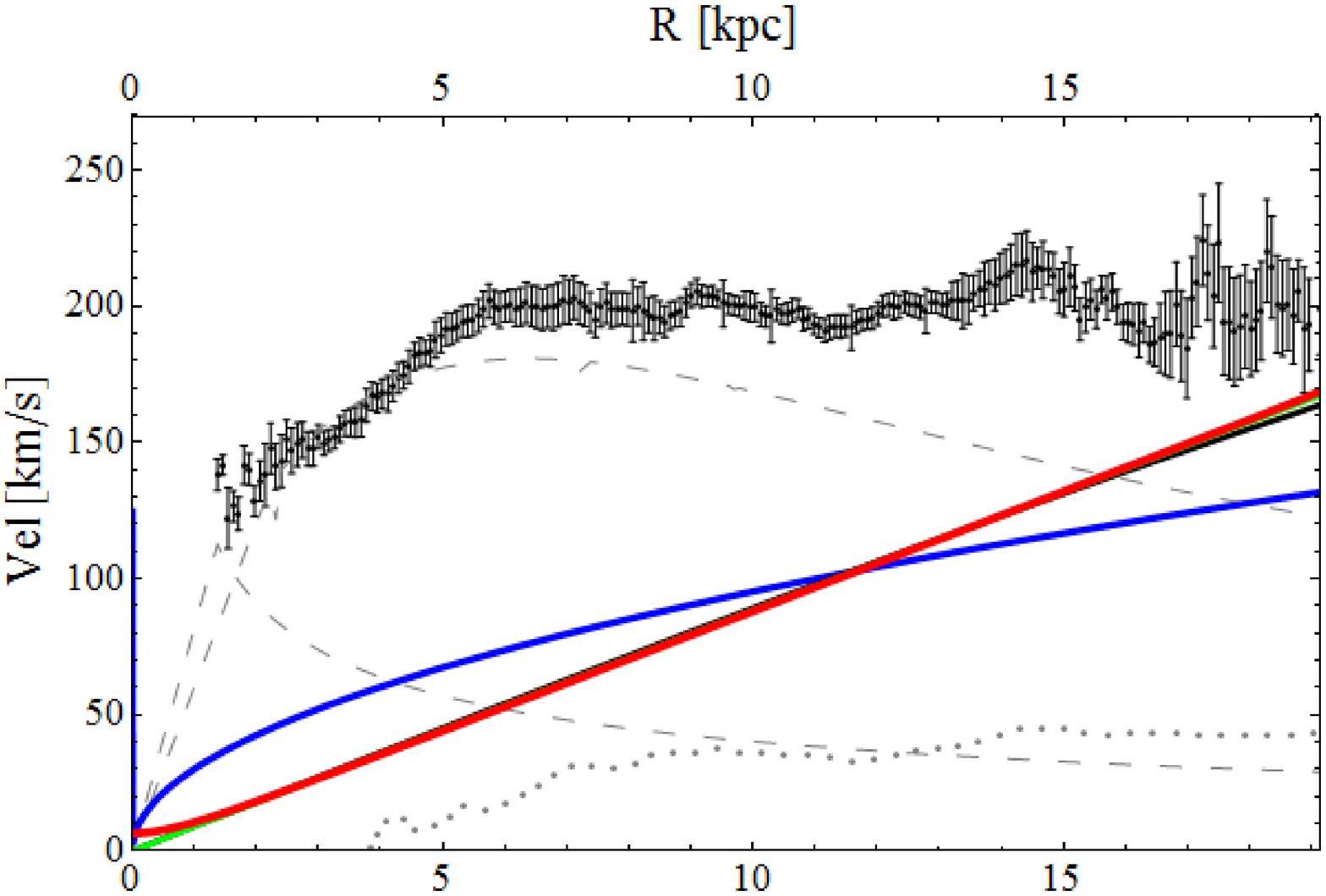} \\
    \includegraphics[width=0.35\textwidth]{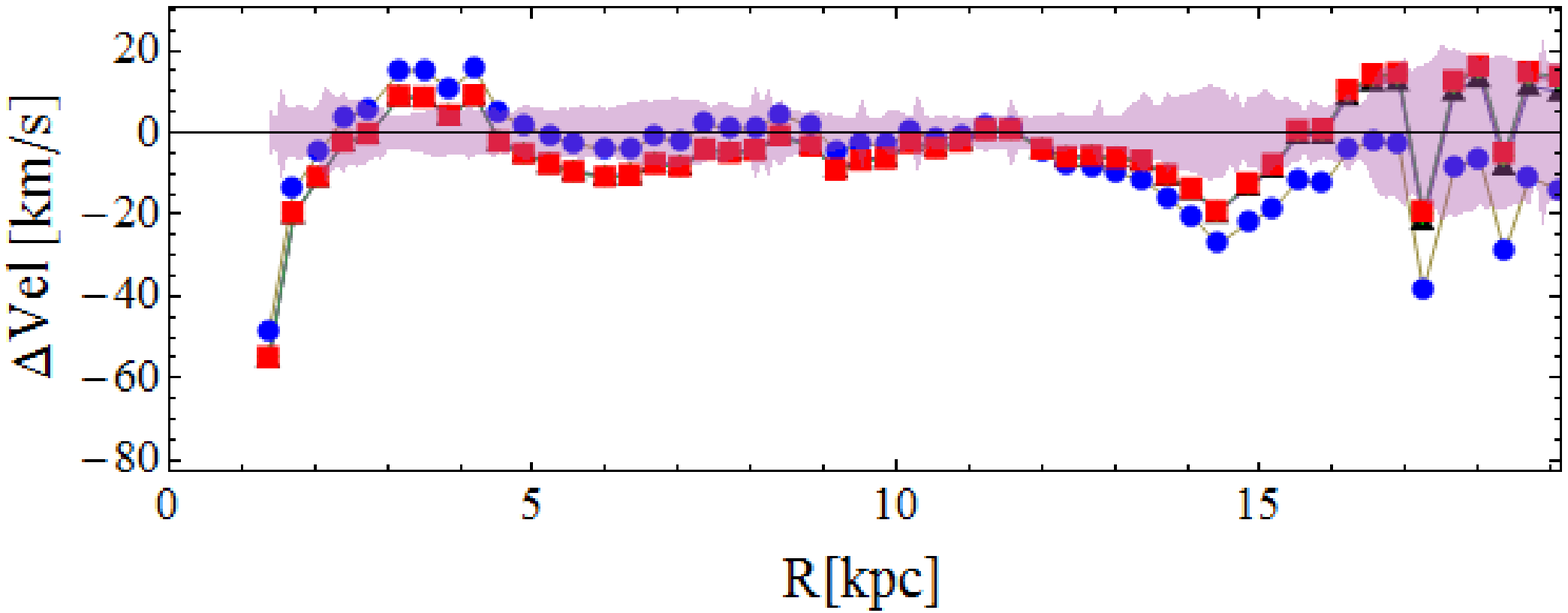}
    \end{tabular}  }
  \caption{\footnotesize{Here we display the rotation curves for the galaxy NGC 6946. Colors and symbols are as in Fig.\ref{fig:DDO154}. The galaxy present a small central object but does not improve remarkably the analysis, so we treat the stellar disk as one component using a mass of $M_\star = 5.8*10^10 M_\sun$ and $R_d=2.97 {\rm kpc}$. From left to right, images from the fit of the galaxy NGC 6946 considering minimal disk, minimal disk+gas, Kroupa, diet-Salpeter.}}
  \label{fig:NGC6946}
\end{figure}

\begin{figure}[h!]
    \subfloat[\footnotesize{Minimal disk}]{
    \begin{tabular}[b]{c}
    \includegraphics[width=0.35\textwidth]{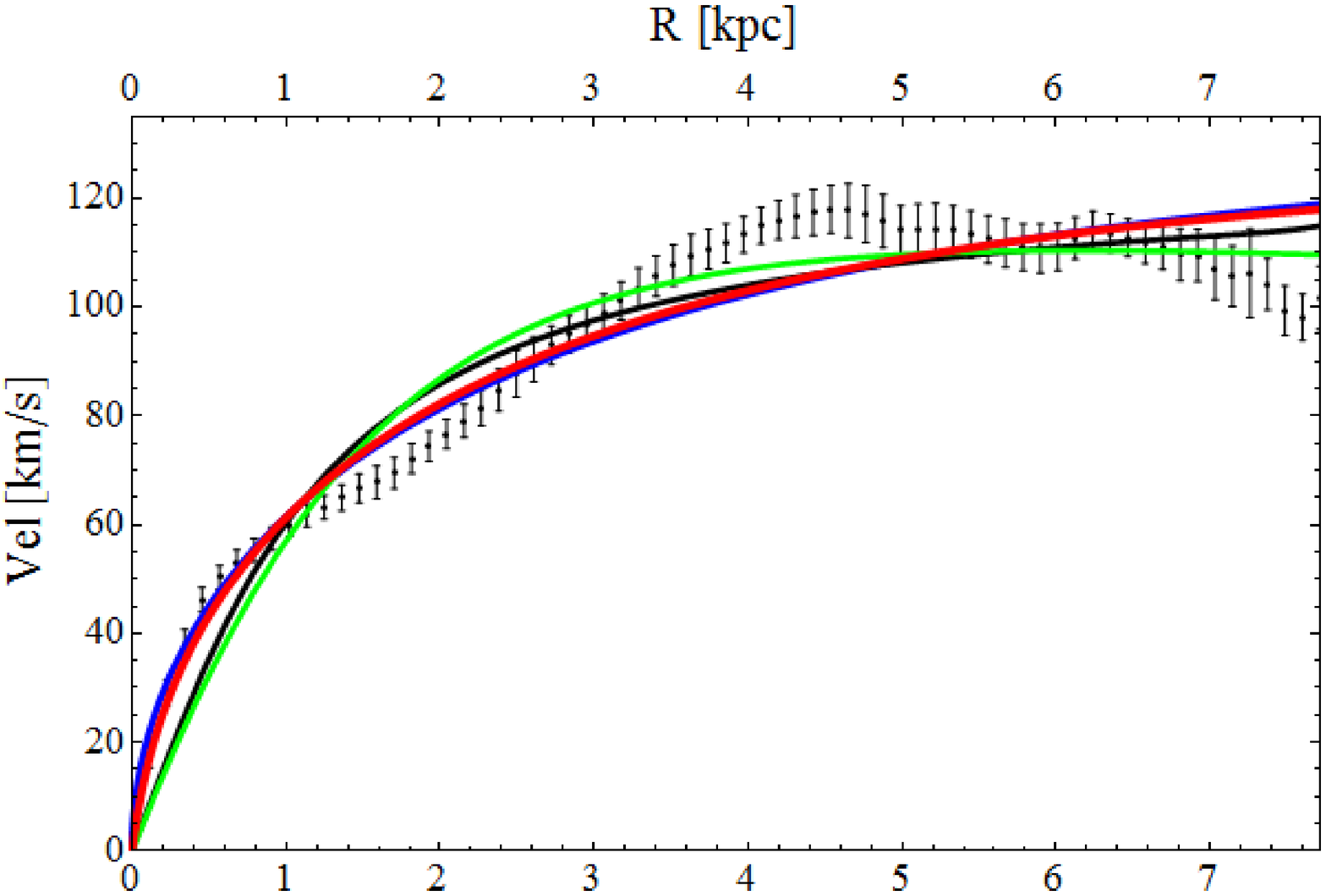} \\
    \includegraphics[width=0.35\textwidth]{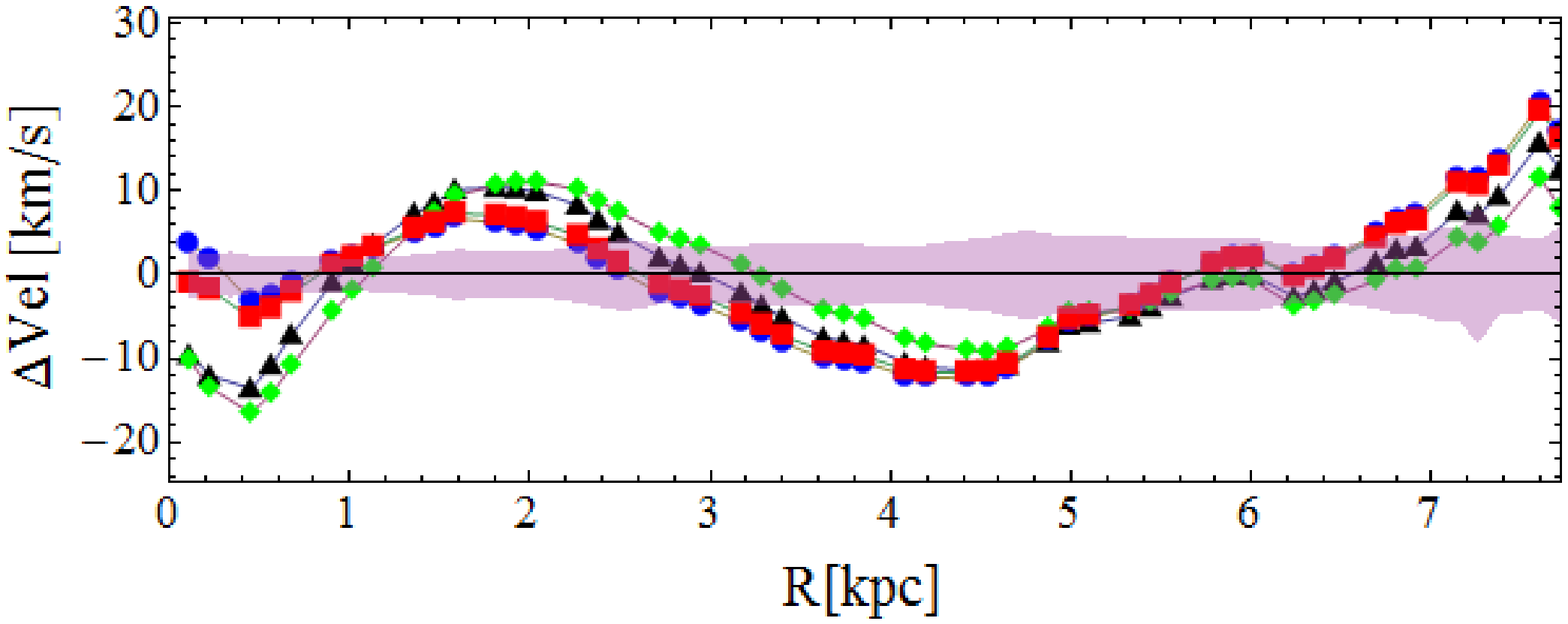}
    \end{tabular}  }
    \subfloat[\footnotesize{Min. disk + Gas}]{
    \begin{tabular}[b]{c}
    \includegraphics[width=0.35\textwidth]{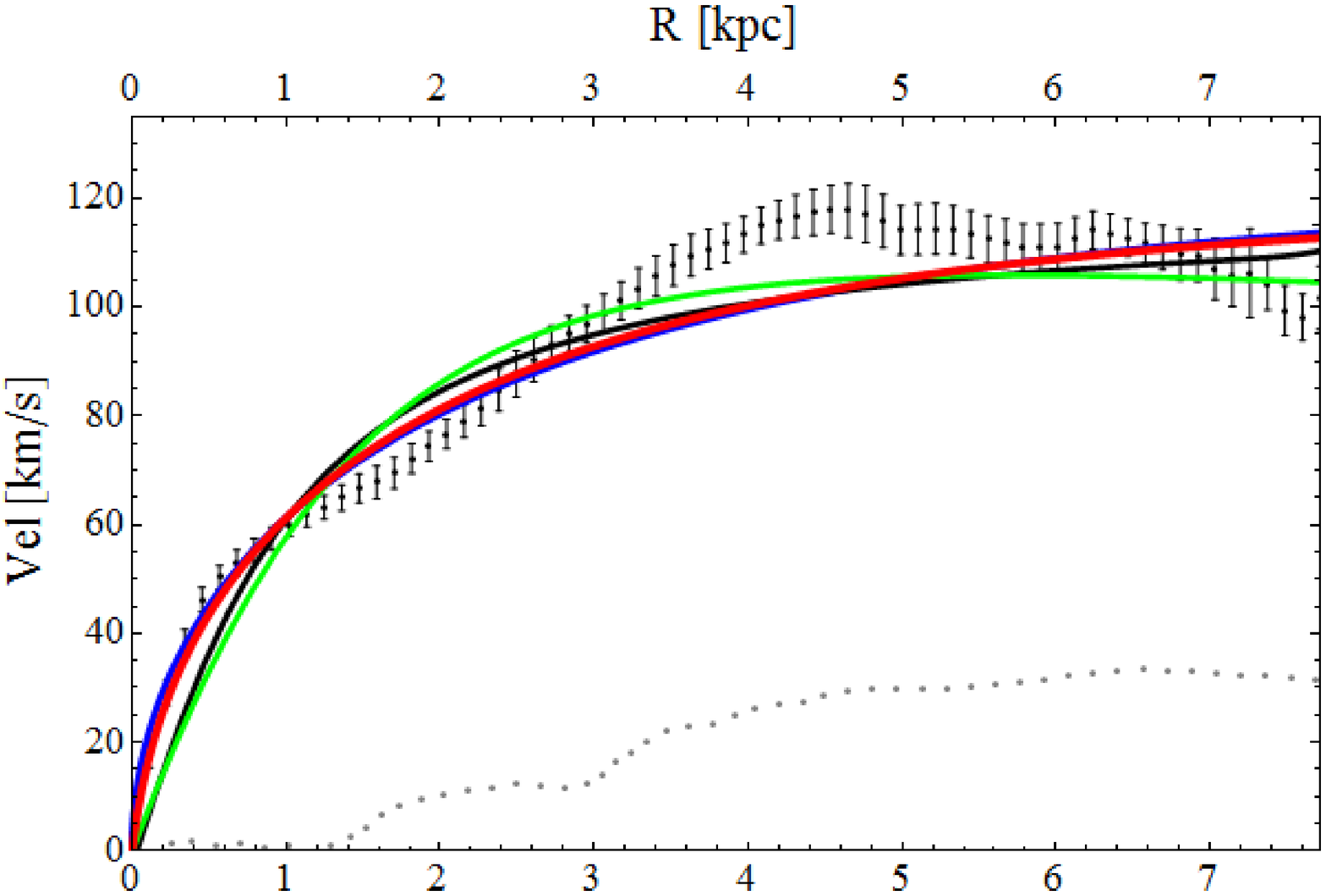} \\
    \includegraphics[width=0.35\textwidth]{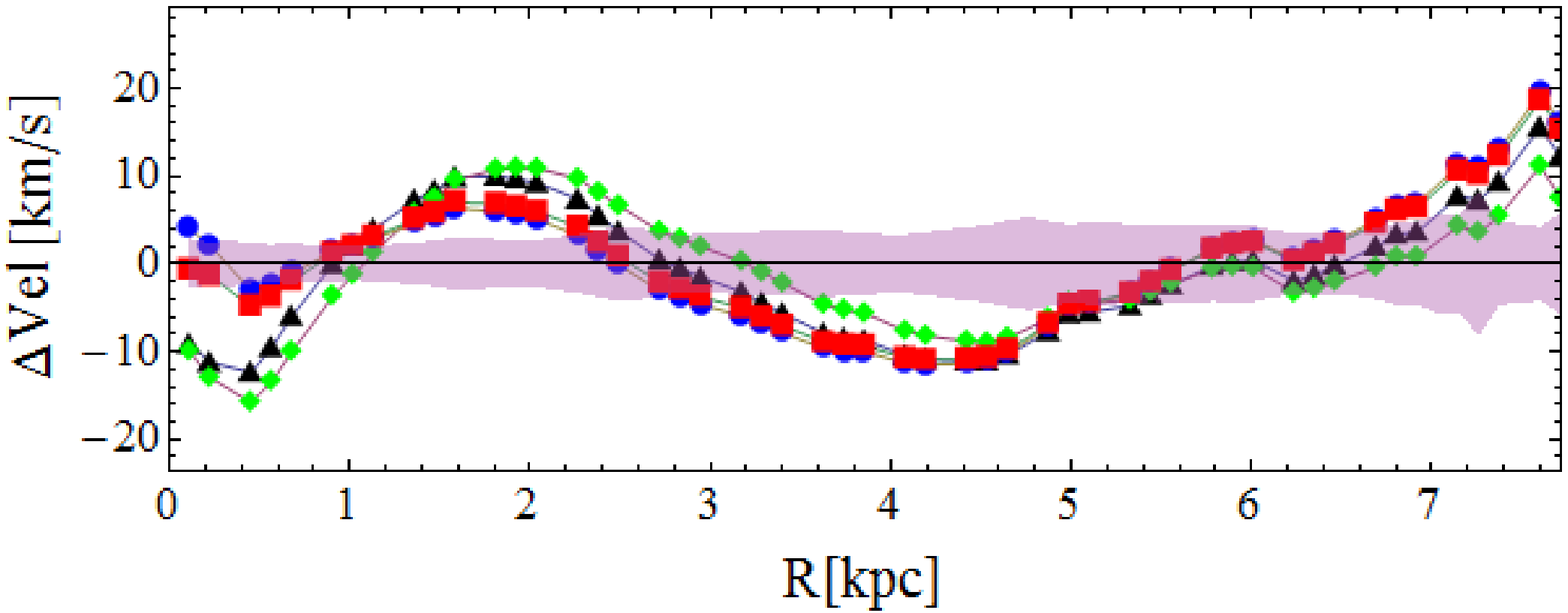}
    \end{tabular}  }  \\
    \subfloat[\footnotesize{Kroupa}]{
    \begin{tabular}[b]{c}
    \includegraphics[width=0.35\textwidth]{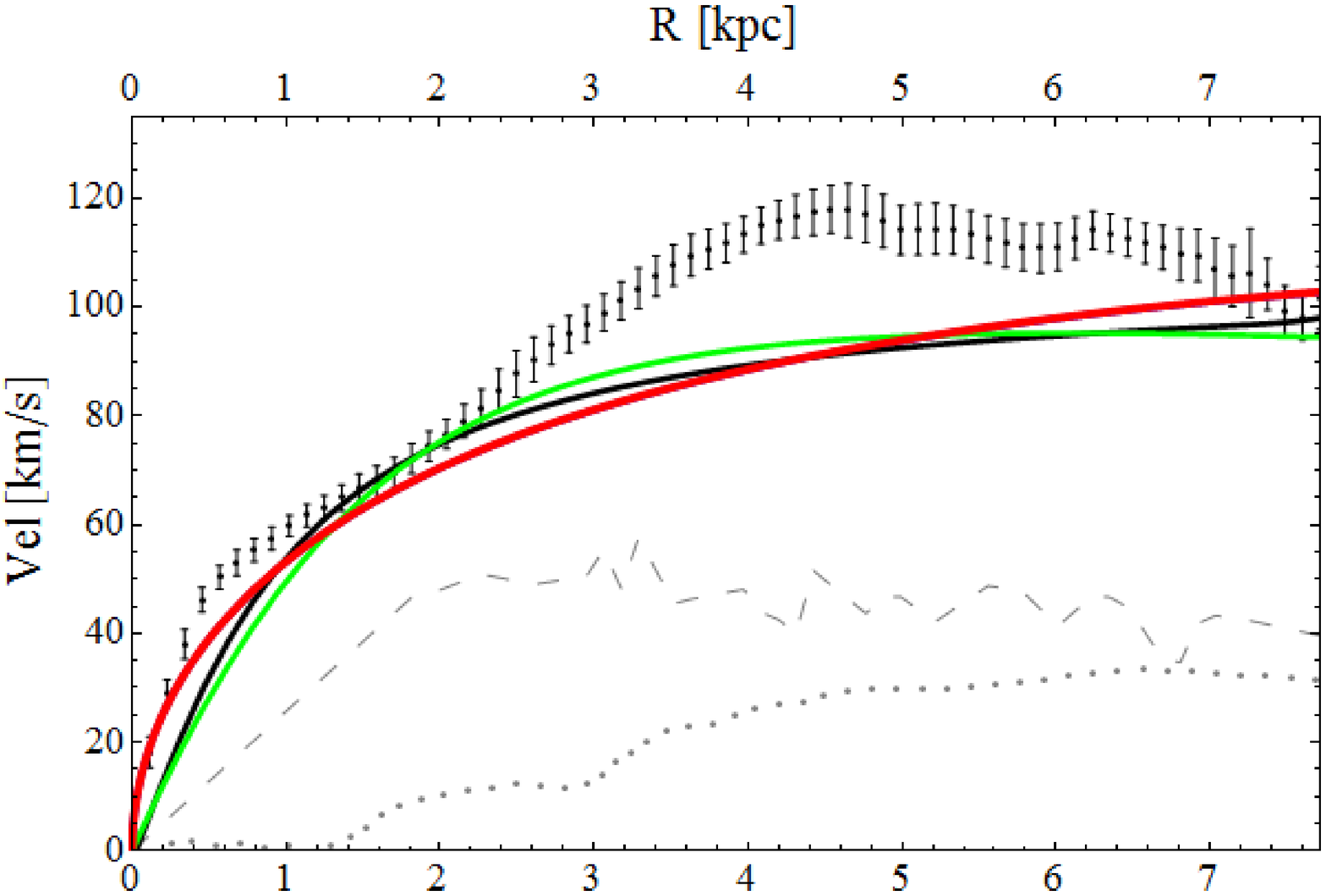} \\
    \includegraphics[width=0.35\textwidth]{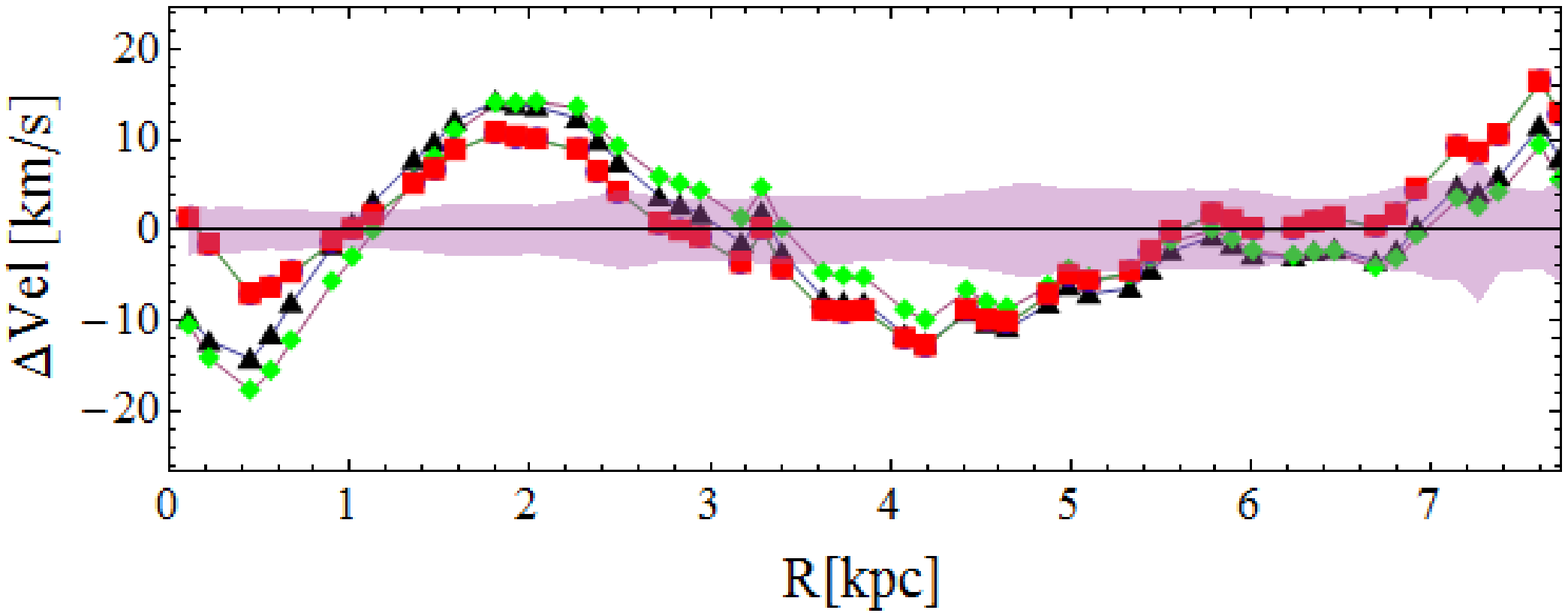}
    \end{tabular}  }
    \subfloat[\footnotesize{diet-Salpeter}]{
    \begin{tabular}[b]{c}
    \includegraphics[width=0.35\textwidth]{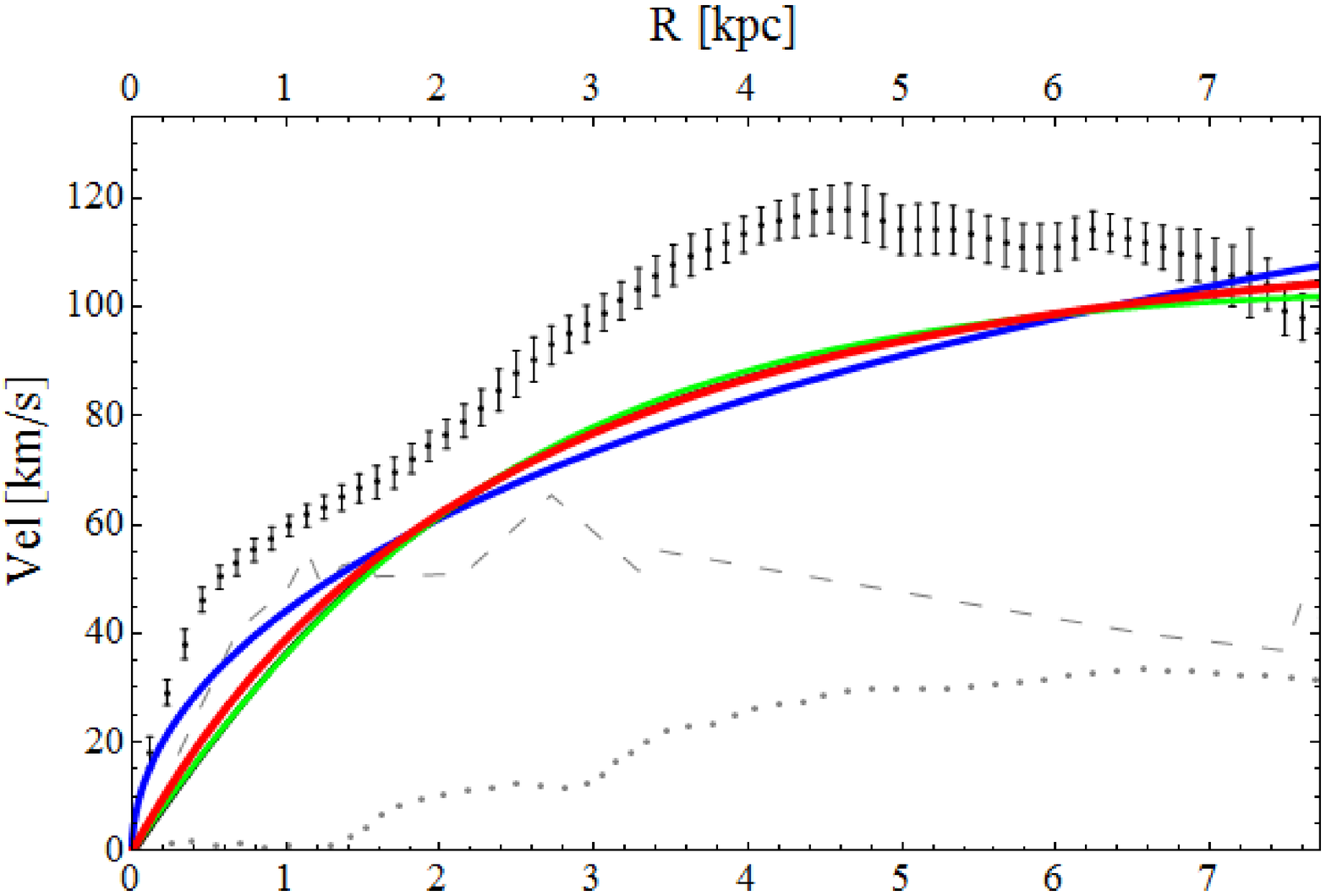} \\
    \includegraphics[width=0.35\textwidth]{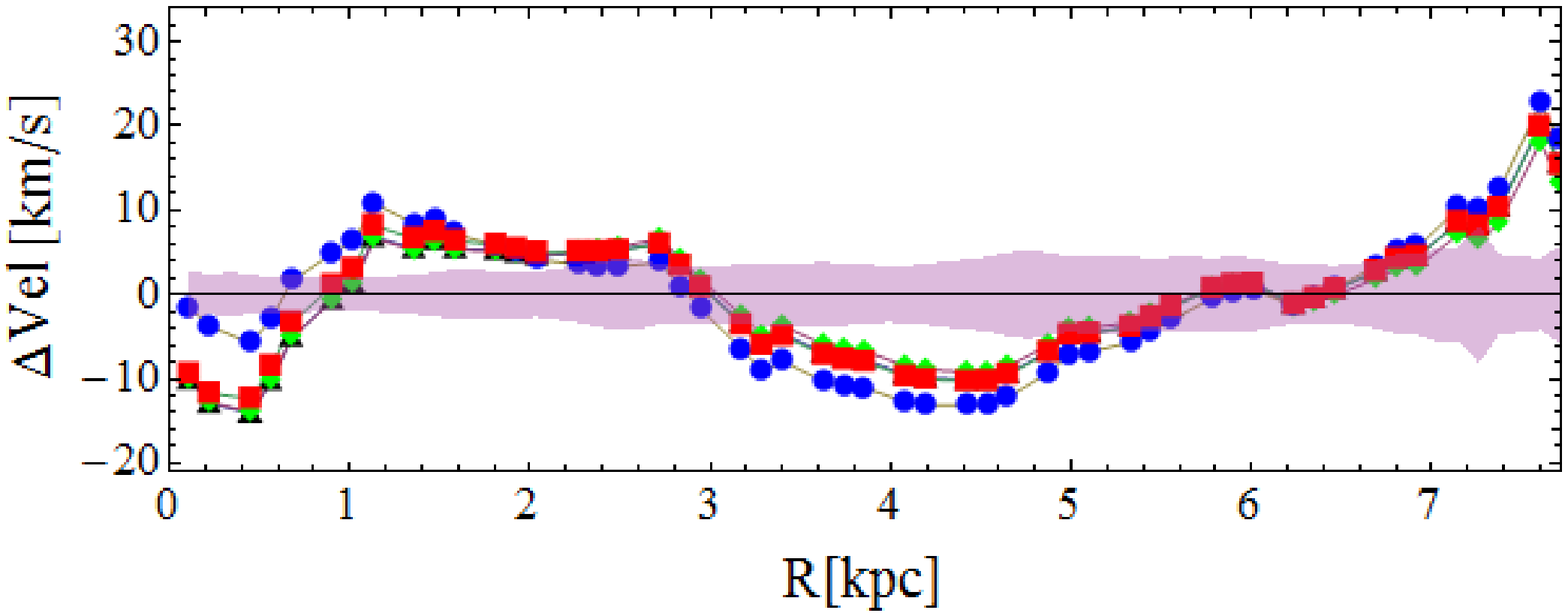}
    \end{tabular}  }
   \caption{\footnotesize{Here we display the rotation curves for the galaxy NGC 7793. Colors and symbols are as in Fig.\ref{fig:DDO154}. This galaxy is a late-type Sd spiral galaxy. It shows a nuclear star cluster, nevertheless its photometric and dynamical importance has become negligible. The (J-K) profile has a small color gradient, so we proceed taking an average of the $\gs$ for the inner and outer parts of the stellar disk and assume a constant value for $\gs = 0.31$. All the profiles roughly predict equal results, close to Kroupa's value. even so, BDM offers the best fit for these galaxy with physical plausible parameters values. From left to right, images from the fit of the galaxy NGC 7793 considering minimal disk, minimal disk+gas, Kroupa, diet-Salpeter. }}
  \label{fig:NGC7793}
\end{figure}
\newpage
\clearpage
\subsection{GROUP B}\label{apendix:gb}
In this section we present the considerations made the galaxies with fitted values of $r_c = r_s$ when analyzed with the minimal disk model, we have named this set of galaxies as G.B. The conclusion are presented in Sec. \ref{conclusion}. At the end of the section can be see the confidence level for each galaxy in every mass model.

\begin{figure}[h]
    \subfloat[\footnotesize{Minimal disk}]{
    \begin{tabular}[b]{c}
    \includegraphics[width=0.35\textwidth]{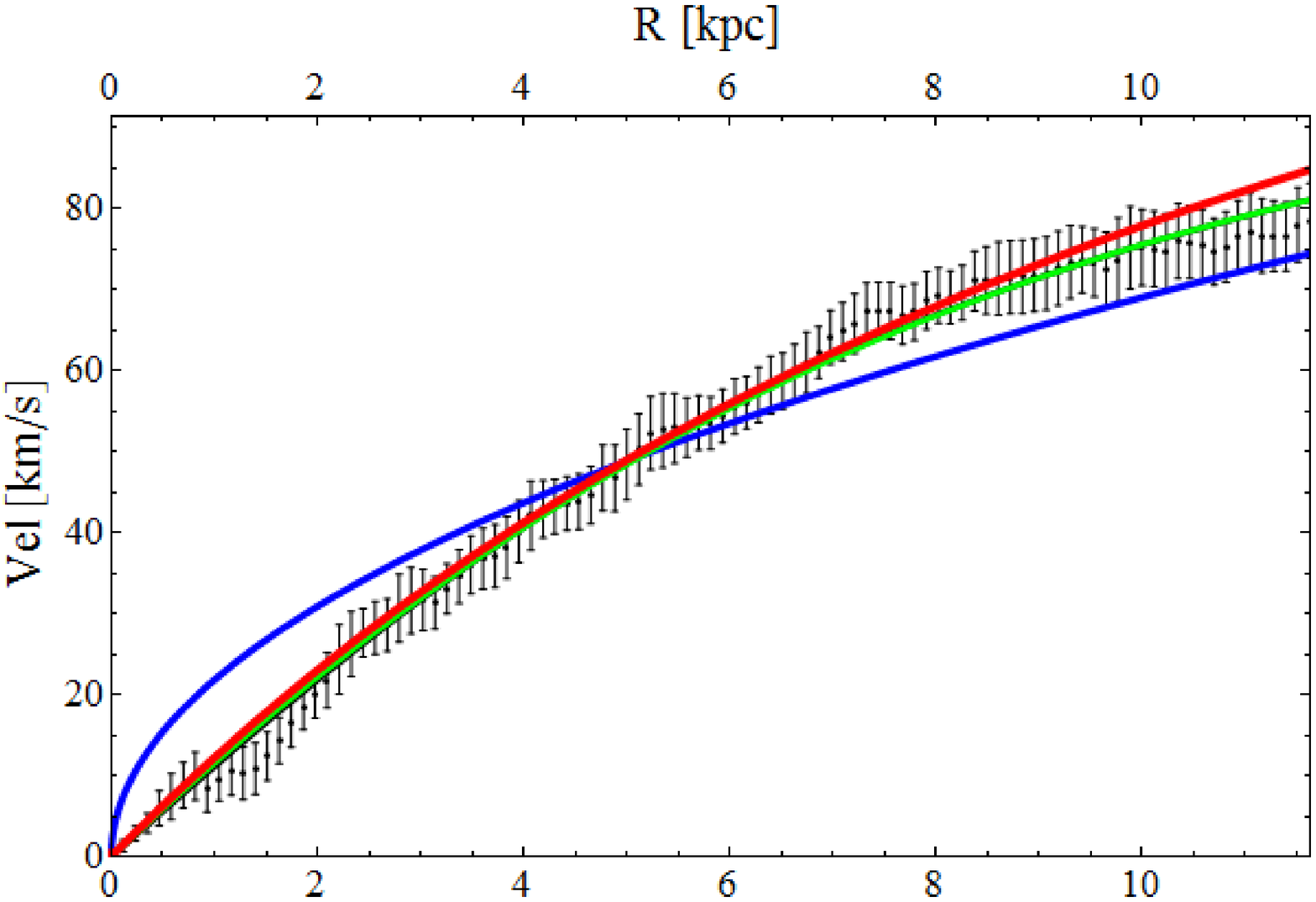} \\
    \includegraphics[width=0.35\textwidth]{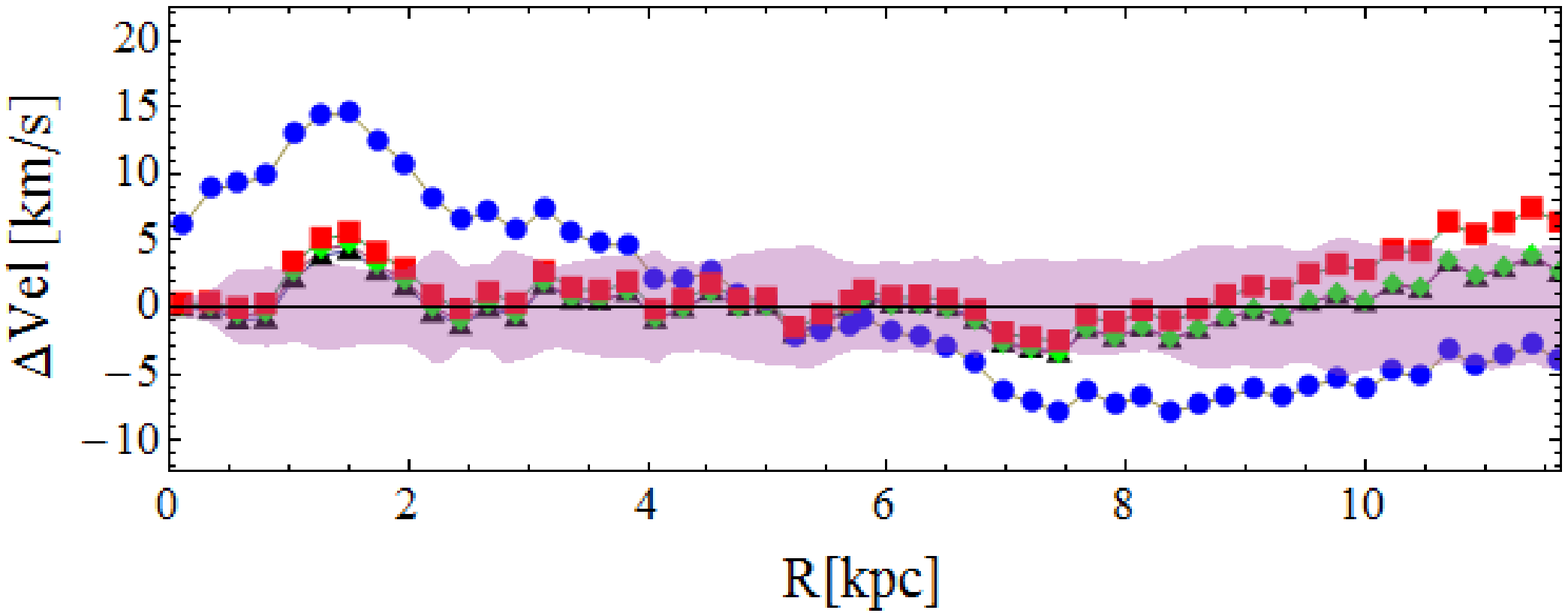}
    \end{tabular}  }
    \subfloat[\footnotesize{Min. disk + Gas}]{
    \begin{tabular}[b]{c}
    \includegraphics[width=0.35\textwidth]{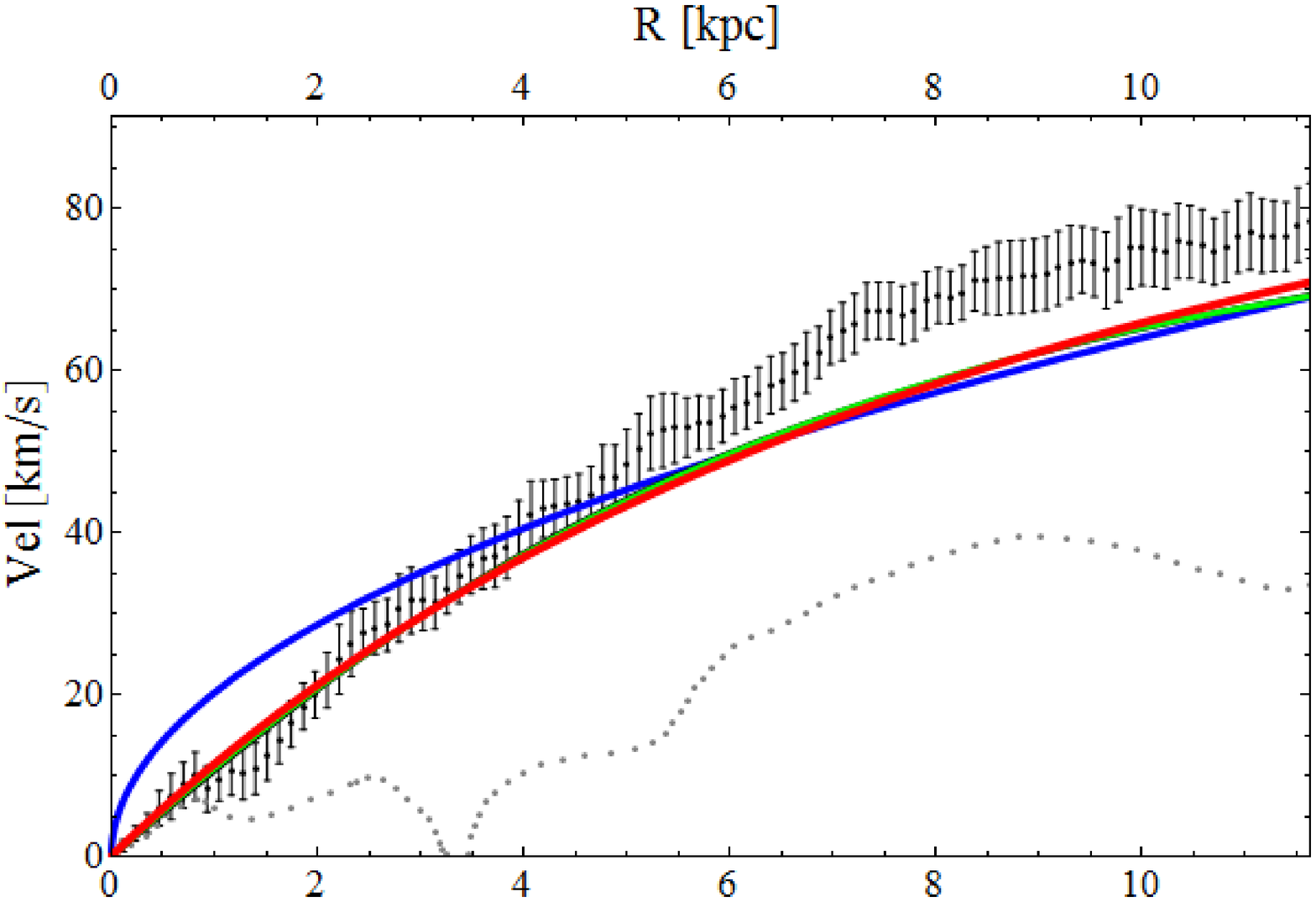} \\
    \includegraphics[width=0.35\textwidth]{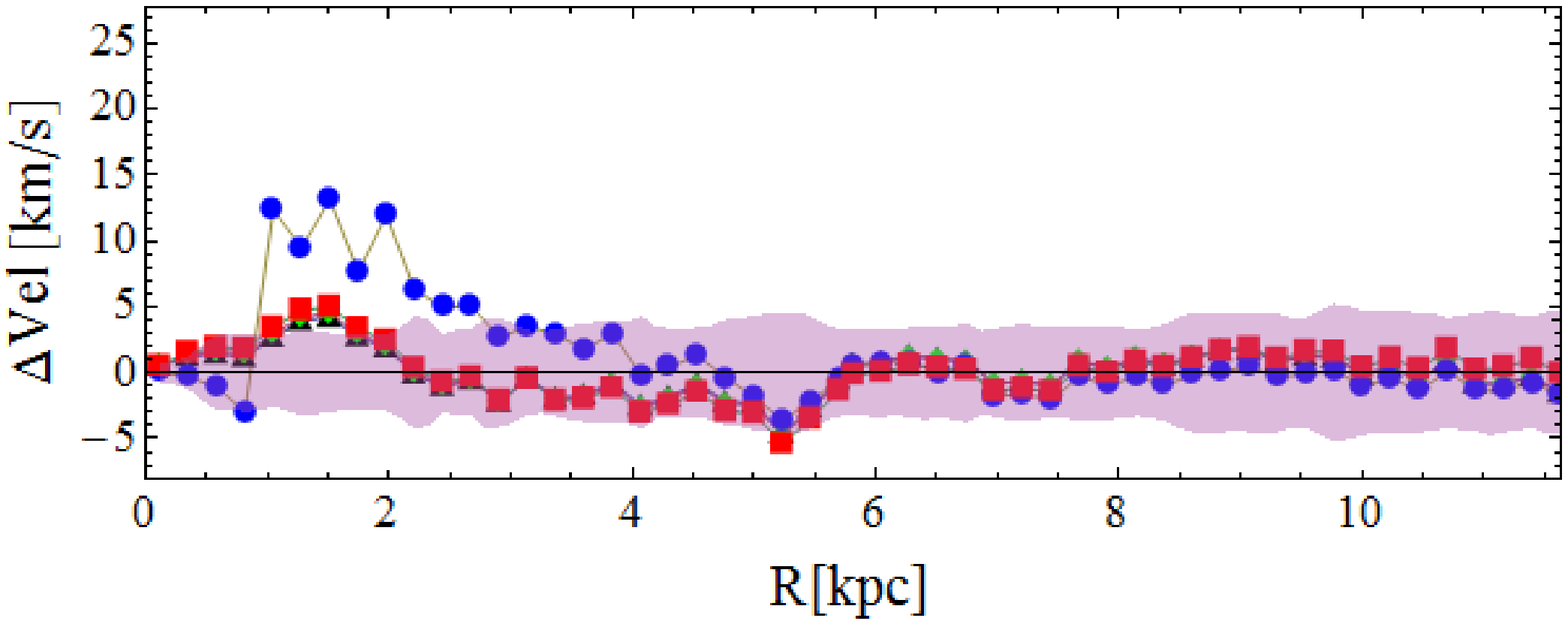}
    \end{tabular}  }  \\
    \subfloat[\footnotesize{Kroupa}]{
    \begin{tabular}[b]{c}
    \includegraphics[width=0.35\textwidth]{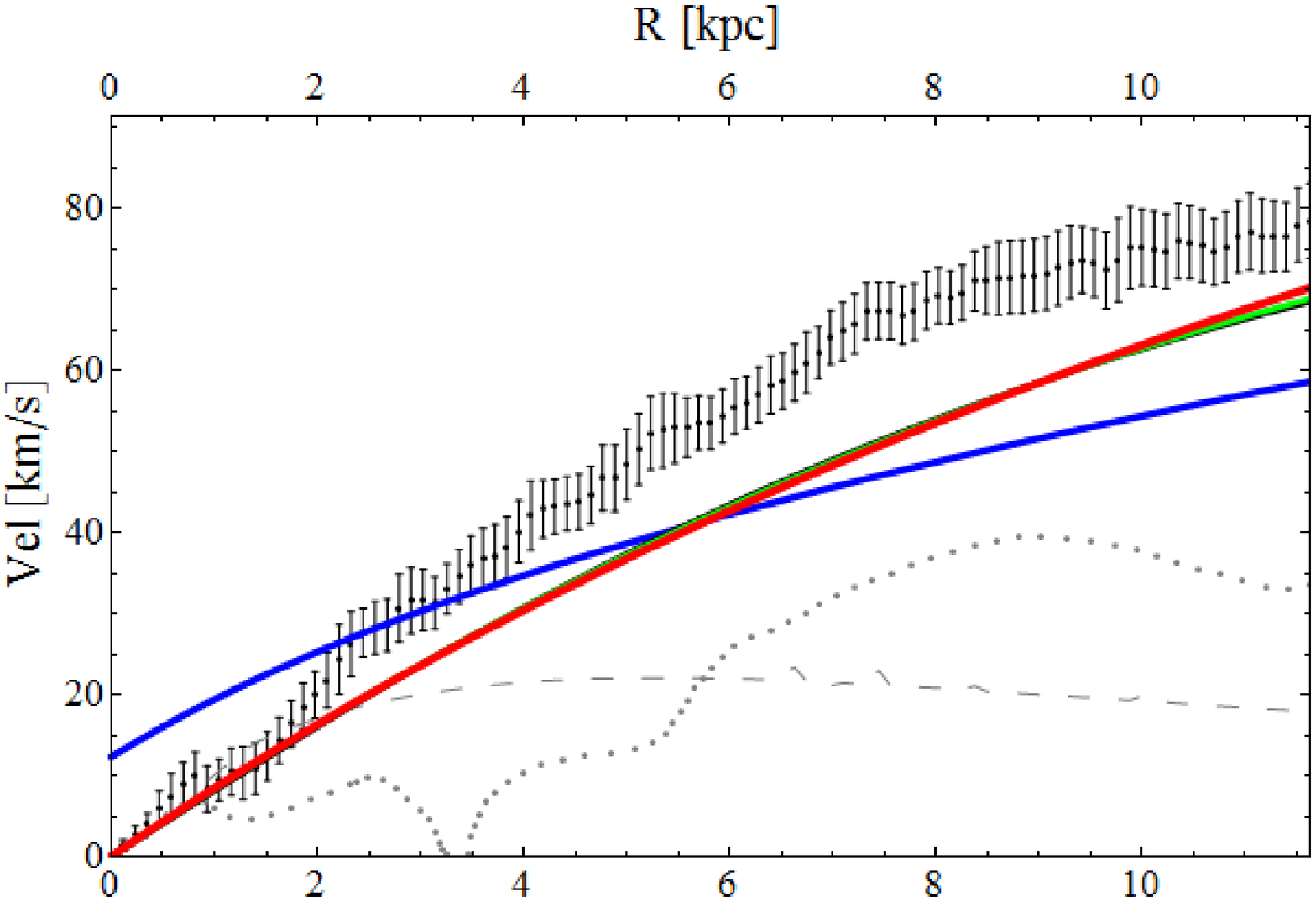} \\
    \includegraphics[width=0.35\textwidth]{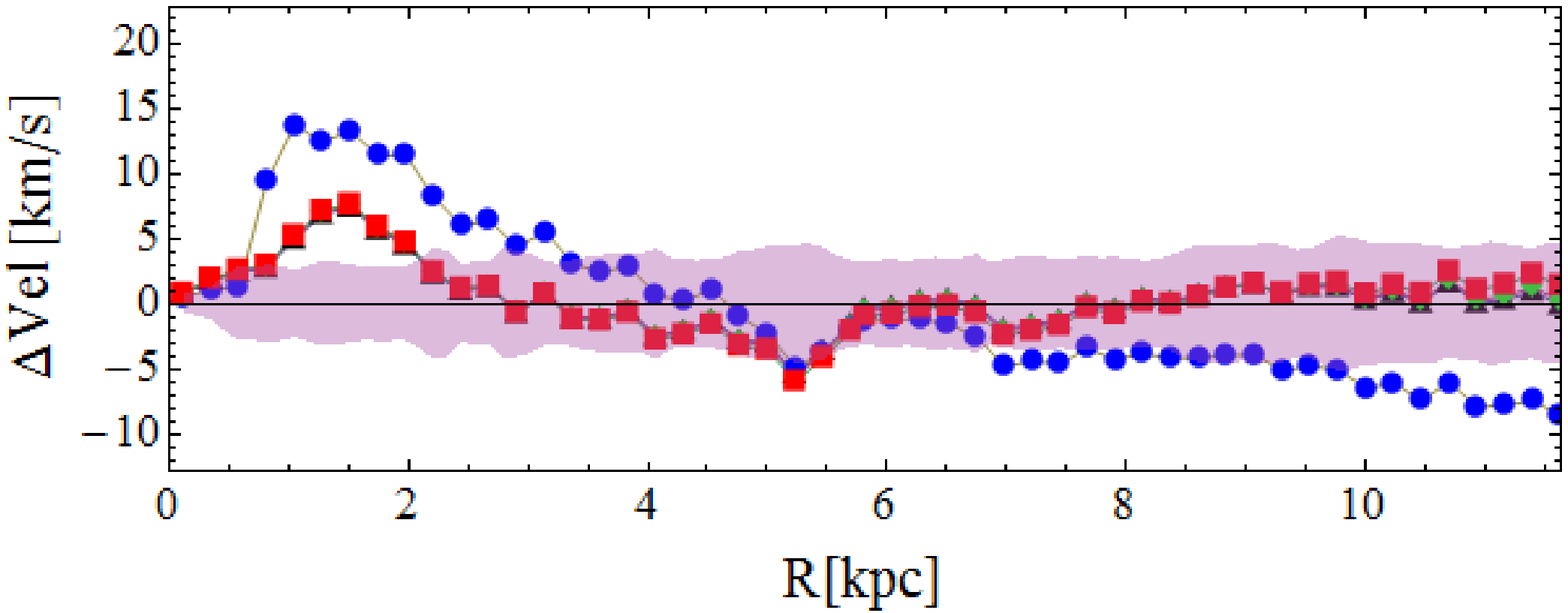}
    \end{tabular}  }
    \subfloat[\footnotesize{diet-Salpeter}]{
    \begin{tabular}[b]{c}
    \includegraphics[width=0.35\textwidth]{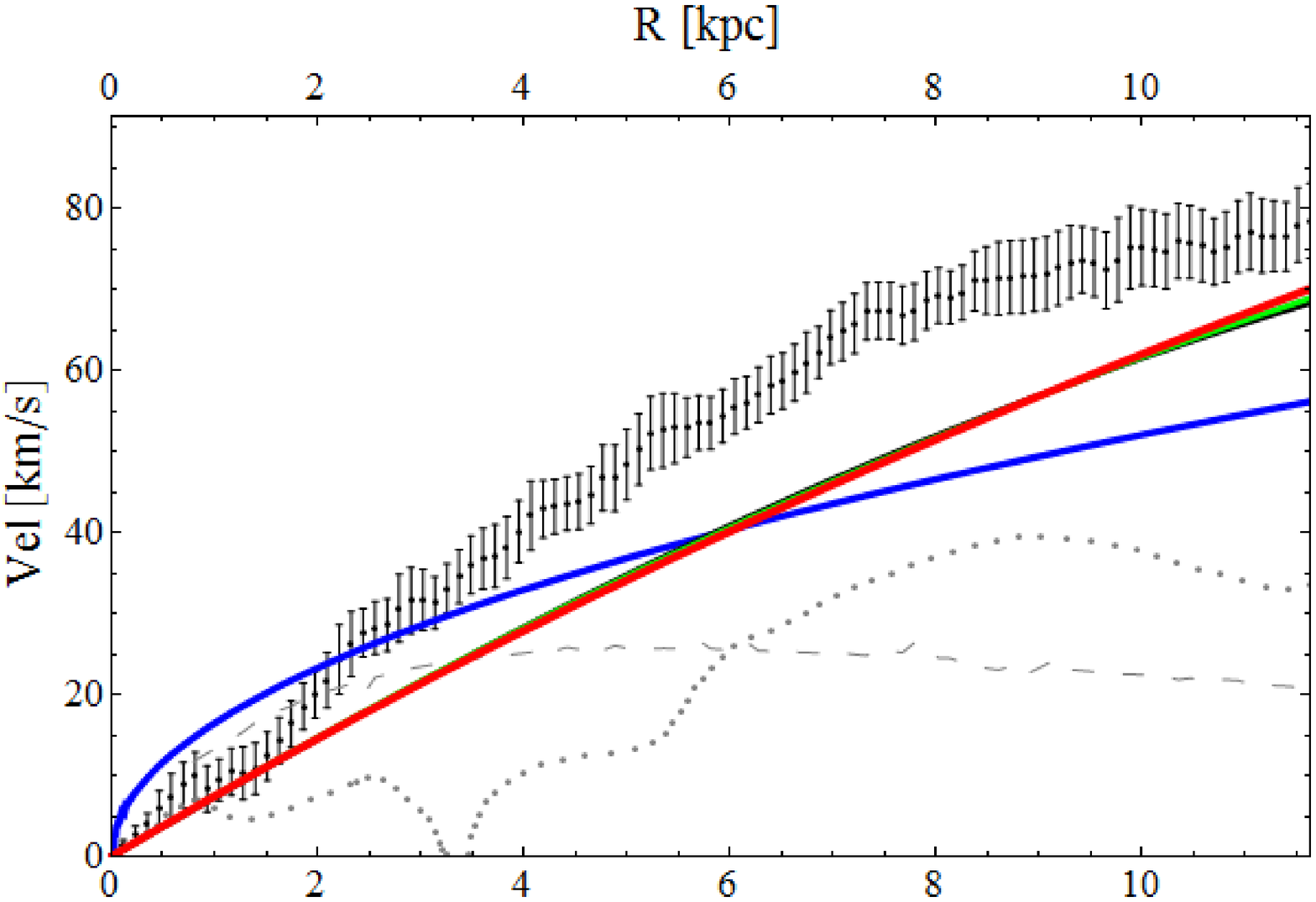} \\
    \includegraphics[width=0.35\textwidth]{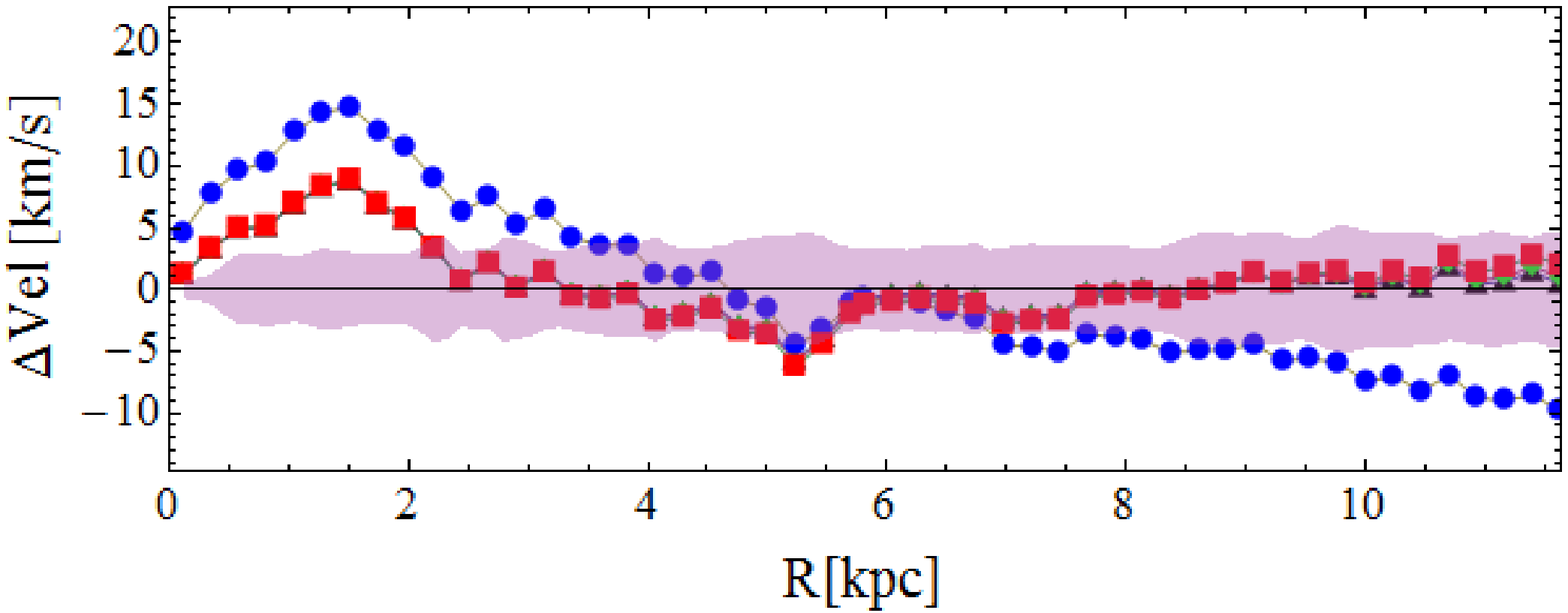}
    \end{tabular}  }
   \caption{\footnotesize{The rotation curves for the galaxy IC 2574. Colors and symbols are as in Fig.\ref{fig:DDO154}. A dwarf galaxy. We analyze this galaxy as a one component exponential disk with mass $M_\star=1.04*10^9 M_\sun$ and $R_d=2.56$ {\rm kpc}. We assume a constant mass-to-light ratio with value $\gs=0.44$. All the profiles predict the minimal plausible value for $\gs$. All the cored profile adjust equally well and NFW fit specially bad. In these kind of galaxy we have the best fit for $r_c = r_s$ even so we do an analysis using the inner behavior of the BDM profile. These galaxy has a very smooth slope and we can not distinguish where the maximum could it be, so we take all the points and adjust it to the inner BDM profile in order to find a plausible value for the core and the central density $r_c = 3.2 {\rm kpc}$ and $\rho_c = 3.4*10^7 M_\sun/{\rm kpc}^3$ for the minimal analysis. Since we take all the points to do de inner analysis it is very difficult to break the degeneracy between the parameters $\rho_0$ and $r_c$. }}
  \label{fig:IC2574}
\end{figure}

\begin{figure}[h!]
    \subfloat[\footnotesize{Minimal disk}]{
    \begin{tabular}[b]{c}
    \includegraphics[width=0.35\textwidth]{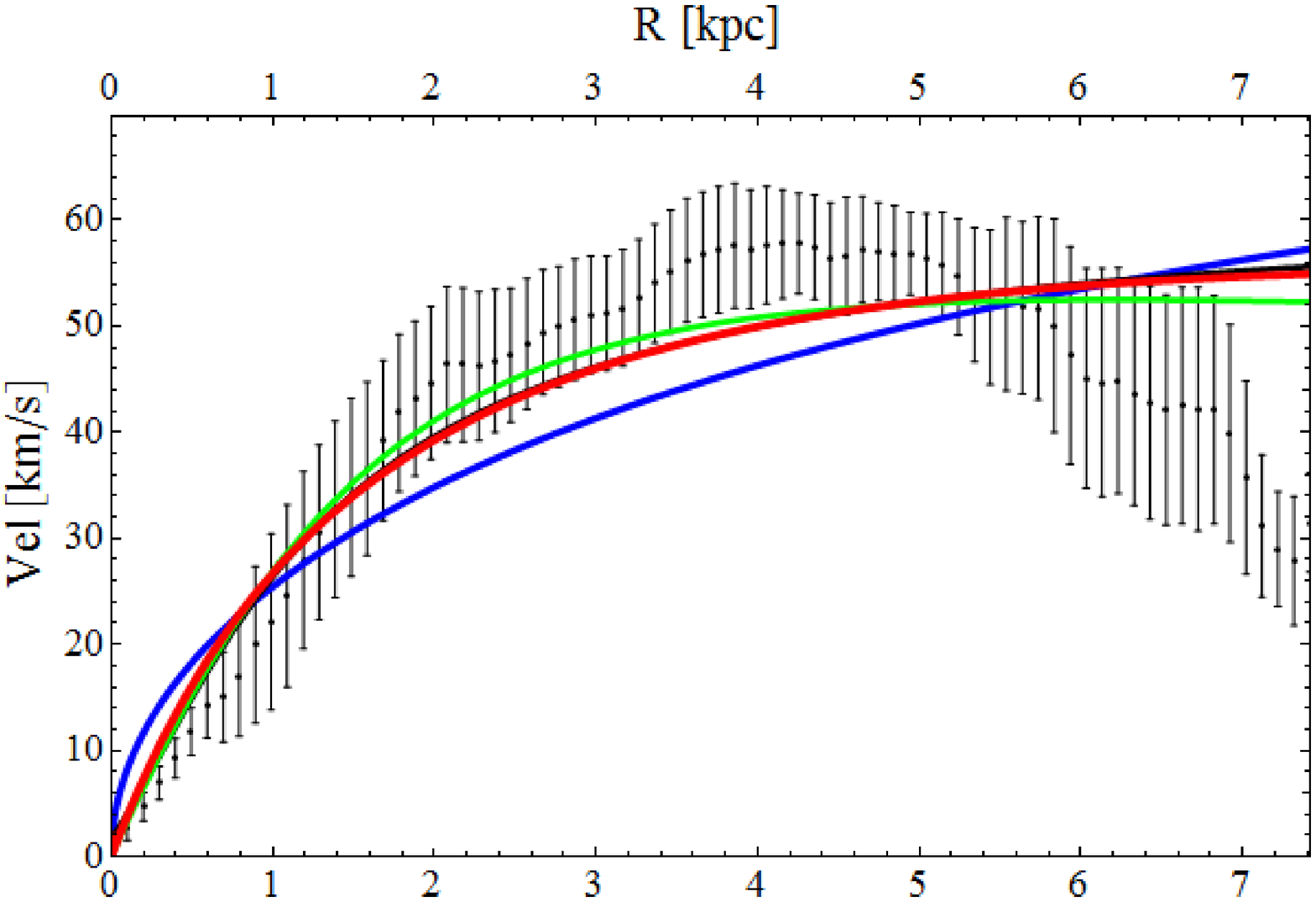} \\
    \includegraphics[width=0.35\textwidth]{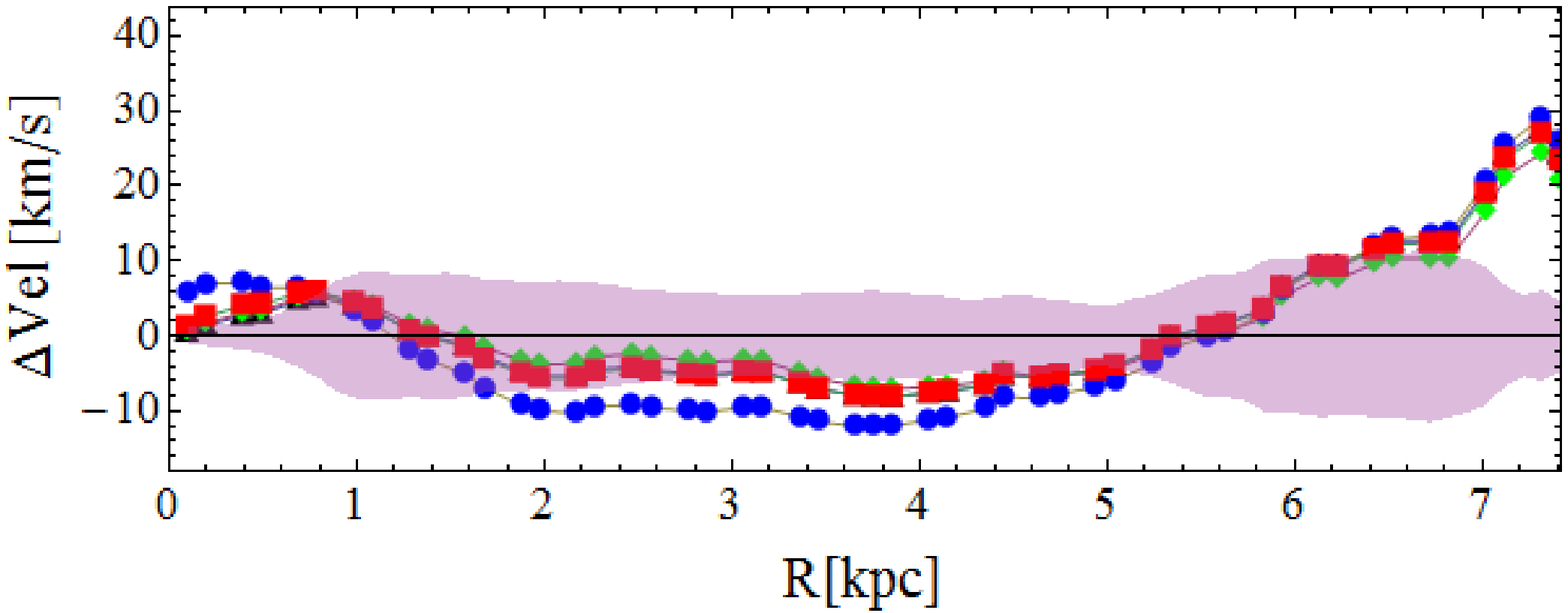}
    \end{tabular}  }
    \subfloat[\footnotesize{Min. disk + Gas}]{
    \begin{tabular}[b]{c}
    \includegraphics[width=0.35\textwidth]{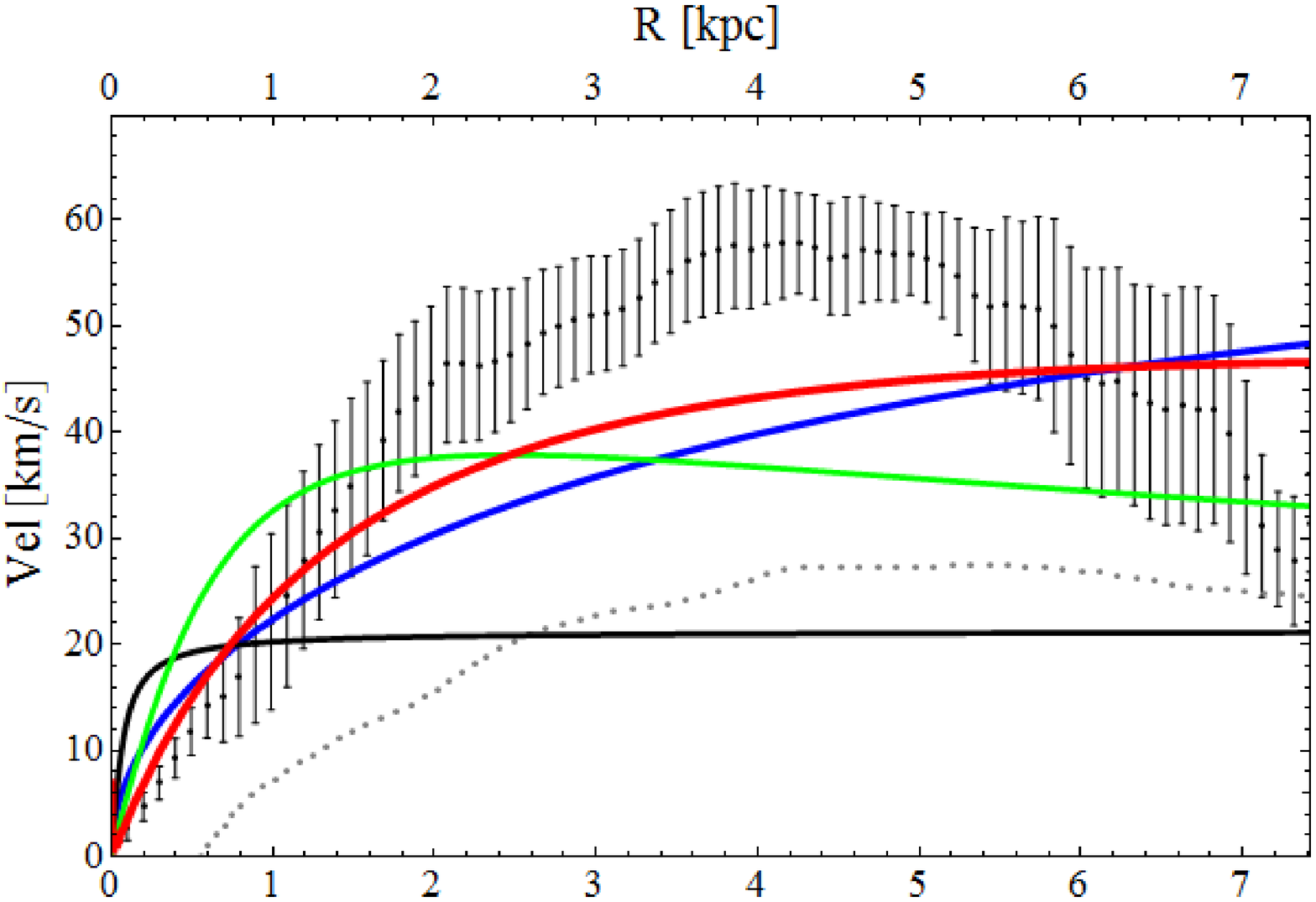} \\
    \includegraphics[width=0.35\textwidth]{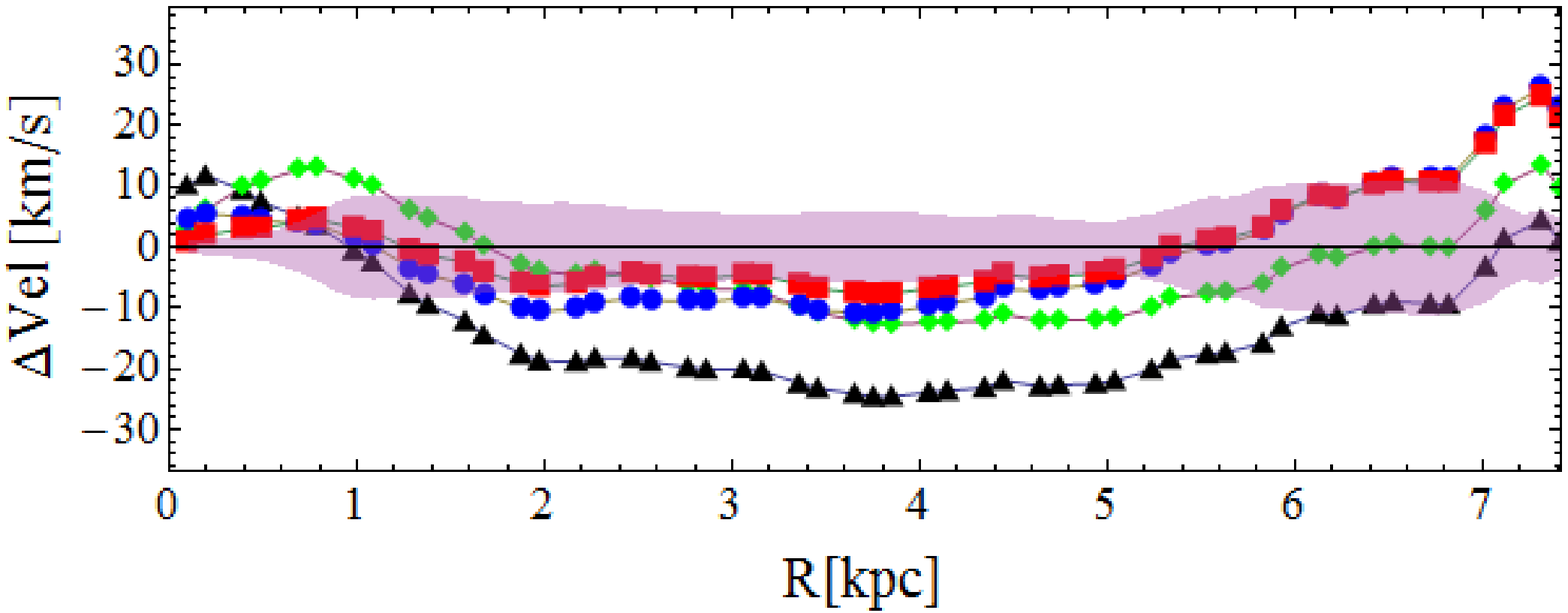}
    \end{tabular}  }  \\
    \subfloat[\footnotesize{Kroupa}]{
    \begin{tabular}[b]{c}
    \includegraphics[width=0.35\textwidth]{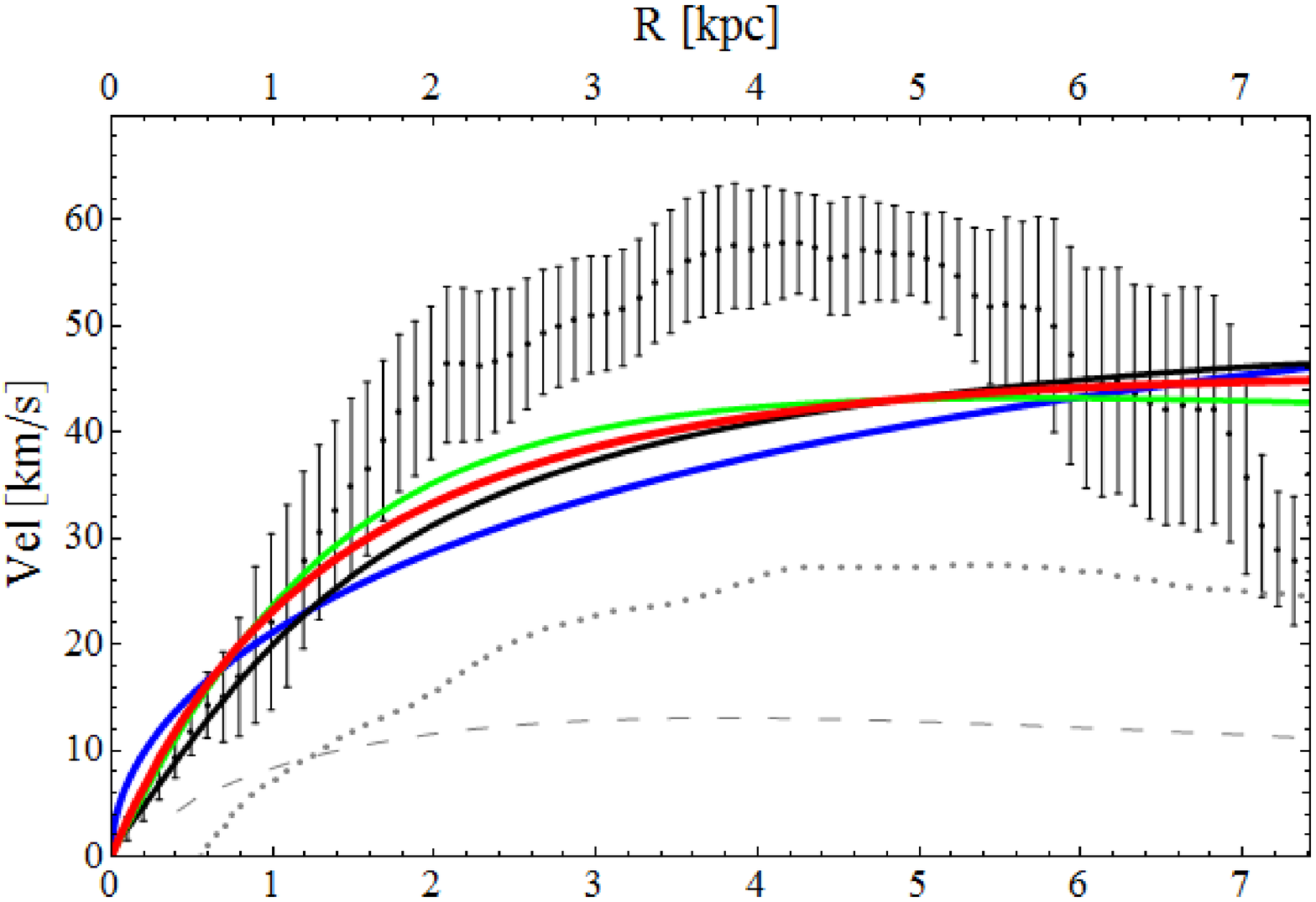} \\
    \includegraphics[width=0.35\textwidth]{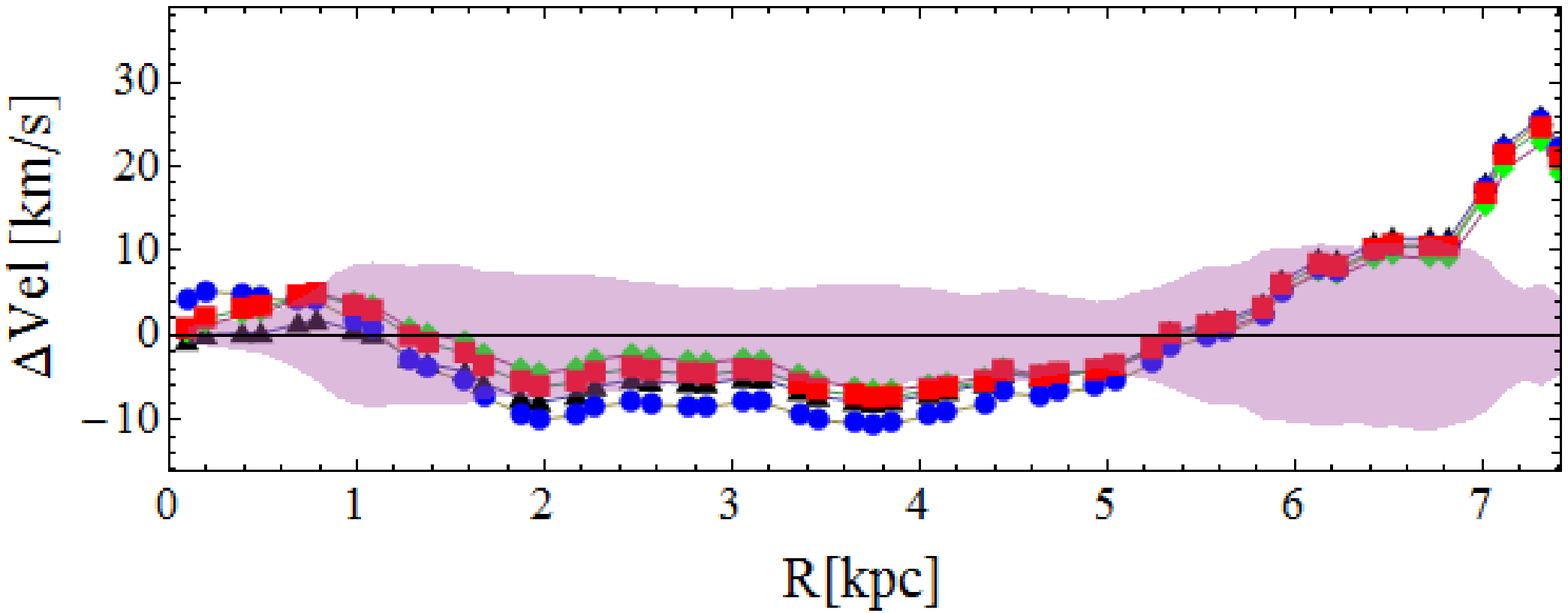}
    \end{tabular}  }
    \subfloat[\footnotesize{diet-Salpeter}]{
    \begin{tabular}[b]{c}
    \includegraphics[width=0.35\textwidth]{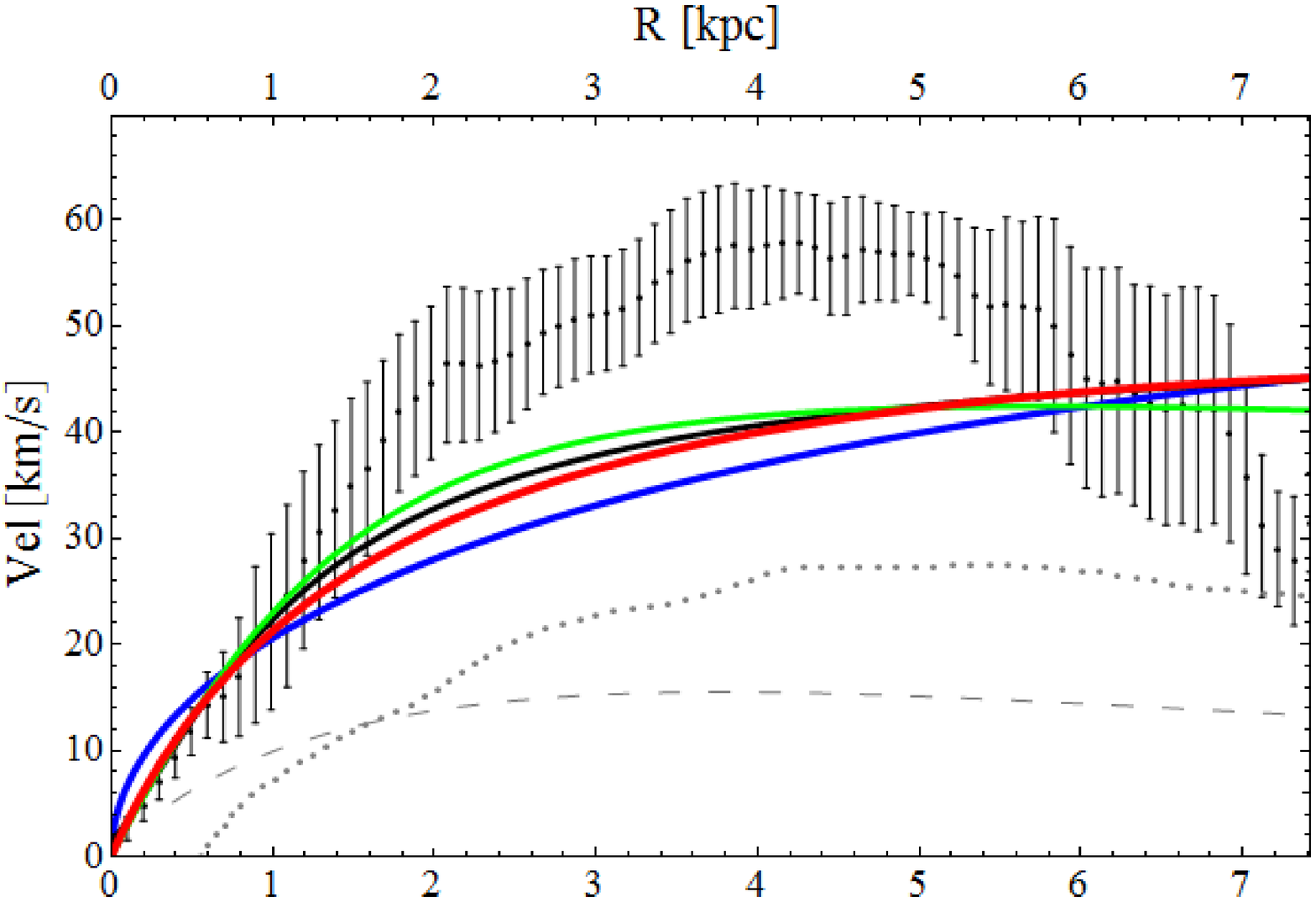} \\
    \includegraphics[width=0.35\textwidth]{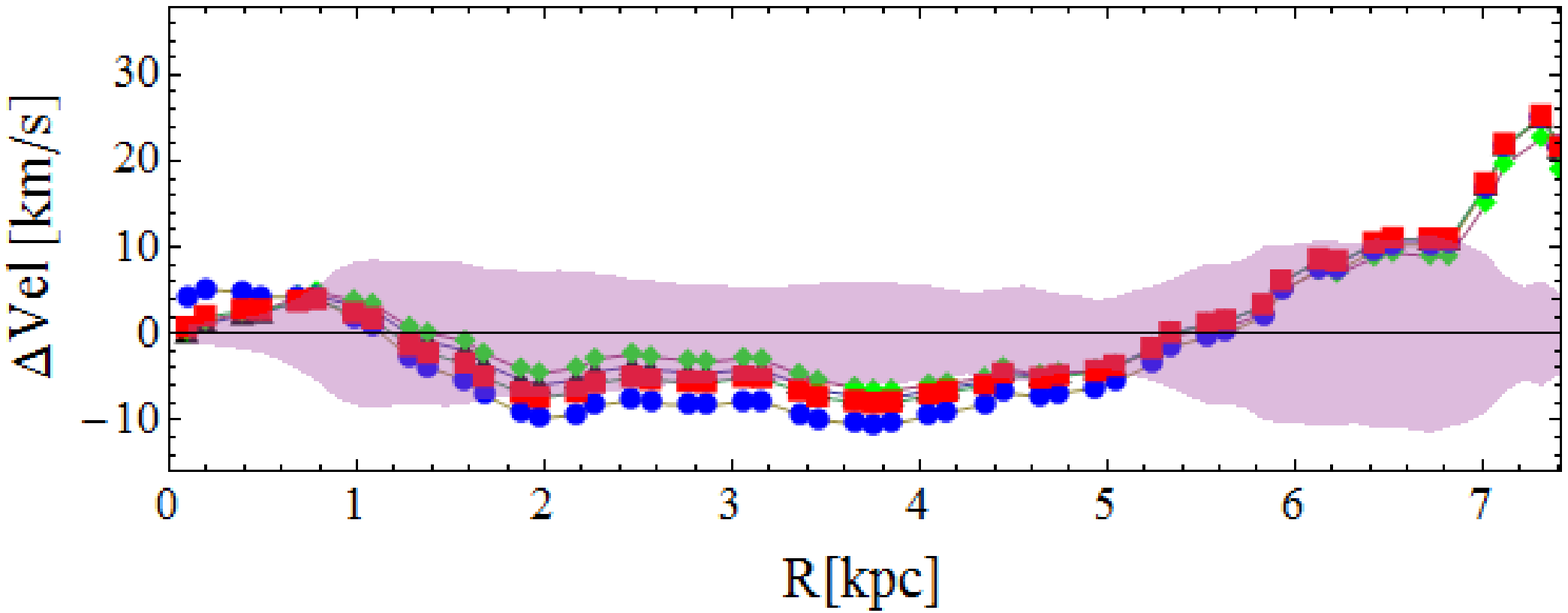}
    \end{tabular}  }
   \caption{\footnotesize{We display the rotation curves for the galaxy NGC 2366. Colors and symbols are as in Fig.\ref{fig:DDO154}. For these galaxy is clear that non circular velocities are affecting the rotation curve. We do inner analysis taken data below the 2.2 {\rm kpc} obtaining values for the core $r_c = 7.4 {\rm kpc}$ and the central density $\rho_c = 2.4 * 10^8 M_\sun/{\rm kpc}^3$. even if the parameters are not absolutely correct can be taken into account in order to do predictions of the behavior of the curve. For the stellar disk we use a $R_d = 1.76$ {\rm kpc} and mass $M_\star = 2.5*10^8 M_\sun$. }}
  \label{fig:NGC2366}
\end{figure}

\begin{figure}[h!]
    \subfloat[\footnotesize{Minimal disk}]{
    \begin{tabular}[b]{c}
    \includegraphics[width=0.35\textwidth]{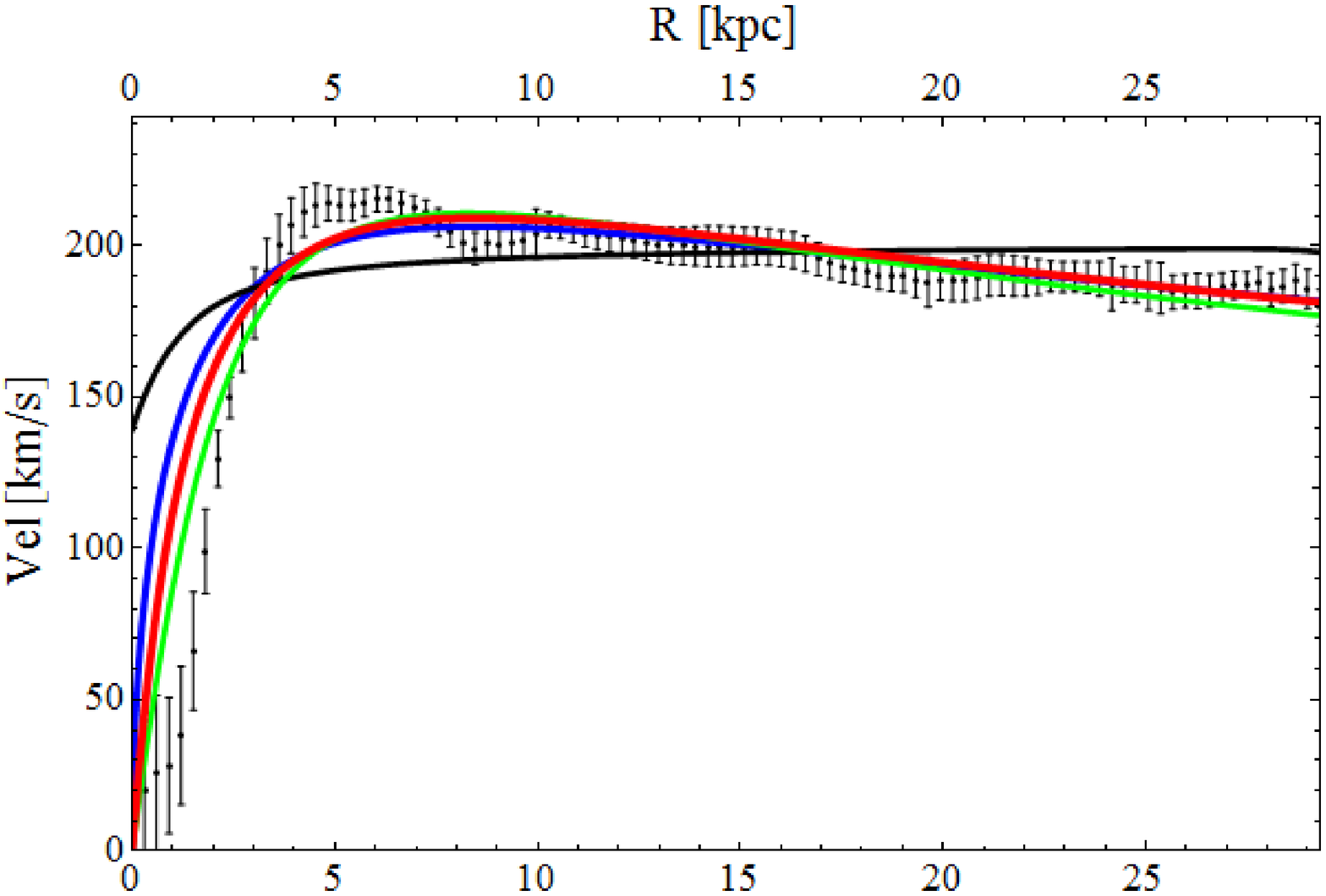} \\
    \includegraphics[width=0.35\textwidth]{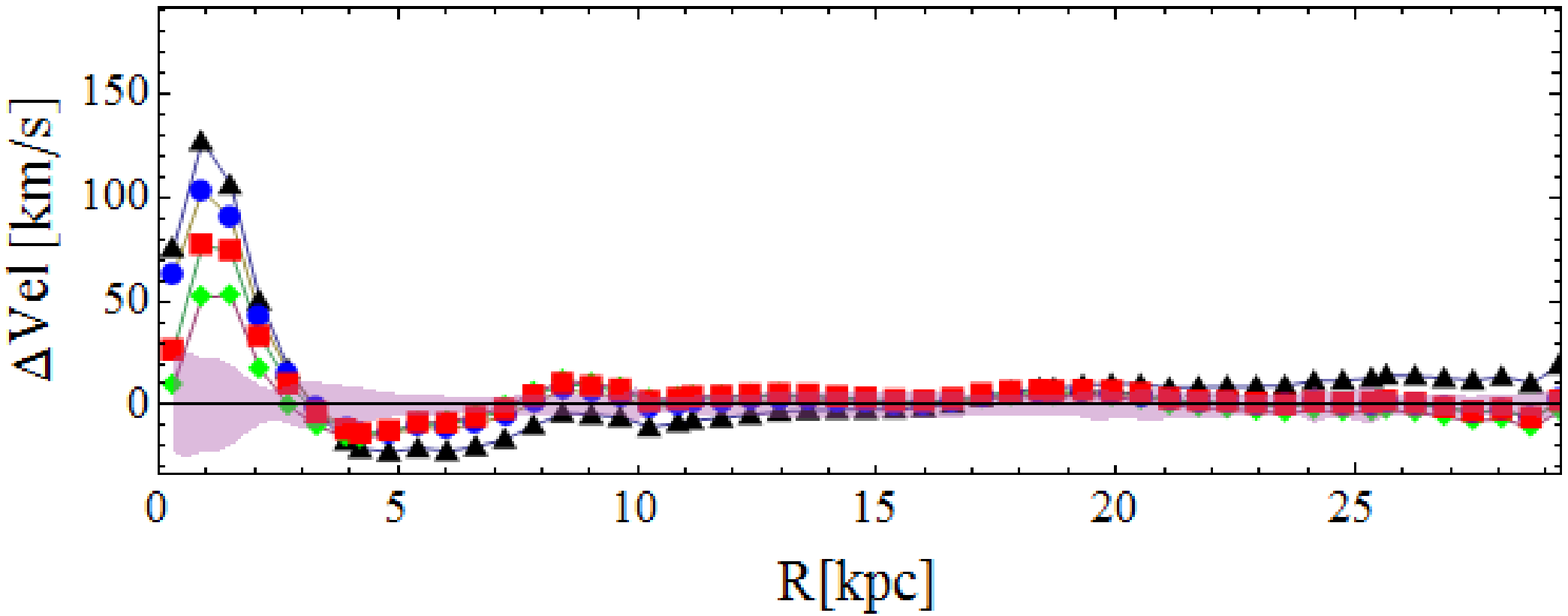}
    \end{tabular}  }
    \subfloat[\footnotesize{Min. disk + Gas}]{
    \begin{tabular}[b]{c}
    \includegraphics[width=0.35\textwidth]{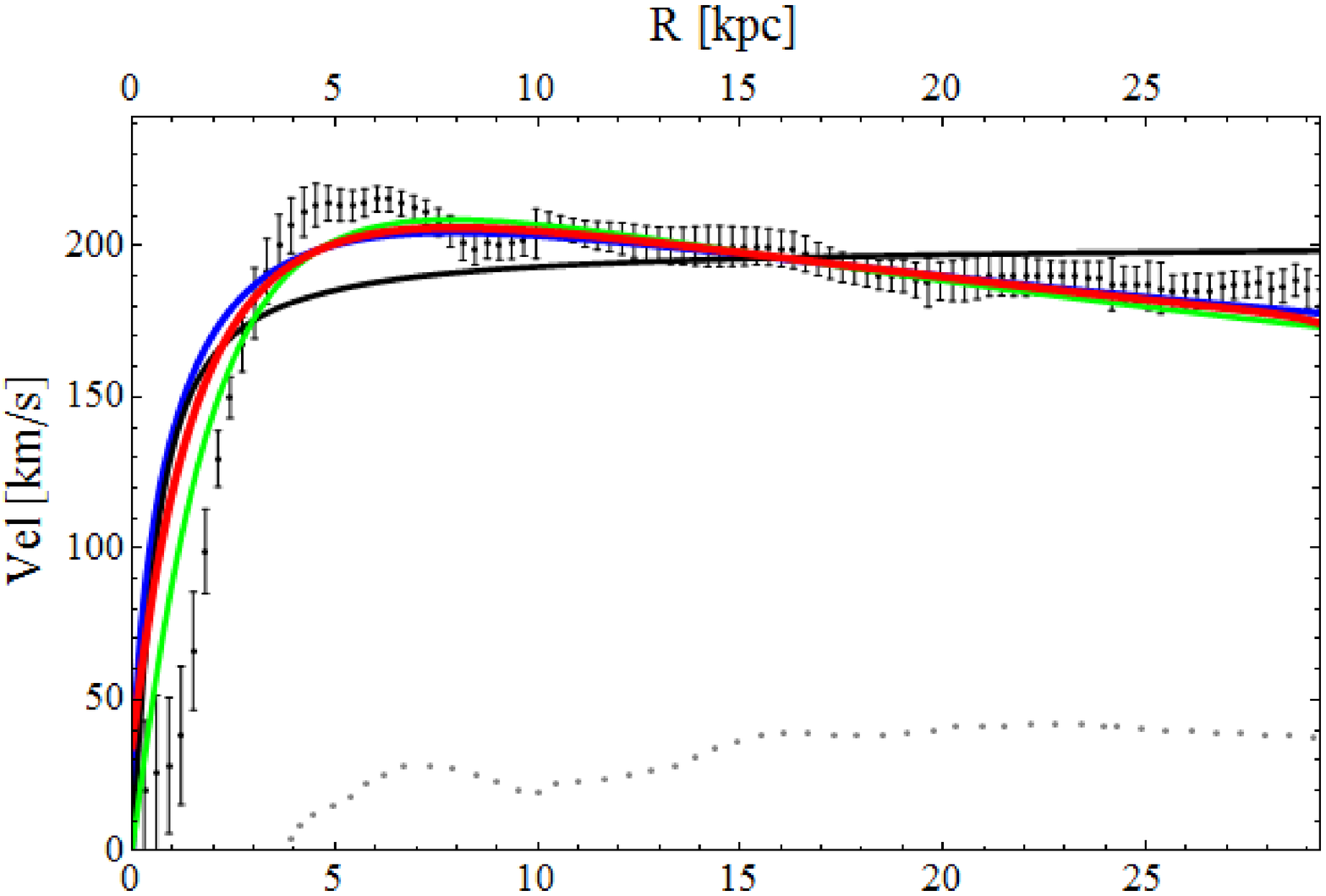} \\
    \includegraphics[width=0.35\textwidth]{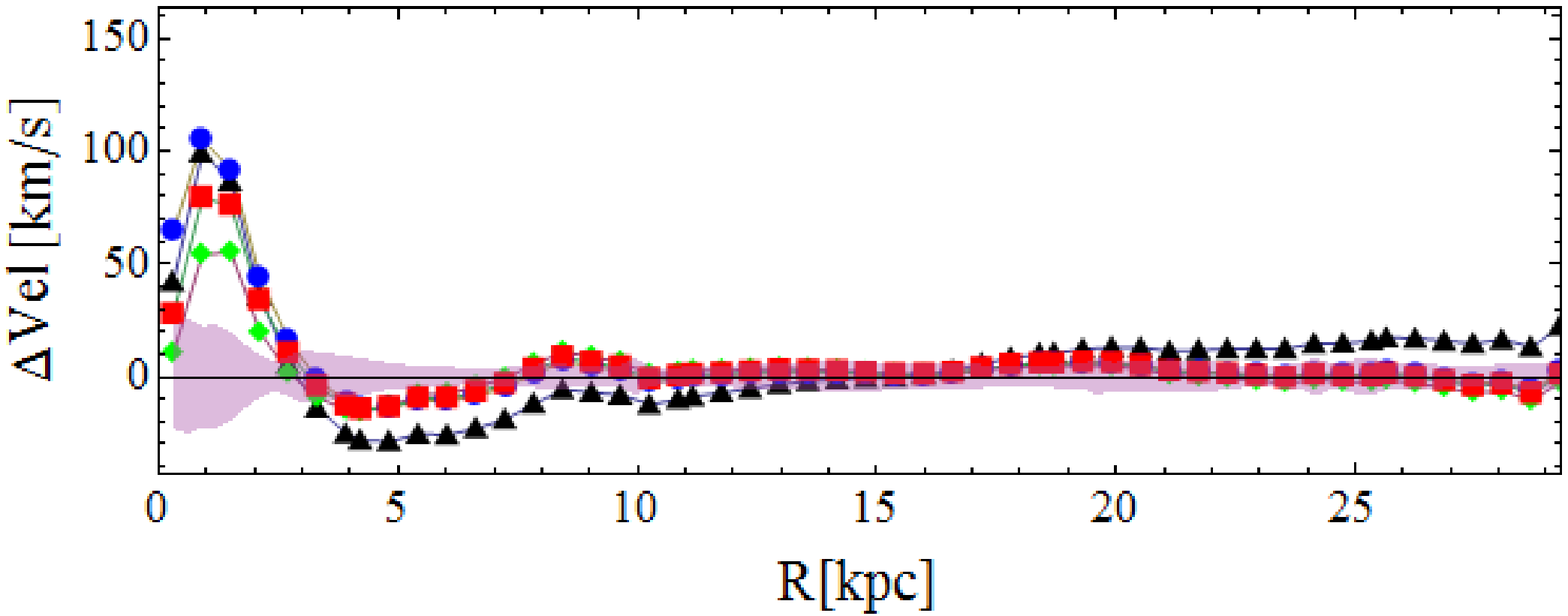}
    \end{tabular}  }  \\
    \subfloat[\footnotesize{Kroupa}]{
    \begin{tabular}[b]{c}
    \includegraphics[width=0.35\textwidth]{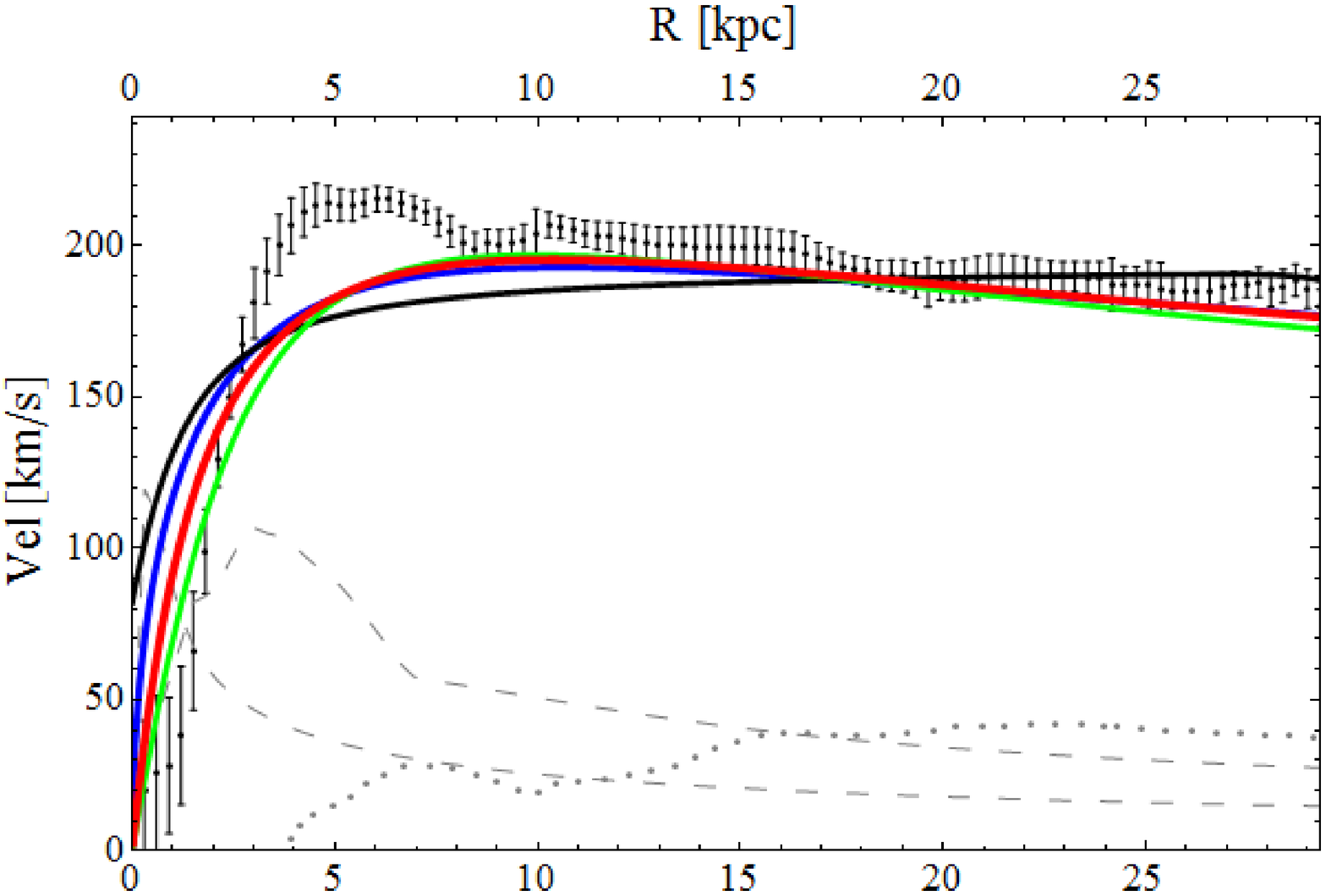} \\
    \includegraphics[width=0.35\textwidth]{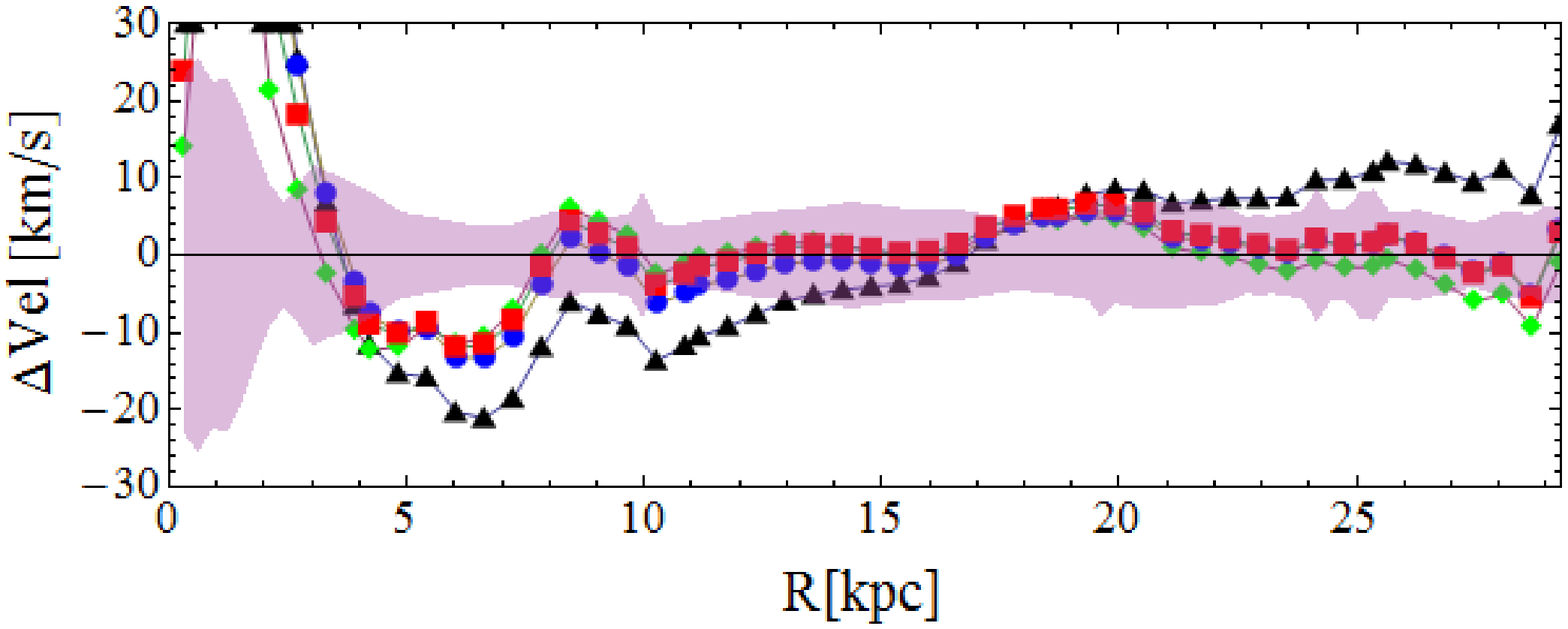}
    \end{tabular}  }
    \subfloat[\footnotesize{diet-Salpeter}]{
    \begin{tabular}[b]{c}
    \includegraphics[width=0.35\textwidth]{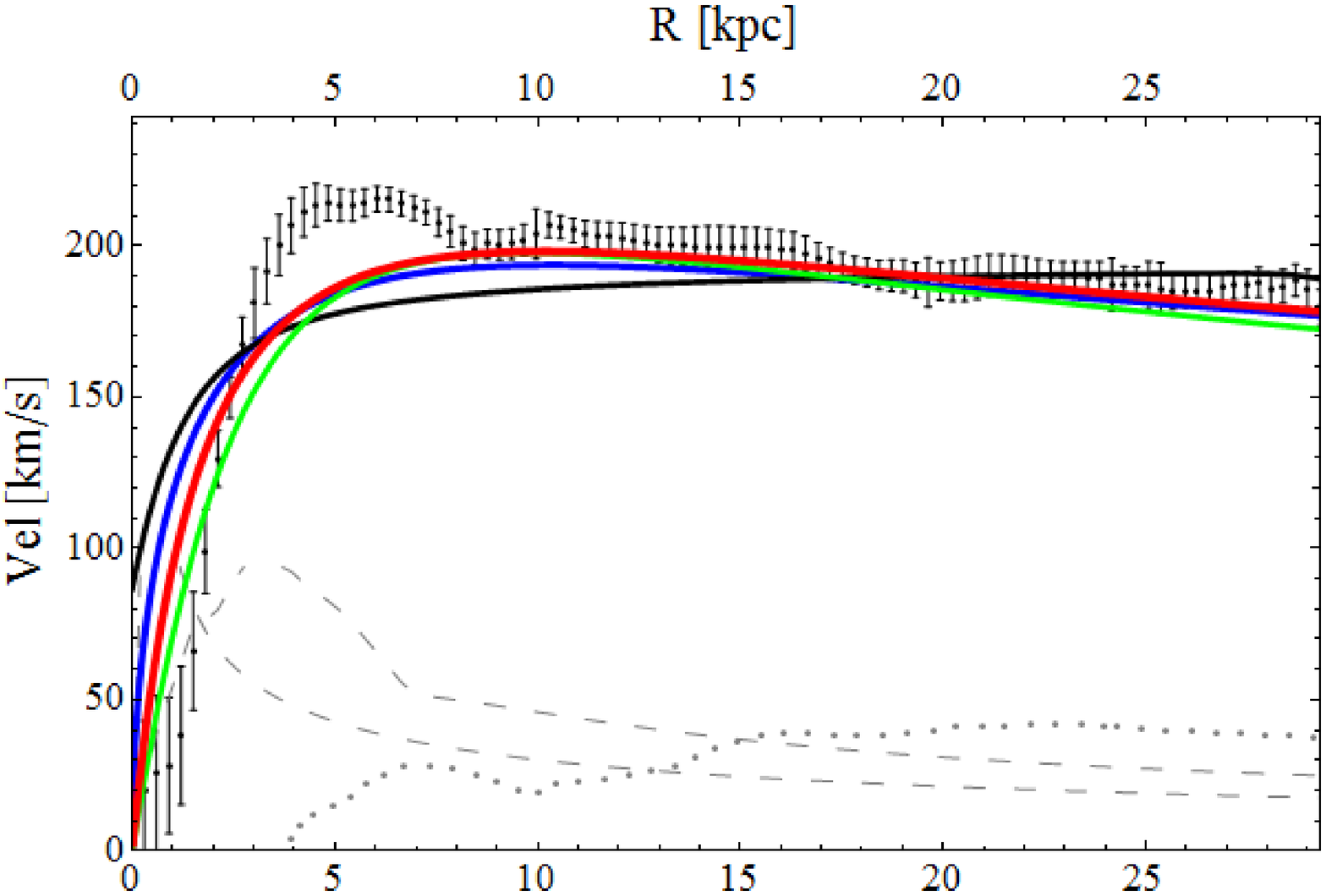} \\
    \includegraphics[width=0.35\textwidth]{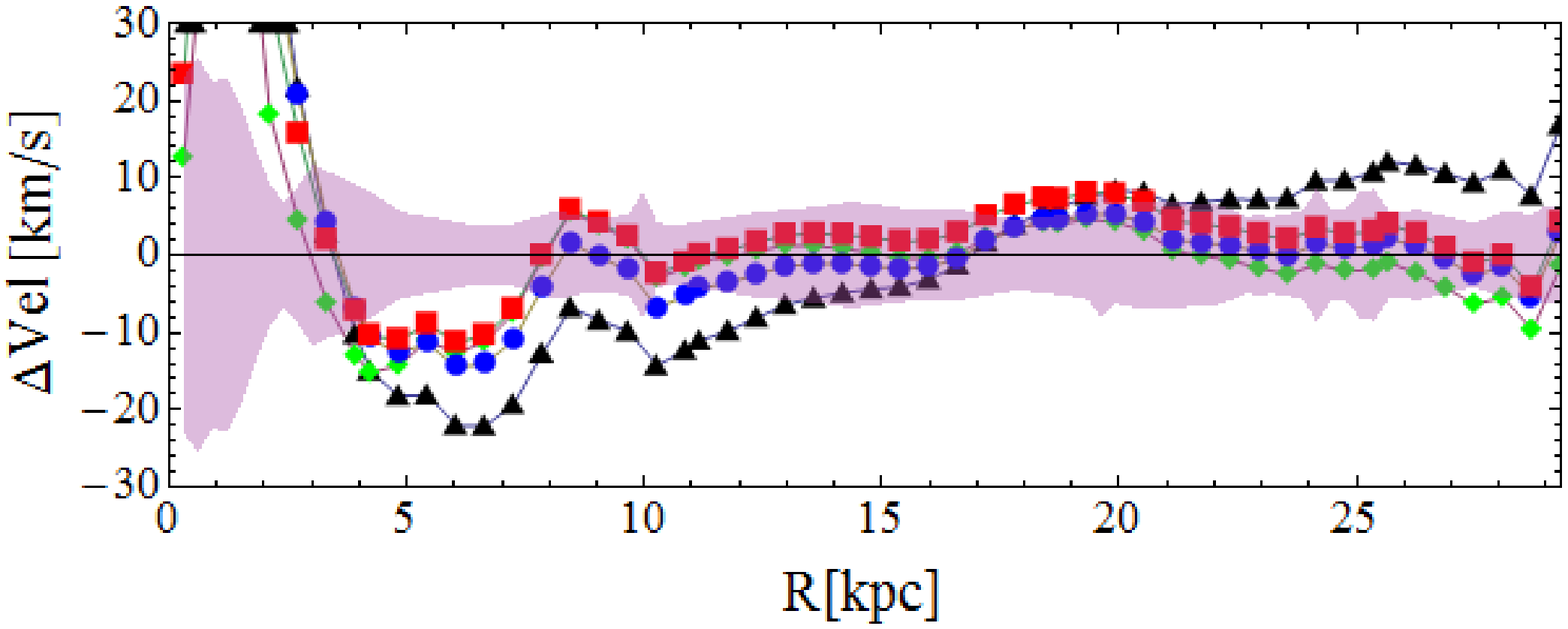}
    \end{tabular}  }
   \caption{\footnotesize{We present the rotation curves for the galaxy NGC 2903. Colors and symbols are as in Fig.\ref{fig:DDO154}. The profile of this galaxy can be decomposed into two components, but the central object provides a small effect on the rotation curve fit and it is ignored. For the outer disk we take values of $M_\star = 3.1*10^10 M_\sun$ and $R_d = 2.40$ {\rm kpc}. The inner analysis is carried out taking data points below 4.5 {\rm kpc} and obtaining values for core and central density of $r_c = 1.7 {\rm kpc}$ and $\rho_c = 6.7*10^8 M_\sun/{\rm kpc}^3$. }}
  \label{fig:NGC2903}
\end{figure}

\begin{figure}[h!]
    \subfloat[\footnotesize{Minimal disk}]{
    \begin{tabular}[b]{c}
    \includegraphics[width=0.35\textwidth]{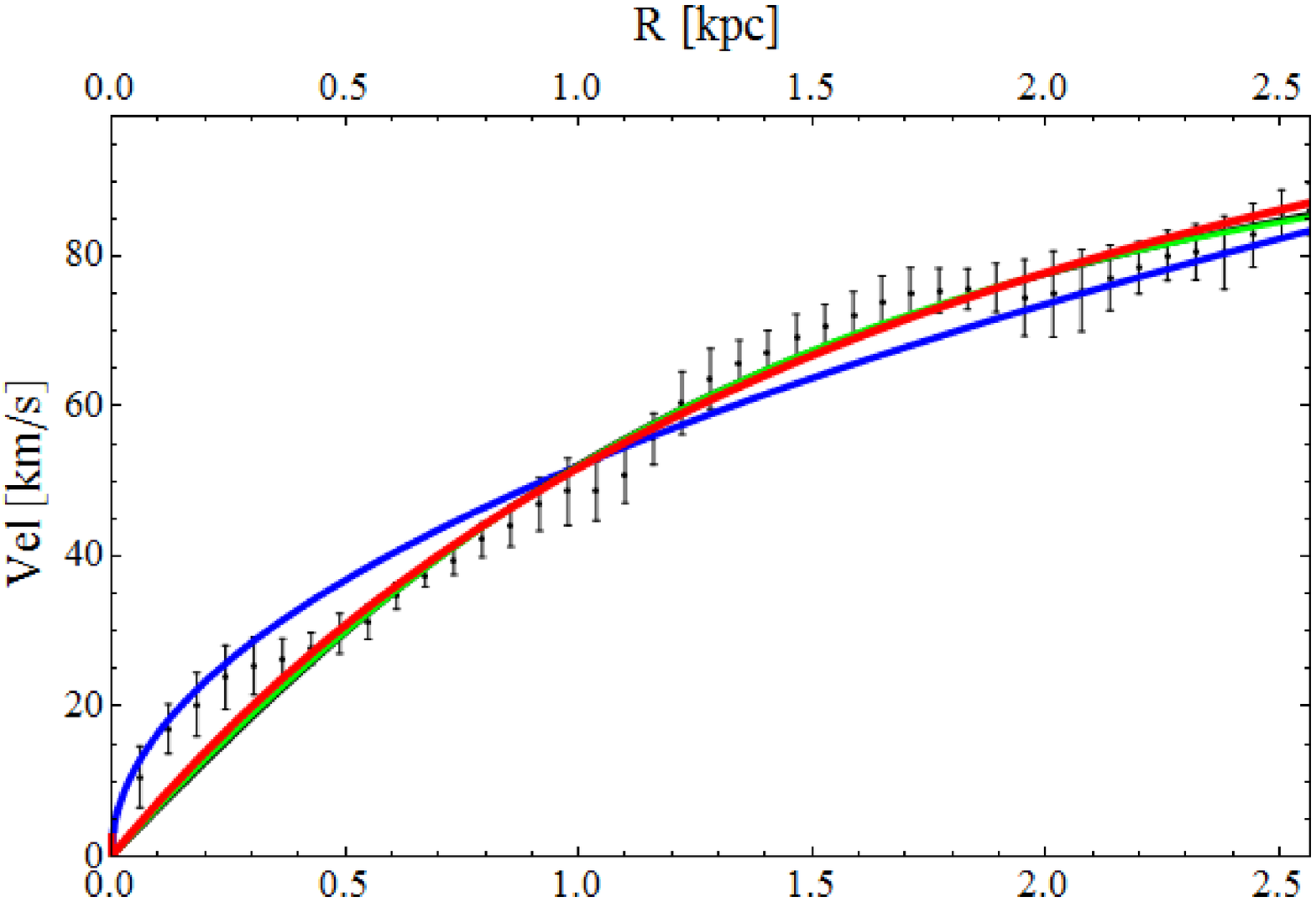} \\
    \includegraphics[width=0.35\textwidth]{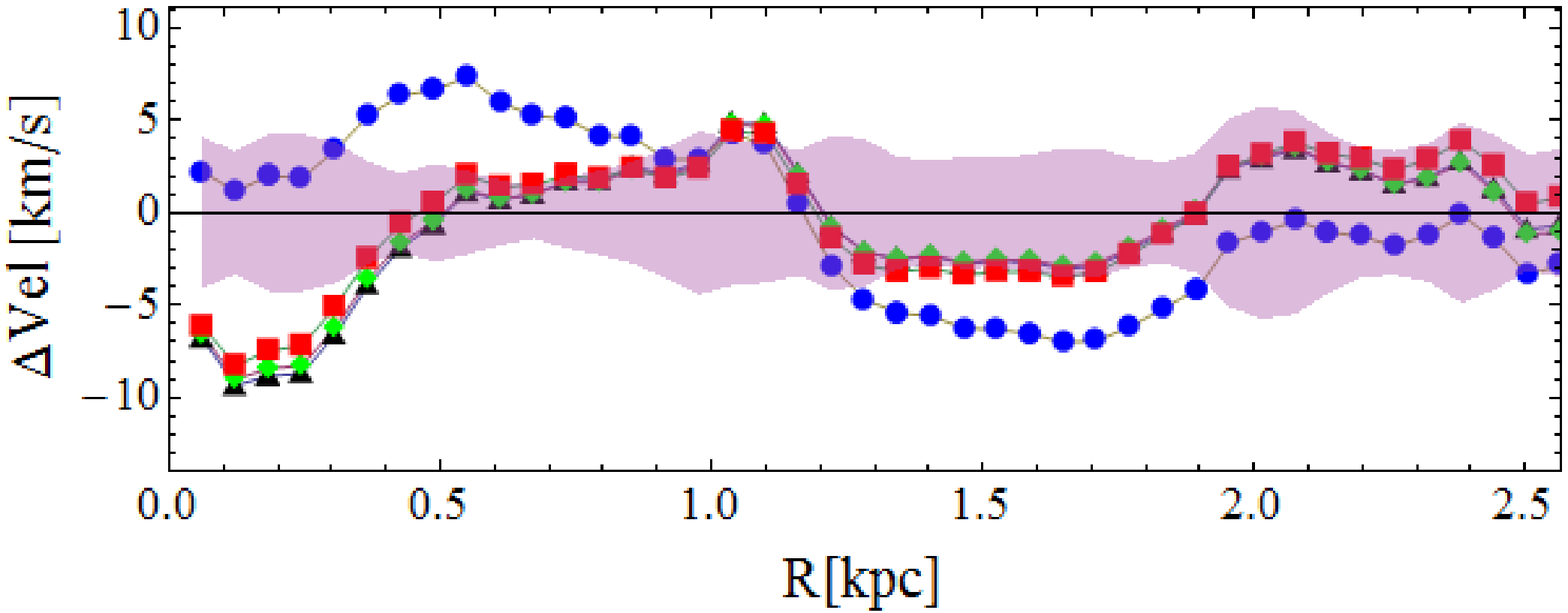}
    \end{tabular}  }
    \subfloat[\footnotesize{Min. disk + Gas}]{
    \begin{tabular}[b]{c}
    \includegraphics[width=0.35\textwidth]{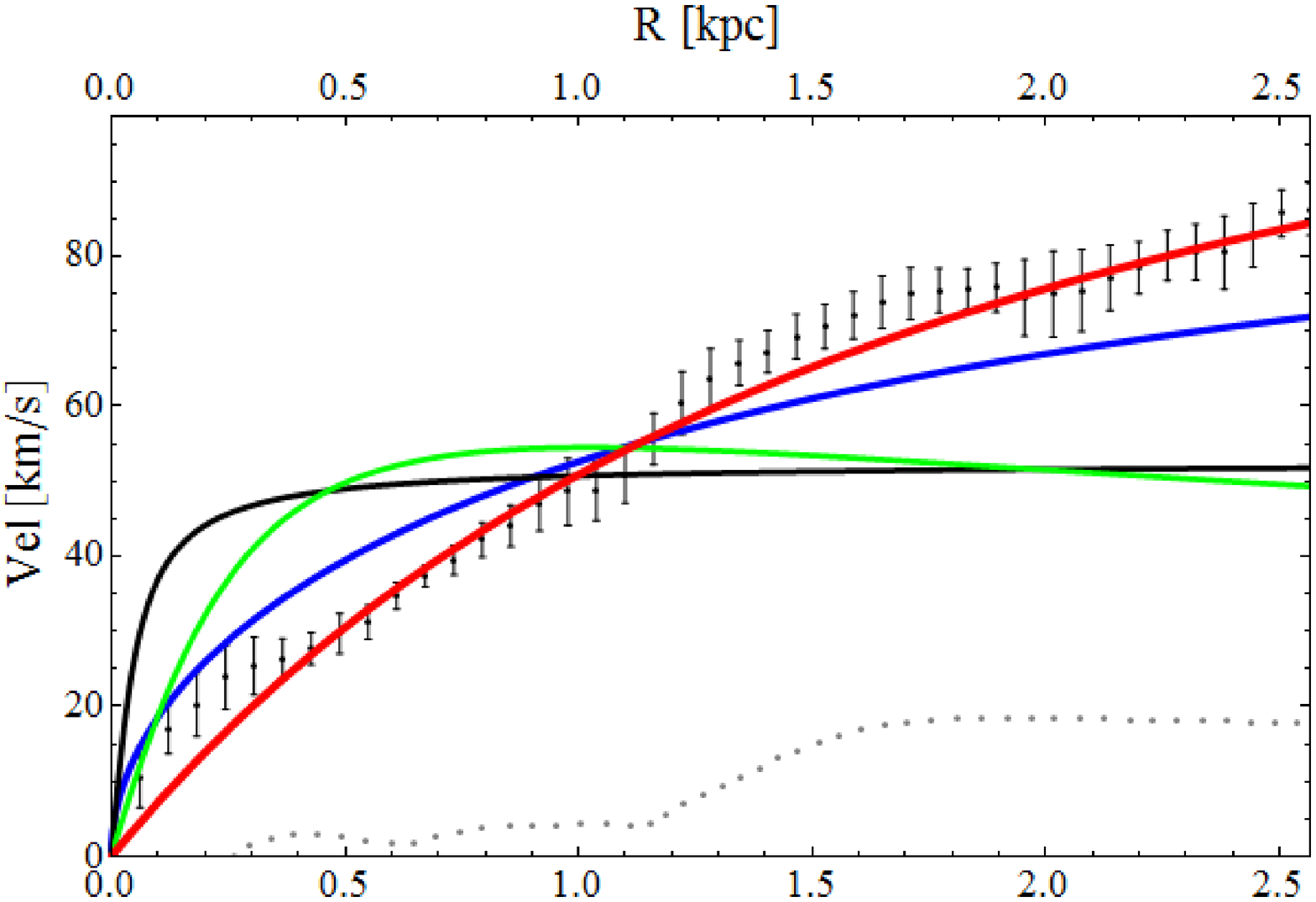} \\
    \includegraphics[width=0.35\textwidth]{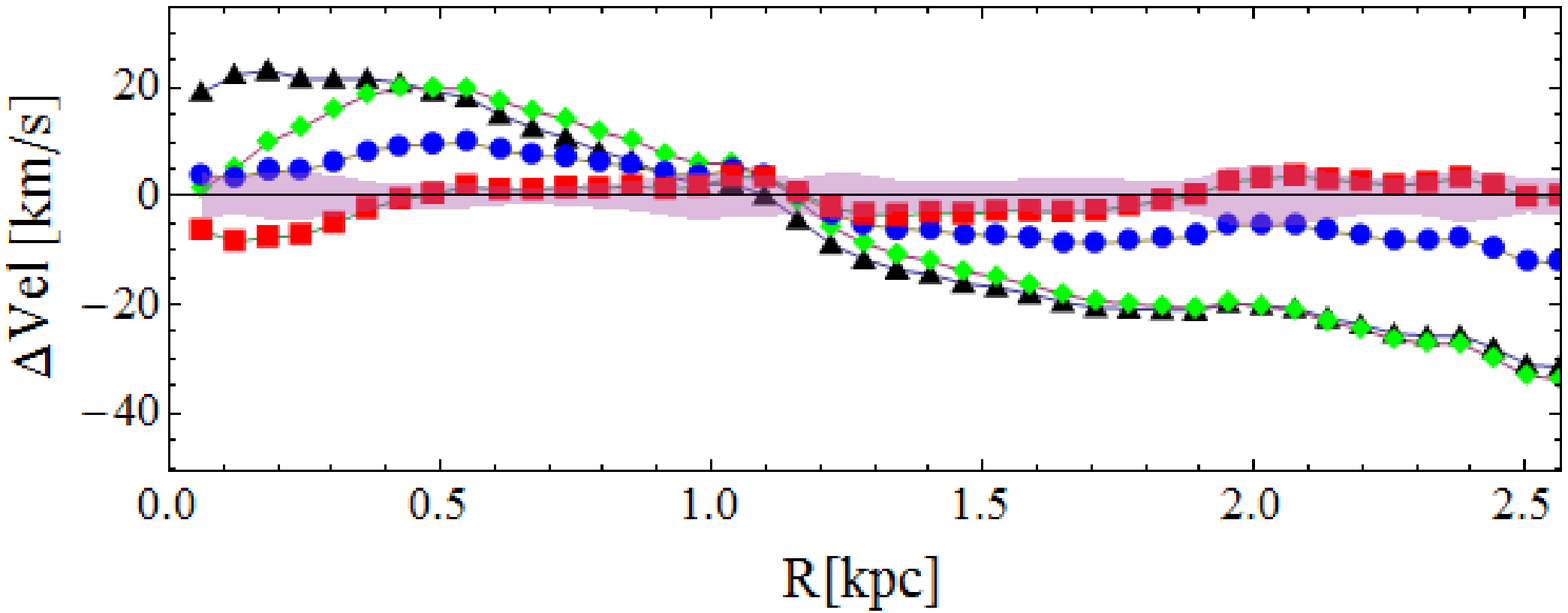}
    \end{tabular}  }  \\
    \subfloat[\footnotesize{Kroupa}]{
    \begin{tabular}[b]{c}
    \includegraphics[width=0.35\textwidth]{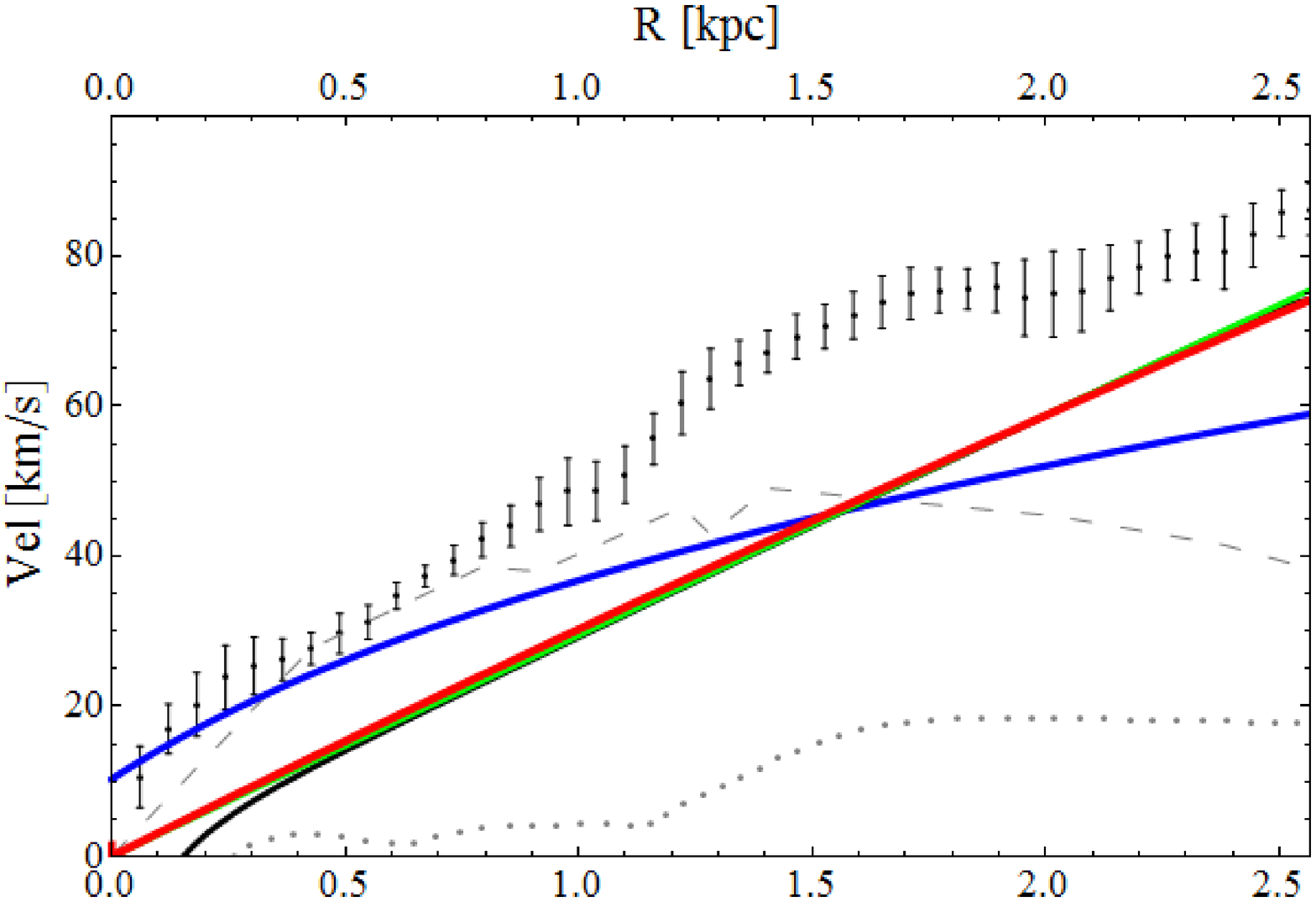} \\
    \includegraphics[width=0.35\textwidth]{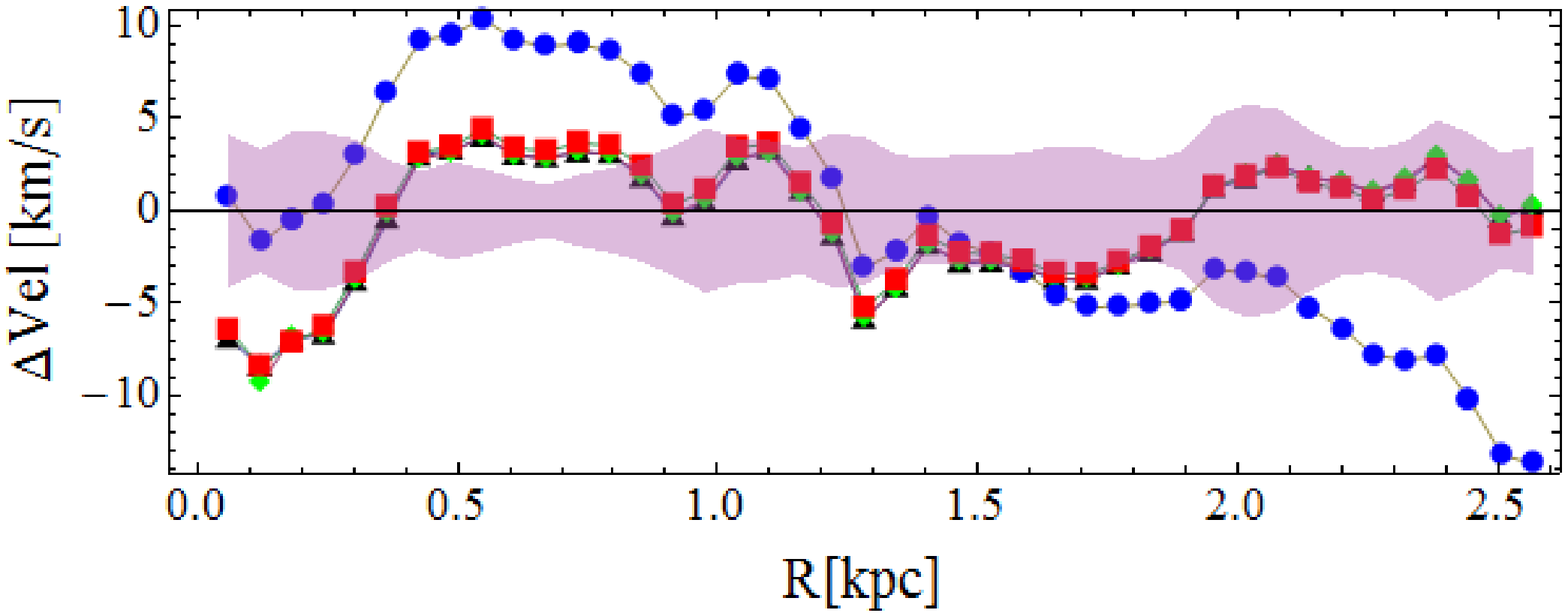}
    \end{tabular}  }
    \subfloat[\footnotesize{diet-Salpeter}]{
    \begin{tabular}[b]{c}
    \includegraphics[width=0.35\textwidth]{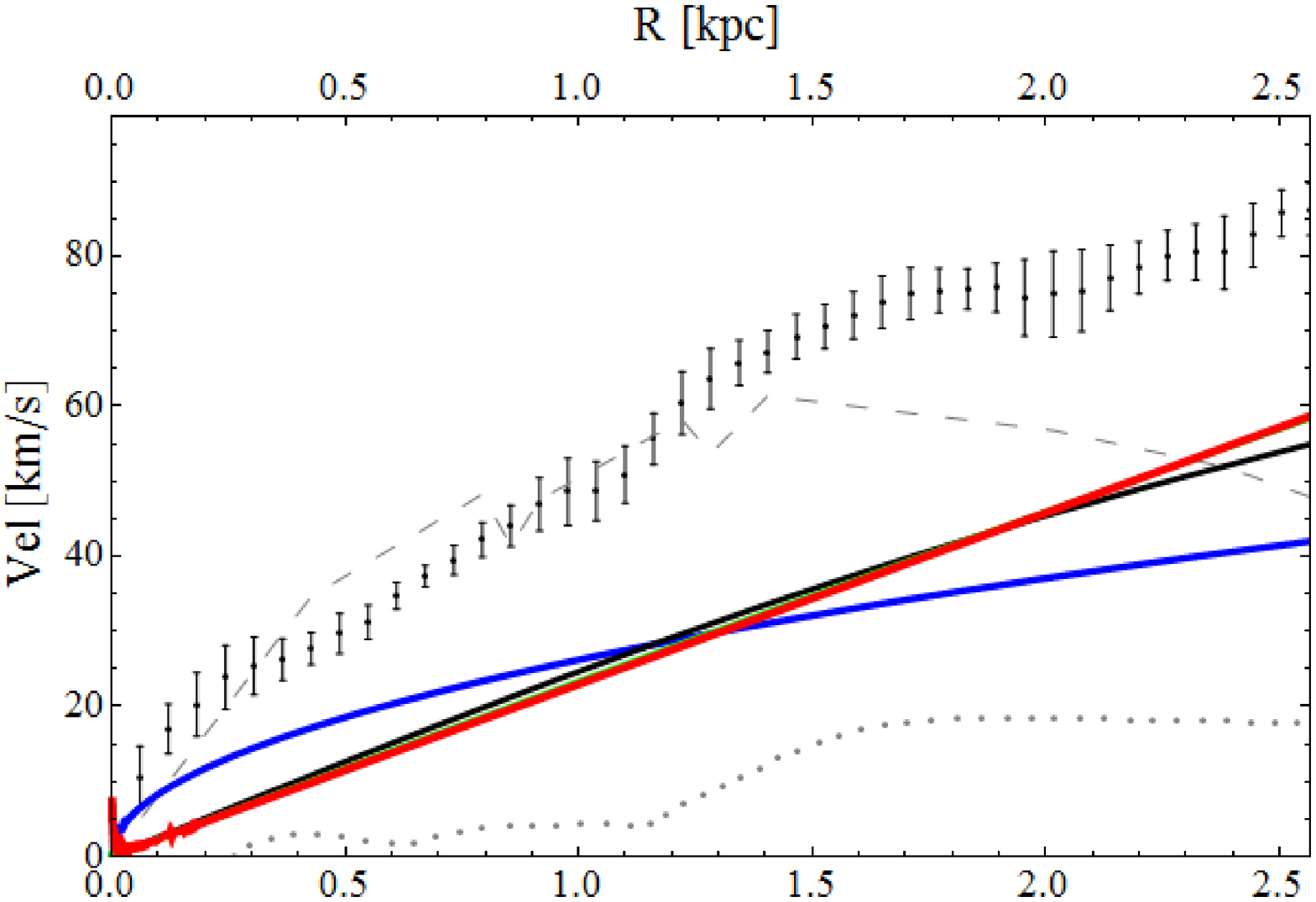} \\
    \includegraphics[width=0.35\textwidth]{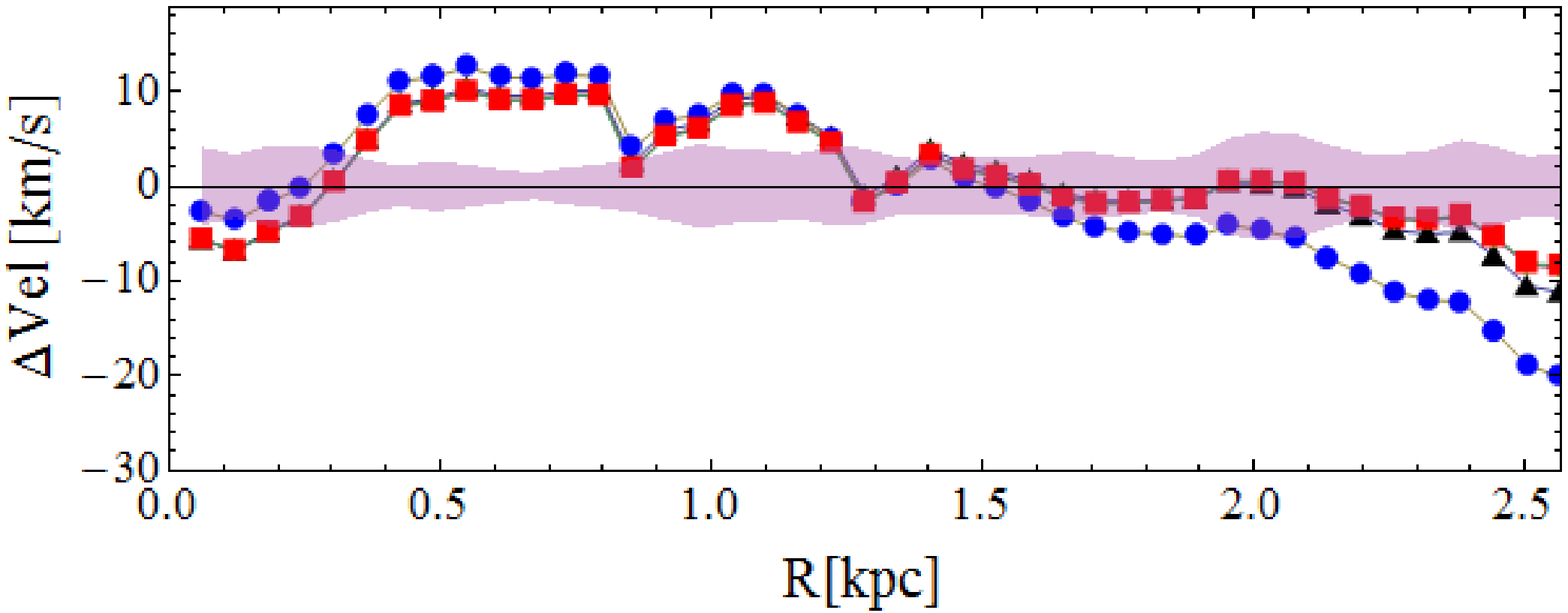}
    \end{tabular}  }
   \caption{\footnotesize{We display the rotation curves for the galaxy NGC 2976. Colors and symbols are as in Fig.\ref{fig:DDO154}. The galaxy show a color gradient, for sake of simplicity we take $\gs = 0.52$ for all radii. We assume the value $\mu_0 = 17.78$ mag arcsec$^{-2}$ and $R_d = 0.91$ {\rm kpc}. For the inner analysis we take all the observational data, up to 2.5 {\rm kpc}, since the slope of the rotation curve is very smooth and can not be distinguished where the maximum velocity could be located. The fitted values obtained from the inner analysis are $r_c = 0.25 {\rm kpc}$ and $\rho_c = 1.3*10^8 M_\sun/{\rm kpc}^3$. It is difficult to break the degeneracy between $r_c$ and $\rho_0$ since we have no more data. }}
  \label{fig:NGC2976}
\end{figure}

\begin{figure}[h!]
    \subfloat[\footnotesize{Minimal disk}]{
    \begin{tabular}[b]{c}
    \includegraphics[width=0.35\textwidth]{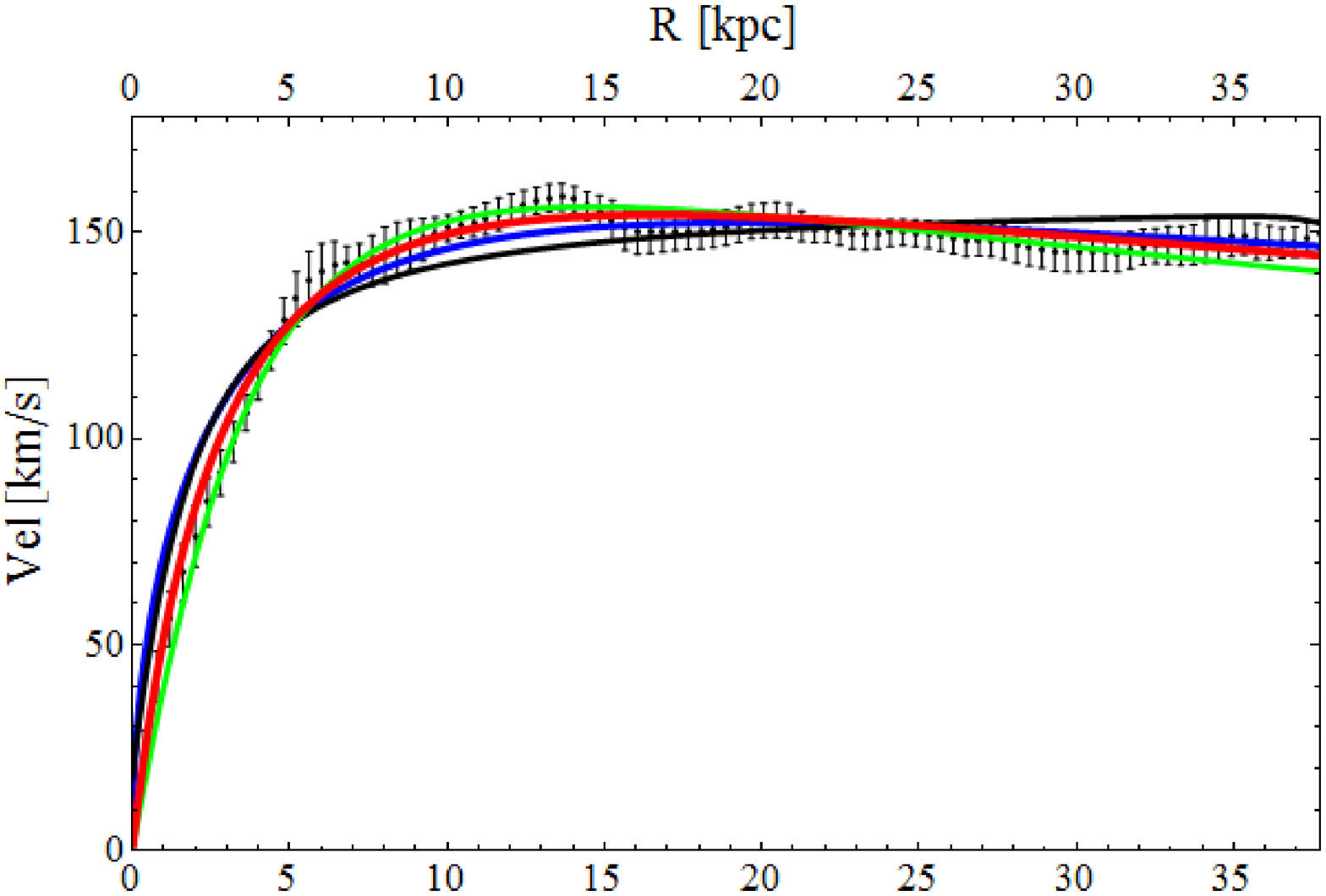} \\
    \includegraphics[width=0.35\textwidth]{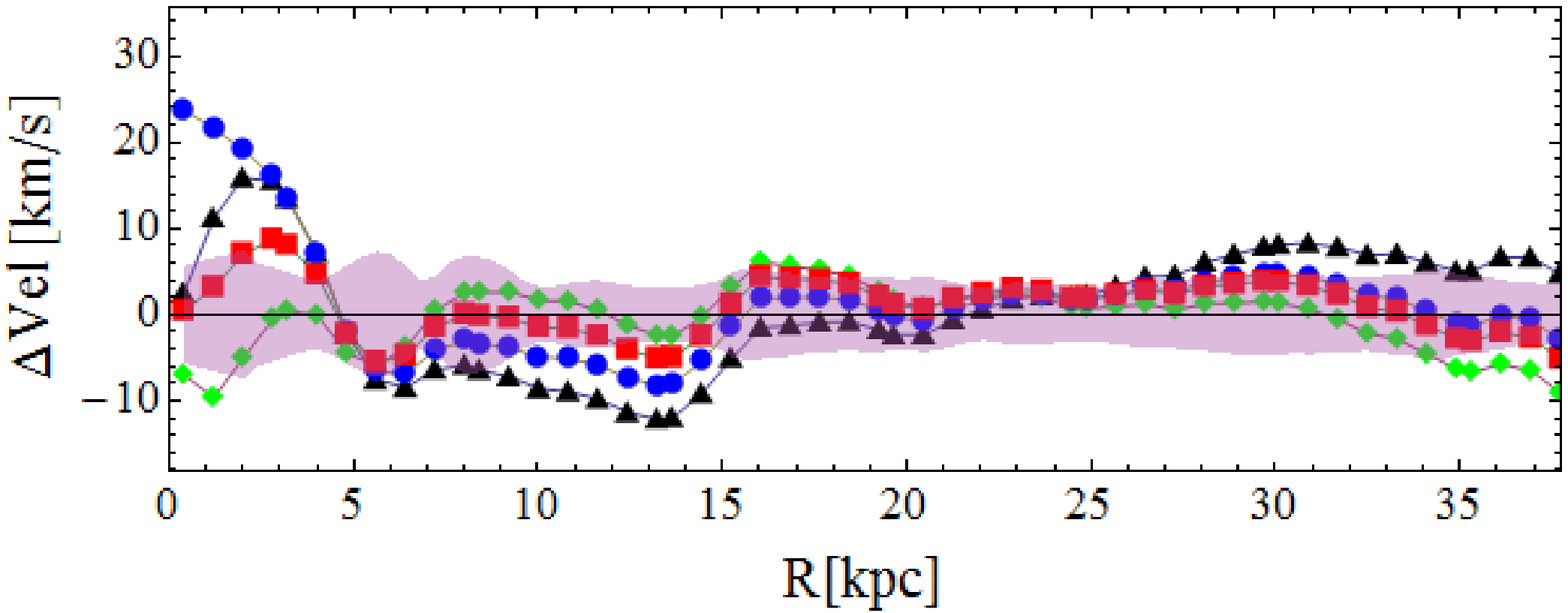}
    \end{tabular}  }
    \subfloat[\footnotesize{Min. disk + Gas}]{
    \begin{tabular}[b]{c}
    \includegraphics[width=0.35\textwidth]{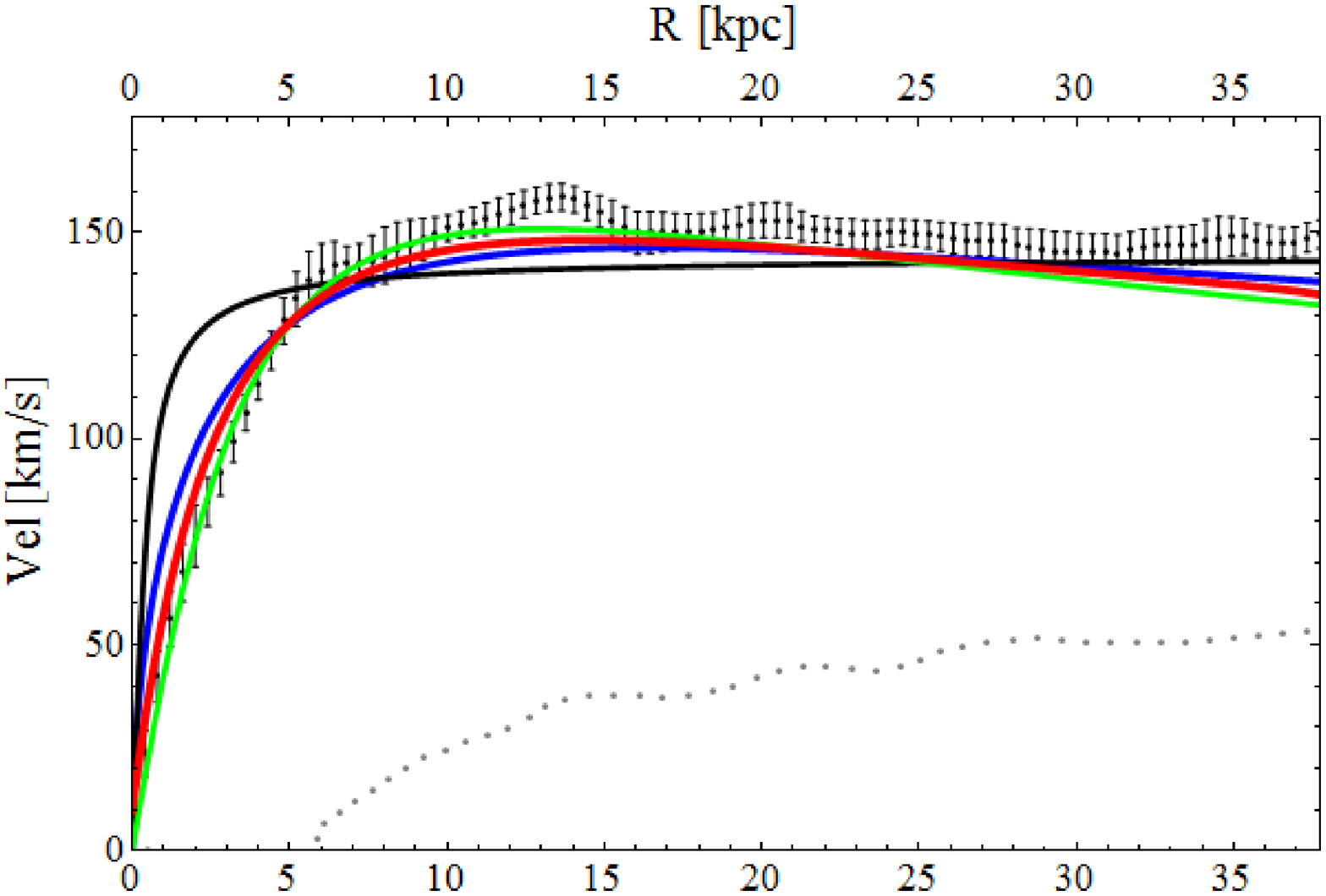} \\
    \includegraphics[width=0.35\textwidth]{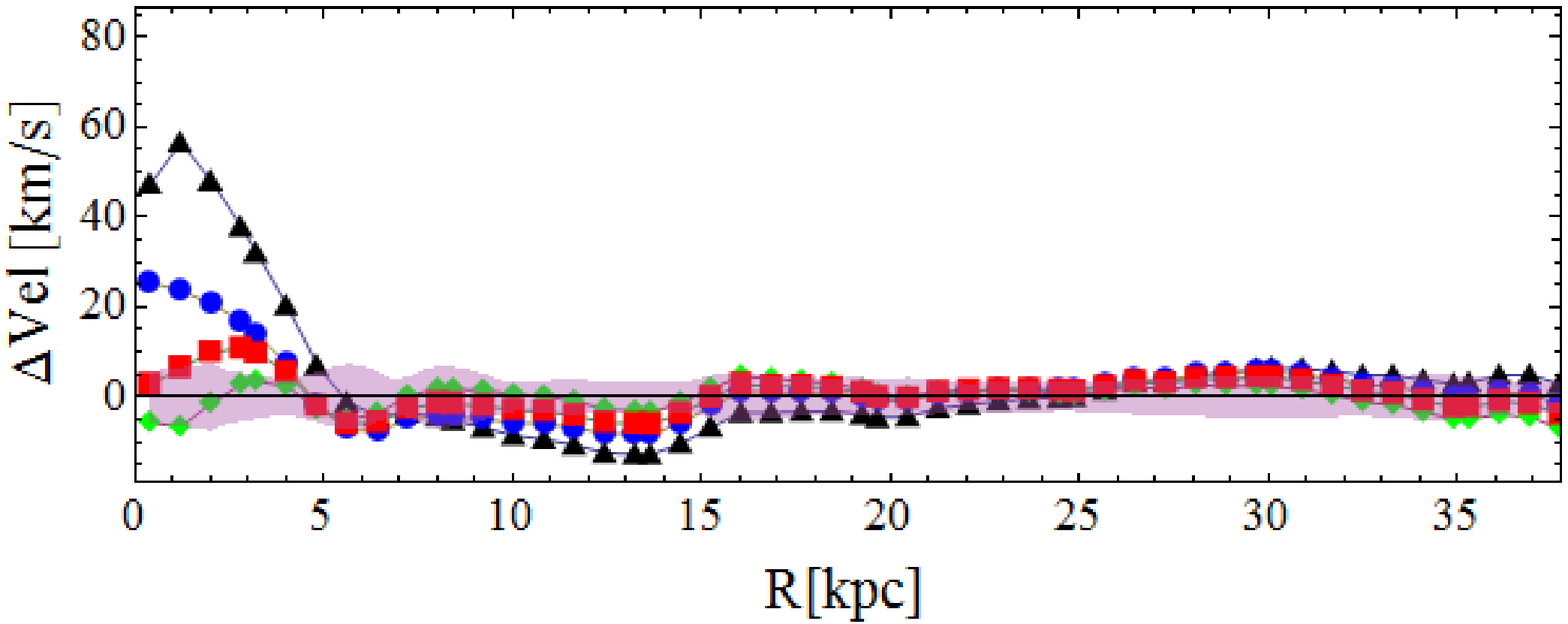}
    \end{tabular}  }  \\
    \subfloat[\footnotesize{Kroupa}]{
    \begin{tabular}[b]{c}
    \includegraphics[width=0.35\textwidth]{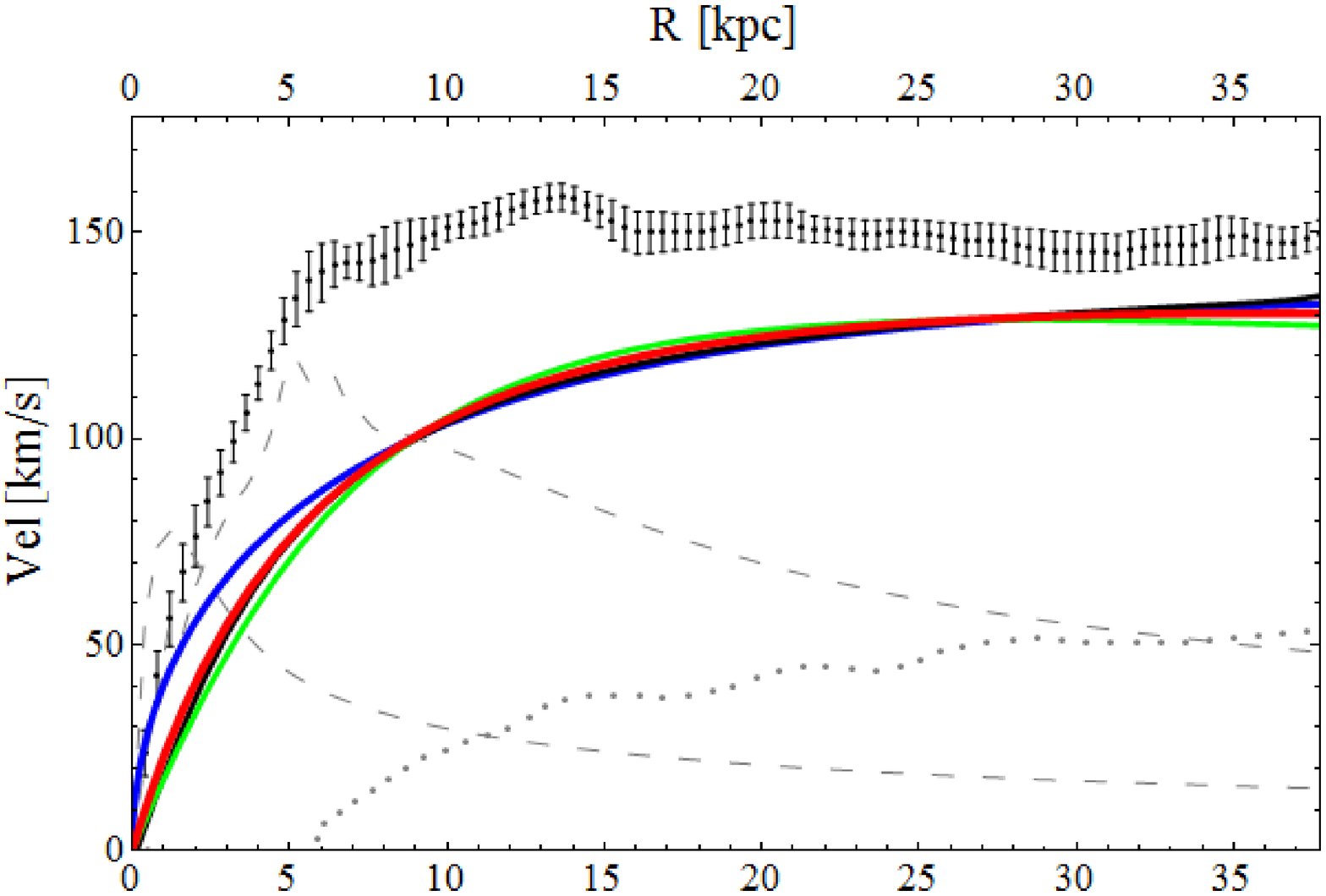} \\
    \includegraphics[width=0.35\textwidth]{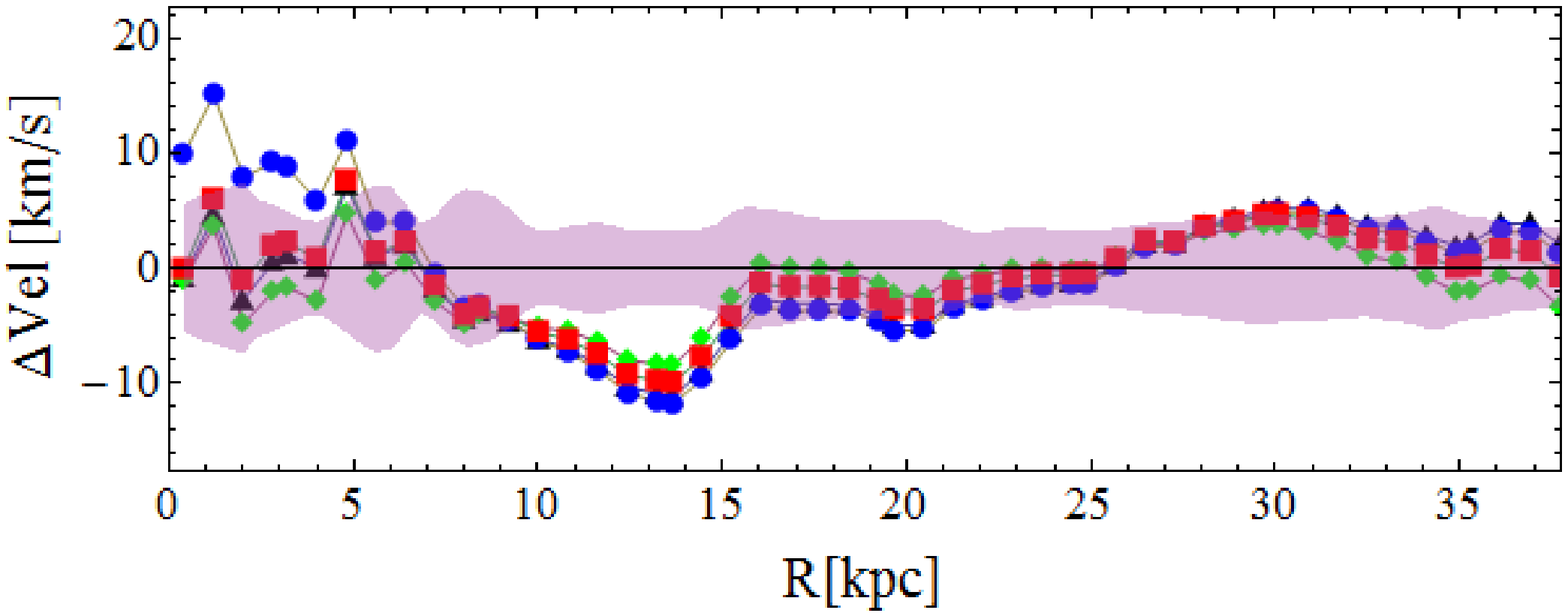}
    \end{tabular}  }
    \subfloat[\footnotesize{diet-Salpeter}]{
    \begin{tabular}[b]{c}
    \includegraphics[width=0.35\textwidth]{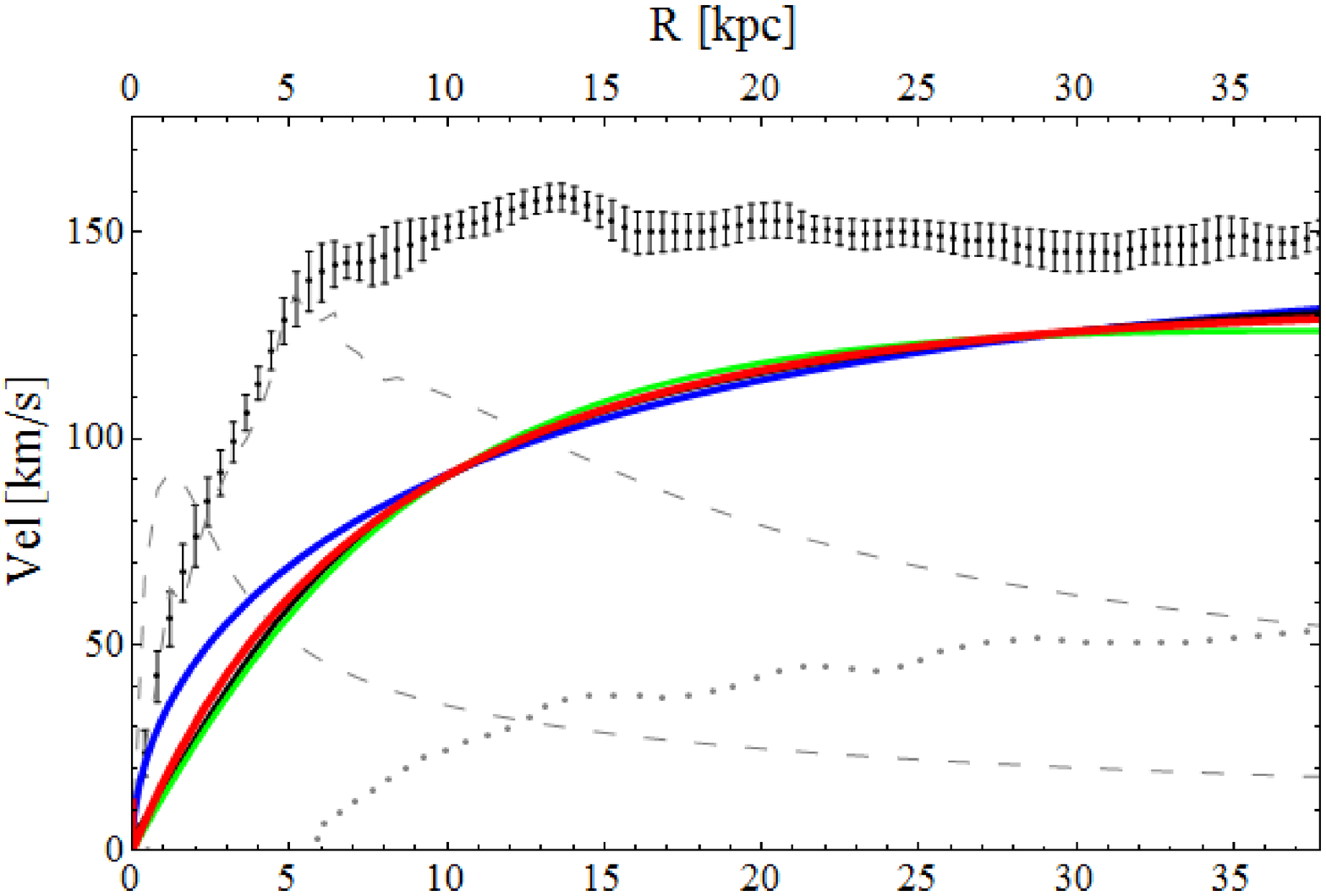} \\
    \includegraphics[width=0.35\textwidth]{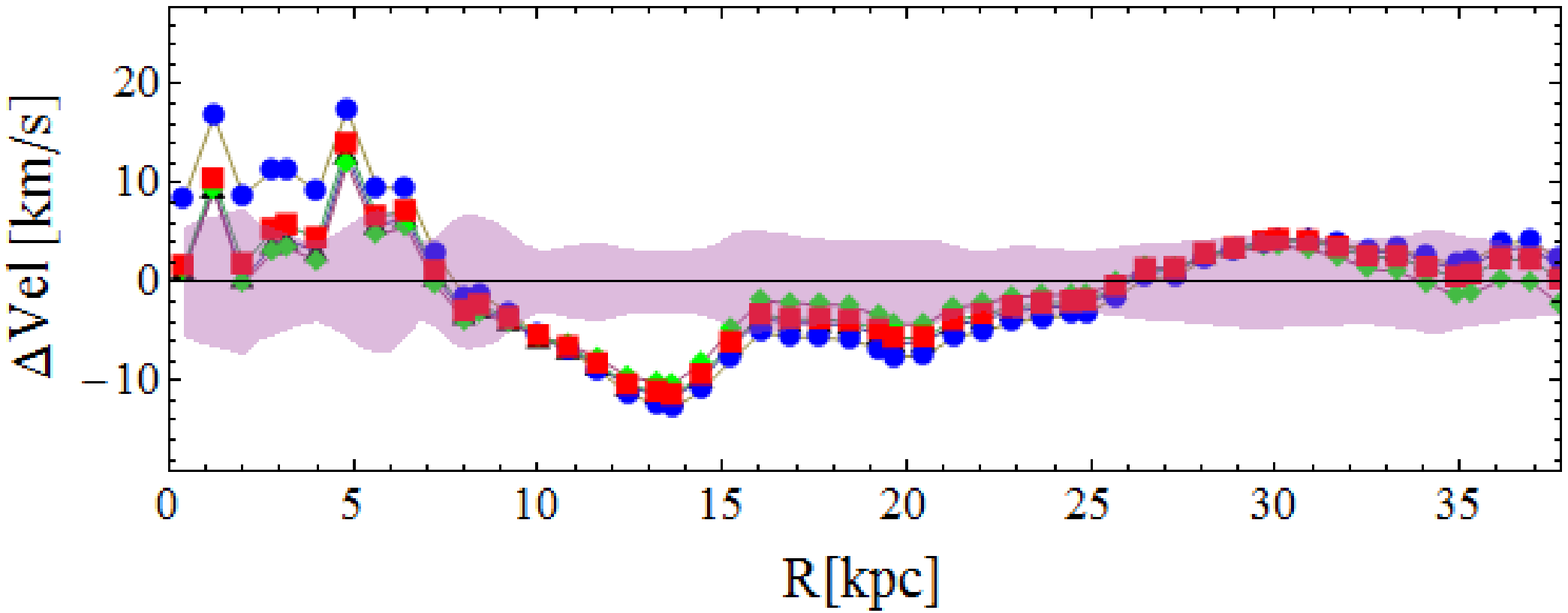}
    \end{tabular}  }
   \caption{\footnotesize{We present the rotation curves for the galaxy NGC 3198. Colors and symbols are as in Fig.\ref{fig:DDO154}. This galaxy shows a complicated stellar disk in the center, it could be assume two component but for simplicity and because it provides a small effect on the rotation curve we analyze it with only one component with parameters, $\mu_0=17.1 mag arcsec^{-2}$ and $R_d=3.18 {\rm kpc}$. For the inner analysis we take data below 6.8 {\rm kpc}, and obtained values of $r_c = 0.21 {\rm kpc}$, $r_s = 16.64$ and $\rho_0 = 7.7 *10^6 M_\sun/{\rm kpc}^3$ for the minimal disk.  }}
  \label{fig:NGC3198}
\end{figure}

\begin{figure}[h!]
    \subfloat[\footnotesize{Minimal disk}]{
    \begin{tabular}[b]{c}
    \includegraphics[width=0.35\textwidth]{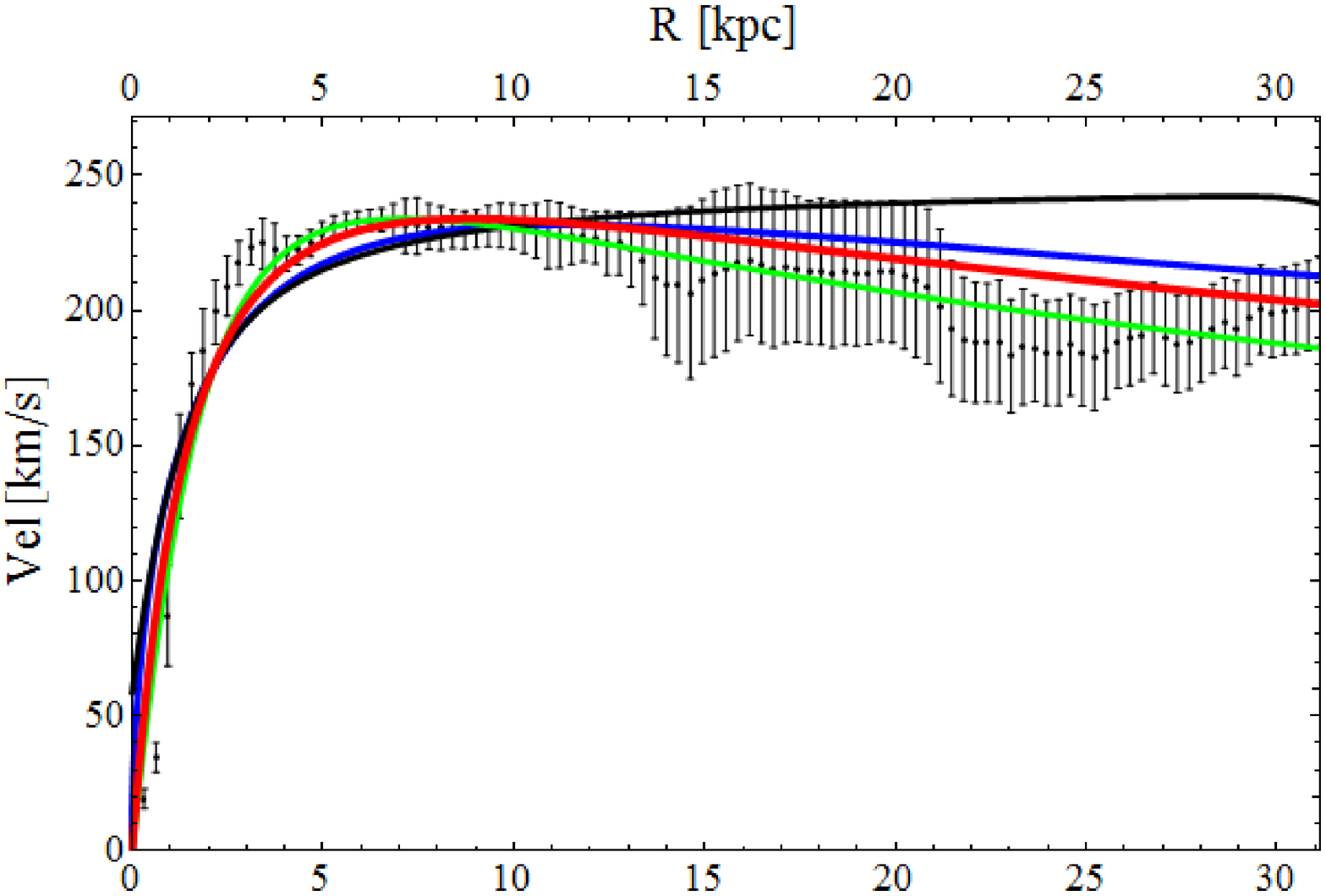} \\
    \includegraphics[width=0.35\textwidth]{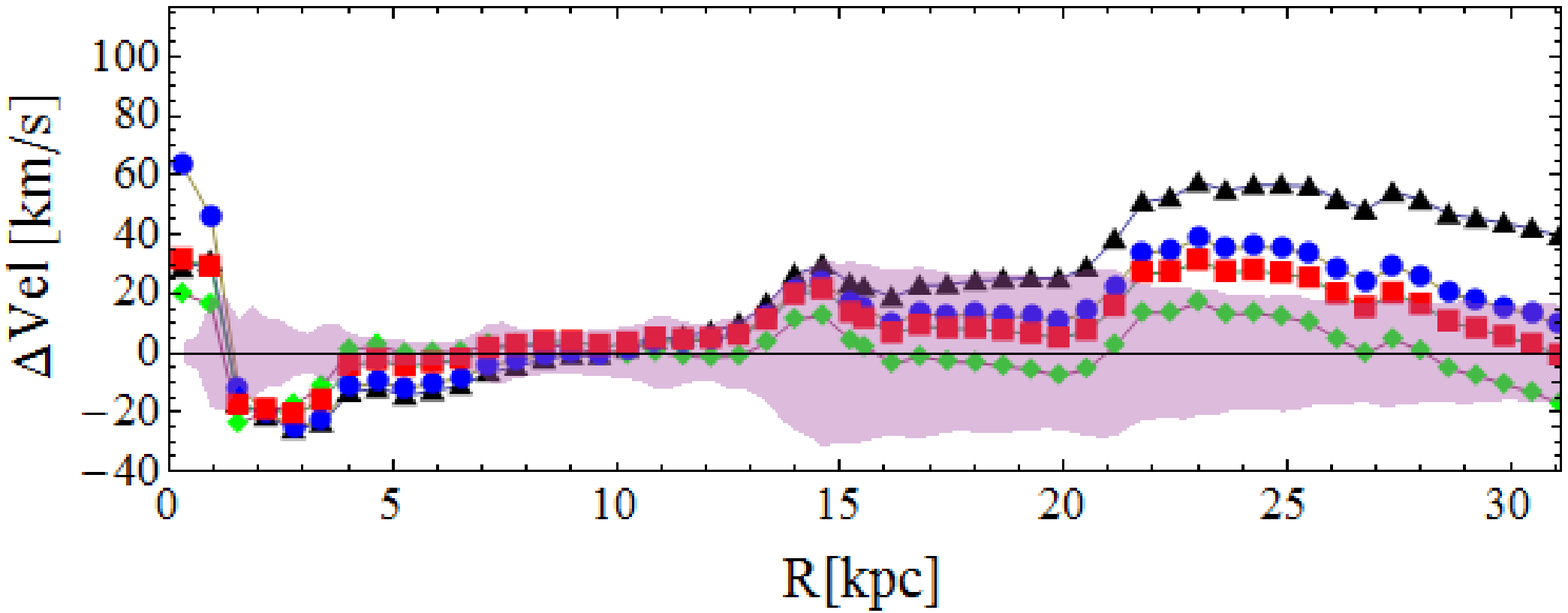}
    \end{tabular}  }
    \subfloat[\footnotesize{Min. disk + Gas}]{
    \begin{tabular}[b]{c}
    \includegraphics[width=0.35\textwidth]{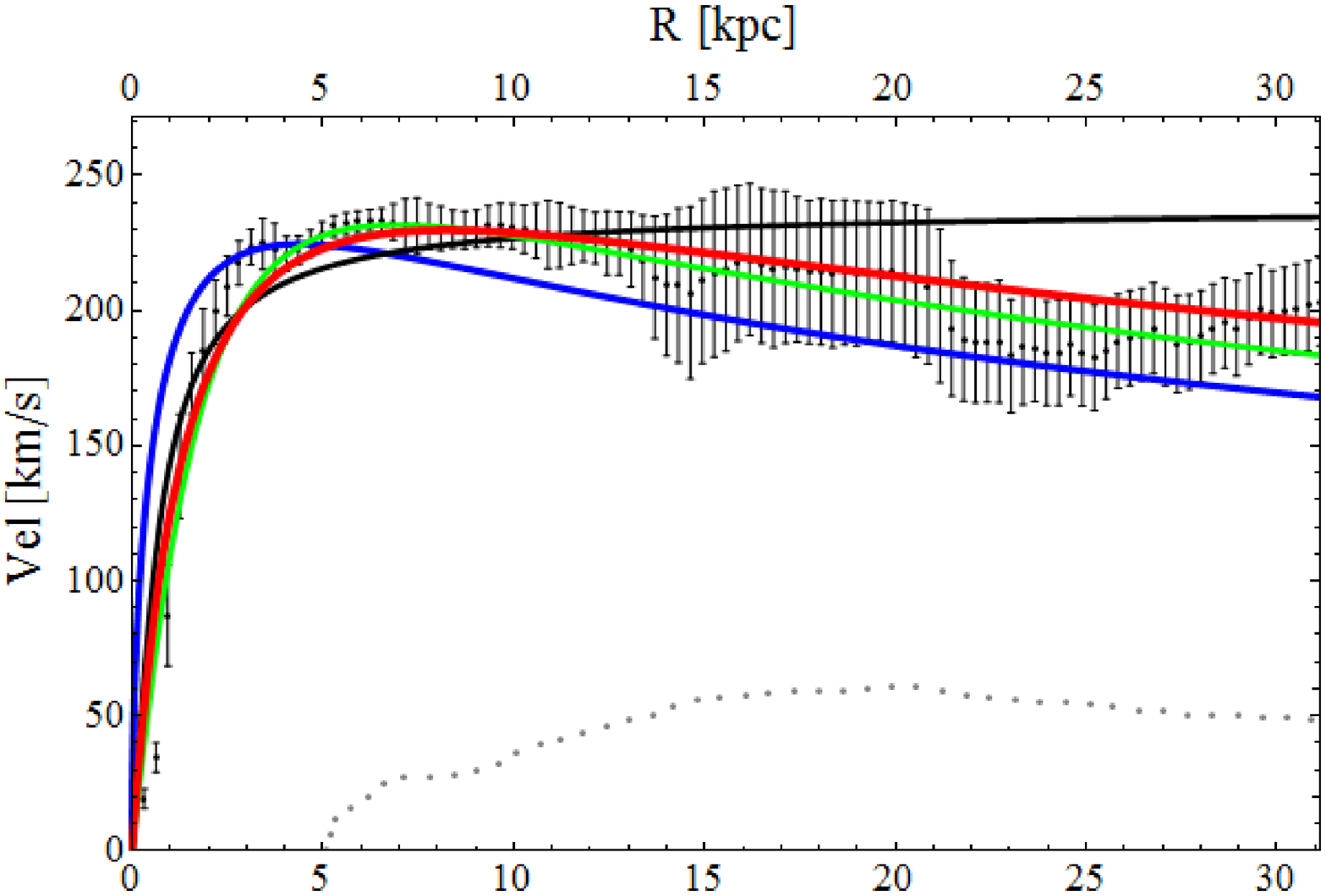} \\
    \includegraphics[width=0.35\textwidth]{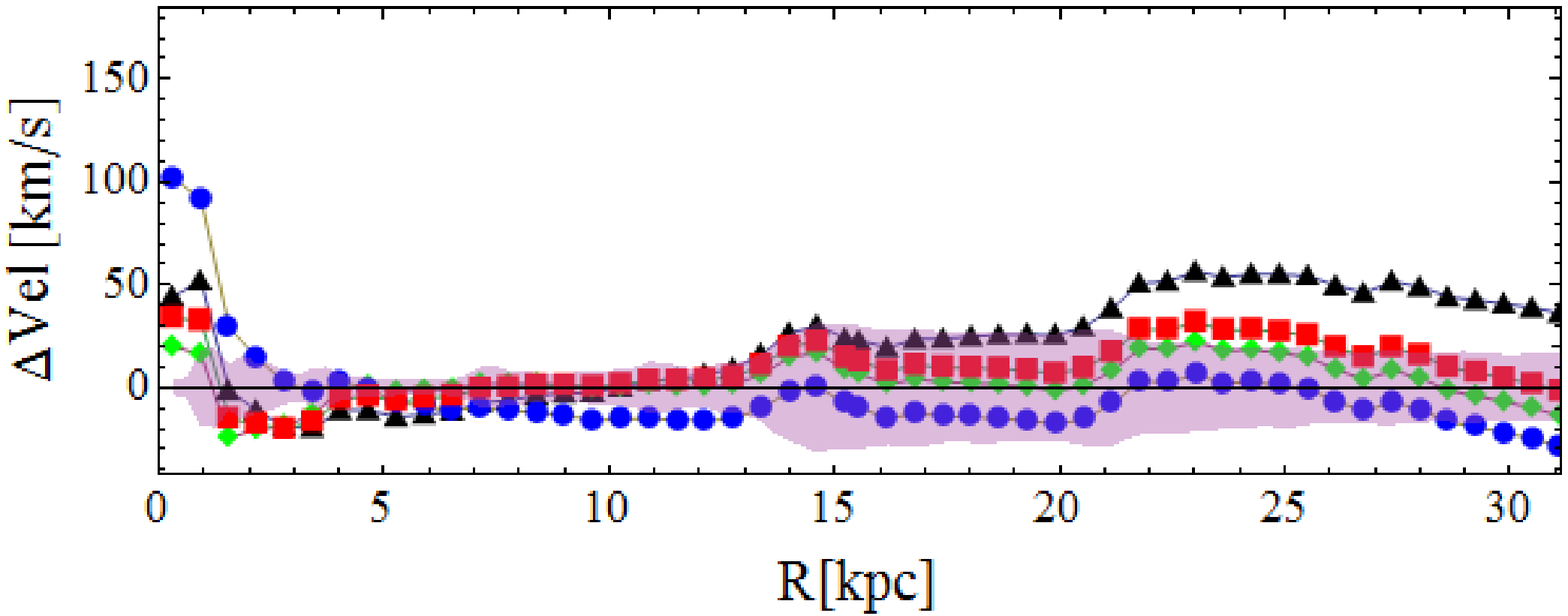}
    \end{tabular}  }  \\
    \subfloat[\footnotesize{Kroupa}]{
    \begin{tabular}[b]{c}
    \includegraphics[width=0.35\textwidth]{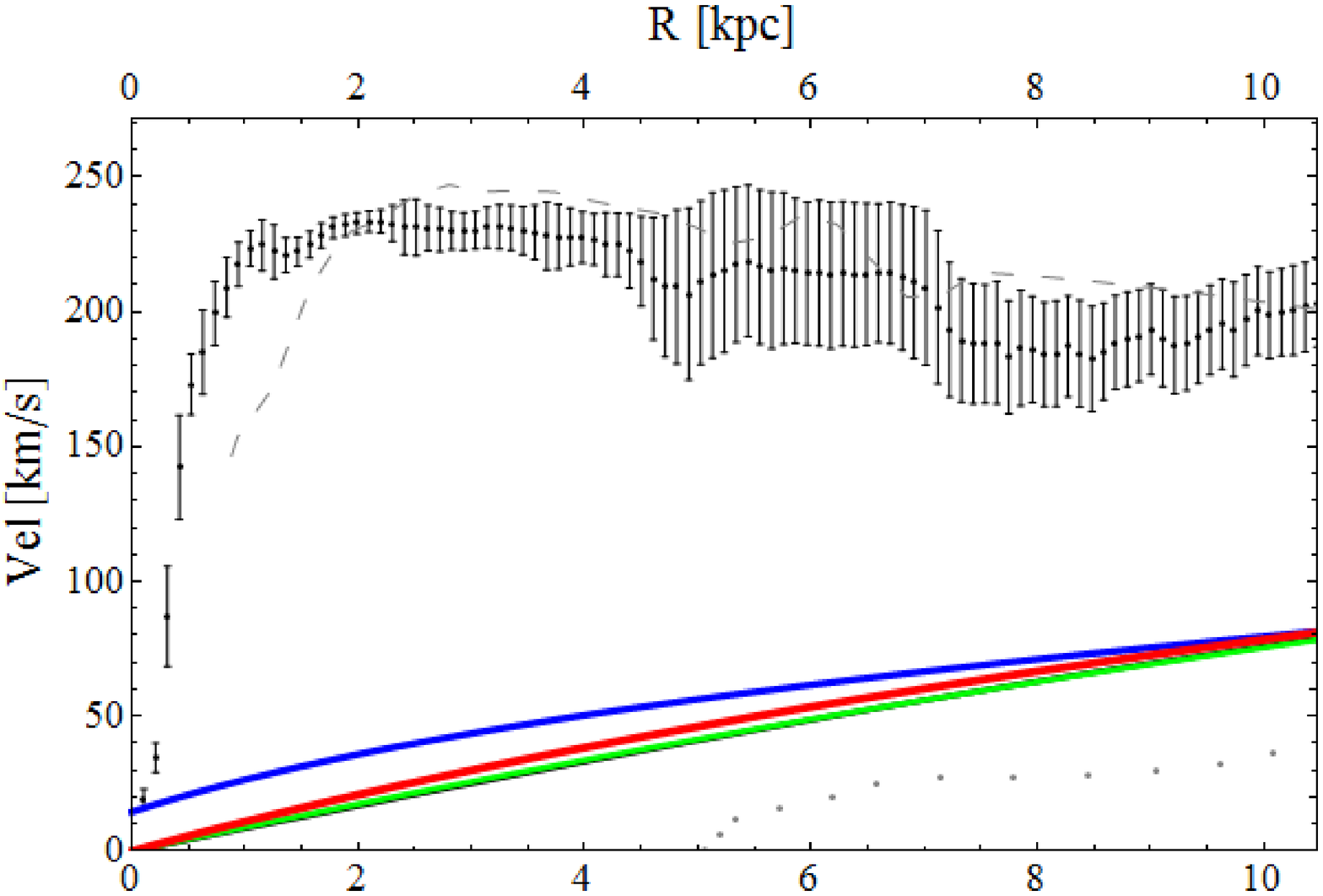} \\
    \includegraphics[width=0.35\textwidth]{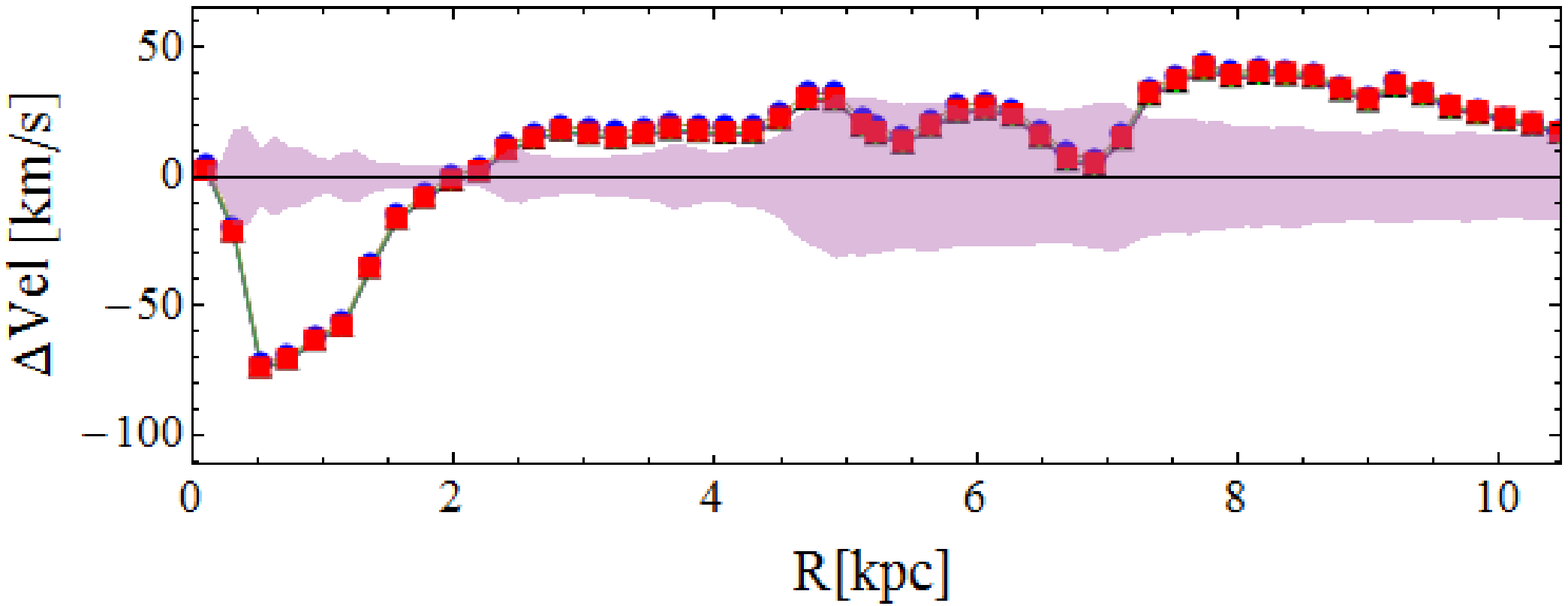}
    \end{tabular}  }
    \subfloat[\footnotesize{diet-Salpeter}]{
    \begin{tabular}[b]{c}
    \includegraphics[width=0.35\textwidth]{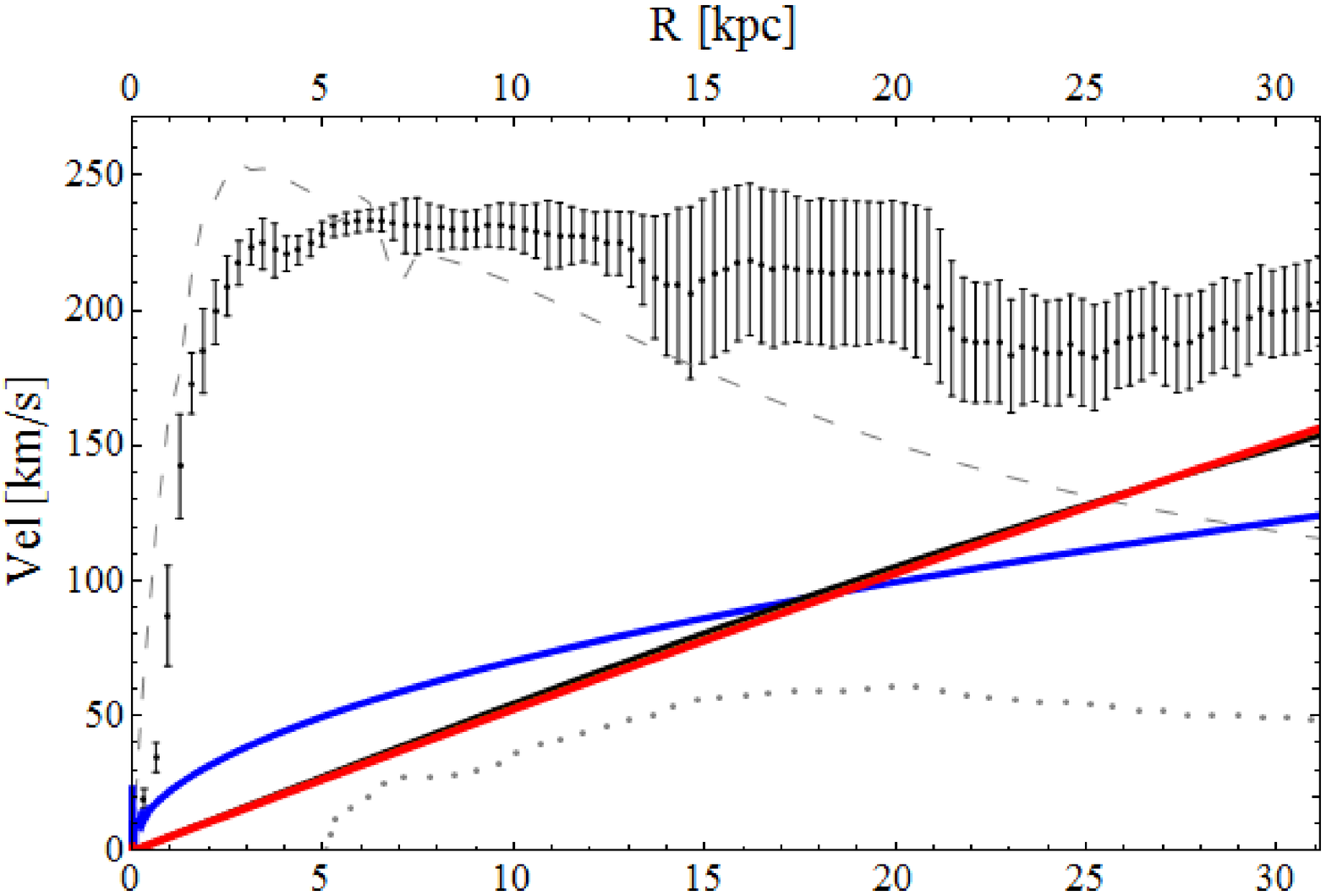} \\
    \includegraphics[width=0.35\textwidth]{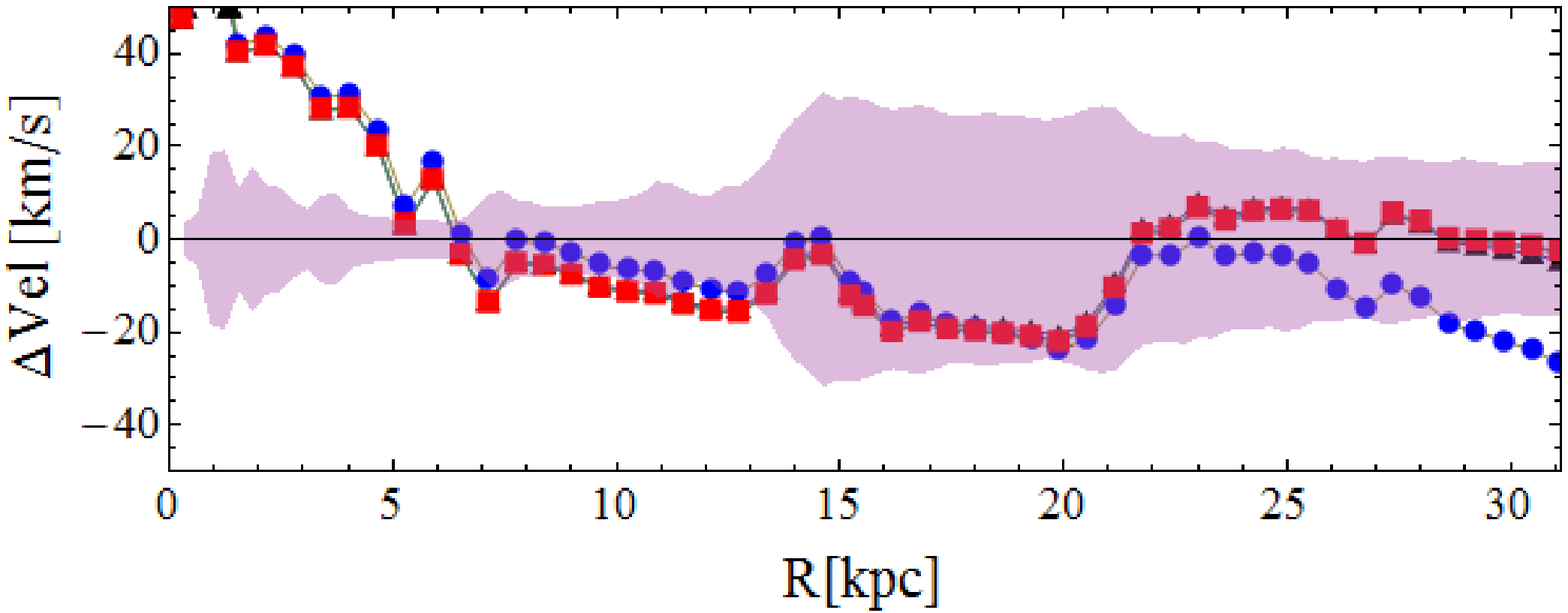}
    \end{tabular}  }
   \caption{\footnotesize{The rotation curves for the galaxy NGC 3521. Colors and symbols are as in Fig.\ref{fig:DDO154}. Because the nature of the galaxy it is difficult to obtain reliable data from the stellar component (see deBlok08). even so, we treat the stellar disk as a single component with mass $M_\star = 1.2 *10^{11} M_\sun$ and $R_d = 3 {\rm kpc}$. A color gradient is present. The inner analysis is carried out with data below the 3.7 {\rm kpc} and obtaining values for the core and central density $r_c = 2.4$ {\rm kpc} and $r_s = 5.5$ {\rm kpc} and $\rho_0 = 2.4 * 10^8 M_\sun/{\rm kpc}^3$ for the minimal disk.  }}
  \label{fig:NGC3521}
\end{figure}

\begin{figure}[h!]
    \subfloat[\footnotesize{Minimal disk}]{
    \begin{tabular}[b]{c}
    \includegraphics[width=0.35\textwidth]{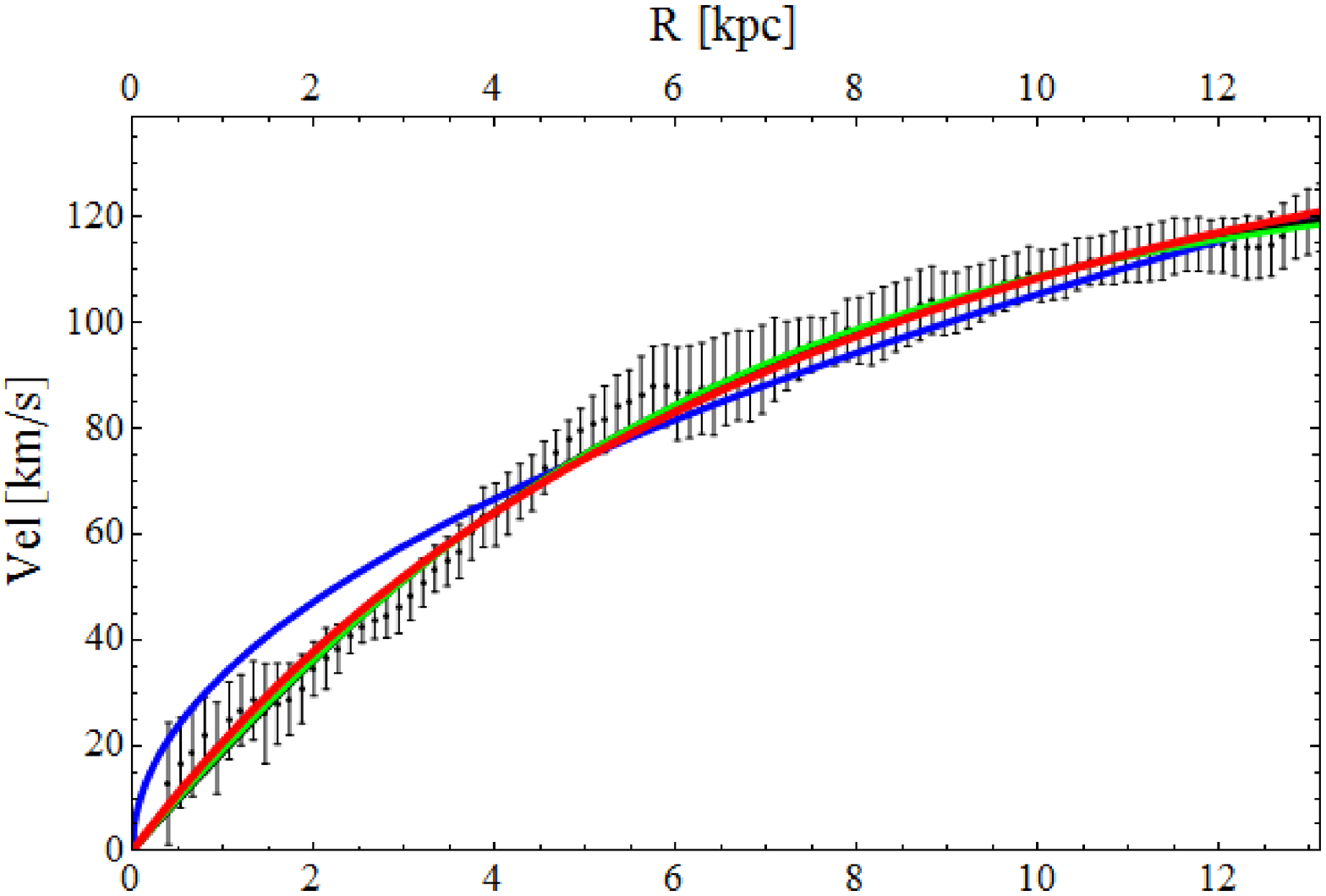} \\
    \includegraphics[width=0.35\textwidth]{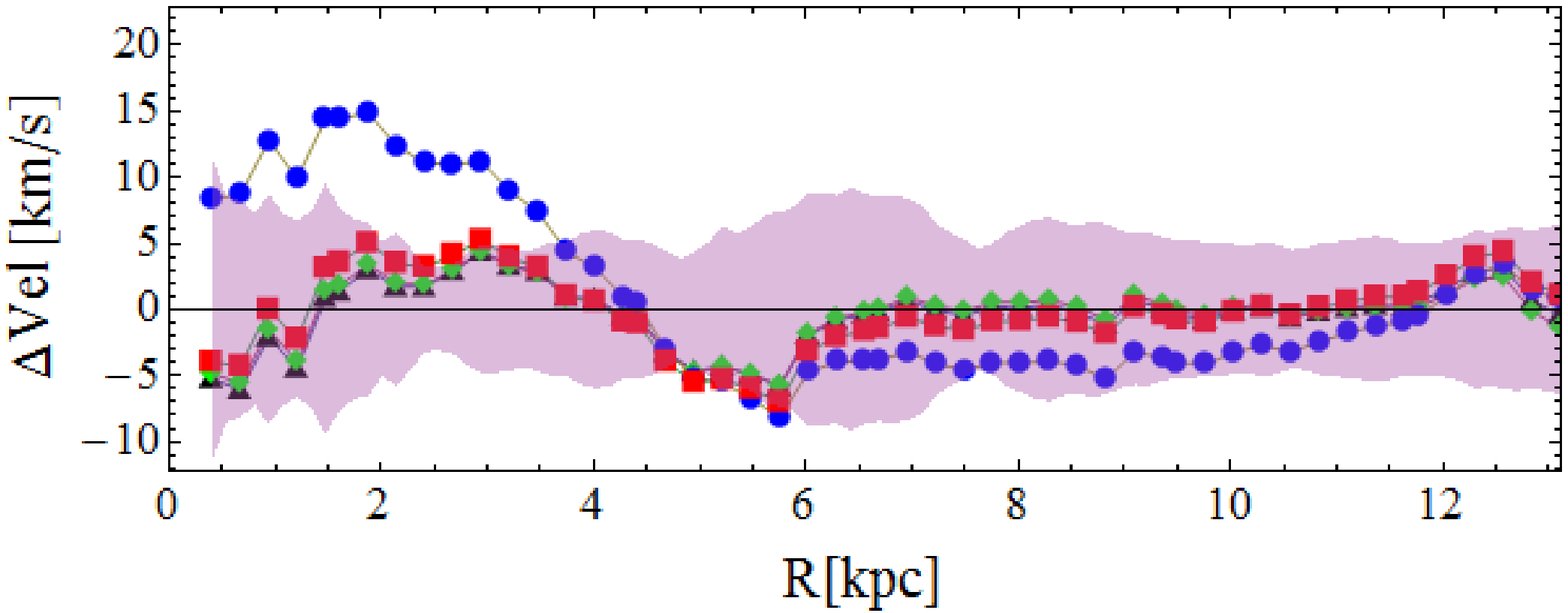}
    \end{tabular}  }
    \subfloat[\footnotesize{Min. disk + Gas}]{
    \begin{tabular}[b]{c}
    \includegraphics[width=0.35\textwidth]{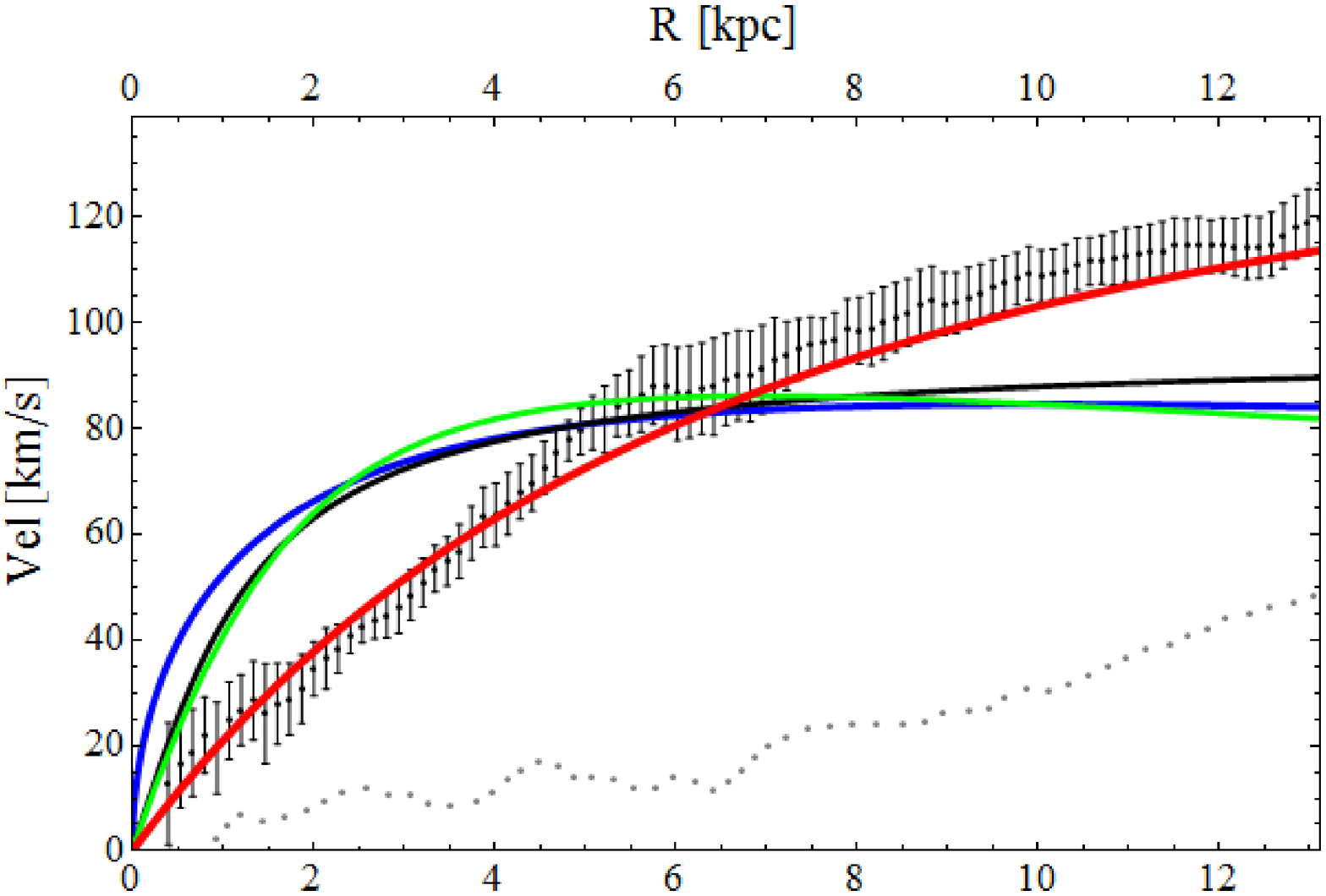} \\
    \includegraphics[width=0.35\textwidth]{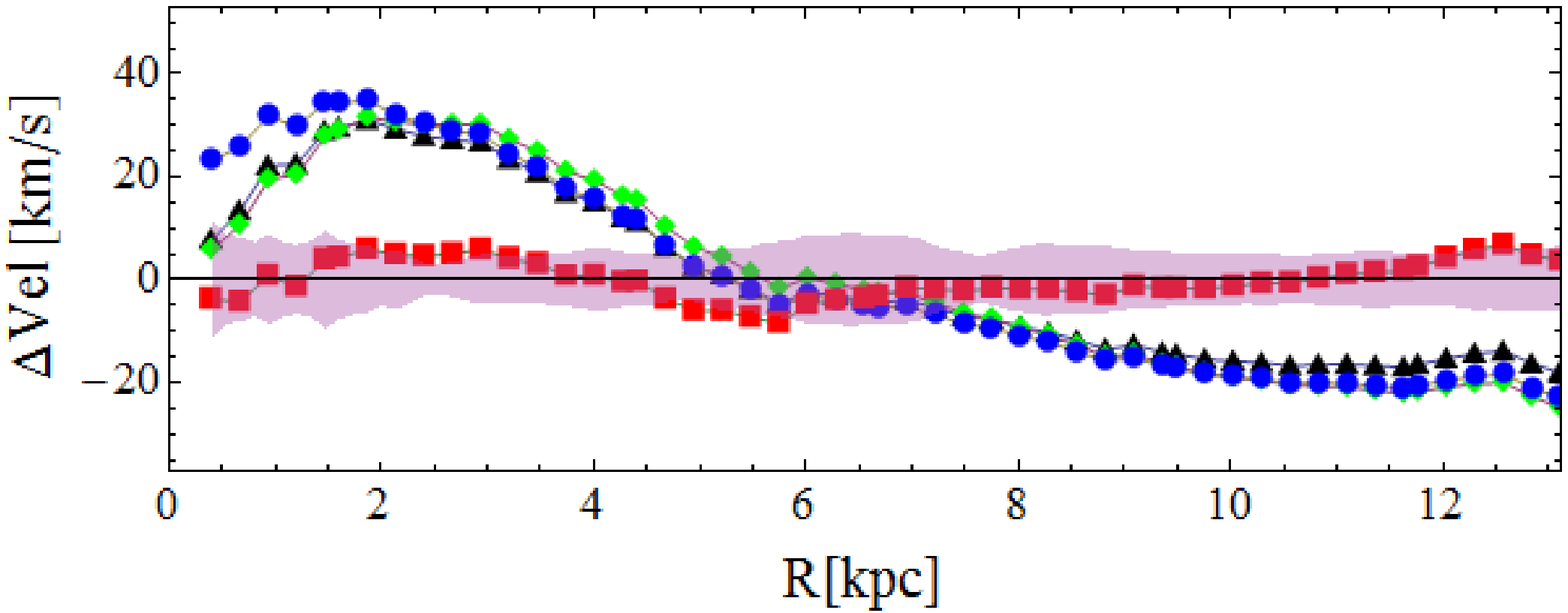}
    \end{tabular}  }  \\
    \subfloat[\footnotesize{Kroupa}]{
    \begin{tabular}[b]{c}
    \includegraphics[width=0.35\textwidth]{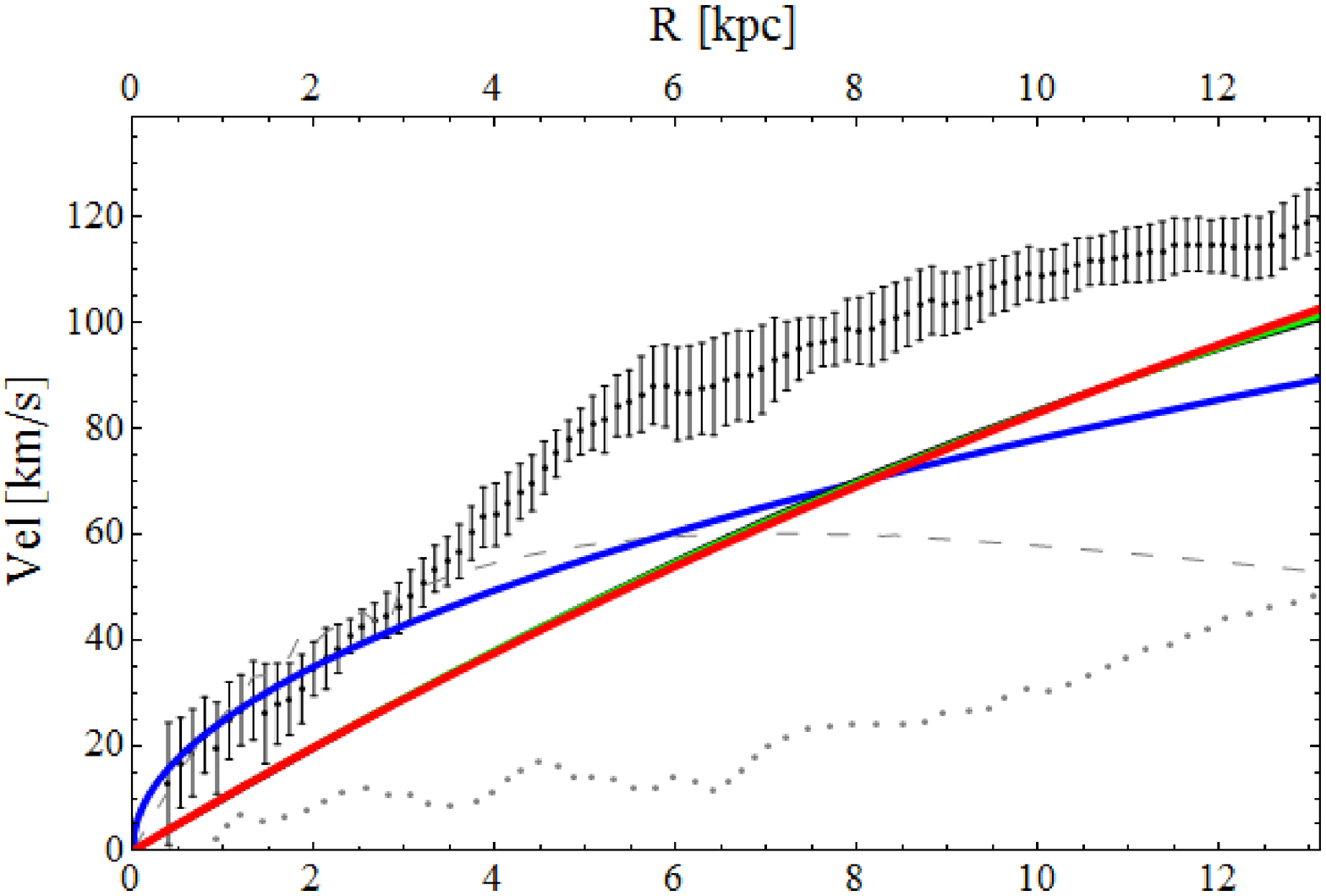} \\
    \includegraphics[width=0.35\textwidth]{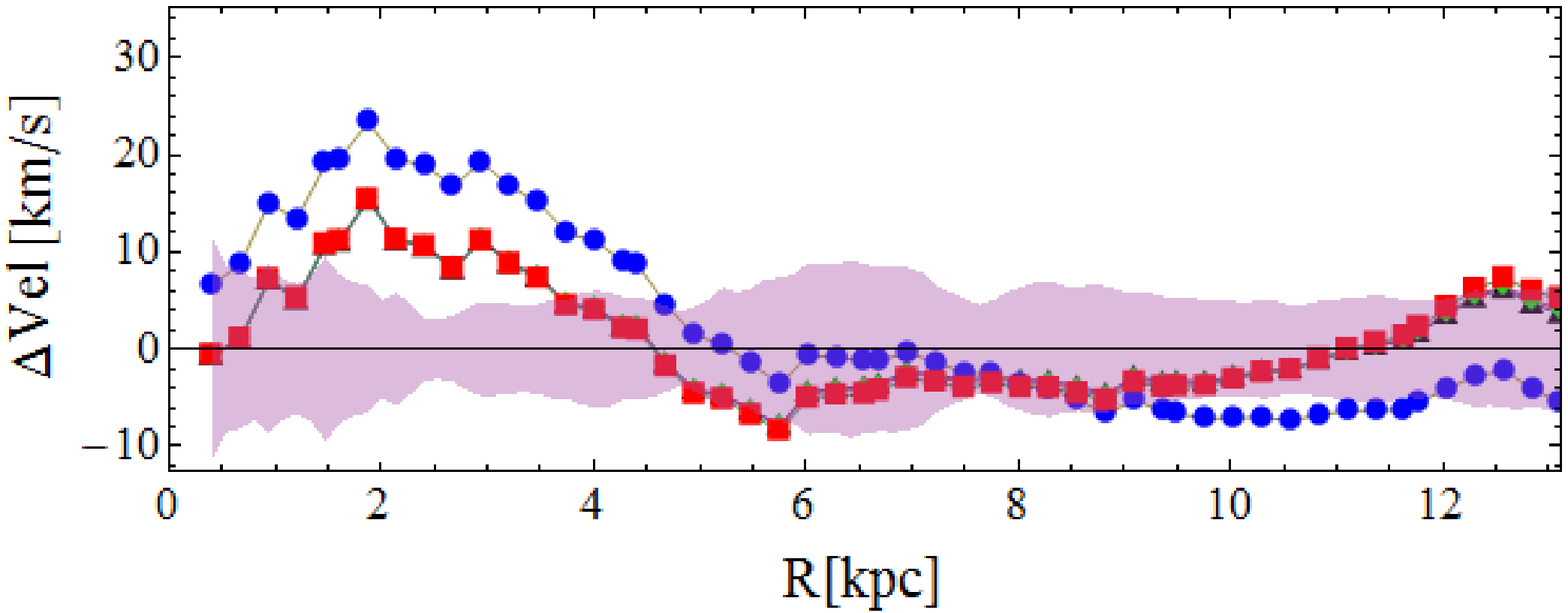}
    \end{tabular}  }
    \subfloat[\footnotesize{diet-Salpeter}]{
    \begin{tabular}[b]{c}
    \includegraphics[width=0.35\textwidth]{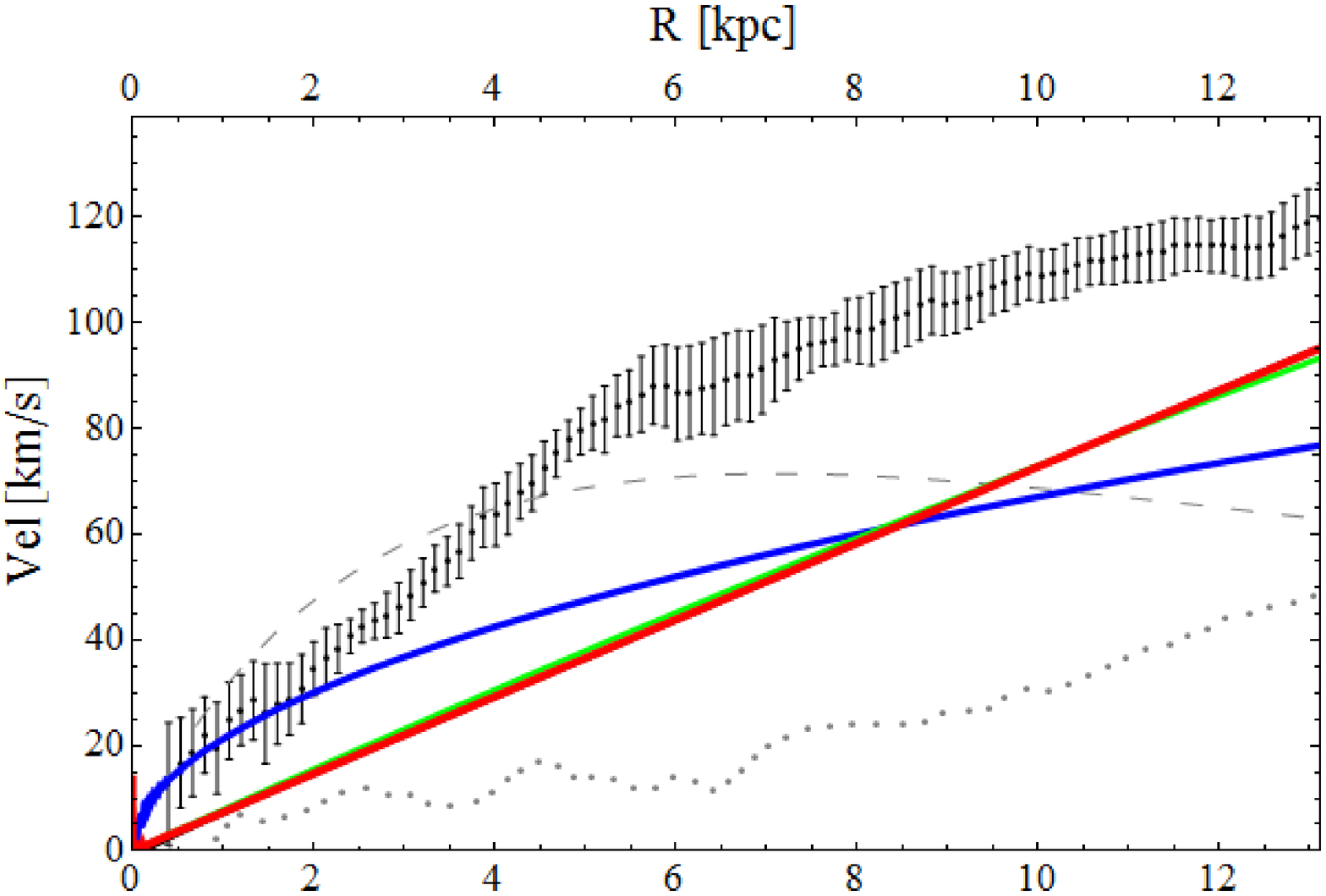} \\
    \includegraphics[width=0.35\textwidth]{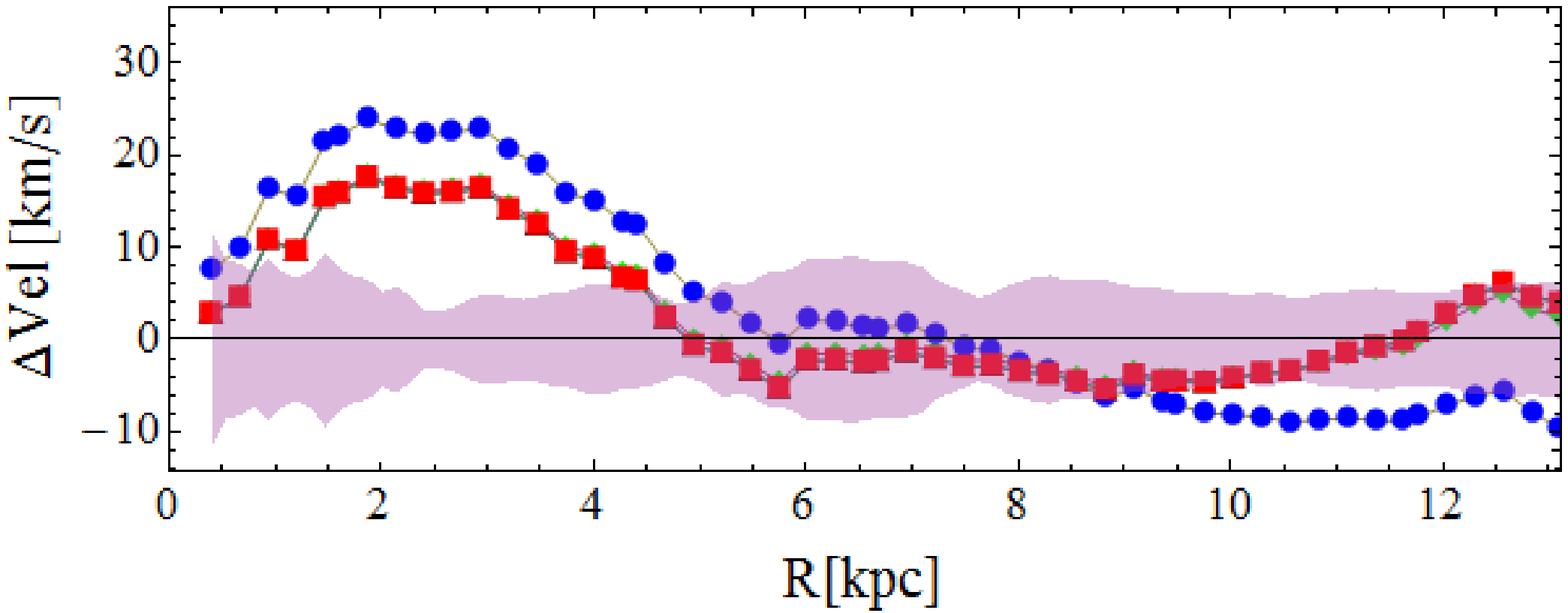}
    \end{tabular}  }
   \caption{\footnotesize{We display the rotation curves for the galaxy NGC 925. Colors and symbols are as in Fig.\ref{fig:DDO154}. It is classified as a late-type barred spiral. Shows no evidence for a bright central component. There is also no evidence for a strong color gradient and we assume a $\gs=0.65$ for the stellar disk. The stellar disk is described by a central surface brightness $\mu_0=18.9$ mag$arcsec^{-2}$ and a scale length $R_d=3.31$ {\rm kpc}. This galaxy is specially difficult to fit with reasonable values specially because it present a very smooth slope. For inner analysis we take data below 6 {\rm kpc}, obtaining values for the minimal disk analysis of $r_c = 10.8 {\rm kpc}$ for the core and $\rho_c = 2*10^8 M_\sun/{\rm kpc}^3$ for the central density. Unlike the previous IC 2574 and NGC 2976 galaxies we have few data points after the 6 {\rm kpc} so we can break the degeneracy and obtain values for $r_s = 24.6$ {\rm kpc} and $\rho_0 = 8.1*10^6 M_\sun/{\rm kpc}^3$ for the minimal disk.  }}
  \label{fig:NGC925}
\end{figure}

\newpage
\clearpage

\newpage
\clearpage
\subsection{GROUP C}\label{apendix:gc}
In this section we present the considerations made for the galaxies with fitted values of $r_c = 0$ when analyzed with the min.disk mass model. We have named this set of galaxies as G.C . The conclusion are presented in Sec. \ref{conclusion} for further discussion. At the end of the section can be see the confidence level for each galaxy in every mass model.

\begin{figure}[h!]
    \subfloat[\footnotesize{Minimal disk}]{
    \begin{tabular}[b]{c}
    \includegraphics[width=0.35\textwidth]{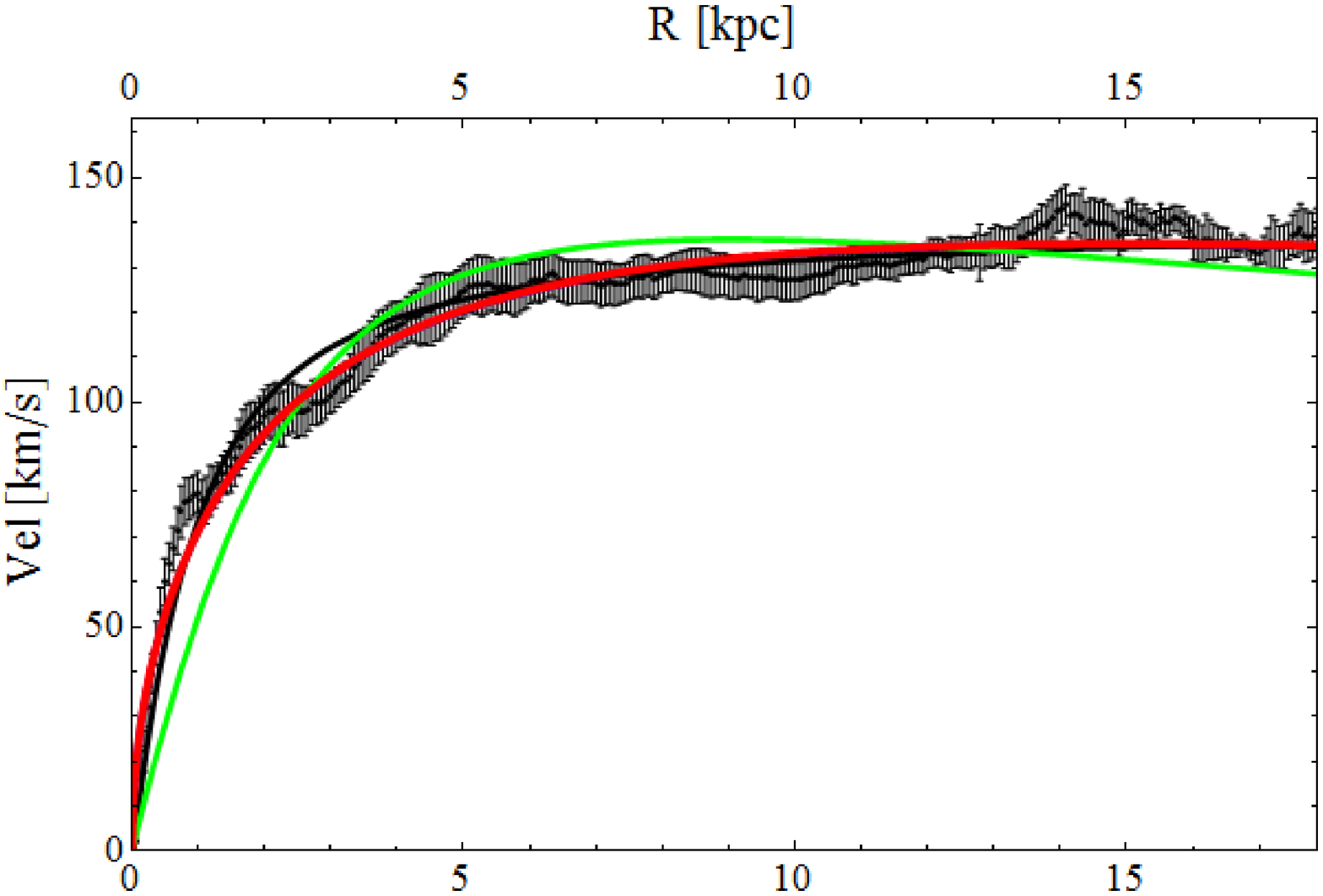} \\
    \includegraphics[width=0.35\textwidth]{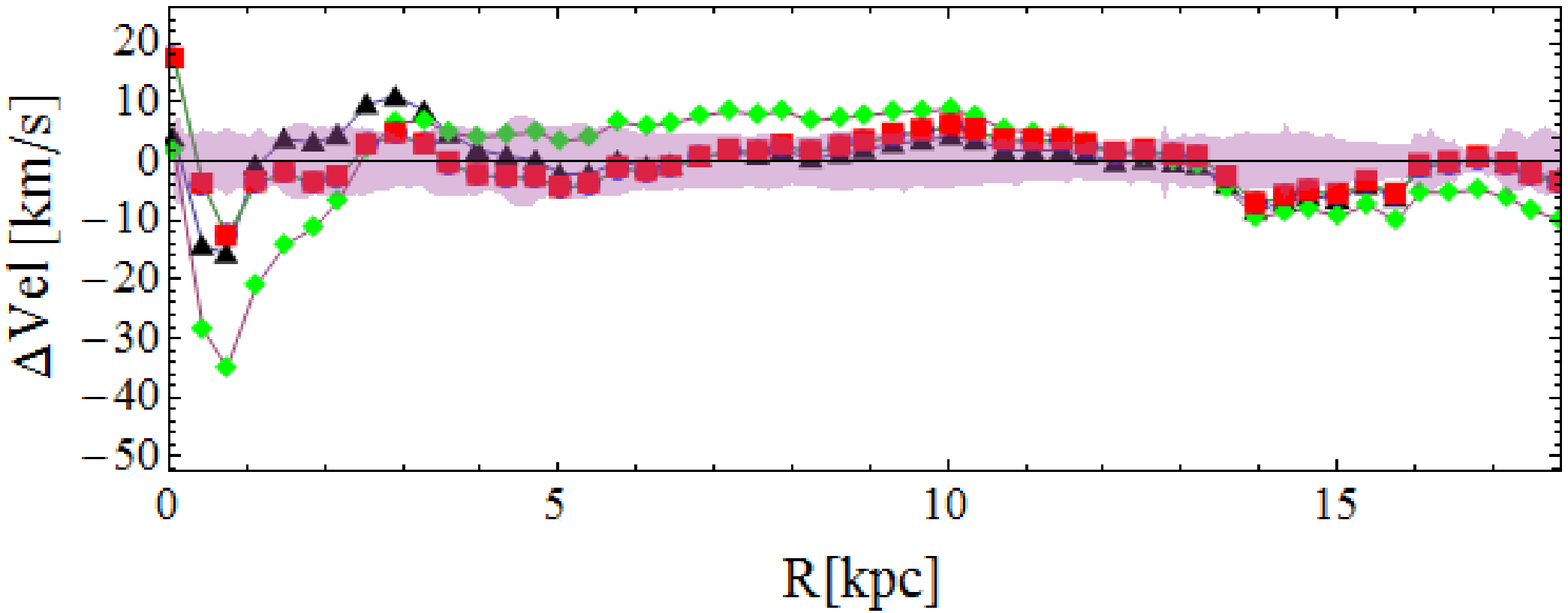}
    \end{tabular}  }
    \subfloat[\footnotesize{Min. disk + Gas}]{
    \begin{tabular}[b]{c}
    \includegraphics[width=0.35\textwidth]{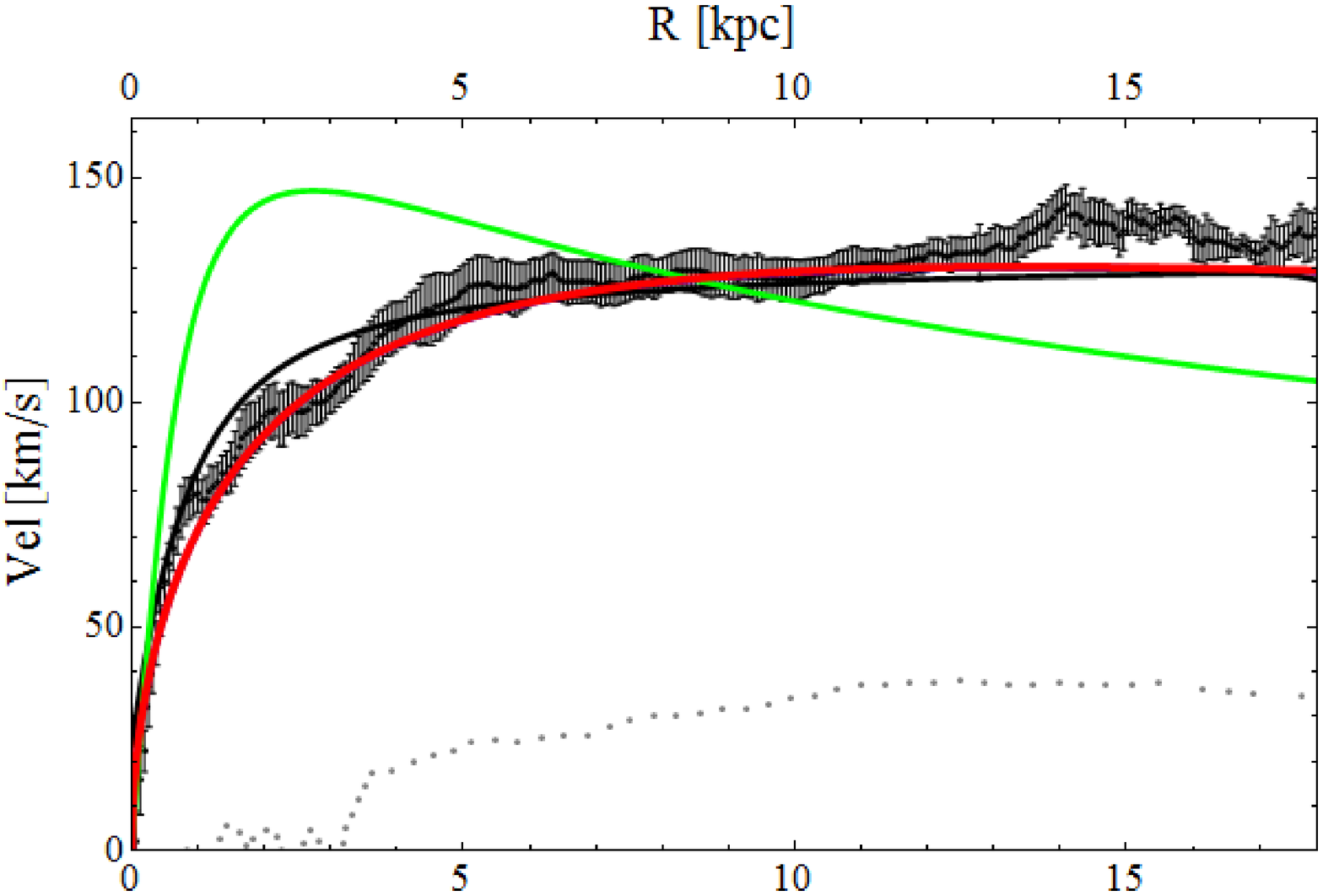} \\
    \includegraphics[width=0.35\textwidth]{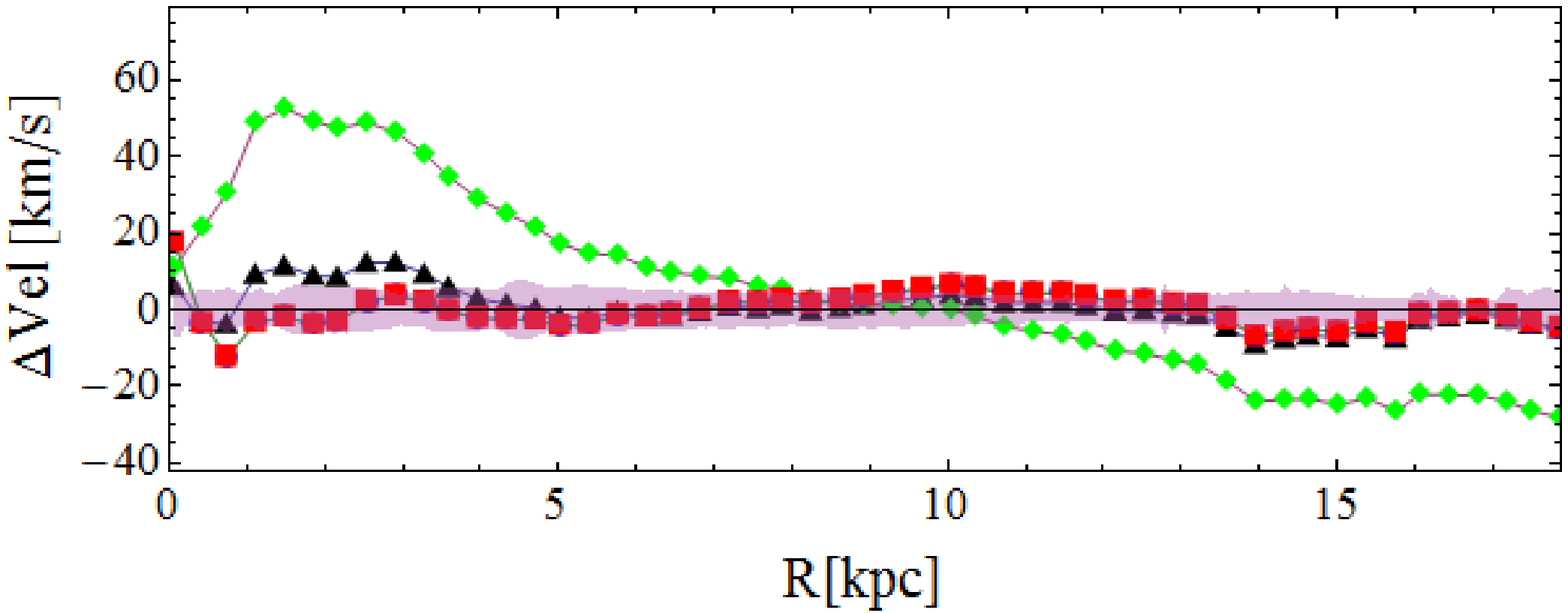}
    \end{tabular}  }  \\
    \subfloat[\footnotesize{Kroupa}]{
    \begin{tabular}[b]{c}
    \includegraphics[width=0.35\textwidth]{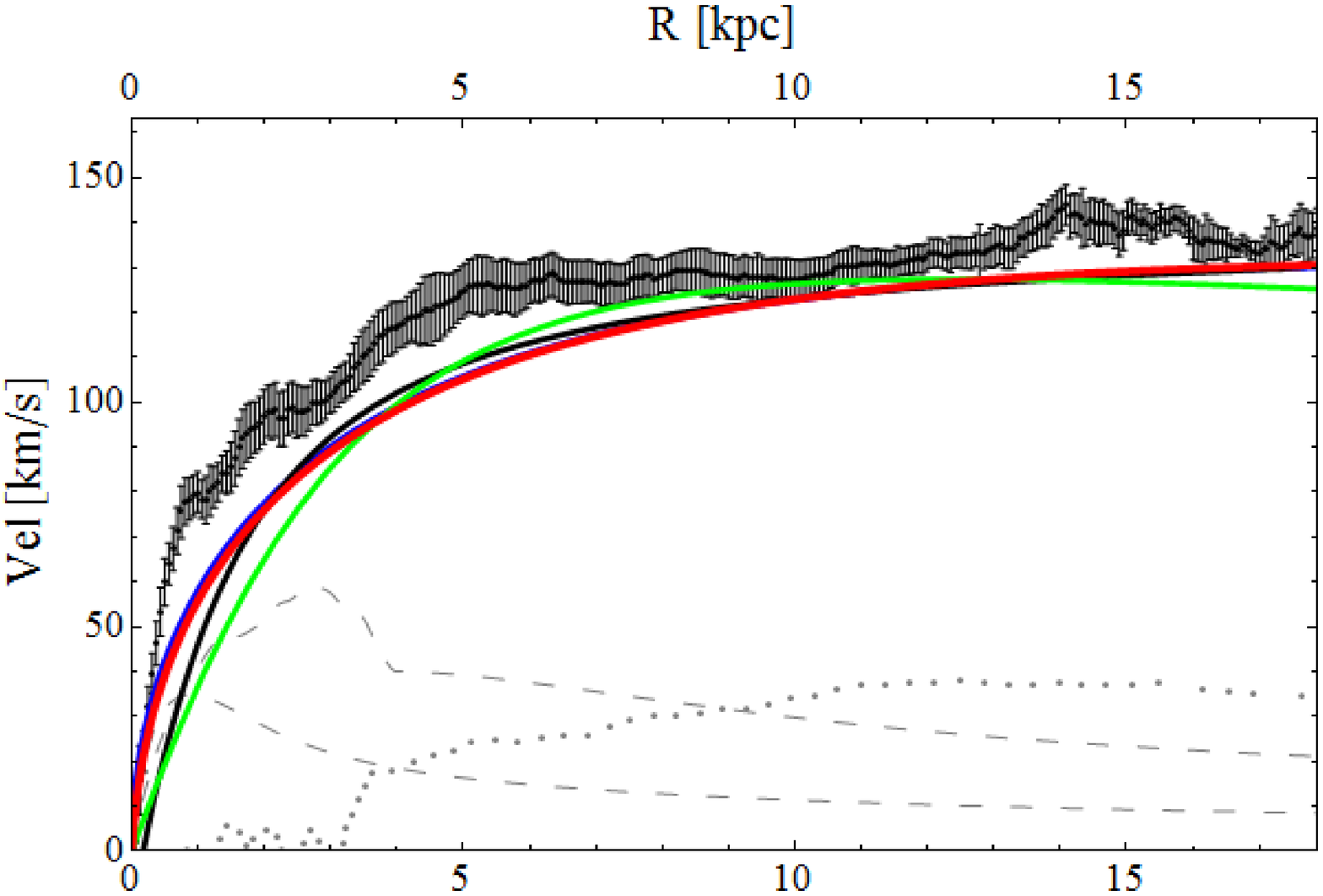} \\
    \includegraphics[width=0.35\textwidth]{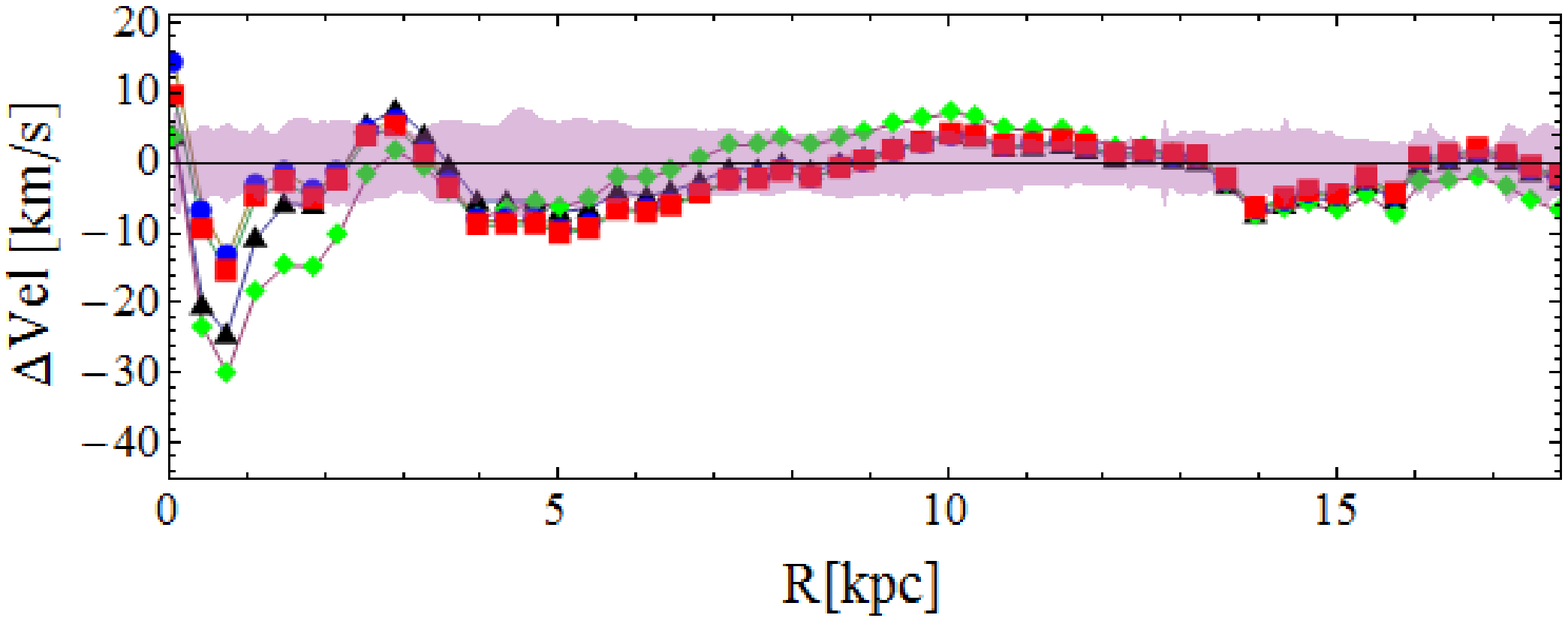}
    \end{tabular}  }
    \subfloat[\footnotesize{diet-Salpeter}]{
    \begin{tabular}[b]{c}
    \includegraphics[width=0.35\textwidth]{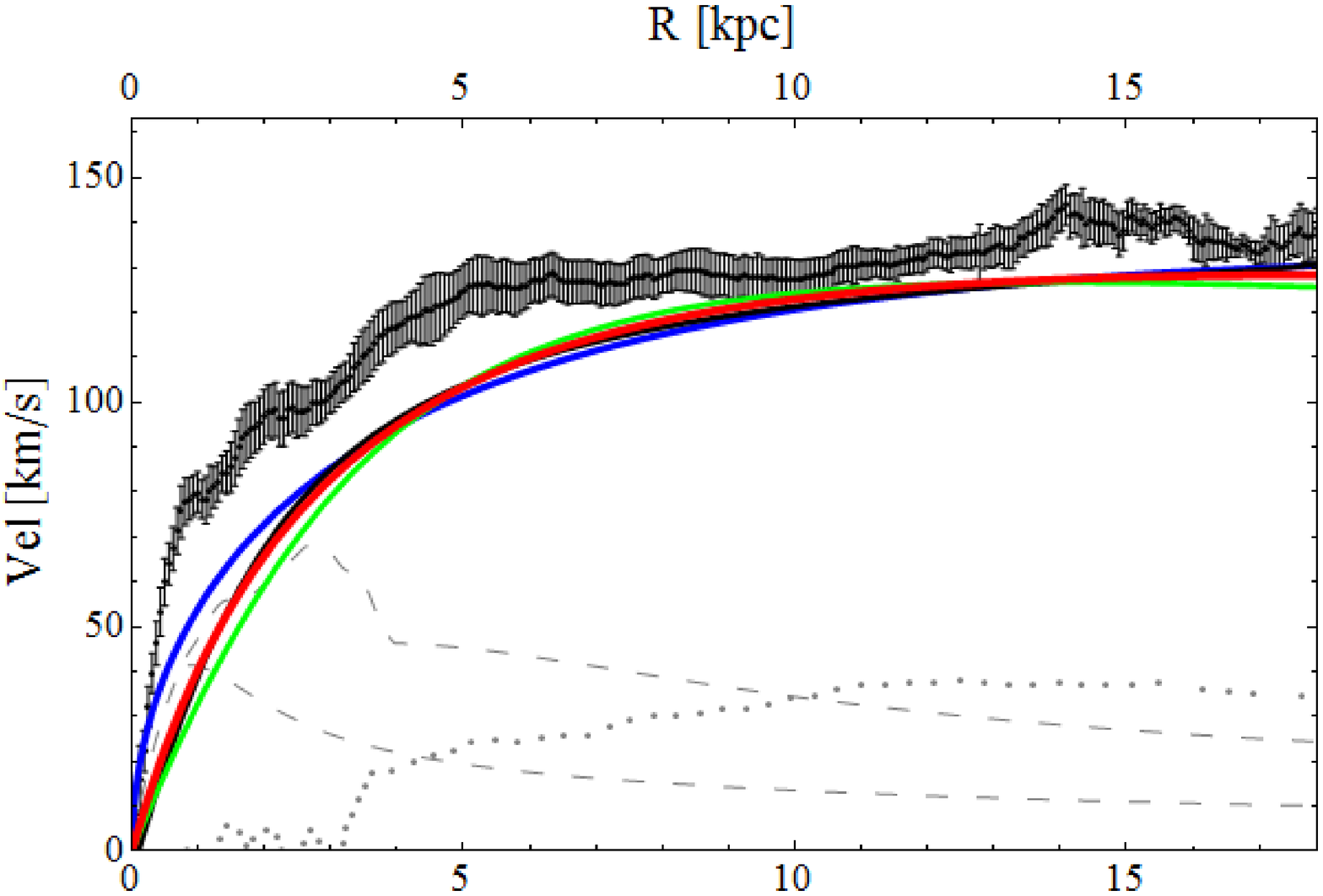} \\
    \includegraphics[width=0.35\textwidth]{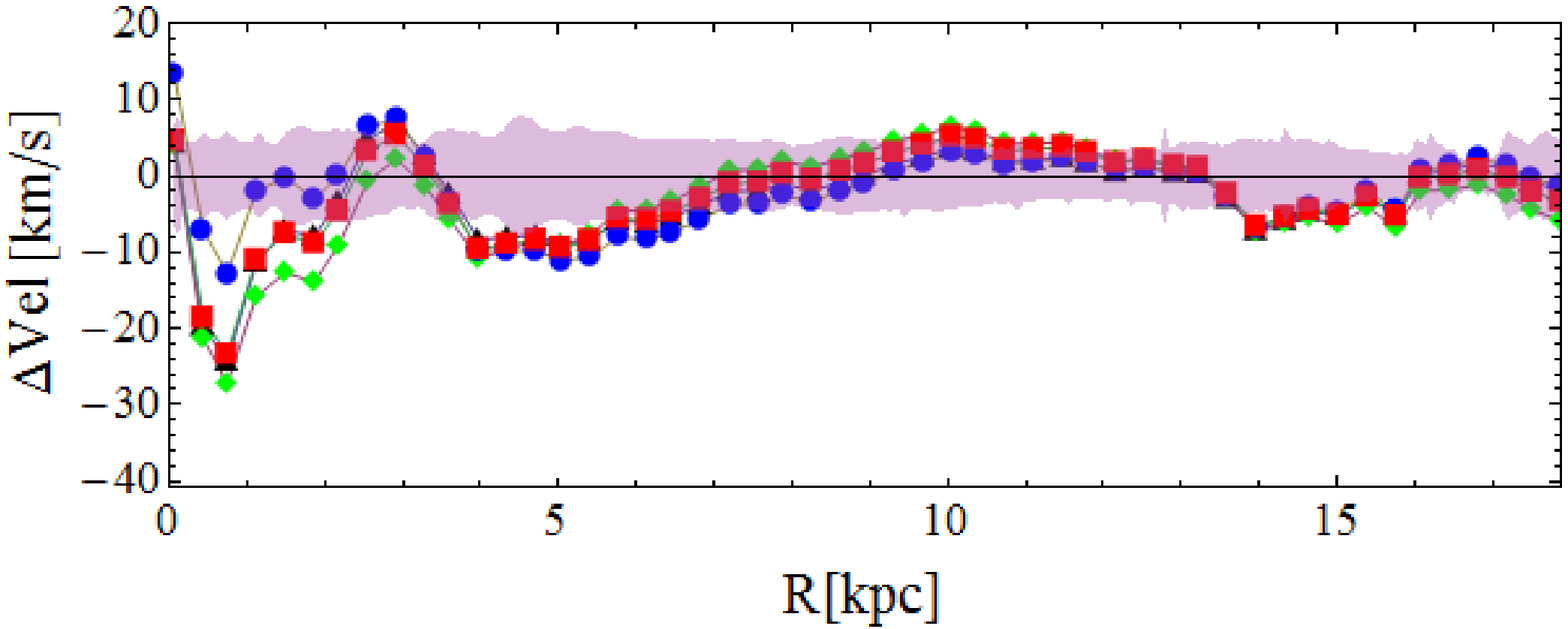}
    \end{tabular}  }
  \caption{\footnotesize{The rotation curves for the galaxy NGC 2403. Colors and symbols are as in Fig.\ref{fig:DDO154}. It is a late-type Sc spiral. The galaxy shows a (J-k) color gradient that produce a small difference in terms of $\gs$ for the inner and outer part of the stellar component. Since the color gradient is small we considerer a constant $\gs=0.41$. The surface brightness profile shows evidence for a second component in the inner parts but it is uncertain if this constitutes a dynamically separate component. For our purpose we consider a single-component in the stellar disk described by a central surface brightness $\mu_0=16.90${\rm  mag arcsec${^{-2}}$} and a scale length $R_d = 1.81$ {\rm kpc}. For the number of inner data points less than 2.1 {\rm kpc} it is clearly that we can do a inner analysis as for B group galaxies, in which the results are $r_c = 0.004 {\rm kpc}$ and $\rho_c = 2.1*10^8$ wich is consistant until $1\sigma$ for the confidence level with zero. From left to right, images from the fit of the galaxy NGC 2403 considering minimal disk, minimal disk+gas, Kroupa, diet-Salpeter. }}
  \label{fig:NGC2403}
\end{figure}

\begin{figure}[h!]
    \subfloat[\footnotesize{Minimal disk}]{
    \begin{tabular}[b]{c}
    \includegraphics[width=0.35\textwidth]{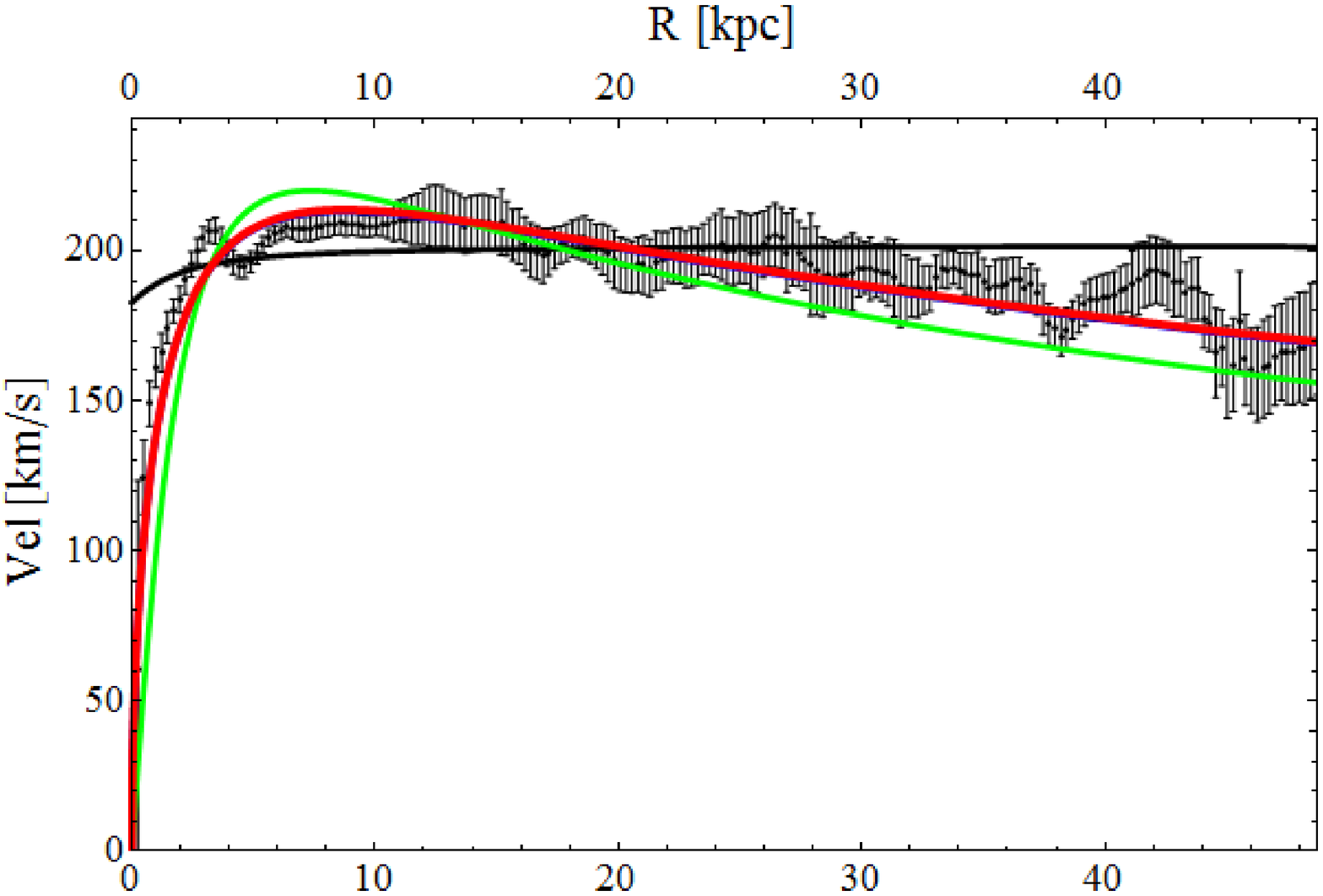} \\
    \includegraphics[width=0.35\textwidth]{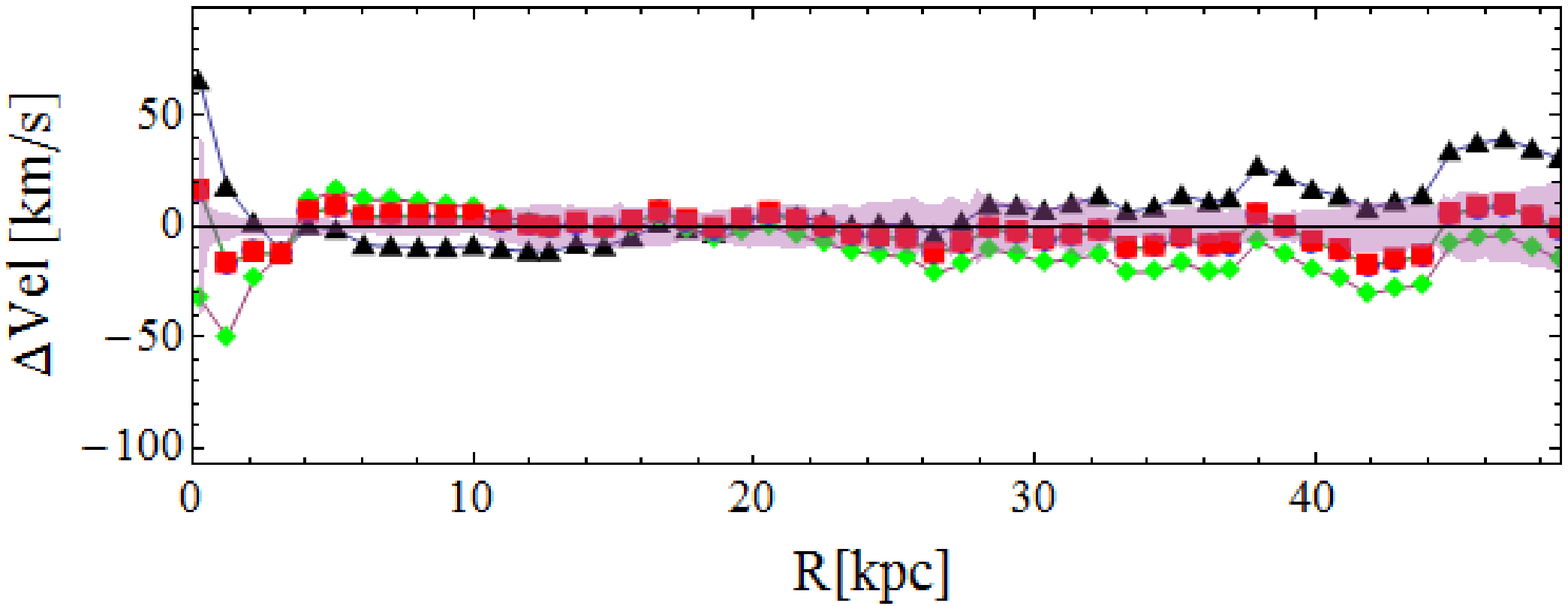}
    \end{tabular}  }
    \subfloat[\footnotesize{Min. disk + Gas}]{
    \begin{tabular}[b]{c}
    \includegraphics[width=0.35\textwidth]{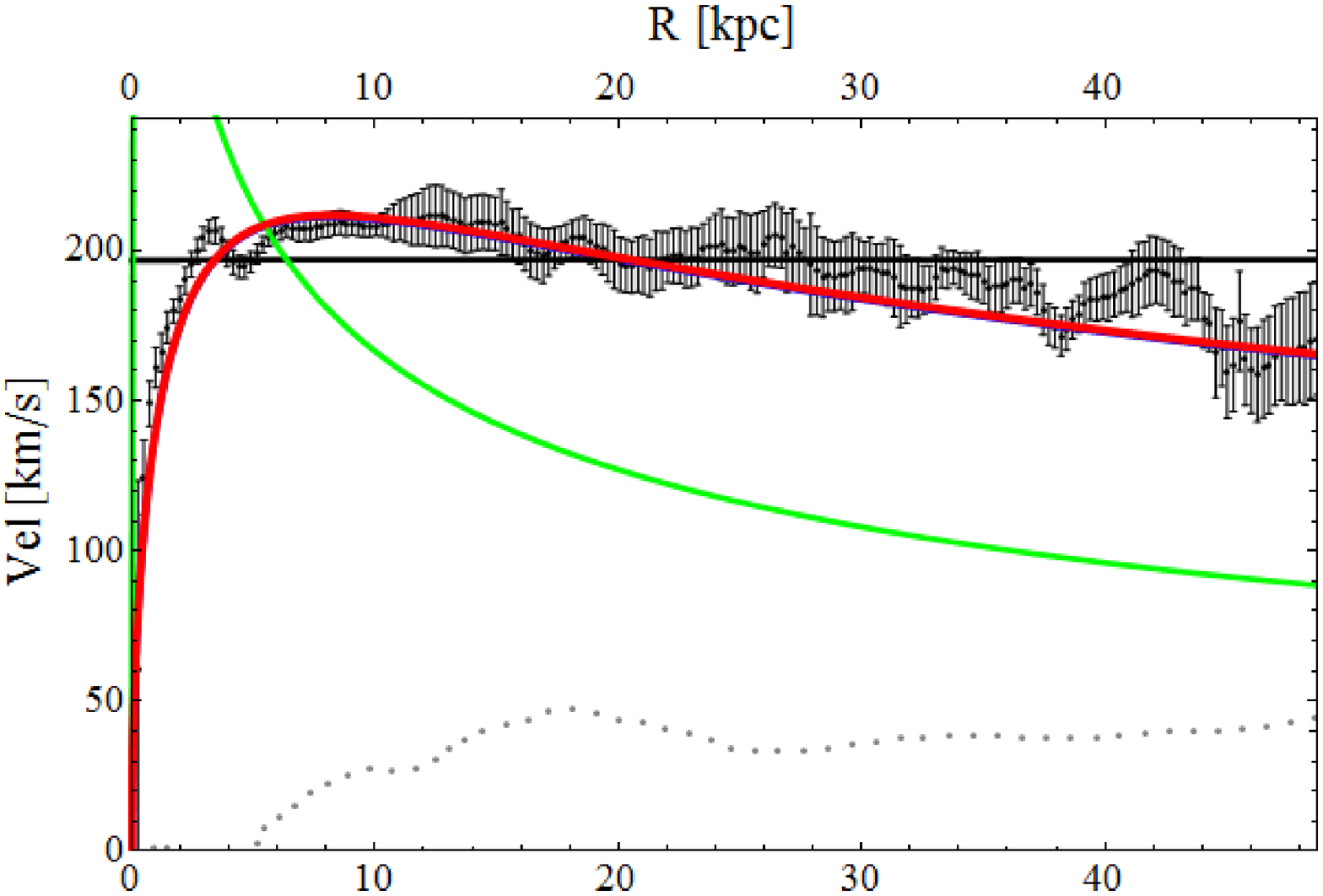} \\
    \includegraphics[width=0.35\textwidth]{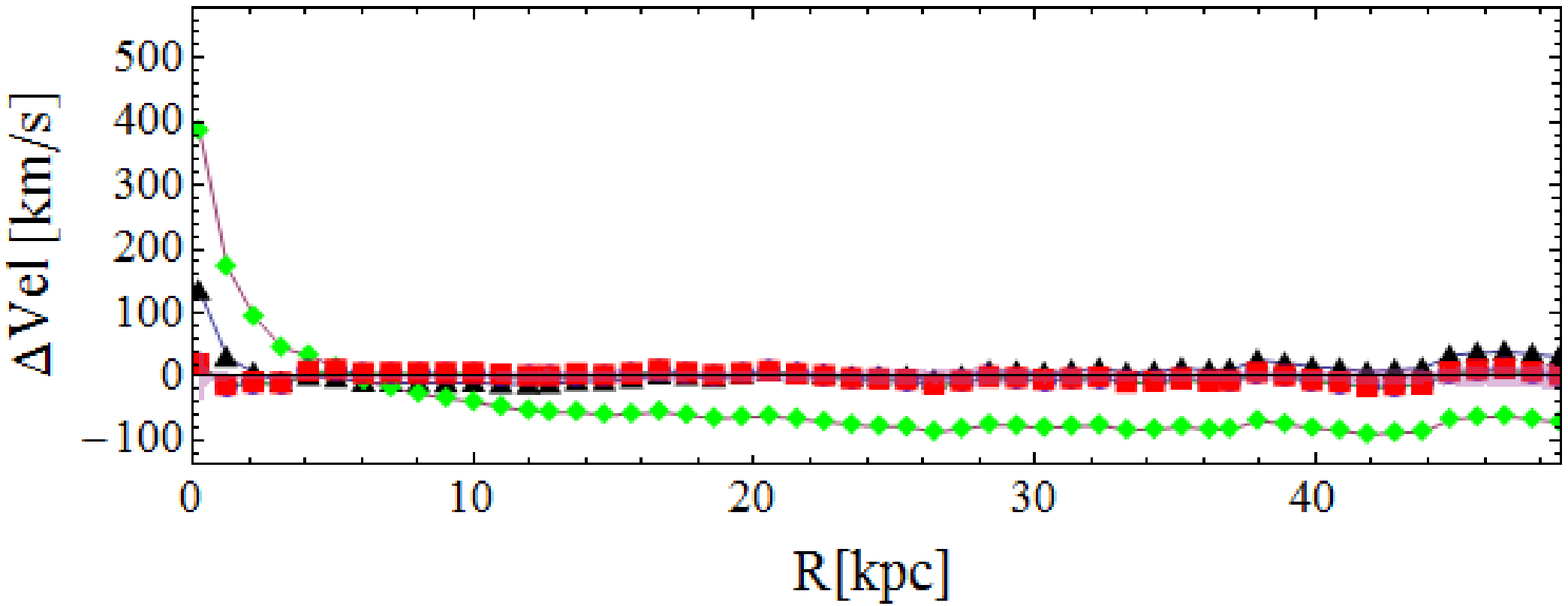}
    \end{tabular}  }  \\
    \subfloat[\footnotesize{Kroupa}]{
    \begin{tabular}[b]{c}
    \includegraphics[width=0.35\textwidth]{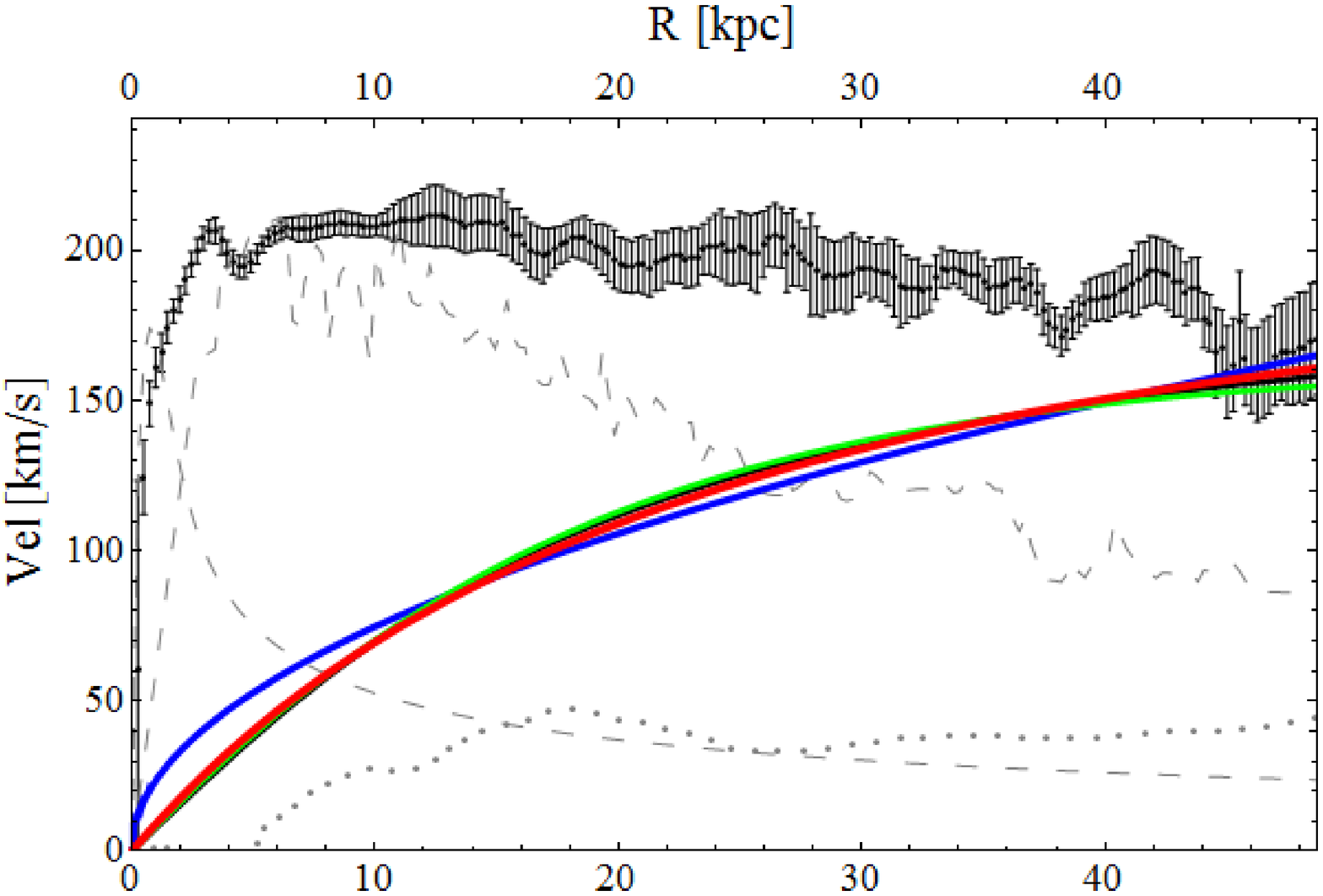} \\
    \includegraphics[width=0.35\textwidth]{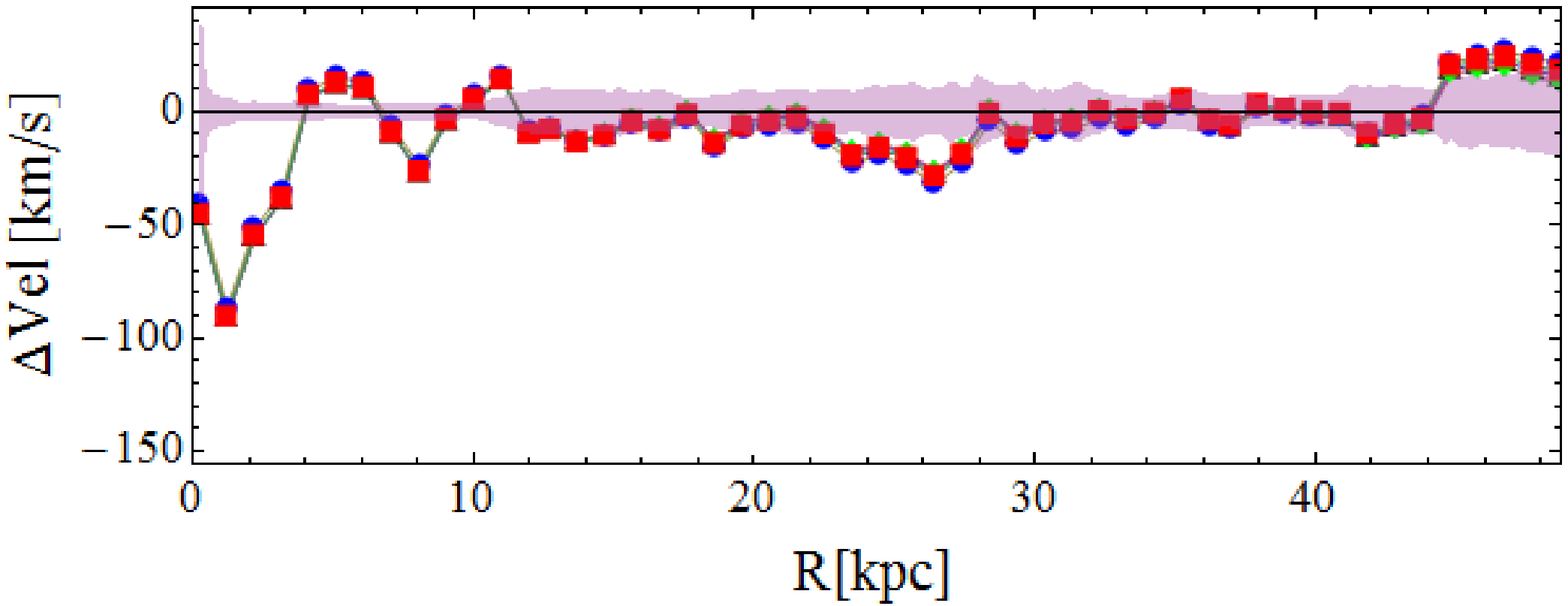}
    \end{tabular}  }
    \subfloat[\footnotesize{diet-Salpeter}]{
    \begin{tabular}[b]{c}
    \includegraphics[width=0.35\textwidth]{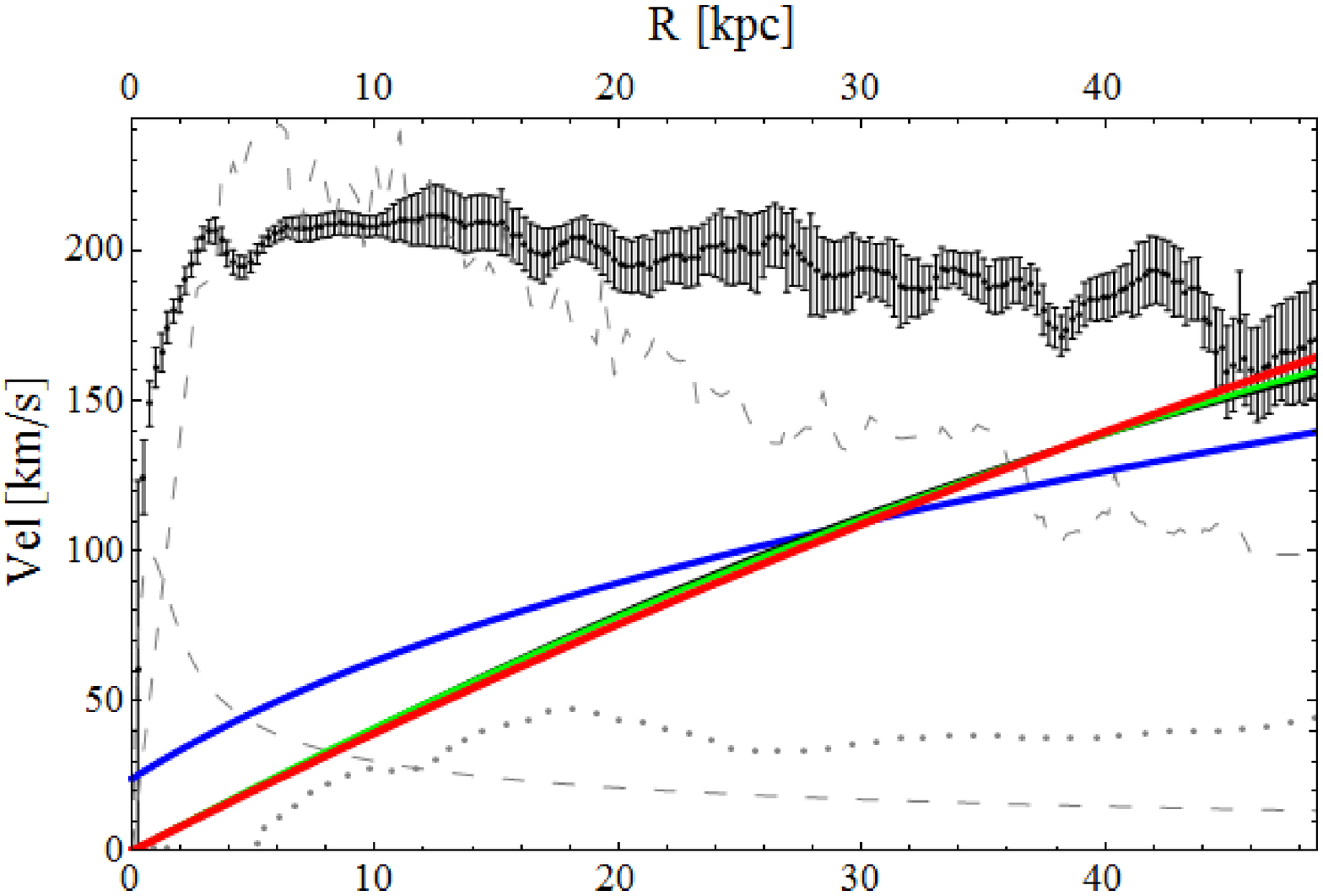} \\
    \includegraphics[width=0.35\textwidth]{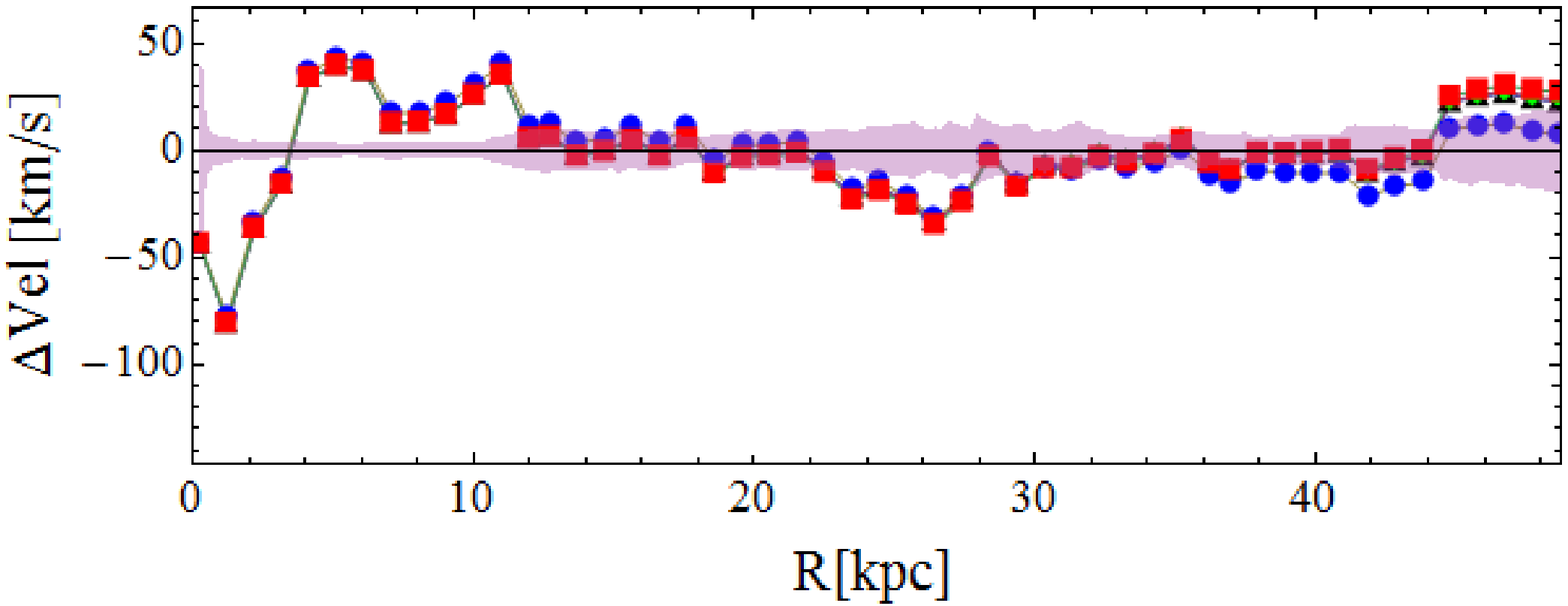}
    \end{tabular}  }
  \caption{\footnotesize{Here we display the rotation curves for the galaxy NGC 5055. Colors and symbols are as in Fig.\ref{fig:DDO154}. We take the parameter for this galaxy as $\mu_0 = 16.7$ mag arcsec$^{-2}$ and $R_d = 3.68$ {\rm kpc} for the stellar disk. Inner analysis is in agreement with the zero value for the core until $1\sigma$ of confidence level. From left to right, images from the fit of the galaxy NGC 5055 considering minimal disk, minimal disk+gas, Kroupa, diet-Salpeter. Colors and symbols are as in Fig.\ref{fig:DDO154}. }}
  \label{fig:NGC5055}
\end{figure}

\begin{figure}[h!]
    \subfloat[\footnotesize{Minimal disk}]{
    \begin{tabular}[b]{c}
    \includegraphics[width=0.35\textwidth]{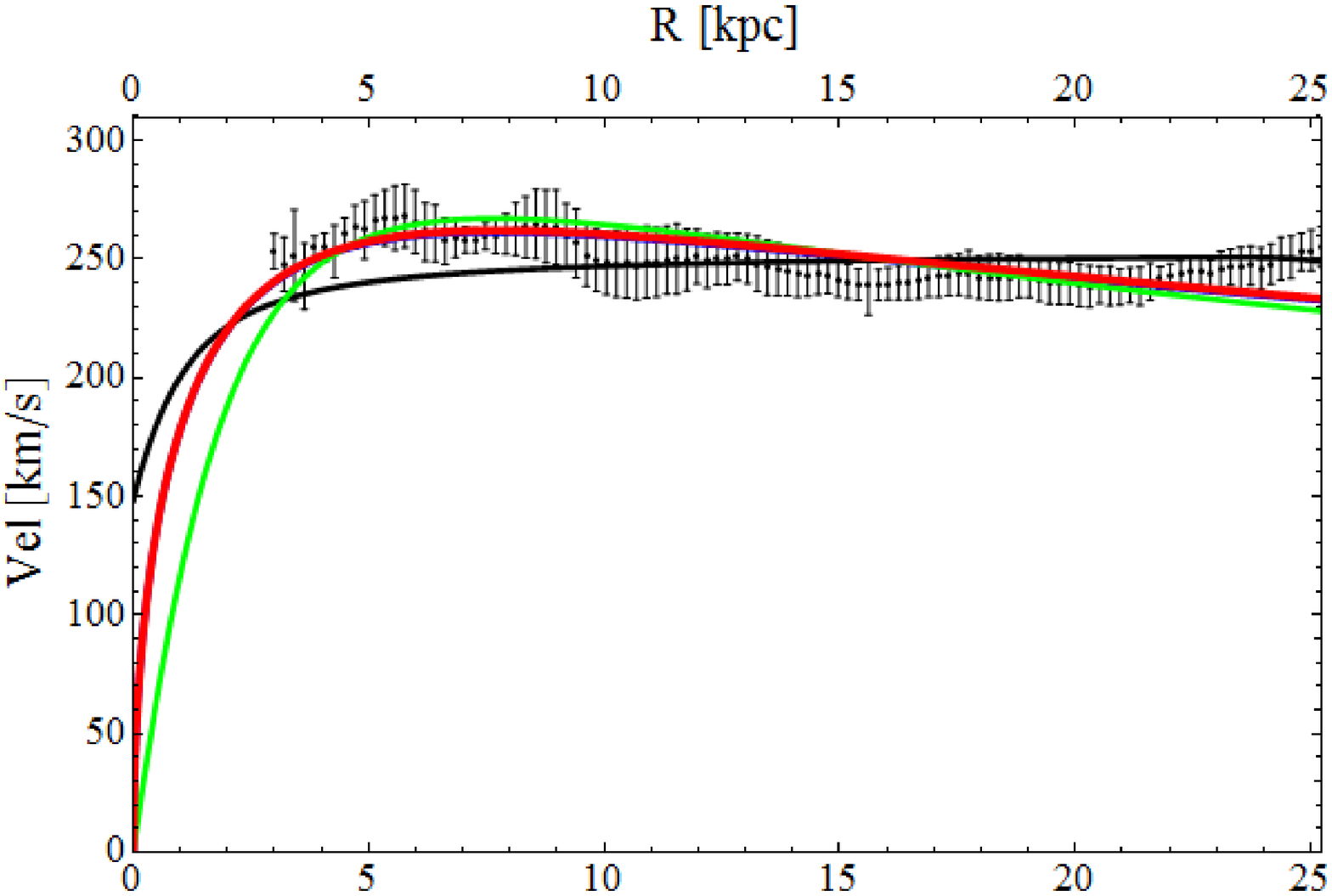} \\
    \includegraphics[width=0.35\textwidth]{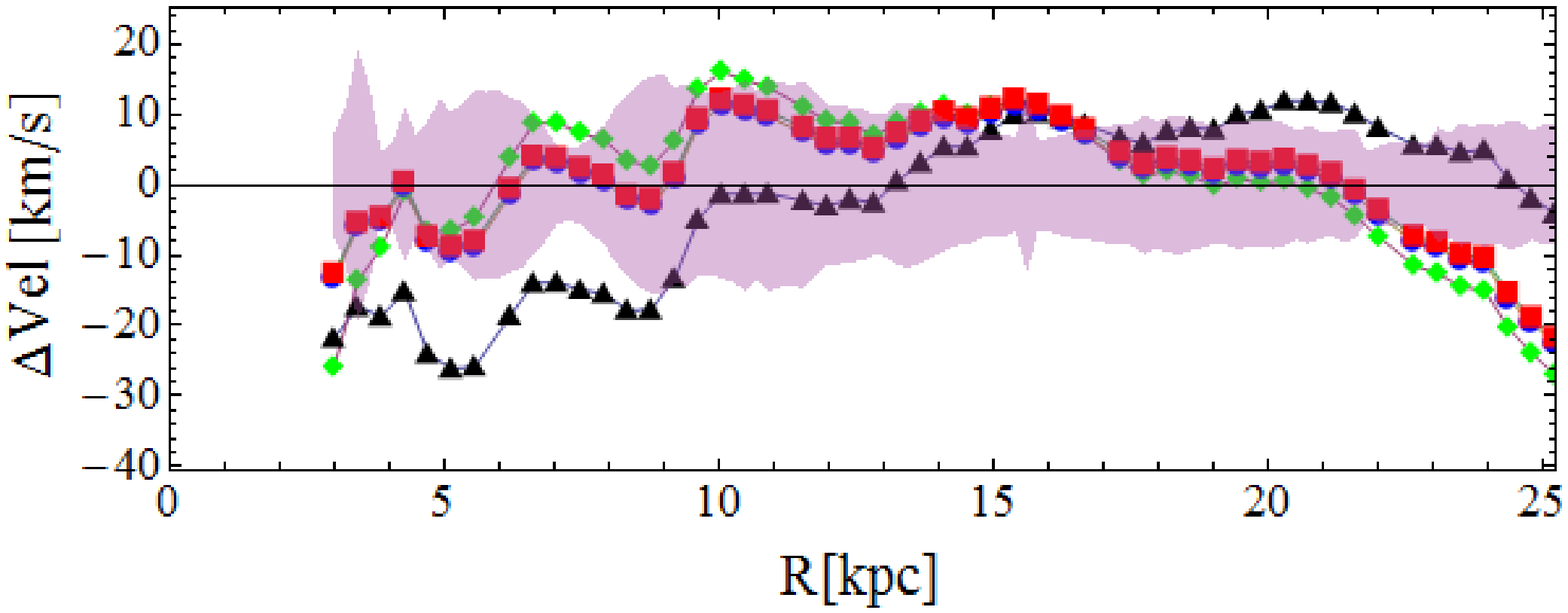}
    \end{tabular}  }
    \subfloat[\footnotesize{Min. disk + Gas}]{
    \begin{tabular}[b]{c}
    \includegraphics[width=0.35\textwidth]{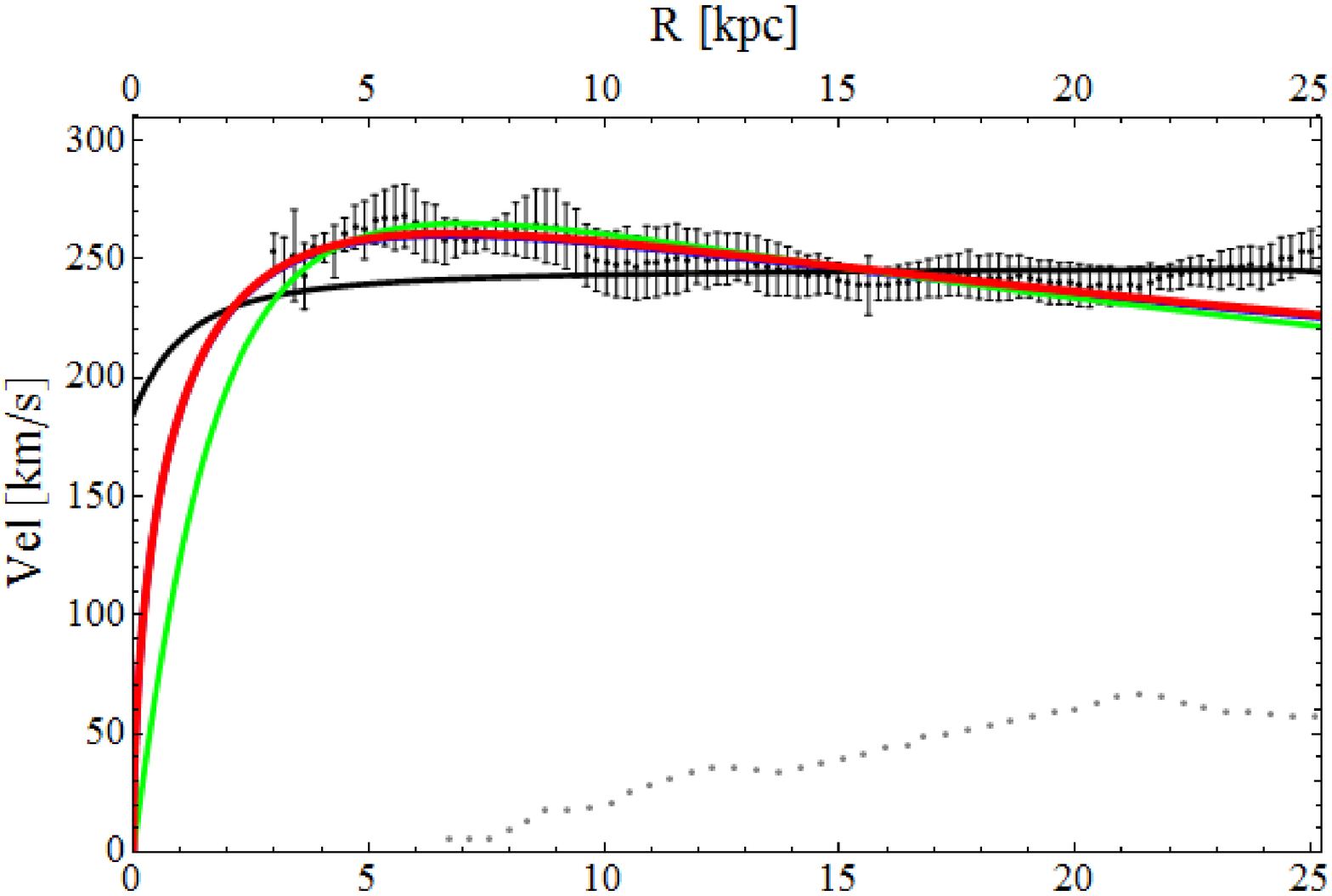} \\
    \includegraphics[width=0.35\textwidth]{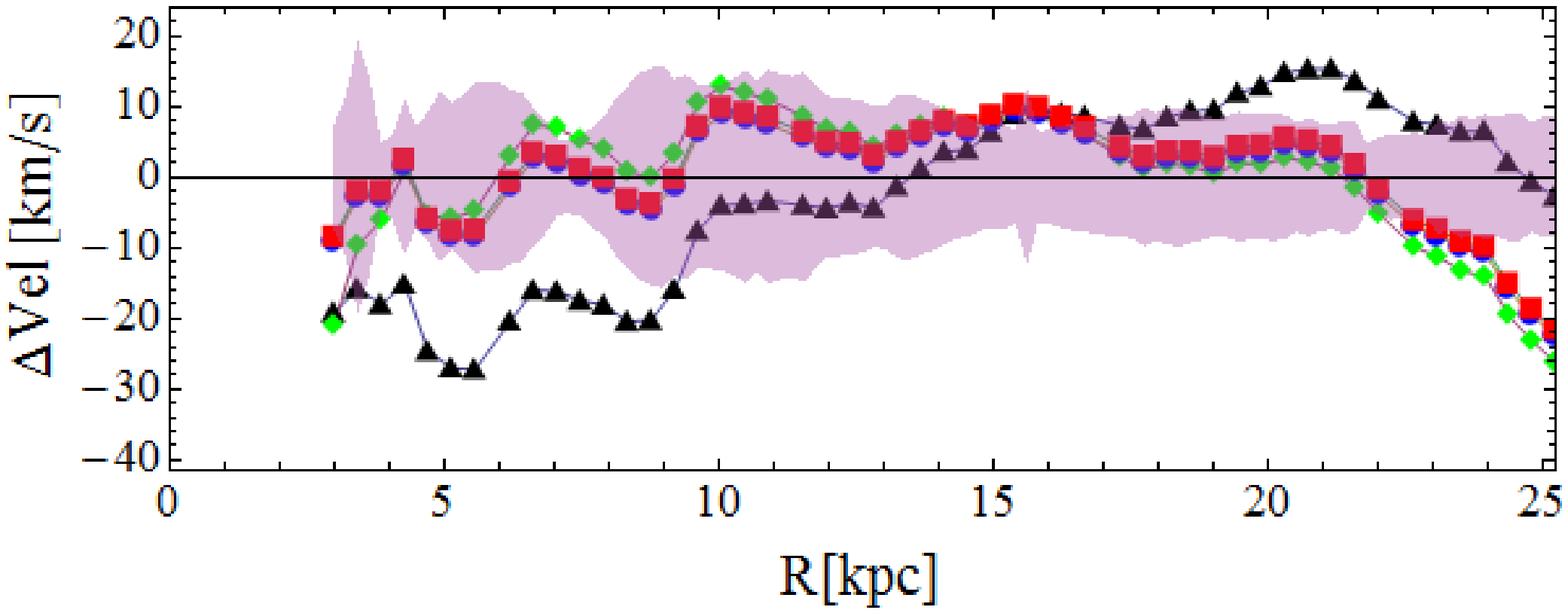}
    \end{tabular}  } \\
    \subfloat[\footnotesize{Kroupa}]{
    \begin{tabular}[b]{c}
    \includegraphics[width=0.35\textwidth]{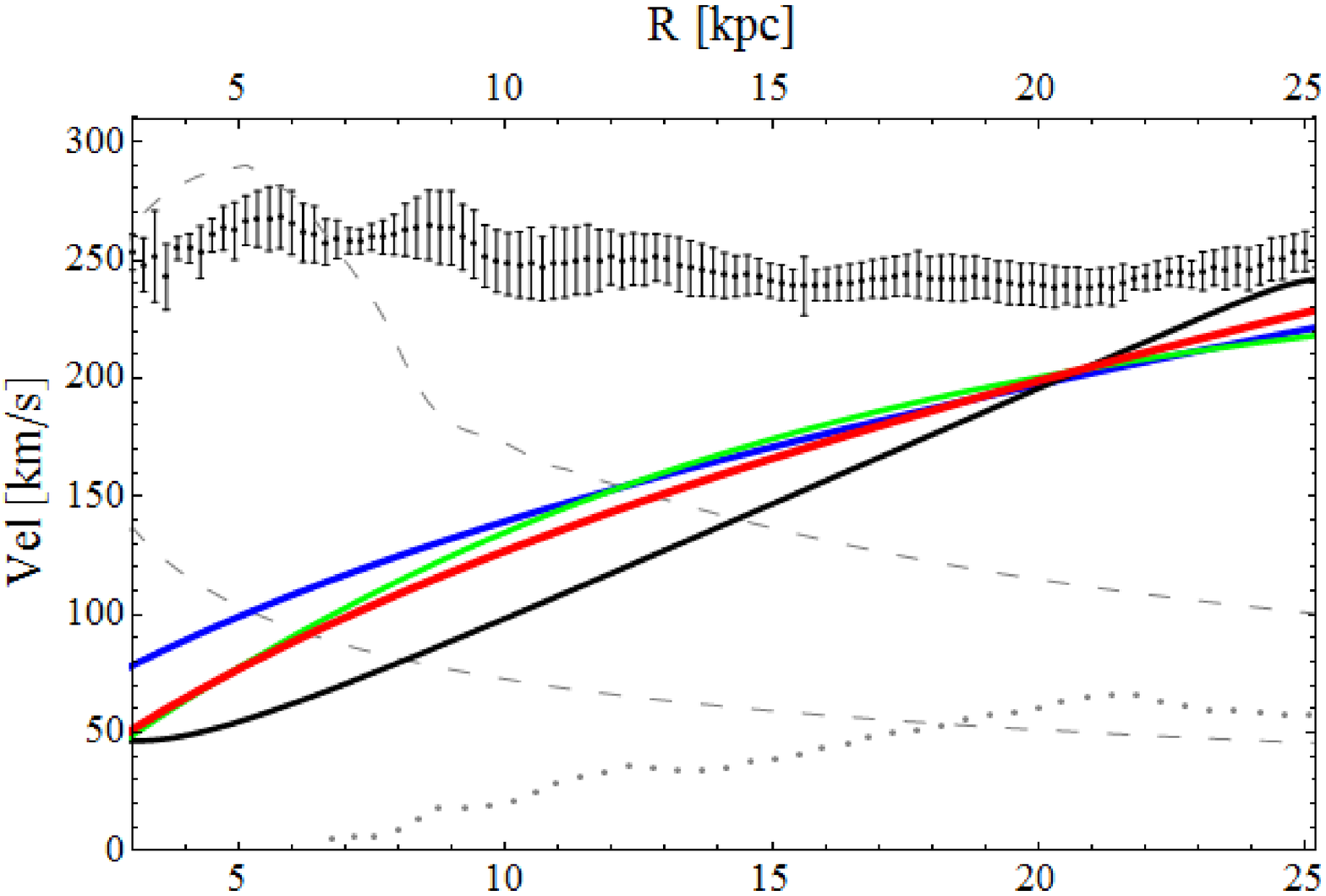} \\
    \includegraphics[width=0.35\textwidth]{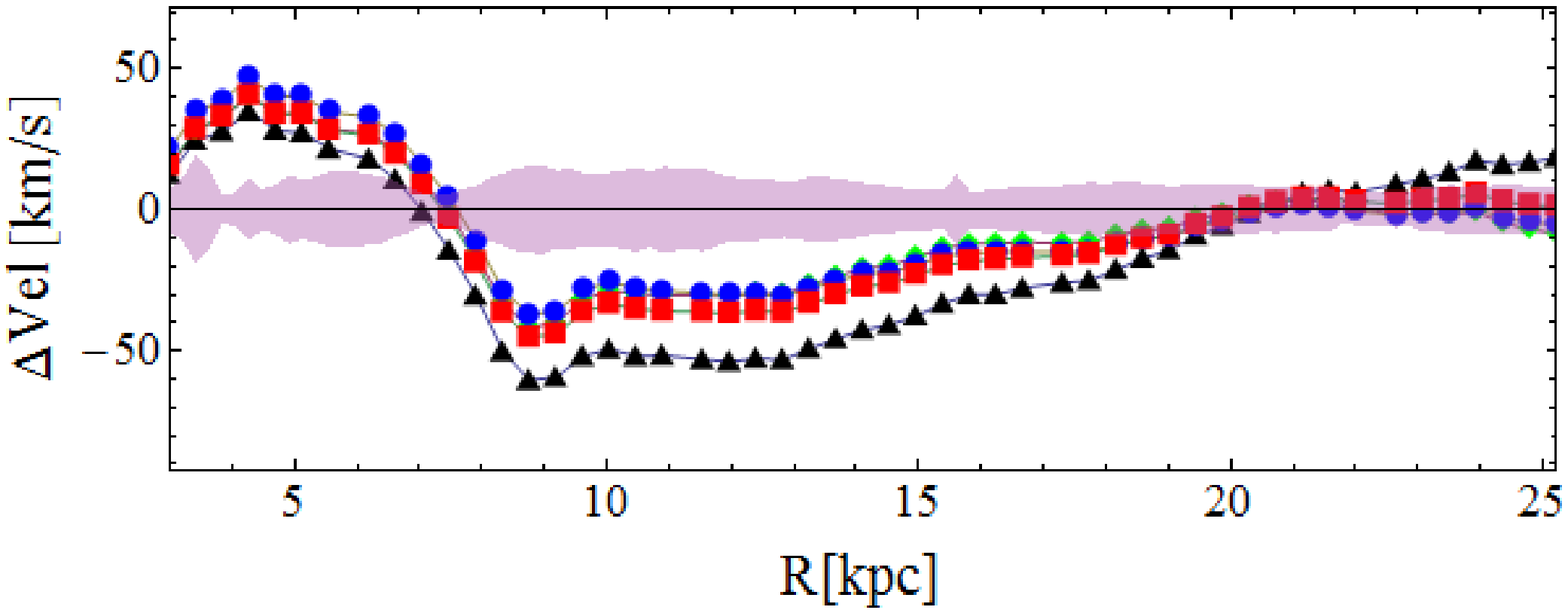}
    \end{tabular}  }
    \subfloat[\footnotesize{diet-Salpeter}]{
    \begin{tabular}[b]{c}
    \includegraphics[width=0.35\textwidth]{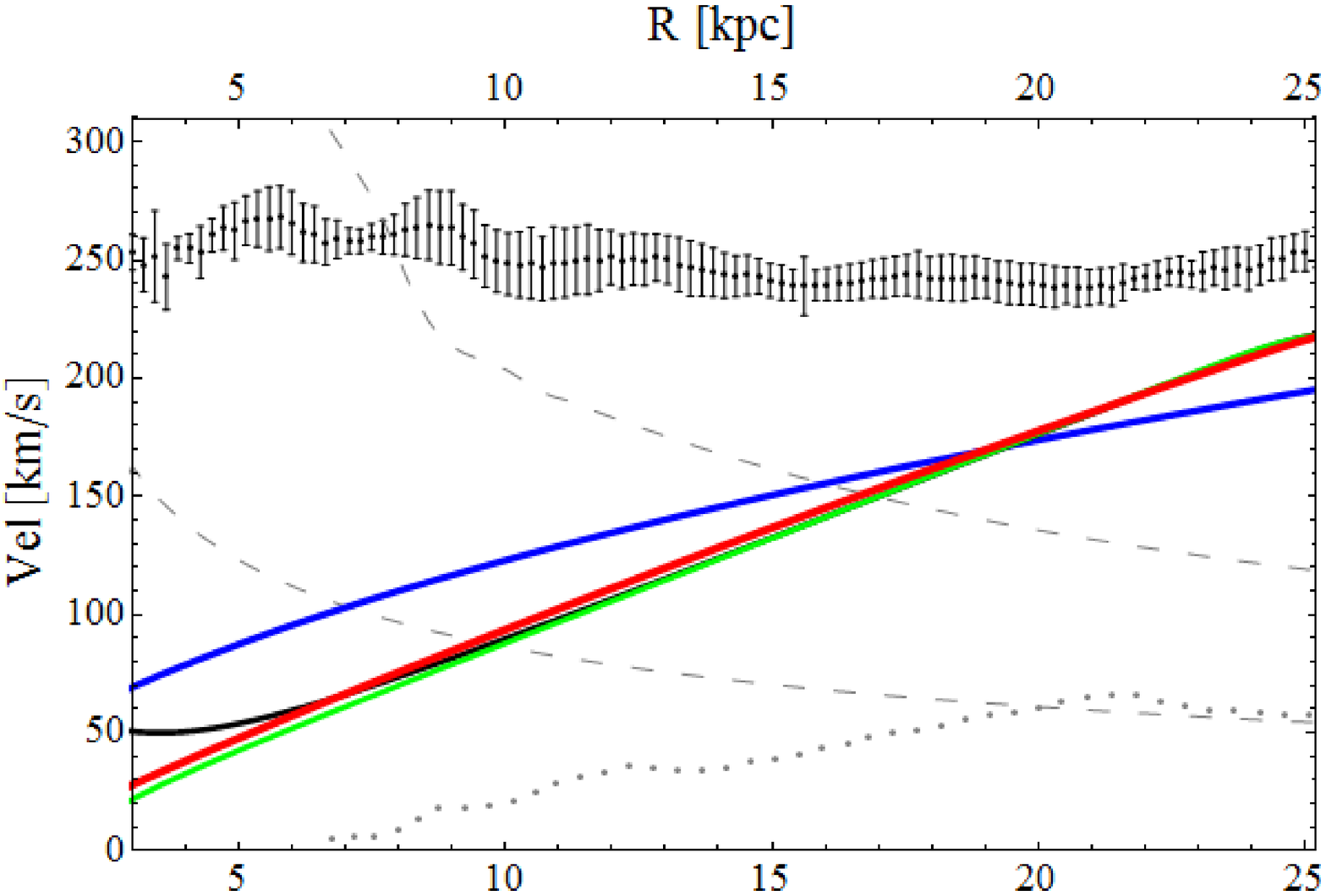} \\
    \includegraphics[width=0.35\textwidth]{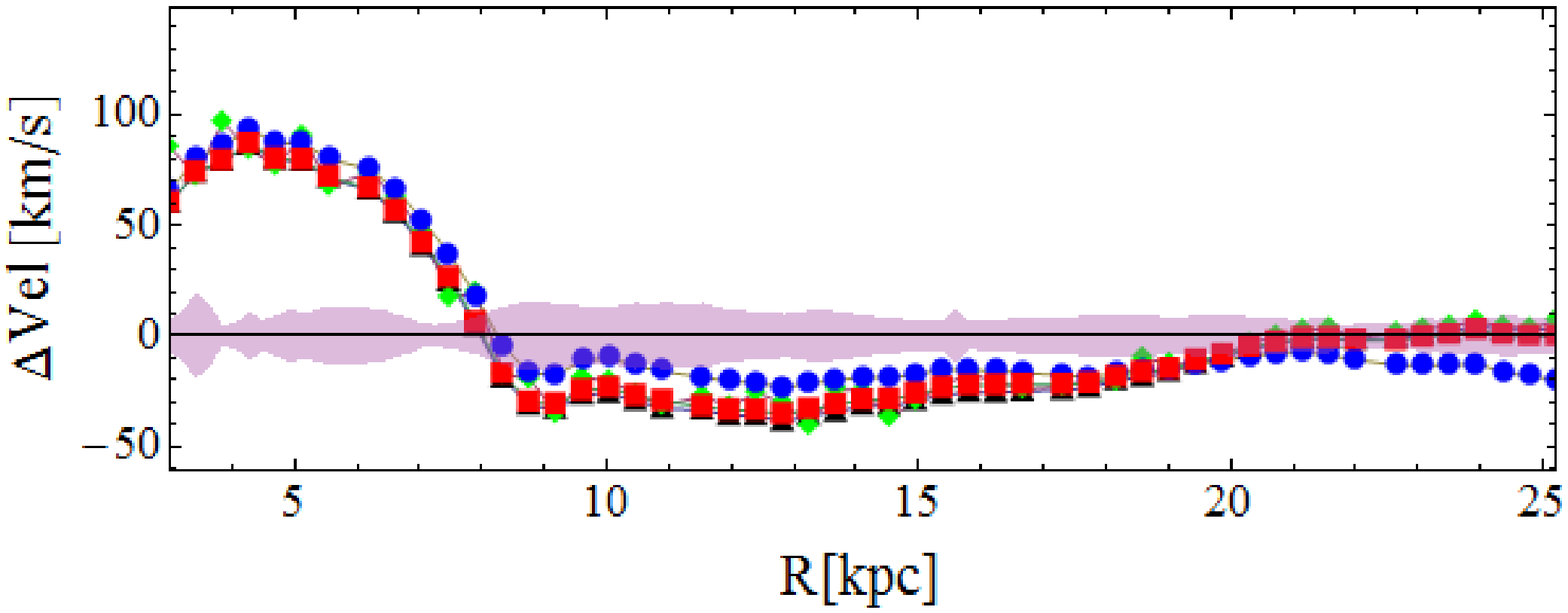}
    \end{tabular}  }
  \caption{\footnotesize{The rotation curves for the galaxy NGC 7331. Colors and symbols are as in Fig.\ref{fig:DDO154}. We assume values for the parameters $\mu_0 = 15.3$ mag arcsec$^{-2}$ and $R_d = 2.42$ {\rm kpc} for the outer disk, respectively. We can not extract relevant information from these galaxy because the observational data is not sufficiently close to center. From left to right, fits considering minimal disk, minimal disk+gas, Kroupa, diet-Salpeter. Colors and symbols are as in Fig.\ref{fig:DDO154}. }}
  \label{fig:NGC7331}
\end{figure}
\end{subappendices}

\end{appendices}

\newpage
\clearpage
\thebibliography{}
 \footnotesize{

\bibitem{delaMacorra:2009yb}
  A. de la Macorra,
Astropart.\ Phys.\  {\bf 33}, 195 (2010)
  [arXiv:0908.0571 [astro-ph.CO]].

\bibitem{KrAl78} P. C. van der Kruit and R. J. Allen,  Annu. Rev. Astron. Astrophys.   \textbf{16}, 103 (1978).

\bibitem{Tr87} V. Trimble,  Annu. Rev. Astron. Astrophys.   \textbf{25}, 425 (1987).

\bibitem{SoRu01} Y. Sofue and V. Rubin,  Annu. Rev. Astron. Astrphys.   \textbf{39}, 137 (2001).

\bibitem{deBlok10}   W.J.G. de Blok,
Ad.  Astron.\  {\bf 2010},  Article ID 789293 (2010)

\bibitem{Navarro:1995iw}
  J.~F.~Navarro, C.~S.~Frenk and S.~D.~M.~White,
  Astrophys.\ J.\  {\bf 462}, 563 (1996)
  [arXiv:astro-ph/9508025].

\bibitem{Navarro:1996gj}
  J.~F.~Navarro, C.~S.~Frenk and S.~D.~M.~White,
  Astrophys.\ J.\  {\bf 490}, 493 (1997)
  [arXiv:astro-ph/9611107].

\bibitem{Bu95} A. Burkert,  Astrophys. J.   \textbf{447}, L25 (1995).

\bibitem{KuMcBl08}
 R. Kuzio de Naray, S. S. McGaugh, and W.J.G. de Blok, Astrophys.  J.  {\bf 676}, 920 (2008).

\bibitem{KuMcMi09}
R. Kuzio de Naray, S. S. McGaugh, and C. Mihos, Ap.  J.  {\bf 692}, 1321 (2009).

\bibitem{vandenBosch:1999ka}
  F.~C.~van den Bosch, B.~E.~Robertson, J.~J.~Dalcanton and W.~J.~G.~de Blok,
  Astron.\ J.\  {\bf 119}, 1579 (2000)
  [arXiv:astro-ph/9911372].

\bibitem{Swaters:2000nt}
  R.~A.~Swaters, B.~F.~Madore and M.~Trewhella,
  Astrophys. J. \textbf{531} L107 (2000)

\bibitem{Simon:2003xu}
  J.~D.~Simon, A.~D.~Bolatto, A.~Leroy and L.~Blitz,
  Astrophys.\ J.\  {\bf 596}, 957 (2003)
  [arXiv:astro-ph/0307154].

\bibitem{Rhee:2003vw}
  G.~Rhee, A.~Klypin and O.~Valenzuela,
  Astrophys.\ J.\  {\bf 617}, 1059 (2004)
  [arXiv:astro-ph/0311020].

\bibitem{Swater99}
Swaters, R. R. 1999, PhD thesis, Univ. of Groningen

\bibitem{deBlok:2008wp}
  W.~J.~G.~de Blok, F.~Walter, E.~Brinks, C.~Trachternach, S.~H.~Oh and R.~C.~.~Kennicutt,
  Astrophys.\ J.\  {\bf 136}, 2648 (2008).

\bibitem{SpGiHa05}
K. Spekkens, R. Giovanelli, M. P. Haynes,  Astron. J.   \textbf{129}, 2119 (2005).

\bibitem{Oh:2010mc}
  S.~H.~Oh {\it et al.},
  arXiv:1011.2777 [astro-ph.CO].

\bibitem{Pontzen:2011}
  A. Pontzen, F. Governato,
  ``How Supernova Feedback Turns Dark Matter Cusps Into Cores''
  arXiv:1106.0499 [astro-ph.CO].

\bibitem{Ogiya:2011ta}
  G.~Ogiya and M.~Mori,
  arXiv:1106.2864 [astro-ph.CO].

\bibitem{delaMacorra:2011df}
  A.~de la Macorra, J.~Mastache and J.~L.~Cervantes-Cota,
  Phys.\ Rev.\  D {\bf 84}, 121301 (2011)
  [arXiv:1107.2166 [astro-ph.CO]].

\bibitem{Walter:2008wy}
  F.Walter {\it et al.},
  Astron.\ J.\  {\bf 136}, 2563 (2008)
  [arXiv:0810.2125 ].

\bibitem{Gentile:2004tb}
  G.~Gentile, P.~Salucci, U.~Klein, D.~Vergani and P.~Kalberla,
  Mon.\ Not.\ Roy.\ Astron.\ Soc.\  {\bf 351}, 903 (2004)
  [arXiv:astro-ph/0403154].

\bibitem{Salucci:2007tm}
  P.~Salucci, A.~Lapi, C.~Tonini, G.~Gentile, I.~Yegorova and U.~Klein,
  Mon.\ Not.\ Roy.\ Astron.\ Soc.\  {\bf 378}, 41 (2007)
  [arXiv:astro-ph/0703115].

\bibitem{Salucci:2010pz}
F. Donato, G. Gentile, P. Salucci, C. Frigerio Martins, M.I. Wilkinson, et al  Mon.\ Not.\ Roy.\ Astron.\ Soc.\ {\bf 397}, 1169 (2009);
G. Gentile, P. Salucci, U. Klein, D. Vergani, P. Kalberla,  Mon.\ Not.\ Roy.\ Astron.\ Soc.\ {\bf 351},903 (2004).
 P.~Salucci,
 arXiv:1008.4344 [astro-ph.CO].

\bibitem{vanAlbada}
  van Albada, T.S, Sancisi, R.,
  1986, Phil. Trans. R. Soc. Lond. A, 320, 447

\bibitem{Salpeter:1955it}
  E.~E.~Salpeter,
  Astrophys.\ J.\  {\bf 121}, 161 (1955).

\bibitem{Kroupa:2000iv}
  P.~Kroupa,
  Mon.\ Not.\ Roy.\ Astron.\ Soc.\  {\bf 322}, 231 (2001)
  [arXiv:astro-ph/0009005].

\bibitem{Bottema:1997qe}
  R.~Bottema,
  ``The maximum rotation of a galactic disc'',
  [astro-ph/9706230].

\bibitem{NBDM}
A.~ de la Macorra and J.~ Mastache, in preparation

\bibitem{Larson:2010gs}
  D.~Larson {\it et al.},
  Astrophys.\ J.\ Suppl.\  {\bf 192}, 16 (2011)
  [arXiv:1001.4635 [astro-ph.CO]].

\bibitem{Komatsu:2010fb}
  E.~Komatsu {\it et al.}  [WMAP Collaboration],
  Astrophys.\ J.\ Suppl.\  {\bf 192}, 18 (2011)
  [arXiv:1001.4538 [astro-ph.CO]].

\bibitem{Dunkley:2010ge}
  J.~Dunkley {\it et al.},
  arXiv:1009.0866 [astro-ph.CO].

\bibitem{Macorra.DE}
  A.~De la Macorra  and C.~Stephan-Otto,
  Phys.\ Rev.\ Lett.\  {\bf 87}, 271301 (2001)
  [arXiv:astro-ph/0106316];
  JHEP {\bf 0301}, 033 (2003)
  [arXiv:hep-ph/0111292];  A.~de la Macorra,
  Phys.\ Rev.\  D {\bf 72}, 043508 (2005)
  [arXiv:astro-ph/0409523]

\bibitem{Macorra.DEDM}
 A.~de la Macorra,
  Phys.\ Lett.\  B {\bf 585}, 17 (2004)
  [astro-ph/0212275]

\bibitem{Adams:2005dq}
  J.~Adams {\it et al.}  [STAR Collaboration],
  Nucl.\ Phys.\  A {\bf 757}, 102 (2005)
  [arXiv:nucl-ex/0501009].

\bibitem{Bazavov:2009zn}
  A.~Bazavov {\it et al.},
  Phys.\ Rev.\  D {\bf 80}, 014504 (2009)
  [arXiv:0903.4379 [hep-lat]].

\bibitem{Simon:2005}
  J.D. Simon, A.D. Bolatto, A. Leroy, L. Blitz,
   Astrophys.\ J.\  {\bf 621}, 757-776 (2005)

\bibitem{Trachternach:2008wv}
  C.~Trachternach, W.~J.~G.~de Blok, F.~Walter, E.~Brinks and R.~C.~.~Kennicutt,
  arXiv:0810.2116 [astro-ph].

\bibitem{Oh:2008ww}
  S.~H.~Oh, W.~J.~G.~de Blok, F.~Walter, E.~Brinks and R.~C.~.~Kennicutt,
  arXiv:0810.2119 [astro-ph].

\bibitem{Kennicutt:2003dc}
  R.~C.~.~Kennicutt {\it et al.},
  Publ.\ Astron.\ Soc.\ Pac.\  {\bf 115}, 928 (2003)
  [arXiv:astro-ph/0305437].

\bibitem{McGaugh:2006vv}
  S.~S.~McGaugh, W.~J.~G.~de Blok, J.~M.~Schombert, R.~K.~de Naray and J.~H.~Kim,
  Astrophys.\ J.\   {\bf 659}, 149 (2007)
  [arXiv:astro-ph/0612410].

\bibitem{Portas:2009}
  A.~Portas, E.~Brinks, A.~Usero, F.~Walter, W.~J.~G. de Blok, Jr., R.~C.~Kennicutt,
  in The Galaxy Disk in Cosmological Context, Proceedings of IAU Symposium No. 254, eds. J. Andersen, J. Bland-Hawthorn \& B. Nordström (Cambridge: Cambridge
University Press), p 52

\bibitem{Freeman:1970mx}
  K.~C.~Freeman,
  Astrophys.\ J.\  {\bf 160}, 811 (1970).

\bibitem{Burlak:1997}
  Burlak, A.N., Gubina, V.A., Tyurina, N.V.,
  Astro.\ Lett.\ {\bf 23}, 522 (1997).

\bibitem{Bell:2000jt}
  E.~F.~Bell and R.~S.~de Jong,
  Astrophys.\ J.\  {\bf 550}, 212 (2001)
  [arXiv:astro-ph/0011493].

\bibitem{Jarrett:2003uy}
  T.~H.~Jarrett, T.~Chester, R.~Cutri, S.~E.~Schneider and J.~P.~Huchra,
  Astron.\ J.\  {\bf 125}, 525 (2003).

\bibitem{Taylor:2005sf}
  V.~A.~Taylor, R.~A.~Jansen, R.~A.~Windhorst, S.~C.~Odewahn and J.~E.~Hibbard,
  Astrophys.\ J.\  {\bf 630}, 784 (2005)
  [astro-ph/0506122].

\bibitem{de Blok:1996ns}
  W.~J.~G.~de Blok, S.~S.~McGaugh and J.~M.~van der Hulst,
  Mon.\ Not.\ Roy.\ Astron.\ Soc.\  {\bf 283}, 18 (1996)
  [arXiv:astro-ph/9605069].

\bibitem{Moore:1994yx}
  B.~Moore,
  Nature {\bf 370}, 629 (1994).

\bibitem{Gnedin:2004cx}
  O.~Y.~Gnedin, A.~V.~Kravtsov, A.~A.~Klypin and D.~Nagai,
  Astrophys.\ J.\  {\bf 616}, 16 (2004)
  [astro-ph/0406247].

\bibitem{Kormendy:2004se}
  J.~Kormendy and K.~C.~Freeman,
  [astro-ph/0407321].

}

\end{document}